\documentclass[12pt, a4paper]{article}
\usepackage{hyperref}
\usepackage{graphicx}
\usepackage{mathrsfs}
\usepackage{amssymb}
\usepackage{color}
\usepackage{amsmath}
\usepackage{ascmac}
\usepackage{amsthm}
\usepackage{booktabs}
\usepackage{lscape}
\usepackage{dcolumn}
\usepackage{longtable}
\usepackage{dcolumn}
\usepackage{arydshln}
\usepackage{bm}
\usepackage{caption}
\usepackage{subfig}
\usepackage{setspace}
\usepackage{cases}
\usepackage{comment}
\usepackage{multirow}
\usepackage{booktabs}
\usepackage[square, sort,comma,numbers]{natbib}
\bibpunct[, ]{(}{)}{;}{a}{}{,}
\usepackage{xurl}

\usepackage{tikz}
\usetikzlibrary{positioning,calc,fit}
\usepackage{graphicx} 
\definecolor{inC}{HTML}{DCE8F5}   
\definecolor{exC}{HTML}{F8E1CC}

\makeatletter
\newcommand{\subsubsubsection}{\@startsection{paragraph}{4}{\z@}%
{1.5\baselineskip \@plus.5\dp0 \@minus.2\dp0}%
{.5\baselineskip \@plus2.3\dp0}%
{\reset@font\normalsize\bfseries}
}
\newcommand{\subsubsubsubsection}{\@startsection{subparagraph}{5}{\z@}%
{1.5\baselineskip \@plus.5\dp0 \@minus.2\dp0}%
{.5\baselineskip \@plus2.3\dp0}%
{\reset@font\normalsize\itshape}
}
\makeatother
\begin{document}

\title{
{\LARGE Urban Reconstruction and Population Redistribution:}\\
\vspace{0.2cm}
{\fontsize{16}{19}\selectfont Evidence from Tokyo after the Great Kanto Earthquake}
}

\author{Kota Ogasawara\thanks{Department of Industrial Engineering and Economics, School of Engineering, Institute of Science Tokyo, 2-12-1, Ookayama, Meguro-ku, Tokyo 152-8552, Japan (E-mail: ogasawara.k.ab@m.titech.ac.jp).\newline
I would like to thank the seminar participants at the University of Tokyo for their helpful comments.
I am also grateful to the CIRJE at the University of Tokyo for granting me the opportunity to be a short-term visitor in 2026.
This study was supported by JSPS KAKENHI (23KK0221).
There are no conflicts of interest to declare.
All errors are my own.
}
}
\date{
Institute of Science Tokyo\\[2ex]
\today}
\maketitle

\begin{abstract}
\begin{spacing}{0.85}
This study examines the impact of the 1923 Great Kanto Earthquake on population distribution within Tokyo City.
Using newly digitized census statistics and variation in fire damage across areas, I show that land readjustment under the reconstruction project reduced residential land in the burned area, and rents rose throughout the city, particularly in the burned area.
Although residential floor area recovered through the construction of multi-story dwellings, the number of households in the burned area remained below its pre-earthquake level, whereas the population had returned to roughly its pre-earthquake level by 1935.
The pre-earthquake zoning system had little effect, suggesting that post-disaster population distribution was shaped by market-based price adjustments rather than regulation.
The population in the burned area thus reflected the net effect of two opposing forces: fewer households due to the loss of residential land, partly offset by more households per dwelling through lodging, and larger households through the inclusion of non-kin members, such as servants and employees.
\end{spacing}
\bigskip

\bigskip

\end{abstract}

\newpage
\section{Introduction}

A capital city serves as the economic hub of a nation, with the government, numerous businesses, and many households investing vast amounts of capital there.
As a result, buildings are concentrated in the capital, and land use is closely linked to economic activity and population distribution.
Understanding changes in population distribution within the capital is essential to grasping shifts in resource allocation at the heart of the economy.\footnote{In 1920, buildings accounted for approximately $50$\% of Japan's total net capital stock, with residential buildings comprising $63.5$\% and non-residential buildings $36.5$\% of that total (at 1934--1936 prices). The remaining $50$\% consisted almost entirely of producers' durable equipment and structures other than buildings. Of the net worth of housing, approximately $54$\% was non-farm housing. See \citet[pp.~149; 251]{ltes1966v3} for these statistics.}

This study examines the impact of the unprecedented disaster of 1923, the Great Kanto Earthquake, on the population distribution within Tokyo City.
While there is a long history of economic theory research on urban land use, the empirical population distribution within the city has been understudied.
In particular, it remains largely unclear how the distribution of the urban population changes after a large-scale disaster occurs.
Is the post-disaster distribution of the urban population consistent with theoretical predictions?
This paper shows that the population in the affected area is influenced more by land-market adjustments resulting from land readjustment after the disaster than by land-use regulations shaped by path dependence.
In addition, I examine how the affected area accommodated its population, including a margin not considered in the standard model: an increase in household size through the inclusion of non-kin members, such as servants and employees.

The earthquake not only caused buildings to collapse due to the tremors but also triggered massive fires.
Strong winds from a typhoon near the Noto Peninsula spread the fires unpredictably, so that the extent of fire damage across areas was largely beyond human control.
Exploiting this variation, I first assessed the consistency between observed phenomena and predictions derived from the monocentric model.
Using official census reports, I constructed ward-level panel data on housing, rents, population, and households, and conducted an event-study analysis.
I next analyzed whether the impact of the disaster depended on the land-use zoning system, which was designed before the earthquake based on the initial geographical features.
To do so, I digitized the census statistics on population and households at the smallest available grid unit, the town block.
I then identified burned and industrial blocks from official documents and estimated their interaction effects within an event-study framework.
Finally, I analyzed the two margins through which the burned area accommodated its population: the number of households per dwelling and average household size.
I examined lodging using cross-sectional block-level data on rents and lodging rents digitized from a comprehensive household survey, and household size using block-level panel data.

The results are summarized as follows.
First, the post-earthquake changes are consistent with the predictions of the model.
Before the earthquake, both average rent and household density were higher in the burned area.
After the earthquake, land readjustment reduced the amount of residential land in the burned area, and real rents rose throughout the city, with a larger increase in the burned area.
When the reconstruction plan was completed in 1930, total floor area in the burned area had almost recovered to its pre-earthquake level owing to the construction of multi-story dwellings, but both the number of households and the population remained lower.
The number of households in the burned area remained below its pre-earthquake level through 1935, while the population in the city's outlying areas grew gradually.
The zoning system designed before the earthquake had little effect on the population in the burned area.

Second, the population in the burned area recovered faster than the number of households, returning to roughly its pre-earthquake level by 1935.\footnote{This recovery is in line with studies showing that the spatial distribution of economic activity is persistent even after large shocks, owing to locational fundamentals \citep{Davis2002-ng, Bosker2017-bx, Michaels2017-vs, Fluckiger2020-ml}; see \citet{Hanlon2022-ad} for a review. My analysis, however, focuses on the redistribution of households and population within a city over about a decade, rather than on long-term development across cities.}
A decomposition of population density shows that the persistent relative shortfall in floor area was more than offset by a more intensive use of the housing stock: about $70\%$ of this offset is accounted for by more households per unit of floor area, which corresponds to smaller housing per household in the model, and about $30\%$ by larger households, which lies outside the model.
The first margin operated through lodging.
Lodger households, usually single workers or married couples, rented rooms in other people's dwellings and were counted as separate households.
In the burned area, the unit rent paid by lodgers relative to that paid by landlords was lower than in the unburned area, so that landlords bore a larger share of housing costs.
The second margin operated through the inclusion of non-kin members.
As rents rose, the number of independent kinship households declined, while domestic servants and business employees increasingly lived in their employers' houses and were counted as members of those households.
Overall, the population in the burned area reflected the net effect of two opposing forces: fewer households due to the loss of residential land, partly offset by lodging, and larger households through the inclusion of non-kin members.
These results suggest that, even when the population of a devastated city center recovers, a disaster can leave lasting changes in the composition of its households.

This study makes the following three contributions.
First, it adds novel empirical evidence on population evolution in a capital city after a disaster.
Theoretical research on urban land use has evolved to account for heterogeneous residents, housing durability, the endogenous emergence of central points, and mixed land use.
Compared to these studies, however, the body of empirical research on urban structure remains limited.\footnote{\citet{Fujita1989-zu} provides a review of early theoretical research. \citet{Duranton2015-mm} offers a comprehensive review of both recent theoretical and empirical research on urban land use.}
Existing empirical tests of the monocentric model have mainly exploited changes in commuting costs.
For example, \citet{BaumSnow2007} shows that new highways in U.S. metropolitan areas reduced the population of central cities, consistent with the model's prediction that lower commuting costs lead to suburbanization.
This study instead exploits a shock to the supply of residential land: land readjustment after the earthquake reduced the share of land available for housing in the burned area, holding commuting costs largely unchanged.
This study shows that the post-disaster changes are consistent with a closed-city monocentric model, in which a reduction in residential land raises rents throughout the city and reduces the number of households in the affected area.
It also shows that the population declined by less than the number of households because both the number of households per dwelling and average household size increased.

Second, this study contributes to the literature on the consequences of regulations on urban structure.
Generally, land-use regulations negatively affect the economy by constraining flexibility in land demand \citep{Gyourko2015-ir}.
This study analyzed the impact of a zoning system that was designed based on initial geographical features, using a large-scale disaster in a metropolis as a natural experiment.
The results show that the zoning had little effect on the population distribution in the burned area.
Market adjustments to the loss of residential land, reflected in rising rents, had a more pronounced influence on population distribution than institutional regulation.
This is consistent with the findings of \citet{Hornbeck2017-wv}, who argued that the large-scale rebuilding of the urban area following the Great Boston Fire caused land prices--which had been suppressed by older buildings--to rise.\footnote{\citet{Siodla2017-jy}  also found that the residential land in the burned area shifted to nonresidential use after the San Francisco Fire of 1906, suggesting that the disaster had increased the flexibility of land use and boosted land productivity.}
Their study focused on land and building prices, and the demographic trend after the disaster is outside its scope.
In this light, my results provide new evidence that disasters affect not only real estate but also population structure.

Third, it broadens the horizons of economic history research focused on Tokyo.
While there is a study providing an overview of suburbanization in Tokyo Prefecture based on estimates of residential floor area \citep[pp.~126; 127]{ltes1966v3}, little is known about the impact of the 1923 earthquake on the city population.
Furthermore, conventional urban social history research suggests that low-income households that migrated to Tokyo City were able to form households and sustain their livelihoods by the 1920s \citep{Nakagawa1985}.
However, it has not been sufficiently clarified why the working-class households were able to sustain their livelihoods during this period.\footnote{\citet{Ogasawara2026-uc} presents evidence from the perspective of the financial system. Around 1920, workers used credit extensively to smooth consumption in response to idiosyncratic income shocks.}
My results suggest that, as rents rose after the earthquake, housing sharing may have served as a means of coping with higher housing costs in the metropolis.
In urban housing literature, previous studies have documented the rental housing and room-rental markets in Tokyo after the earthquake \citep{Ono2006, Ono2014}.
This study complements them by quantifying how these markets shaped the spatial distribution of households and population.

The remainder of the study is organized as follows. 
Section~\ref{sec:sec_hb} provides brief overviews of the historical context.
Section~\ref{sec:sec_tf} introduces the theoretical framework for this paper.
Section~\ref{sec:sec_data} describes the data and Section~\ref{sec:sec_ea} summarizes empirical analysis.
Section~\ref{sec:sec_con} concludes.

\section{Historical Background} \label{sec:sec_hb}

\subsection{Central Business District and the Zoning System Designed before the Disaster} \label{sec:sec_hb_cbd}

\begin{figure}[htbp]
\centering
\captionsetup{justification=centering,margin=1.5cm}
\includegraphics[width=0.50\textwidth]{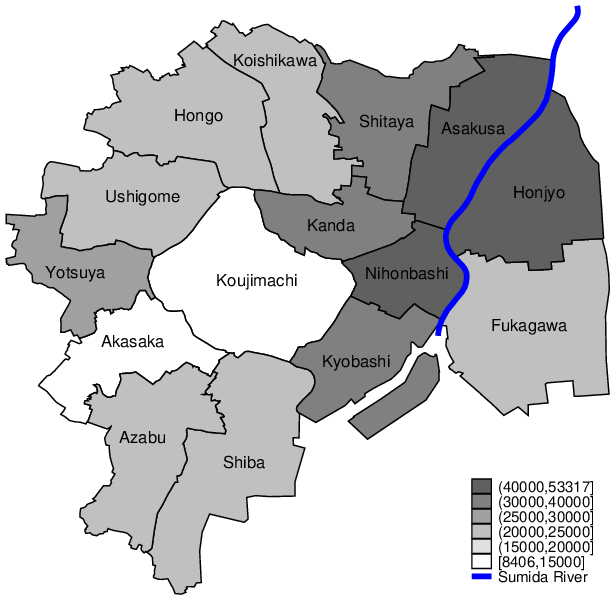}
\caption{15 Administrative Wards, Sumida River, and Population Density ($/\text{km}^{2}$) in 1920 Tokyo City}
\label{fig:map_ward_pden1920_mt}
\scriptsize{\begin{minipage}{450pt}
\setstretch{0.9}
Notes:
Figure~\ref{fig:map_ward_pden1920_mt} shows the 15 administrative wards in Tokyo City and the Sumida River (shown in a thick blue line).
The Nihonbashi, Kyobashi, Kanda, and Asakusa Wards include many commercial districts.
Parts of the Shiba and Shitaya Wards also have commercial districts.
The manufacturing zones were mainly in Honjo and Fukagawa Wards, as well as a part of Kyobashi Ward.
Most of the Koishikawa, Hongo, Ushigome, Yotsuya, Akasaka, and Azabu Wards were residential districts.
The population density ($/\text{km}^{2}$) calculated using the 1920 Population Census is illustrated in the same figure.
99\% of Kojimachi Ward is occupied by the Imperial Palace and government-owned land.
Sources: Created by the author using the Tokyo City Office (1922; 1922a).
\end{minipage}}
\end{figure}

According to the 1920 census report, Tokyo City was the largest city in Japan with a population of $2.17$ million, accounting for approximately $4$\% of the total population.
The majority of the city's population was concentrated in the wards east of the Imperial Palace.
Figure~\ref{fig:map_ward_pden1920_mt} illustrates the spatial distribution of the 15 wards and the population density as of 1920.
The Imperial Palace was located in Kojimachi Ward, while the commercial districts were situated in the Nihonbashi, Kyobashi, Kanda, and Asakusa Wards west of the Sumida River, as well as in parts of Shiba and Shitaya Wards.
Most of these areas were the castle town (\textit{shitamachi}) near the Imperial Palace, where various commercial functions--such as government offices, banks, wholesalers, and retail stores--were densely concentrated.
Nihonbashi, in particular, had been the starting point of the Five Highways (\textit{gokaid\=o}) since the Edo era (from the 17th century) and was lined with financial institutions, including the Bank of Japan, and major \textit{kimono} shops (\textit{gofukuten}).
The manufacturing zones were primarily located in Honjo and Fukagawa Wards on the east side of the Sumida River, where various factories were concentrated.
These areas constituted the central business district (CBD) and served as the industrial hub of the time.
Hereafter, these commercial (\textit{sh\=ogy\=o}) and manufacturing (\textit{k\=ogy\=o}) districts are collectively referred to as the industrial (\textit{sangy\=o}) districts.

The areas located west of the Imperial Palace (Koishikawa, Hongo, Ushigome, Yotsuya, Akasaka, and Azabu Wards) were mostly residential.
Additionally, although not included on the map, the areas to the north and east of the city, bordering the industrial districts, were also mostly residential.
In other words, industrial districts had developed along both banks of the Sumida River, with residential areas spreading out to surround them.
Workers commuted to these industrial districts using railways and electric tramways \citep[pp.~44--48]{Horiuchi1978}.\footnote{See also Tokyo City (1928) for this point. Almost the entire Kojimachi Ward (Figure~\ref{fig:map_ward_pden1920_mt}) is occupied by the Imperial Palace. As a result, it is impossible to accurately calculate the center of gravity of the working population. When I forcibly calculate the center of gravity of the employed population recorded in the 1920 census, it falls within Nihonbashi Ward and the northern part of Kanda Ward. If Kojimachi Ward were excluded, therefore, the center of gravity would lie farther east, specifically around the border among the Nihonbashi, Asakusa, and Honjo Wards. Taking into account the residential areas adjacent to the north and east of Tokyo City, the center must lie around the center of Asakusa, Nihonbashi, and Honjo Wards. \citet[p.~76]{Horiuchi1978} manually illustrates the circular spread of the population centered on these industrial areas.}

Historical records indicate that the distribution of these industrial districts was shaped during the Edo era in accordance with geographical features \citep{Horiuchi1978}.
At that time, ships were the primary means of transportation, and the Sumida River was a key transportation route.
Thus, industrial districts developed in coastal zones and regions bordering rivers (Figure~\ref{fig:map_ward_pden1920_mt}).
With the onset of the Meiji era (from the late 19th century), the first railway line was laid between Shimbashi (in Kyobashi Ward) and Yokohama City.
This means that the location of industrial districts was determined by an initial geographical feature suitable for water transport, but was not created by the introduction of railways.

\begin{figure}[htbp]
\centering
\captionsetup{justification=centering}
\subfloat[Residential (green), Commercial (red), Undesignated (yellow), and Manufacturing (blue) Zones]{\label{fig:zoning_1925_mt}\includegraphics[width=0.525\textwidth]{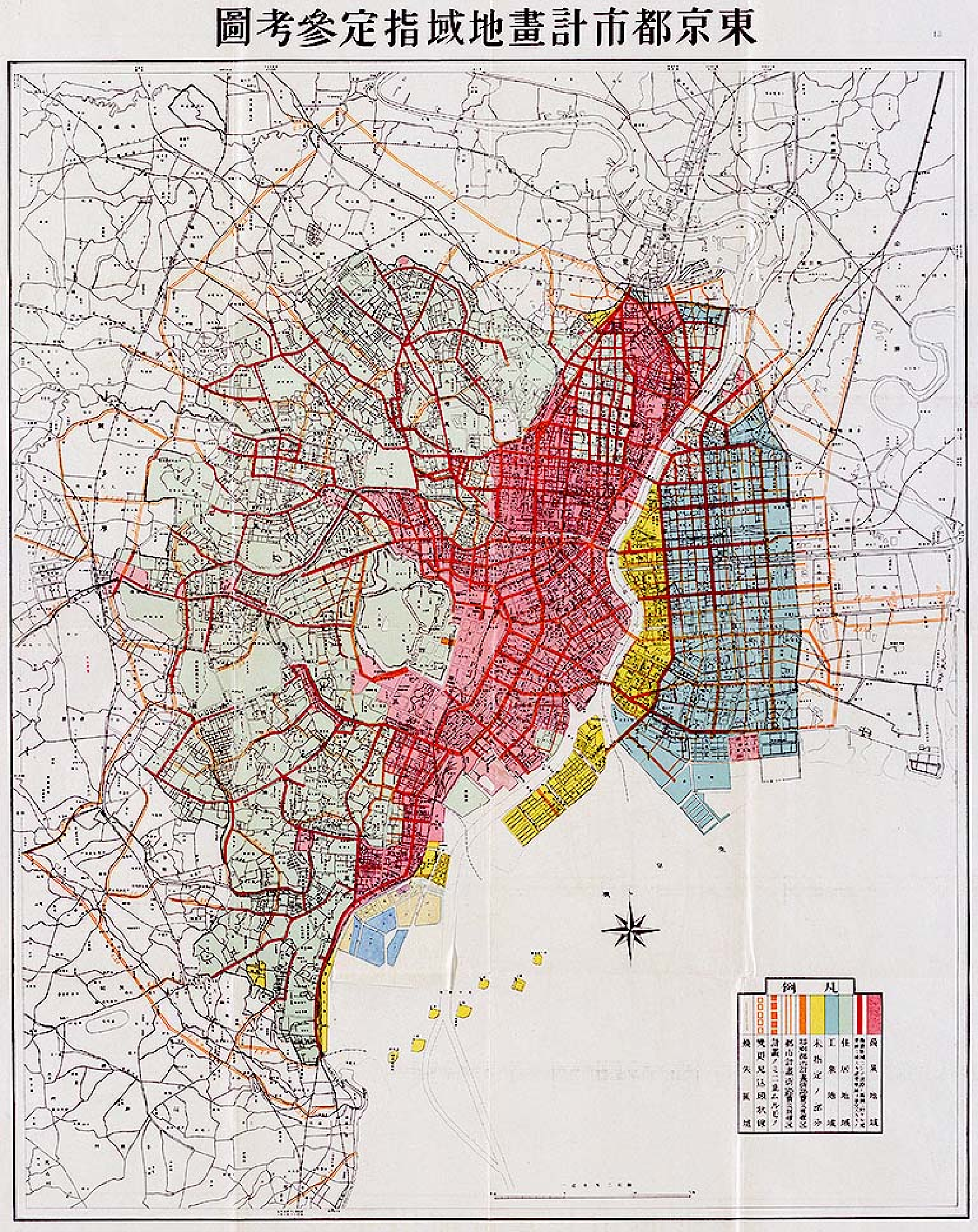}}
\hspace{5pt}
\subfloat[Burned Blocks (orange and pink): Wind (blue arrow) and Spread (red arrow) Directions]{\label{fig:fire_mt}\includegraphics[width=0.45\textwidth]{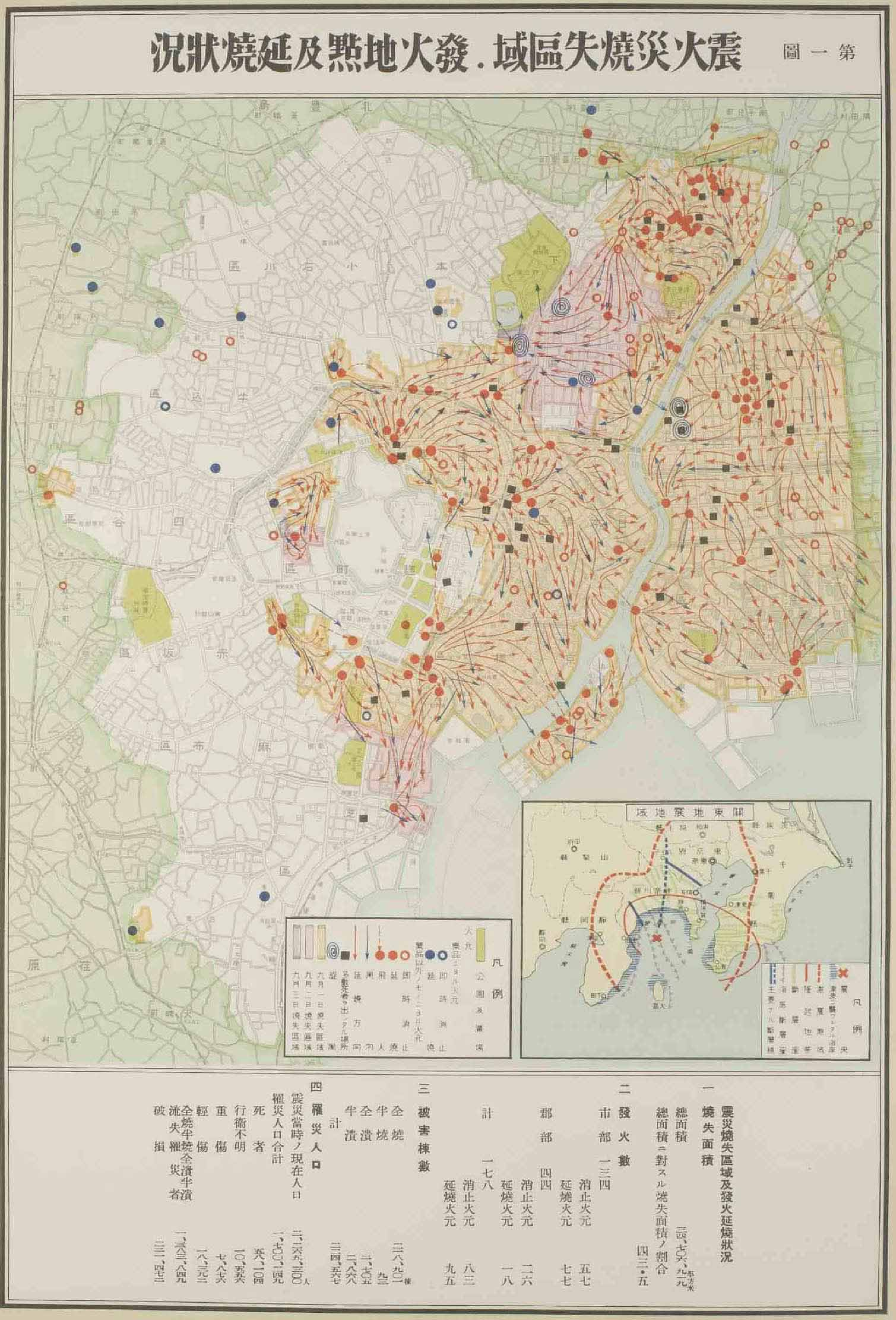}}
\caption{Land-use Zones and Burned Blocks in Tokyo City}
\label{fig:map_mt}
\scriptsize{\begin{minipage}{450pt}
\setstretch{0.9}
Notes:
Figure~\ref{fig:zoning_1925_mt} illustrates the residential (green), commercial (red), and manufacturing (blue) zones within the Tokyo City Planning Area (\textit{T\=oky\=o Toshi Keikaku Kuiki}), which was designed before the 1923 earthquake and enacted in 1925.
Although the undesignated (yellow) zone is designed as a buffer between the city's commercial center (red) and the manufacturing area (blue), it encompasses principal manufacturing districts and thus constitutes a de facto manufacturing zone.
Figure~\ref{fig:fire_mt} illustrates the burned blocks (orange and pink) in Tokyo City.
The blue arrow indicates the wind direction, whereas the red arrow indicates the direction of the spread of fire.
The red circles show the origin of fire, whereas the blue circles show the origin of fire caused by chemicals.
Black squares indicate locations where multiple deaths occurred.\\
Sources:
The map of land-use zones is obtained from the National Archive of Japan (Fu B02002100), website.
The map of the burned blocks is obtained from the Tokyo City Office (1930, Figure 1).
\end{minipage}}
\end{figure}

Driven by industrial development and urbanization following World War I, the government began to recognize the need for systematic urban planning.
The population of residential areas located outside the industrial districts had begun to grow due to the development of public transportation.
In March 1920, the Ministry of Interior began drafting a zoning system proposal, based on the City Planning Act (\textit{Toshi Keikaku H\=o}) of 1919, which was approved in August 1923.
Although the promulgation of the zoning system was delayed by the Great Kanto Earthquake in September 1923, a system with virtually identical content was promulgated in January 1925.\footnote{A detailed summary of the historical background for the introduction of the zoning system is provided in the Online Appendix~\ref{sec:seca_zoning}. The only changes from the original draft were that a small number of towns in the undesignated area were reclassified as industrial zones (footnote~\ref{fn:fn_a}). The statement of reasons for zoning and the supporting documents were also attached to the revised draft, exactly as they appeared in the original draft \citet[pp.50--51]{Horiuchi1978}. The city planning before the disaster is summarized in Online Appendix~\ref{sec:seca_before}.}
Figure~\ref{fig:zoning_1925_mt} illustrates the spatial distribution of these land-use zones (\textit{y\=otochiiki}) designed within the Tokyo City Planning Area (\textit{T\=oky\=o Toshi Keikaku Kuiki}).
The explanatory memorandum for the zoning system clearly states that zoning districts were determined based on geographical conditions.
Consequently, based on the initial geography, commercial zones were designated west of the Sumida River and industrial zones east of it.
The zoning system prohibited the construction of factories exceeding a certain scale or equipped with certain facilities in residential and commercial zones.\footnote{For existing factories, although a grace period was granted, the law was generally applied when they underwent expansion or renovation. In industrial zones, there were generally no restrictions on factory construction. Even within industrial zones,  however, the areas where construction was permitted differed between factories handling explosives such as gunpowder and ether, and those handling substances like sulfuric acid and nitric acid that posed a risk of water pollution or other health hazards \citep[p.~51]{Horiuchi1978}. Since it is practically impossible to distinguish between these categories based on the available data, this paper treats them uniformly as industrial zones.}
Importantly, however, it did not impose restrictions on the construction of new homes.
The yellow areas in the figure were designated as undesignated (\textit{mishitei}) zones to limit potential negative externalities of industrial land, but they consist largely of industrial areas.
Thus, they are treated as industrial districts for analysis in the following analysis.\footnote{\label{fn:fn_a} Tsukishima, an island at the mouth of the Sumida River, was an example. It covered a large part of the undesignated zone (Figure~\ref{fig:zoning_1925_mt}) but included a major industrial area with numerous machinery factories. See \citet{Ogasawara2026-uc} for the details of this island in the early 1920s. Additionally, there are cases where areas designated as undesignated zones in the original plan were included in industrial zones upon implementation (e.g., the Shibaura area).}

To summarize, the zoning system assigned institutional labels to the distribution of early industrial districts, which had been shaped by initial geographical features.
Although the zoning system underwent a few revisions throughout the 1920s and 1930s, there were virtually no changes to the zoning districts within Tokyo City (Online Appendix~\ref{sec:seca_zoning}).
Consequently, it did not fundamentally transform the distribution of industrial and residential areas within a span of a dozen or so years.
As existing historical research has pointed out, the institutional binding power of the zoning system had been considerably weak \citep{Numajiri2002}.

\subsection{The Great Kanto Earthquake: Devastation in Tokyo City} \label{sec:sec_hb_gke}

On September 1, 1923, the Great Kanto Earthquake struck, causing the greatest material and human damage in Japan's history of natural disasters.\footnote{The data and references used in this section are summarized in the Online Appendix~\ref{sec:seca_gke}, which provides a more detailed explanation of the content of this section.}
The maximum seismic intensity is estimated at M$8.1 \pm 0.2$, and the economic damage amounted to approximately $5.5$ billion yen, equivalent to about $40$\% of the GDP.
The scale of the damage becomes clear when the national annual budget was $1.4$ billion yen at that time.
According to a report by the Ministry of Interior, approximately $73$\% of households in Tokyo City were affected, and $3$\% of the city's population died or went missing.

The earthquake struck at a time when many families were preparing lunch.
Furthermore, because a typhoon had formed near the Noto Peninsula, winds were reportedly quite strong that day.
Since nearly all homes at the time were wooden, fires that broke out at multiple locations spread rapidly in the wind, burning down more than 40\% of the city.
As a result, $85$\% of the affected households were completely destroyed by fires subsequently caused by the earthquake.
Figure~\ref{fig:fire_mt} shows the distribution of the areas affected by the fires.
As is evident, the majority of the CBD was lost to the fires.

The distribution of these fires was not merely related to ground softness.
Since the Asakusa, Honjo, and Fukagawa Wards were built on alluvial deposits, they experienced severe shaking, and fires from collapsed houses spread flames throughout the area.
However, the Nihonbashi and Kyobashi Wards, built on a submerged wave-cut platform called the Nihonbashi Plateau, were resistant to shaking.
Both wards had the lowest rate of house collapse among all wards and had fewer ignition points.
Despite this advantage, buildings there caught fire from flying embers because the area was downwind of the ignition points in the Kanda and Asakusa Wards.
These mixed external factors, due to the ground softness and winds, introduced randomness into the distribution of damage.

Although the disaster hit the CBD, Tokyo City remained the capital in the end.
As examined in Section~\ref{sec:sec_hb_cbd}, the pre-earthquake business district was distributed in accordance with its geographical characteristics.
This suggests that the city's long-run development was influenced by the locational fundamentals.
Unlike the location of capital, however, the pattern of population distribution within the city after the disaster is not self-evident.
Generally, when zoning regulations restrict housing supply, the number of housing units in an area decreases, and housing prices rise \citep{Gyourko2015-ir}.
In the case of Tokyo, the zoning system imposed constraints on factory construction but did not regulate housing supply (Section~\ref{sec:sec_hb_cbd}).
Instead, the land readjustment project in the reconstruction plan directly reduced the amount of land available for residential use.
In the following section, I provide an overview of the project to consider the potential mechanisms underlying population changes in areas affected by the fire.

\subsection{Reconstruction: Land Readjustment in the Burned Area}\label{sec:sec_hb_rec}

Immediately after the earthquake, the national government and Tokyo City were required to take various measures.
I focus herein on the reconstruction project implemented in the devastated areas.
These emergency responses by the national government and the city are summarized in detail in Online Appendix~\ref{sec:seca_gov}.\footnote{\citet{Hunter2019-yt} provides a review of these emergency measures from the perspective of price controls. Finer details of the history of city planning in Tokyo before and after the earthquake are summarized in Online Appendix~\ref{sec:seca_plan}. The supplies of wood via procurement and markets after the disaster are summarized in Online Appendix~\ref{sec:seca_wood}.}

On September 19, 1923, the government established the Imperial Capital Reconstruction Council (\textit{Teito Fukk\=o Shingikai}).
The Imperial Capital Reconstruction Agency (\textit{Teito Fukk\=o-in}) was then established on the 27th.
Through both bodies, the government formulated the Imperial Capital Reconstruction Plan (\textit{Teito Fukk\=o Keikaku}), which accounted for $50$\% of the national budget.
The main project of the plan was land readjustment.
This was a large-scale undertaking in which the government and Tokyo City jointly carried out land readjustment on $91$\% of the burned area.
Details of the way in land readjustment are examined in Online Appendix~\ref{sec:seca_rec}.

The land readjustment project aimed to increase the amount of public space, such as roads and parks, to improve fire prevention.
Before the earthquake, the city had roads laid out during the Edo era.
These old roads were narrow and winding, designed to prepare for urban warfare.
The project involved straightening and widening these roads.
In areas where land readjustment was implemented, the average road width increased from $9.7$ meters to $13.7$ meters, whereas in areas where it was not implemented, the average width remained virtually unchanged (Online Appendix Table~\ref{tab:res_area}).
The most significant result from road widening and the creation of new parks was the reduction in residential areas.
The proportion of residential land within the total area of the burned area decreased from $75.3$\% to $63.8$\%, while public land, such as roads and parks, increased from $24.6$\% to $36.2$\% (Online Appendix Table~\ref{tab:road}).
This decline in residential land suggests that housing costs in the city were likely to increase.

The land readjustment has further shaped the structure of modern-day Tokyo.
Although Tokyo was reduced to ashes by air raids during World War II, many of the major roads constructed as part of the post-earthquake land readjustment still exist today.
Not only that, but many of the bridges and parks remain as well.
Online Appendix Figure~\ref{fig:road_bridge} shows an example of the legacy roads and bridges.
In this light, the land readjustment carried out in the burned area created long-term path dependence in the land use.

\section{Theoretical Framework}\label{sec:sec_tf}

In the empirical analysis, I aim to interpret the empirical distribution of the population in Tokyo City by employing a simplified monocentric city model.\footnote{I considered \citet{Wheaton1974-tx} as a reference model. Since the theory of land use was proposed by \citet{Alonso1964}, a body of research has accumulated on theories regarding residential location choice within cities in urban economics \citep[e.g.,] []{Mills1967-ks, Beckmann1969-sr, Montesano1972-iy}. \citet{Straszheim1987-hw} and \citet{Fujita1989-zu} provide comprehensive reviews of the foundational theory. A comprehensive review, including recent theoretical research, is provided by \citet{Duranton2015-mm}.}
This section provides a brief overview of the theoretical predictions regarding the population distribution.
A detailed explanation of the derivations is available in Online~\ref{sec:secb}.

Consider a closed city populated by $n$ households (or consumers) with identical preferences.
They commute to the city center, where they earn income ($y$).
Given their income, each household consumes a composite good ($x$) and housing ($s$).
Both goods are necessities and normal goods, and the utility function is represented by $u(x, s)$.
The utility function is strictly increasing in each good and is assumed to be twice-differentiable and strictly quasi-concave.
Commuting from a residence location ($m \geq 0$) to the city center incurs a commuting cost ($C(m)$), which is strictly increasing with respect to distance ($\frac{\partial C}{\partial m} > 0$).
The unit housing price depends on the residential location ($R(m)$), and the composite good is assumed to be a num\'eraire.
Under this setting, the household's budget constraint is given by $sR(m) = y - C(m) - x$.
The household then maximizes utility by choosing a residential location that affects both rent and commuting cost, subject to a trade-off between the size of the residence and its proximity to the city center.
Since preferences are identical across households, all households achieve the same level of utility ($\hat{u}$) at the residential equilibrium, regardless of their housing location.

The function that derives the maximum rent per unit of housing (land) that a household is willing to pay under a given utility level ($u$) is called the bid rent function \citep{Alonso1964} and is defined as follows:
	\begin{align}
	R(m, u, y) = \max_{x, s} \left[\frac{y-C(m)-x}{s} \middle| u(x, s) = u \right]
	\end{align}
This means that, for a household with consumption $(x, s)$ that achieves a given utility level $u$ at a distance $m$, the payable rent per unit area is represented by $(y-C(m)-x)/s$.
Let $(x^{*}, s^{*}) = (x(m, \hat{u}, y), s(m, \hat{u}, y))$ denote the optimal consumption that achieves maximum utility ($\hat{u}$) at distance $m$, and let $\hat R(m,\hat u,y) \equiv R(m,\hat u,y)$ denote the market rent at the equilibrium utility level for $m \le m^*$.
One can then derive the following relationships between market rent, housing size, and distance:\footnote{By the envelope theorem, $d\hat R/dm = -C'(m)/s < 0$; because compensated housing demand decreases in its own price, $ds/dm > 0$.}
	\begin{align}\label{theory_prediction}
	\frac{d\hat{R}(m, \hat{u}, y)}{d m} < 0 \\
	\frac{d s(m, \hat{u}, y)}{d m}  > 0.
	\end{align}
This result implies that as the distance from the city center to the residence increases, market rental prices decrease and housing size increases.

Assume that land beyond the city boundary ($m^{*}$) is used for agricultural production that generates an agricultural rent ($R^{*} \geq 0$) per unit area.
At the city boundary, therefore, $R(m^{*}, \hat{u}, y) = R^{*}$ holds.
Let $\phi(m)\in(0,1]$ denote the share of land at distance $m$ available for residential use.
Since all residential land within the city is consumed, the total number of households is expressed as follows:
	\begin{align}\label{fixed_pop}
	\int_0^{m^*}\frac{\phi(m)\,2\pi m}{s(m,\hat u,y)}\,dm = n.
	\end{align}
The standard monocentric model corresponds to the case of $\phi(m)=1$.
Note that $\phi(m)$ affects only this land-market condition, and not the household problem or Equations (2) and (3).
At equilibrium, the residential density ($1/s(m, \hat{u}, y)$) decreases toward the boundary, as housing size increases toward the city boundary.
This means that, for a given $\phi$, household density within the city is high near the city center and low in the suburbs.

Land readjustment following the earthquake reduced $\phi(m)$ in the burned area (Section~\ref{sec:sec_hb_rec}).\footnote{Land readjustment also widened and straightened roads, which could have reduced commuting costs. I abstract from this channel by holding $C(m)$ fixed, because transportation fares were uniform within the city and remained largely unchanged throughout the sample period (Section~\ref{sec:sec_rob}).}
Because the number of households is fixed in a closed city, the equilibrium utility $\hat u$ must decline, which shifts the bid rent function upward at every location and moves the city boundary outward (Online Appendix Figure~\ref{fig:brf_shift}).
In other words, the reduction in residential land area resulting from land readjustment increases market rental prices within the city and leads to suburbanization.
Consequently, the number of households within the pre-earthquake city boundary decreases.
The same holds for the burned area itself: although smaller housing allows more households to reside per unit of residential land, households outside the burned area also consume less housing and the city boundary expands, so the number of households in the burned area declines (Online Appendix~\ref{sec:secb4}).

I summarize the following two predictions made by the model for the empirical analysis:

\begin{itemize}
\item[] \textbf{Prediction 1:} Before the land readjustment, the city center could have a higher average rent and a higher household density (the number of households per unit of land) than the surrounding areas.

\textbf{Prediction 2:} After the land readjustment, the number of households in the burned area could decline relative to other areas. The reduction in residential land directly decreases the number of households in the burned area, while smaller housing per household, which arises citywide as the equilibrium utility declines, partly offsets this decline.
\end{itemize}

Because the model treats each household as a single consumption unit, both predictions concern households.
Population, which I use as the main outcome in the empirical analysis, equals the number of households multiplied by the average household size.
Population and households therefore move together only if average household size remains unchanged.
Section~\ref{sec:sec_ea_block} examines changes in household size, which lie outside the model.
The zoning system was a path-dependent institution that did not constitute a new regulation on residential areas (Section~\ref{sec:sec_hb_cbd}).\footnote{I analyzed the impact of the earthquake on population in the burned area, while controlling for the influence of zoning on population in the city center. As I will summarize in Section ~\ref{sec:sec_ea_het}, the impacts of zoning are negligible in the statistical sense.}
This supports the validity of the market-based framework for the empirical analysis rather than the regulation-based framework.
The above predictions are then used to interpret the quantitative results from datasets digitized from Tokyo City census statistics.\footnote{Several studies have tested the prediction of the monocentric model that population density and land and housing prices decline with distance from the city center \citep{BaumSnow2007}. Rigorous tests of model validity are, however, difficult in practice due to the limited availability of land and housing prices statistics and the diversity of city center definitions \citep[pp.~522--530]{Duranton2015-mm}. The current study also does not aim to verify the model from multiple perspectives. Instead, it presents evidence that the population distribution following the earthquake is consistent with a few model predictions.}

\section{Data}\label{sec:sec_data}

I digitized census statistics to create ward- and block-level datasets on housing, rents, population, and households.
This section provides details on variable construction across the datasets and defines the sample used in the analysis.
Finer details are available in Online~\ref{sec:secc}.

\subsection{Ward-level Census Statistics}\label{sec:sec_data_ward}

The ward-level data summarized in Panel A of Table ~\ref{tab:sum} are used in Section~\ref{sec:sec_ea_ward} to provide an overview of the disaster's impacts on housing, rents, population, and households.

\begin{table}[htbp]
\def\arraystretch{0.95}
\centering
\captionsetup{justification=centering}
\begin{center}
\caption{Summary Statistics}
\label{tab:sum}
\scriptsize
\scalebox{0.97}[1]{
{\setlength\doublerulesep{2pt}
\begin{tabular}{lrrrrrr}
\toprule[1pt]\midrule[0.3pt]
\textbf{Panel A: Ward-level Dataset}
&\multicolumn{3}{c}{Burned}&\multicolumn{3}{c}{Unburned}\\
\cmidrule(rrr){2-4}\cmidrule(rrr){5-7}
&Mean&Std. Dev.&Obs.&Mean&Std. Dev.&Obs.\\
\cmidrule(rrrrrrr){1-7}
\multicolumn{7}{l}{Panel A-1: Buildings, floor space, and rent}\\
\hspace{10pt}Buildings (counts $/\text{km}^{2}$)			&3,868	&1,825	&153	&3,509	&423		&102		\\
\hspace{10pt}Multi-story dwellings (counts $/\text{km}^{2}$)	&2,163	&1,371	&153	&1,711	&489		&102		\\
\hspace{10pt}Floor area (square meters $/\text{km}^{2}$)		&285,141	&130,872	&153	&288,053	&53,231	&102		\\
\hspace{10pt}Average monthly unit rent ($\text{yen}/\text{m}^{2}$)		&0.275	&0.245	&152	&0.138	&0.080	&102		\\
\multicolumn{7}{l}{Panel A-2: Population and households}\\
\hspace{10pt}Population								&140,850	&57,747	&63	&98,011	&34,641	&42		\\
\hspace{10pt}Households								&30,626	&13,327	&63	&21,234	&7,455	&42		\\
\multicolumn{7}{l}{Panel A-3: Burned area share}\\
\hspace{10pt}Burned Area (\%)							&73	&29	&63	&6	&6	&42		\\
&&&&&&\\
\textbf{Panel B: Block-level Dataset}
&\multicolumn{3}{c}{Burned}&\multicolumn{3}{c}{Unburned}\\
\cmidrule(rrr){2-4}\cmidrule(rrr){5-7}
&Mean&Std. Dev.&Obs.&Mean&Std. Dev.&Obs.\\
\cmidrule(rrrrrrr){1-7}
\multicolumn{7}{l}{Panel B-1: Block-year level dataset}\\
\hspace{10pt}Population							&1,222.2	&3,265.9	&5,316	&1,359.1	&1,718.8	&3,138	\\
\hspace{10pt}Household							&269.1	&742.9	&5,316	&302.8	&400.5	&3,138	\\
\hspace{10pt}Average household size (individuals/households)	&5.1		&1.6		&5,136	&4.7		&1.1		&2,937	\\
\hspace{10pt}Average sex ratio (female/male)			&0.8	&0.3	&5,134	&0.9	&0.6	&2,937	\\
\hspace{10pt}Industrial blocks					&0.9 &0.3	&5,316	&0.3	&0.4	&3,138	\\
\multicolumn{7}{l}{Panel B-2: Block-size-year level dataset}\\
\hspace{10pt}1 person				&20.6	&53.1	&5,316	&24.1&52.9&3,138\\
\hspace{10pt}2 people				&87.5	&256.6	&5,316	&95.6&147.0&3,138\\ 
\hspace{10pt}3 people				&151.0	&444.3	&5,316	&165.7&230.5&3,138\\
\hspace{10pt}4 people				&175.0	&510.8	&5,316	&198.5&263.6&3,138\\
\hspace{10pt}5 people				&175.5	&497.9	&5,316	&204.0&270.0&3,138\\
\hspace{10pt}6 people				&153.9	&428.3	&5,316	&182.1&243.4&3,138\\
\hspace{10pt}7 people				&121.8	&331.2	&5,316	&145.3&194.5&3,138\\
\hspace{10pt}8 people				&88.8	&231.4	&5,316	&104.6&138.8&3,138\\
\hspace{10pt}9 people				&63.4	&162.4	&5,316	&70.3&91.6&3,138\\
\hspace{10pt}10 people				&46.3	&120.0	&5,316	&48.2&63.6&3,138\\
\hspace{10pt}11 people				&28.6	&68.3	&5,316	&29.5&39.2&3,138\\
\hspace{10pt}12 people				&22.3	&55.3	&5,316	&20.5&28.4&3,138\\
\hspace{10pt}13 people				&17.0	&42.5	&5,316	&14.5&22.2&3,138\\
\hspace{10pt}14 people				&12.8	&34.9	&5,316	&10.7&17.5&3,138\\
\hspace{10pt}15+ people				&57.7	&143.6	&5,316	&45.5&84.2&3,138\\
\multicolumn{7}{l}{Panel B-3: Block-level household survey dataset}\\
\hspace{10pt}Landlords' average monthly rent (yen/\textit{tatami})		&2.20&0.40&678&1.83&0.24&418\\
\hspace{10pt}Lodgers' average monthly lodging rent (yen/\textit{tatami})	&2.19&0.36&678&2.04&0.22&418\\
\hspace{10pt}Average area per lodger (\textit{tatami})				&4.36&1.04&676&4.76&0.69&416\\
\hspace{10pt}Average monthly lodging rent per lodger (yen)				&9.56&2.81&676&9.69&1.75&416\\
\hspace{10pt}Rent ratio (lodging rent/rent)							&1.01&0.15&678&1.13&0.15&418\\
\midrule[0.3pt]\bottomrule[1pt]
\end{tabular}
}
}
{\scriptsize
\begin{minipage}{435pt}
\setstretch{0.85}
Notes:
\textbf{Panel A}: This panel presents summary statistics for the ward-year level data.
The `Burned' indicates the statistics for the wards where more than a quarter of the administrative area was burned, whereas the `Unburned' denotes the statistics for the other wards.
\underline{Panel A-1} summarizes statistics from the official survey on the number of wooden buildings ($/\text{km}^{2}$), the number of multi-story wooden dwellings ($/\text{km}^{2}$), the floor area of the wooden buildings ($/\text{km}^{2}$), and rent ($/\text{m}^{2}$) from 1919 to 1935.
Rents are deflated by the urban consumer price index excluding rent.
\underline{Panel A-2} summarizes statistics from the population census conducted in 1908, 1920, 1923, 1924, 1925, 1930, and 1935.
\underline{Panel A-3} shows the percentage of burned area relative to the total administrative ward area.
The administrative ward area used as a denominator in the variables in Panels A-1 and A-3 excludes the Imperial Palace grounds, where citizens are unable to live.\\
\textbf{Panel B}:
This panel presents summary statistics for the block-level data collected in the 1908, 1920, 1923, 1924, 1925, and 1930 population censuses.
The `Burned' indicates the statistics for the $886$ burned blocks, whereas the `Unburned' indicates those for the $523$ unburned blocks.
\underline{Panel B-1} shows the summary statistics for the block-year panel data on the number of people, households, average household size, sex ratio, and the industrial zone indicator variable.
Household size is the number of individuals per household.
Sex ratio is the number of females divided by the number of males.
For 381 block-year cells with no individuals, neither the household size nor the sex ratio is defined.
In the sex ratio sample, a few block-year cells that include only females are excluded.
Industrial blocks indicate the commercial and manufacturing zones within the zoning system, which were designed before the earthquake.
\underline{Panel B-2} shows the summary statistics for the block-size-year level data on the number of individuals.
Household bins range from 1 person to 15+ people.
For example, a bin labeled `2 people' shows statistics on households with two individuals.
A bin labeled `1 person' indicates statistics for households with one person, meaning the number of people is equal to the number of households in the bin.
\underline{Panel B-3} presents summary statistics for the variables measured in the Survey on Shared-Dwelling Households (SSDH), conducted between the end of 1929 and the beginning of 1930.
Average monthly rent and lodging rent are reported as a monthly payment per unit area, \textit{tatami} mat ($1.54 \text{m}^{2}$).
Average rented area per lodger indicates the average area per lodger in \textit{tatami} mats.
Average monthly lodging rent per lodger is defined as the average monthly lodging rent per lodger multiplied by the average space per lodger. 
Four blocks with no information on the average rented area per lodger are excluded from the average rented area and the monthly rent samples.
The rent ratio is the lodgers' average monthly lodging rent divided by the landlords' average monthly rent.\\
Sources:
Panel A-1: The statistics on the buildings and rents are digitized using the Tokyo City Office (1921--1937).
The consumer price index is obtained from \citet{ltes1967v8}.
The data on the ward area are obtained from the Tokyo City Office (1911--1937).
Panel A-2: The statistics on the population are digitized from the Tokyo City Office (1909; 1922a; 1925a; 1926a; 1927a; 1932a) and the Statistics Bureau of the Cabinet (1937).
Panel A-3: The statistics on the burned area and total ward area are from the Tokyo City Office (1932c).
The statistics on the Imperial Palace grounds area are from the Tokyo City Office (1922).
Panels B-1 and B-2: The population and household statistics are from the Tokyo City Office (1909; 1922a; 1925a; 1926a; 1927a; 1932a).
Panel B-3: The survey data are from the Tokyo City Social Bureau (1930a).
\end{minipage}
}
\end{center}
\end{table}

\subsubsection*{Burned area}

The total burned area in each administrative ward is documented in an official report for the reconstruction plan, named History of the Imperial Capital Reconstruction and Land Readjustment Project (\textit{Teito Fukk\=o Kukaku Seiri-shi}) published by the Tokyo City Office (1932c).
I calculated the area burned as a percentage of the administrative ward area.
Online Appendix Table~\ref{tab:damage_ward} summarizes the statistics used.

\subsubsection*{Building and Rent}

In prewar Japan, people lived in wooden houses and rarely in other types of housing.
Systematic statistics on wooden houses in Tokyo City are documented in the Tokyo City Statistical Yearbook (\textit{Toky\=o-shi T\=okei Nenpy\=o}, hereafter the TCSY).
This source provides information on the number of residential wooden buildings, their number of stories, and their floor area, reported for the 15 administrative wards.
I have digitized these statistics from 1919 to 1935 and converted them to values per unit ward area ($\text{km}^{2}$).
The TCSY reports average monthly unit rent (per $\text{m}^{2}$), which is rarely available in historical sources \citep{Duranton2015-mm}.
It is described as ``the average price of residential properties transacted over the course of a year.''
However, it would be difficult to comprehensively record all rents across market transactions in a year.
Thus, it is presumably a compilation of prices from all contracts transacted in some areas within each ward.\footnote{The average monthly unit rent fluctuates more significantly than the statistics on the wooden buildings (Panel A-1 of Table~\ref{tab:sum}). I will return to this issue in Section~\ref{sec:sec_ea_ward2} in detail.}
Nevertheless, given the scarcity of rent statistics, these figures have analytical value.
As with the number of dwellings, I have digitized the statistics on the average monthly unit rent from 1919 to 1935.

\subsubsection*{Population and Households}

Data on the number of households and the population residing in the city are available for the regular census years 1920, 1925, 1930, and 1935.
In addition, accurate statistics on the number of households and the population are available for 1908, when the city conducted its pilot census.
Two special censuses were also conducted immediately after the earthquake in November 1923 and October 1924.
Based on the statistics from these complete surveys, I have compiled systematic data covering the seven-year period between 1908 and 1935.
In this study, I refer to these statistical reports as the Original Tables of Tokyo City Statistics (\textit{T\=oky\=o-shi Shisei T\=okei Genpy\=o}, hereafter the OTTCS).\footnote{Although there are slight differences in terminology, there are no practical issues with using a standardized term (Online Appendix~\ref{sec:secc_block1}). As for the population statistics, the TCSY also provides annual population estimates based on official records. However, these estimates contain substantial measurement errors and are not suitable for statistical analysis. This issue is discussed in detail in \citet[p.~125]{ltes1966v3}.}

\subsection{Block-level Population Census Statistics}\label{sec:sec_data_census}

The block-level data listed in Panels B-1 and B-2 of Table~\ref{tab:sum} are considered to analyze population impacts in detail in Sections~\ref{sec:sec_ea_het} and~\ref{sec:sec_ea_block1}.

\subsubsection*{Definition of Blocks}

Information on household composition is recorded in the census reports for six years (1908, 1920, 1923, 1924, 1925, and 1930), except for 1935.
The statistics are broken down by block-level (\textit{ch\=o-ch\=ome}), the smallest administrative unit available.
The cross-sectional unit must remain the same administrative lattice across sample years to investigate within-variation in the dependent variable.
Since there were few changes to administrative blocks, most remained unique throughout the observation period.
In cases where blocks were merged or divided due to land readjustment, I take the following adjustments.\footnote{The division and reorganization took place primarily in three commercial areas--Nihonbashi, Ginza, and Ryogoku--located in the Nihonbashi, Kyobashi, and Honjo Wards.}
First, in merger cases, the merged blocks were aggregated into the grid divisions as they existed in 1930.
Second, when blocks were split into multiple blocks, the data was aggregated by the parent block name.\footnote{A few cases fall in this division. For example, when parts of multiple town blocks were subdivided and consolidated into Marunouchi 1-\textit{ch\=ome} and 2-\textit{ch\=ome}, statistics were compiled using Marunouchi as the parent block. A similar approach applies to large-scale parks developed as part of reconstruction plans. For instance, when parts of multiple town blocks were subdivided to form Hamacho 1-\textit{chome} and Hamacho Park, statistics for Hamacho Park were compiled using the parent block of the original blocks (in this case, Hamacho).}
Changes in grid divisions were identified using the Kadokawa Encyclopedia of Japanese Historical Place Names (\textit{Shinpan Kadokawa Nihon Chimei Daijiten}), the most comprehensive gazetteer available.\footnote{Access to this database was generously provided by the Library of Economics at the University of Tokyo. I am deeply grateful to the library archivists for their support and for granting permission to undertake this research.}
I excluded the Imperial Palace, where ordinary households cannot reside, towns incorporated from outside the city during the observation period, and towns established midway through the period due to land reclamation (26 blocks in total).
The final number of town blocks is $1,409$.

\subsubsection*{Burned Blocks}

All blocks destroyed by the earthquake are recorded in the first volume of the History of the Imperial Capital Reconstruction and Land Readjustment Project (\textit{Teito Fukk\=o Kukaku Seiri-shi}), a compilation of official documents (Tokyo City Office 1932c).
According to this report, almost all affected blocks were completely burned, indicating no systematic difference in treatment intensity across blocks.
This confirms that the effects of the measures are homogeneous, which is an advantage in an empirical setting.
Of the total $1,409$ town blocks, $886$ ($63$\%) were in the burned area.\footnote{Note that this rate is not identical to the percentage of burned area described in Section~\ref{sec:sec_hb_gke} because each town block has a different area.}

\subsubsection*{Industrial Zone}

Of the $1,409$ total blocks, the Ministry of Home Affairs Notification (\textit{Naimush\=o kokuji}) No. 14 indicates that $921$ blocks are designated as industrial (i.e., commercial and manufacturing) zones under the zoning system designed before the earthquake (Section~\ref{sec:sec_hb_cbd}).\footnote{This notification is recorded in an official document by a publisher affiliated with the National Police Agency (Wakamori and Someya 1925). As previously mentioned, the undesignated zones were mostly in the manufacturing districts and some nested commercial districts (Section~\ref{sec:sec_hb_cbd}). I classified these zones as industrial zones. The remaining $488$ blocks ($1409-921$) are defined as residential zones.}
85\% of the industrial blocks were burned, leaving a certain proportion of unburned area.\footnote{Industrial zones account for $88$\% of the burned area, confirming that the affected area has a character as a CBD. At the same time, it is clear that the burned area also includes some residential areas.}
This setting is useful for analyzing whether the impacts of the disaster exhibit zoning-dependent heterogeneity.

\subsubsection*{Population, Households, Household Size, and Sex Ratio}

The number of households and population for each block are documented in the OTTCS, except for 1935.
I created a panel dataset with standardized units, based on block definitions as of the final year.
Thus, this leaves $1,409$ block-level data on households and population for the 1908, 1920, 1923, 1924, 1925, and 1930 census years.
The average household size was calculated as the number of individuals per household, and the average sex ratio was determined as the female-to-male ratio.
The OTTCS further documents the number of individuals by household size for each block.
Although the original statistics provide $19$ categories for household size, the number of households in the category with $15$ or more members is small.
I then combined these upper bins into a single category, leaving $15$ categories ranging from $1$ to $15$ or more members for all the blocks.

\subsection{Block-level Complete Household Survey Statistics}\label{sec:sec_data_survey}

To delve into the mechanisms behind population changes following the disaster, I consider cases in which two or more households live in a single dwelling in Section~\ref{sec:sec_ea_block2}.
The variables used are listed in Panel B-3 of Table \ref{tab:sum}.

A report of the Survey of Shared-Dwelling Households (\textit{D\=okyo Setai ni Kansuru Ch\=osa}, hereafter SSDH) conducted by the Tokyo City Social Bureau remains as a basis for housing policy (Tokyo City Social Bureau 1930a).
This survey was a census-style survey conducted through door-to-door visits from November 1929 to March 1930.\footnote{Because the data compilation period is also included in this period, the statistics are expected to reflect the situation of cohabiting households from the end of 1929 to the beginning of 1930. While this is a census-style survey, hospitals and apartments were excluded from the survey object because their definitions differ from those of general households living in shared accommodations (Tokyo City Social Bureau 1930a, p.~legend). The existence of such a survey at the conclusion of the reconstruction plan indicates that Tokyo City was interested in the actual conditions of cohabitation, which had increased in the 1920s.}
The survey covered landlord households (most of whom were not landowners but were tenants) and lodger households, the former being households that sublet part of their residence, and the latter being independent households renting room(s).

\subsubsection*{Definition of Blocks}

As with the panel data, I exclude the Imperial Palace, blocks established due to land reclamation, and those incorporated from outside the city midway through the process, resulting in $1,409$ blocks.
I excluded $6$ blocks in which the number of households not involved in room renting (i.e., total households minus the combined total of landlord and lodger households) was negative, as well as two blocks in which the average household size of those lodging-related households was abnormally high (e.g., $40$ and $99$ people).\footnote{In other words, I excluded blocks where the combined number of landlord and lodger households was either unnaturally large or small compared to the total number of households reported in the census. The timing of this complete enumeration survey differs slightly from that of the 1930 census. However, since land readjustment projects had almost been completed, large-scale population movements are unlikely; therefore, I cannot rule out the possibility that these discrepancies in the blocks result from errors in data compilation or reporting.}
This resulted in a final total of $1,401$ blocks.
Finally, $7$ blocks with incomplete rent data (in which only rent or rental fees were available) were excluded from the sample used in the regression analysis.\footnote{While I excluded these blocks to make a comparable interpretation across different variables, I confirmed that the result is unchanged if these are included (not reported).}

\subsubsection*{Rent and Lodging Rent}

Using the frequency distribution tables for each block reported in the document, I calculated the average monthly landlord's rent (per unit area), the average monthly lodging rent (per unit area), the average lodging area (per lodger), the average lodging rent (per lodger), and the ratio of average monthly landlord's rent to average lodging unit rent.
The detailed calculation methods are summarized in Online Appendix \ref{sec:secc_block2}.
The unit of area is the \textit{tatami}, which is a traditional mat used on the floors of Japanese homes, with each mat equivalent to $1.54~\text{m}^{2}$.

\section{Empirical Analysis} \label{sec:sec_ea}

The empirical analysis consists of three parts.
Section~\ref{sec:sec_ea_ward} uses ward-level data on housing, rents, population, and households to examine the model's predictions.
Section~\ref{sec:sec_ea_het} uses block-level data to confirm these results and to test whether the zoning system affected the population in the burned area.
Section~\ref{sec:sec_ea_block} examines the two margins through which the burned area accommodated its population: the number of households per dwelling, which corresponds to the decrease in housing size per household in the model, and average household size, which lies outside the model.

\subsection{Ward-level Evidence: Housing and Households} \label{sec:sec_ea_ward}

\subsubsection{Descriptive Analysis}\label{sec:sec_ea_ward1}

\begin{figure}[htbp]
\centering
\captionsetup{justification=centering}
\subfloat[Buildings ($/\text{km}^{2}$)\\ in Burned and Unburned Wards]{\label{fig:ts_ward_tbuild}\includegraphics[width=0.45\textwidth]{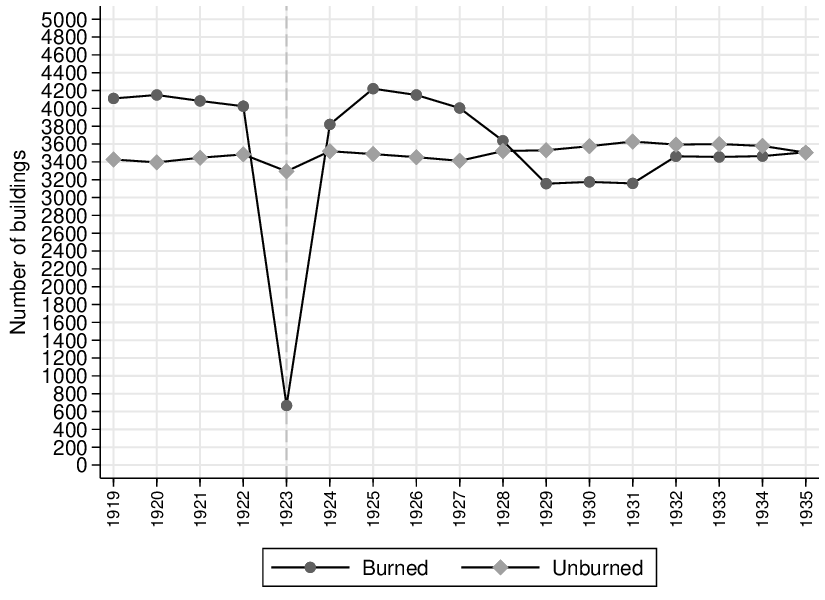}}
\subfloat[Multi-story Dwellings ($/\text{km}^{2}$)\\ in Burned and Unburned Wards]{\label{fig:ts_ward_tbuild_mf}\includegraphics[width=0.45\textwidth]{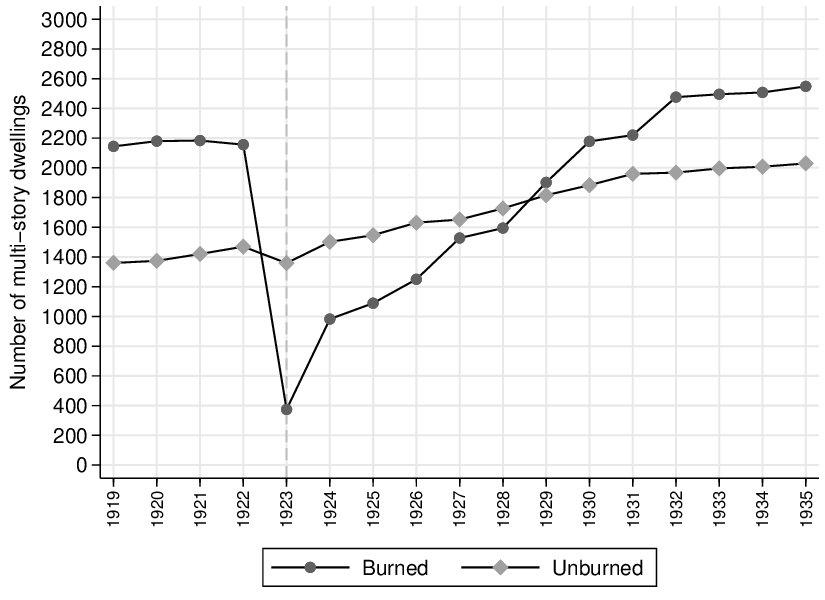}}\\
\subfloat[Floor Area ($\text{m}^{2}/\text{km}^{2}$)\\ in Burned and Unburned Wards]{\label{fig:ts_ward_tbuild_fs}\includegraphics[width=0.45\textwidth]{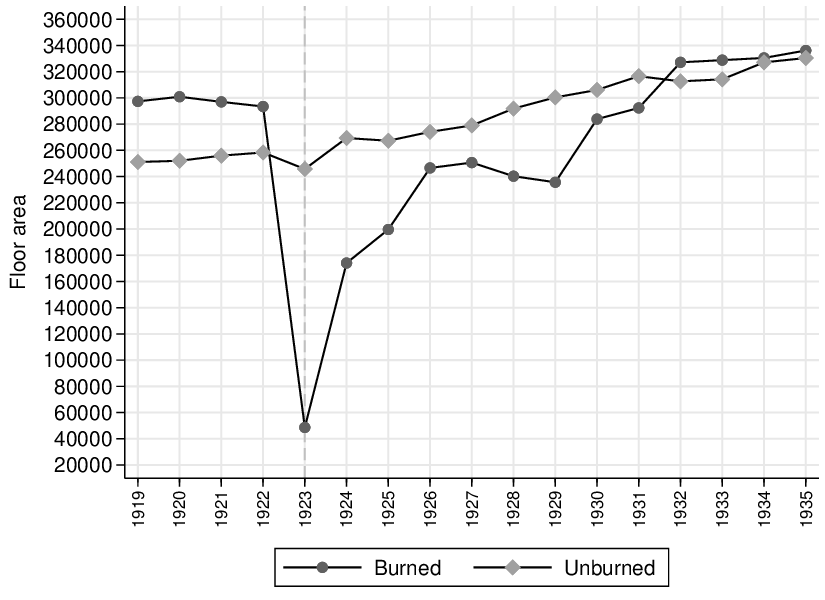}}
\subfloat[Unit Rent (yen / $\text{m}^{2}$)\\ in Burned and Unburned Wards]{\label{fig:ts_ward_mrent}\includegraphics[width=0.45\textwidth]{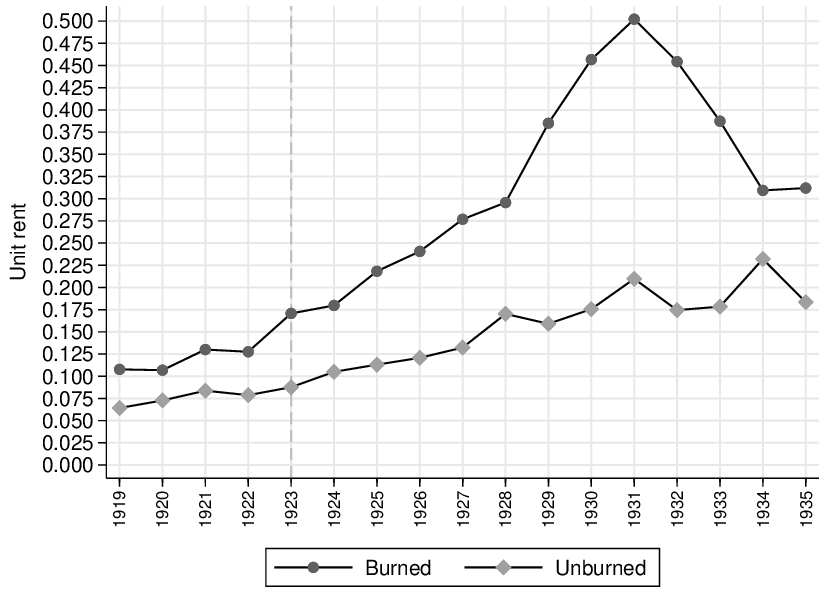}}\\
\subfloat[Population\\ in Burned and Unburned Wards]{\label{fig:ts_ward_pop}\includegraphics[width=0.45\textwidth]{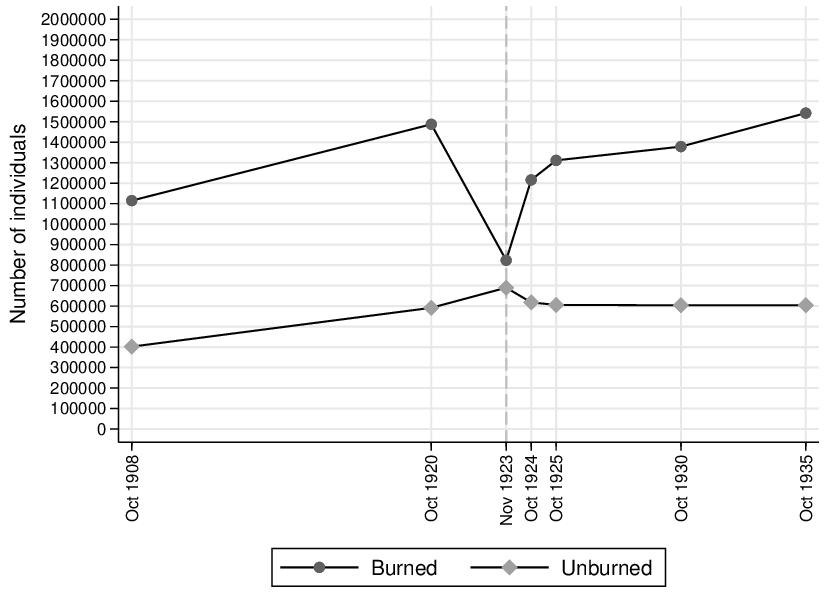}}
\subfloat[Households\\ in Burned and Unburned Wards]{\label{fig:ts_ward_hh}\includegraphics[width=0.45\textwidth]{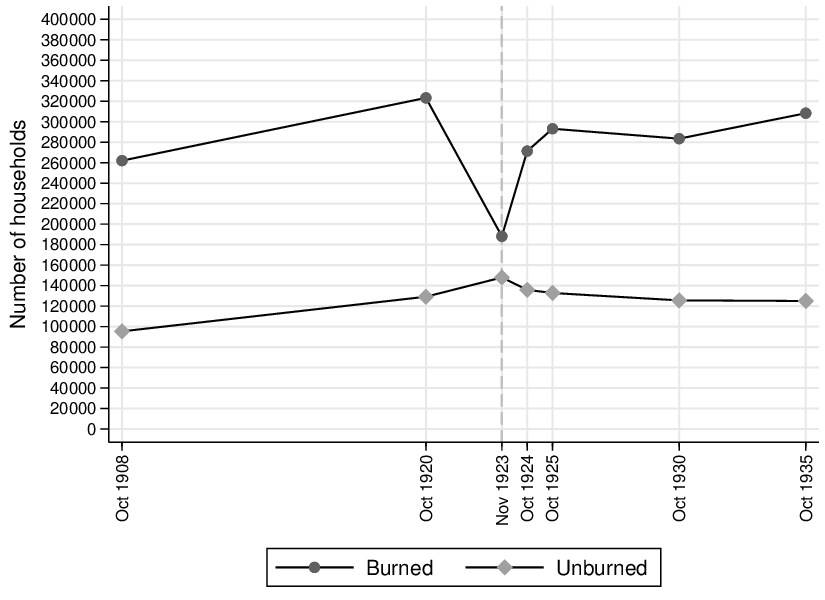}}
\caption{Residencial Buildings, Rents, Population, and Households:\\ Burned v. Unburned Wards}
\label{fig:ts_ward_build_rent}
\scriptsize{\begin{minipage}{450pt}
\setstretch{0.9}
Notes:
Figures~\ref{fig:ts_ward_tbuild}, \ref{fig:ts_ward_tbuild_mf}, and \ref{fig:ts_ward_tbuild_fs} show statistics on the wooden residential buildings between 1919 and 1935 measured in the complete survey by the Tokyo City Office.
Figure~\ref{fig:ts_ward_mrent} shows statistics on the average rent per square meter contracted over a year, surveyed by the Tokyo City Office.
Rents are deflated by the urban consumer price index excluding rent, which corresponds to the price of the composite good in the model (Section~\ref{sec:sec_tf}).
The number of buildings (Figure~\ref{fig:ts_ward_tbuild}), the number of multi-story dwellings (Figure~\ref{fig:ts_ward_tbuild_mf}), and the average floor area (Figure~\ref{fig:ts_ward_tbuild_fs}) are divided by the total area of each ward in $\text{km}^{2}$.
The administrative ward area used as a denominator for the building and rent variables excludes the land for the Imperial Palace, where citizens are not permitted to live.
Figures~\ref{fig:ts_ward_pop} and~\ref{fig:ts_ward_hh} show the population and households in the burned and unburned wards.
The time-series plots of household and population densities (counts per square kilometer of the ward area) show similar results (Online Appendix Figure~\ref{fig:census_hh_pop}).
Burned wards are those in which more than a quarter of the administrative area was burned, whereas unburned wards are the remaining wards.\\
Sources: Created by the author using the ward-level census statistics listed in Panel A of Table~\ref{tab:sum}.
The urban consumer price index is obtained from \citet[pp.~135--136]{ltes1967v8}.
\end{minipage}}
\end{figure}

I begin my analysis by describing the overall transition in the residences and population of the affected wards.

Although it is rational for a landowner or landlord to construct permanent housing rather than costly temporary housing, such construction was prohibited until land readjustments were completed \citep[p.~55]{Ono2014}.
Hence, the construction of temporary housing began in earnest immediately after the earthquake.
These temporary housing included not only private structures but also city-built collective barracks in parks.
Figure~\ref{fig:ts_ward_tbuild} shows that while the number of wooden buildings plummeted immediately after the earthquake, it had recovered to pre-earthquake levels by 1925.
However, temporary housing was much smaller in scale than permanent housing.
Figure~\ref{fig:ts_ward_tbuild_fs} confirms that the floor area of wooden residential buildings in 1925 was significantly below pre-earthquake levels.\footnote{Online Appendix~\ref{sec:seca_wood} provides finer details on post-disaster policies for the procurement of wood, transportation, wood processing and distribution channels, prices, and the construction of policy-based temporary houses (barracks). Online Appendix~\ref{sec:seca_housing} shows the transition of temporary structures, including private housing.}

The collective barracks were dismantled by June 1925, as the city determined that their occupation of public land was undesirable.\footnote{The Tokyo City and D\=ojunkai Foundation (a public corporation established in 1924) also proceeded with the construction of public housing. The number of households that moved into these units, however, was just under 3,000. This is considerably smaller than the number of households in the burned area and was not large enough to affect citywide rent levels \citep[p.~57]{Ono2014}.}
The transition from temporary to permanent housing began around 1925 as prospects for land readjustment took shape.
Figure~\ref{fig:ts_ward_tbuild} shows that the number of buildings declined through around 1930, reflecting the net effect of the decrease in temporary housing and the increase in permanent housing.
Figure~\ref{fig:ts_ward_tbuild_mf} shows that multi-story dwellings began to increase gradually once the prospects for land readjustment became clear, with construction gaining momentum at the end of the 1920s.
Figure~\ref{fig:ts_ward_tbuild_fs} shows fluctuations in floor area during the late 1920s.
This reflects the conflicting trends of a decrease in the number of temporary housing units and an increase in the number of permanent multi-story dwellings.
By the early 1930s, the number of multi-story dwellings had surpassed its pre-earthquake level, and the total floor area had also exceeded its pre-earthquake level.\footnote{Finer details of the reconstruction plan, including the land readjustment project, are summarized in Online Appendix~\ref{sec:seca_rec}. The trends in the buildings and floor area by ward are summarized in Online Appendix~\ref{sec:seca_housing}.}

Landowners who built temporary housing amid shortages of building materials set high rents and security deposits to recoup their high costs.
Figure~\ref{fig:ts_ward_mrent} shows that the average real unit rent in the burned area rose immediately after the earthquake.
As permanent housing replaced temporary housing, rents in the burned area rose further: housing demand in the burned area increased and the vacancy rate declined, so that rents remained high \citep[p.~59]{Ono2014}.\footnote{This trend in rents within the city in the late 1920s is consistent with the report of the housing survey conducted in 1930 (Tokyo City Social Bureau 1931a, p.~11). Online Appendix~\ref{sec:seca_rent} provides additional descriptive analysis of the mechanism behind the rising rents.}
The further increase in real rents after 1930 also reflects the deflation during the depression.
Because nominal rents were downwardly rigid, falling prices raised rents relative to other goods.
This relative rise prompted a nationwide rent-reduction movement (\textit{yachin nesage und\=o}) from late 1929, and many landlords in Tokyo cut nominal rents, typically by 5--20\% for residential houses, between late 1929 and early 1930 \citep[pp.~64--65]{Ono2006}.

Overall, Figure~\ref{fig:ts_ward_mrent} is consistent with the model in two respects.
First, average unit rents in both the burned and unburned wards exceeded their pre-earthquake levels after the completion of the land readjustment, in line with the citywide upward shift of the bid rent function.
Second, the absolute increase was larger in the burned wards, where rents were initially higher.
This reflects the mechanism that the absolute increase in rents tends to be larger near the city center, where housing size is smaller (Online Appendix~\ref{sec:rent_response}).

Figure~\ref{fig:ts_ward_pop} shows population trends derived from the censuses.
In 1920, there was a population difference of $+900,000$ between the burned area and the rest of the city.\footnote{The density difference is $+16,000 / \text{km}^{2}$. The area of each administrative ward remains mostly constant over time. This means that comparing population figures is equivalent to comparing their density. Online Appendix Figure \ref{fig:map_ward_pden} shows the spatial distribution of population density across the wards, which is materially similar to the population distribution.}
Immediately after the earthquake, many of the people living in the burned area fled to unburned area within Tokyo City or sought refuge with relatives and acquaintances outside the city (Section~\ref{sec:sec_hb}).
The temporary population increase in the unburned area in November 1923 reflects the presence of these evacuees.
This phenomenon disappeared around 1925, when the construction of temporary housing was largely complete.

By 1930, the reconstruction plan had been completed, and new residences lined the newly developed blocks.
However, while the population in the unburned area increased slightly, the population in the burned area decreased significantly relative to pre-earthquake levels.
The difference between the two areas had decreased to $800,000$ people in 1930.
The gradual increase in the population of the unburned area is consistent with the development of western residential areas after the earthquake, leading to suburbanization \citep[p.~55]{Horiuchi1978}.
In contrast, the population in the burned area remained considerably lower around 1930.\footnote{Online Appendix~\ref{sec:seca_hh_pop} summarizes in detail the trends in population and households by ward.}

To summarize, in 1920, both average rent and household density were higher in the burned area, consistent with the first prediction of the model.
The land readjustment then restricted the amount of land available for residential use.
Accordingly, real rents rose in the 1920s, particularly in the burned area.
When the reconstruction plan was completed in 1930, the total floor area almost recovered to its pre-earthquake levels.
Despite this, the average unit rent remained higher in the burned area, and the population remained below pre-earthquake levels in 1930.
While total floor area exceeded its pre-earthquake level by 1935, the population in the burned area only approached, but did not exceed, its pre-earthquake level.

\subsubsection{Estimation Strategy}\label{sec:sec_ea_es}

The longitudinal structure in the ward-level data allows for an event-study analysis under a regression framework.
For ward $i$ ($\in \{1,..., N\}$) at period $t$ ($\in \{1,...,T\}$), I specify the model as follows:
	\begin{align}\label{model_es}
	z_{i, t} = \alpha + \sum_{\substack{j \in \{1,...,T\}\\ j \neq F-1}} \beta_{j} \tilde{D}_{i, t}^{j} + \nu_{i} + \lambda_{t} + e_{i, t},
	\end{align}
where $z_{i, t}$ is the dependent variable, $\nu_{i}$ is a ward fixed effect, $\lambda_{t}$ is a time fixed effect, and $e_{i,t}$ is a random error term.
The variable of my interest is the product terms $\tilde{D}_{i, t}^{j}= P_{i} \times I(t = j)$ for $j \in \{1,...,T\}$ and $j \neq F - 1$, where $P_{i}$ is a percentage of burned area and $F$ indicates the period when the earthquake occurs.\footnote{The results are materially similar to those from an alternative specification using a binary treatment variable for the ward where more than a quarter of the administrative area was burned. Online Appendix~\ref{sec:secd_r_ward_alt} summarizes all these results. This confirms that the use of continuous treatment intensity is not arbitrary in modeling.}
The estimated coefficients $\hat{\beta}_{j}$ for $j \in \{F,...,T\}$ suggest the average marginal effect of the disaster in each post-treatment period relative to a reference pre-treatment period ($F-1$).
On the other hand, the estimates for $j \in \{1,...,F-2\}$ serve as a placebo experiment for pre-treatment trends.

The timing of the disaster is exogenously determined.
Evidence indicates that the distribution of burned area is influenced by unpredictable wind patterns on the date of impact as well as the ground softness (Section~\ref{sec:sec_hb_gke}).
This improves the randomness of the assignments represented in $\tilde{D}_{i, t}^{j}$.
Under the systematic reconstruction plan, land readjustment was carried out in accordance with the rule governing the burned area, ensuring that the readjustment policy was homogeneous across the burned area (Section~\ref{sec:sec_hb_rec}).
This institutional homogeneity is a favorable empirical setting because Equation~\ref{model_es} aims to capture the evolution of the disaster's overall impacts, including the reconstruction policy response.

To address heteroskedasticity across cross-sectional units and serial correlation within each cluster, I employ a cluster-robust variance-covariance matrix estimator for statistical inference \citep{Arellano1987-sc}.

\subsubsection{Results}\label{sec:sec_ea_ward2}

The estimates are relative to the year prior to the earthquake, and I focus on their evolution over time.
Because the impact in the earthquake year was extremely large, the estimates from 1924 onward appear relatively small in the figures.

\subsubsection*{Residences}

\begin{figure}[htbp]
\centering
\captionsetup{justification=centering}
\subfloat[Impacts on Buildings ($/ \text{km}^{2}$)]{\label{fig:r_ward_buildr}\includegraphics[width=0.45\textwidth]{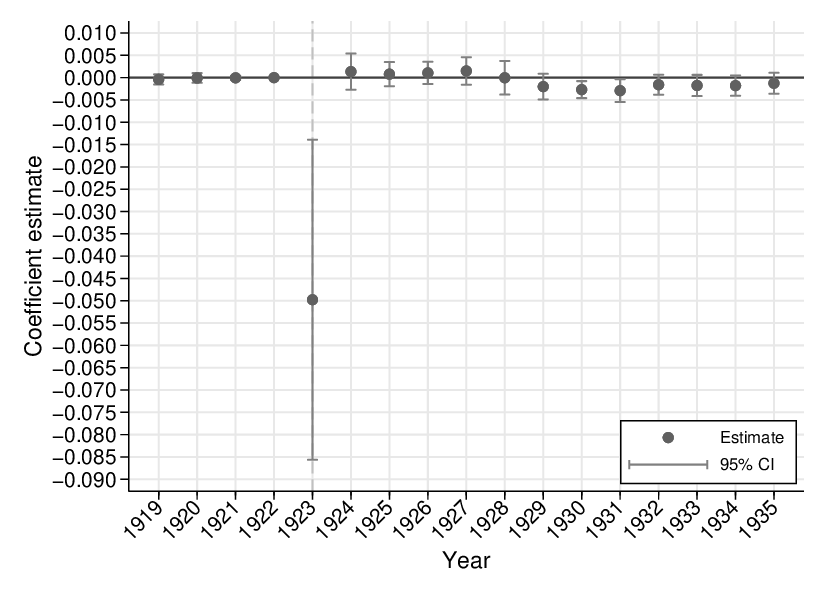}}
\subfloat[Impacts on Multi-story Dwellings ($/ \text{km}^{2}$)]{\label{fig:r_ward_mbuildr}\includegraphics[width=0.45\textwidth]{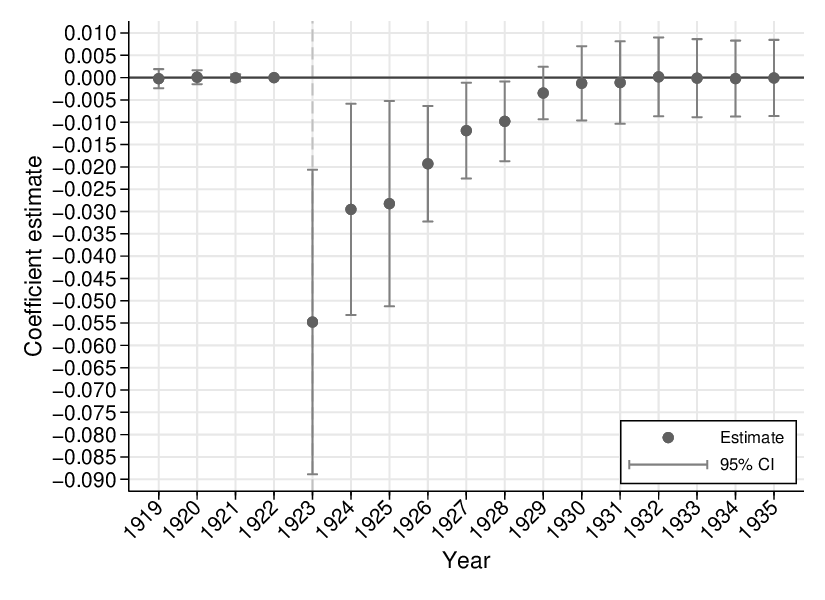}}\\
\subfloat[Impacts on Floor Area ($/ \text{km}^{2}$)]{\label{fig:r_ward_build_arear}\includegraphics[width=0.45\textwidth]
{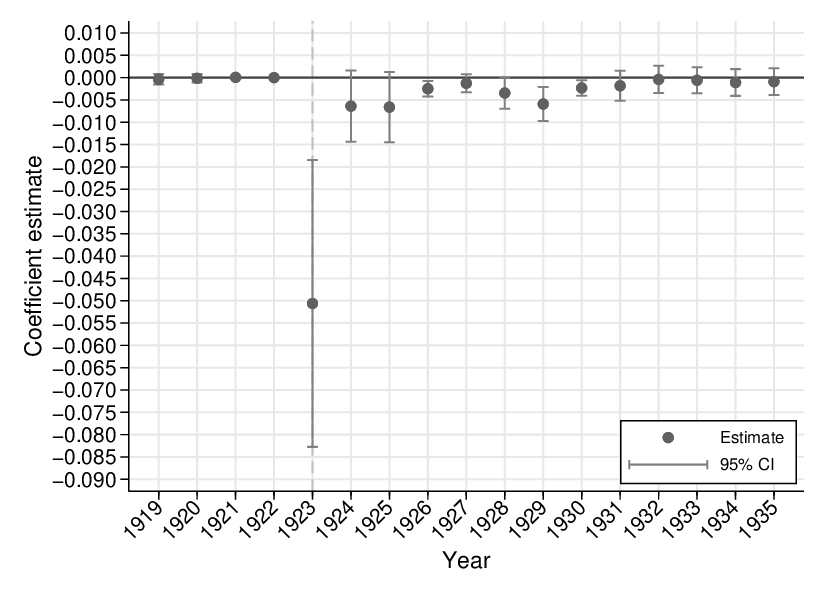}}
\subfloat[Impacts on Rent ($/ \text{m}^{2}$)]{\label{fig:r_ward_rentr}\includegraphics[width=0.45\textwidth]{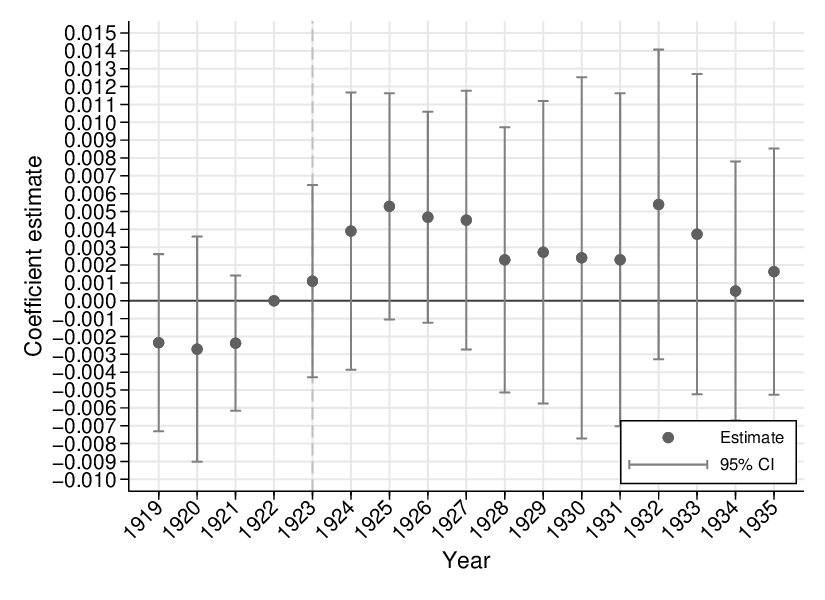}}\\
\subfloat[Impacts on Population]{\label{fig:r_ward_popr}\includegraphics[width=0.45\textwidth]
{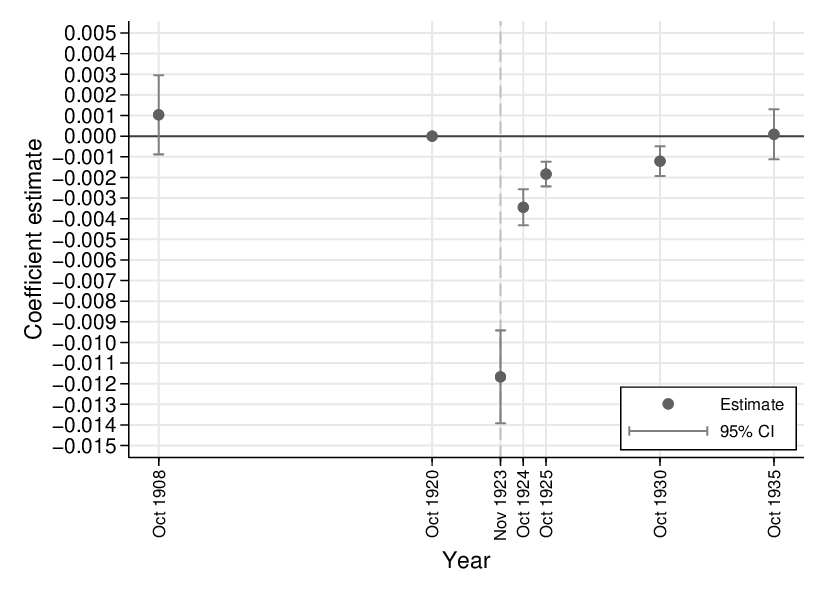}}
\subfloat[Impacts on Households]{\label{fig:r_ward_hhr}\includegraphics[width=0.45\textwidth]
{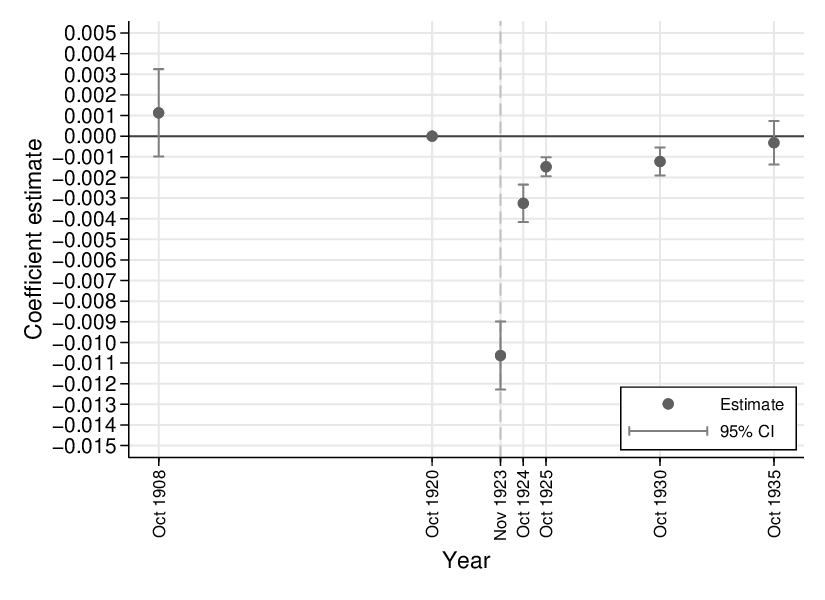}}
\caption{Impacts on the Residences, Rents, and Population:\\ Evidence from the Ward-level Statistics}
\label{fig:r_did_rate}
\scriptsize{\begin{minipage}{450pt}
\setstretch{0.9}
Notes:
This figure shows the estimated coefficients for the interaction terms between the percentage of burned area and the year dummies, as in Equation~\ref{model_es}.
Figure~\ref{fig:r_ward_buildr},~\ref{fig:r_ward_mbuildr}, and~\ref{fig:r_ward_build_arear} illustrate the results for the log-transformed number of wooden buildings (per $\text{km}^{2}$), multi-story wooden dwellings (per $\text{km}^{2}$), and floor area of the wooden buildings (per $\text{km}^{2}$), respectively.
Figure~\ref{fig:r_ward_rentr} shows the result for the log-transformed real rent per square meter.
All these regressions for buildings and rent are weighted by the total ward area, excluding the land for the Imperial Palace, where citizens are unable to live.
Figures~\ref{fig:r_ward_popr} and~\ref{fig:r_ward_hhr} show the results for the log-transformed population and households, respectively.
All the regressions include the ward- and year-fixed effects.
The 95\% confidence intervals are obtained from the cluster-robust variance-covariance matrix estimator using wards as the clustering unit.
The results from an alternative specification using a binary treatment variable are summarized in Online Appendix~\ref{sec:secd_r_ward_alt}.\\
Source: Created by the author using the ward-level census statistics listed in Panel A of Table~\ref{tab:sum}.
\end{minipage}}
\end{figure}

Figures \ref{fig:r_ward_buildr}, \ref{fig:r_ward_mbuildr}, and \ref{fig:r_ward_build_arear} show the results for the number of wooden buildings, the number of multi-story wooden dwellings, and the floor area of wooden buildings per unit area ($\text{km}^{2}$), respectively.
Figure~\ref{fig:r_ward_buildr} confirms that the earthquake caused a sharp decline in the number of dwellings.
The estimated figure for 1923 is extremely high, with a one-percentage-point increase in the burned-area share reducing the number of dwellings by approximately $5$\%.
The rapid construction of temporary housing is reflected in the estimate for the following year.
As land readjustment projects approached completion by the late 1920s, a decline was then observed as the transition from temporary to permanent housing.
Even after the reconstruction plan was completed in 1930, the estimated values remained negative, reflecting the reduction in residential land in the burned area resulting from land readjustment.
The estimated value for 1930 implies that a one-percentage-point increase in the burned-area share corresponds to a decrease of approximately $0.25$\% in the number of buildings.
In other words, the number of dwellings in completely burned wards, such as Nihonbashi and Asakusa, decreased by as much as $22$\%.\footnote{The estimates $\hat{\beta}_j$ represent the change in the log of the dependent variable associated with a one-percentage-point increase in the burned-area share, $P_i$. The percentage effect of a $\Delta P$-point increase is therefore $100 \times (e^{\hat{\beta}_j \Delta P} - 1)$. Throughout this paper, I evaluate the magnitudes for completely burned wards ($\Delta P = 100$) using the exact formula.}

Figure~\ref{fig:r_ward_mbuildr} shows that the estimates after the earthquake are negative because the temporary housing consisted of single-story dwellings.
When the construction of multi-story dwellings accelerated from the late 1920s, the estimates gradually approached zero.

Figure~\ref{fig:r_ward_build_arear} shows the result for the floor area.
Despite the replacement of existing structures with new buildings, the estimate for 1930 is negative.
It indicates that a one-percentage-point increase in the burned-area share resulted in a $2.5$\% decrease in residential floor area.
As multi-story dwellings increased in the early 1930s, the estimates approached zero but remained negative.
In other words, despite the increase in multi-story dwellings, the floor area in the burned area decreased.
This supports the evidence that the reduction in lot size within the burned area due to land readjustment persisted until at least around 1930.

There were technical limitations to building multi-story wooden houses.
In fact, although four-story wooden houses were constructed in the 1930s, they remained uncommon in the city.
Consequently, it was not possible to compensate for the structural reduction in living space resulting from the decrease in residential areas by increasing the floor area.
This means that the earthquake reduced housing supply in the city center, which, in the model, raises rents throughout the city.

\subsubsection*{Rents}

Figure~\ref{fig:r_ward_rentr} shows the result for the average unit rent (per m$^2$).
Immediately after the earthquake, the rental market was in disarray.
After land readjustment projects had begun, the outlook for residential areas improved in some areas by around 1925, which is reflected in the rise in the estimates in the mid-1920s.
The estimates in the late 1920s were approximately $0.0025$, implying that the average unit rent in the completely burned wards rose by about 28\% relative to the unburned area, although the confidence intervals are wide.\footnote{Pooling the post-earthquake years yields a statistically significant relative increase of a similar magnitude (Online Appendix~\ref{sec:secd_r_ward_rent}).}
After 1932, real rents in the burned area declined, and the estimates decreased, which may reflect three factors.
First, multi-story dwellings had increased faster in the burned area than in the unburned area by the early 1930s, leading to a recovery in housing supply.
Second, rents also rose in the unburned area in the early 1930s.
Third, the rent-reduction movement during the depression may have lowered rents more in the burned area, where rents had risen to relatively high levels after the earthquake \citep{Ono2006}.

As shown in Section~\ref{sec:sec_ea_ward1}, real rents in both the burned and unburned wards exceeded their pre-earthquake levels after the completion of the land readjustment.
Because the year fixed effects absorb this common increase, the event-study estimates capture only the differential change between the two areas, which the model does not pin down (Online Appendix~\ref{sec:rent_response}).
The relatively larger increase in the burned area in the late 1920s likely reflects features from which the model abstracts, such as changes in the quality and composition of the rebuilt housing stock and improvements in local amenities resulting from the land readjustment.\footnote{This interpretation is in line with studies on creative destruction, which show that disasters can replace old capital stock with new capital \citep{Heger2019-uq, Imaizumi2016-pc, Okazaki2019-sa}.}
The narrowing of the gap by the mid-1930s, as rents rose in the unburned area, is in line with the citywide increase predicted by the model.

\subsubsection*{Population and Households}

Figure~\ref{fig:r_ward_popr} shows the result for population.\footnote{Since the administrative ward area remains nearly constant over time, changes in density depend solely on changes in the numerator (population size). Therefore, the cross-sectional heterogeneity in area has little effect on the estimates from the within estimator. Online Appendix Figure~\ref{fig:r_ward_den} confirms that the results for population density are materially similar to those obtained for population. The same interpretation holds true in the within estimator for block-level data used in Section~\ref{sec:sec_ea_het}, where area data at the block level is unavailable.}
The impact for 1923 is particularly large ($-0.012$), suggesting that a $1$\% increase in the rate of destruction decreases population by roughly the same magnitude.
This means that the devastated area was uninhabitable immediately after the earthquake.
Even by October 1924, one year after the disaster, the magnitude remained relatively large.
In other words, although the construction of temporary housing had been completed and the majority of evacuees had returned to the city, the population in the burned area remained significantly reduced.
The impact of the earthquake also persisted; as of October 1930, seven years after the disaster, the estimated value was still $-0.0012$.
This indicates that even after the reconstruction plan was completed, the population in the completely burned area had decreased by roughly $11$\%.

In October 1935, 12 years after the earthquake, the estimate became nearly zero.
As examined, the population in the unburned area increased moderately after the earthquake.
The burned area was lined with many multi-story dwellings in the early 1930s.
Consequently, both the difference in average unit rent and the population gap between the burned and unburned areas narrowed around 1935.
Figure~\ref{fig:r_ward_hhr} shows the result for the number of households, which corresponds to the second prediction from the model (Section~\ref{sec:sec_tf}).
Unlike population, the estimate for the number of households remains negative in 1935, indicating that the number of households in the burned area had not returned to its pre-earthquake level.

For both population and households, the estimates for 1908 are not statistically significant.
This supports the finding that the burned area exhibits a demographic trend practically similar to that of the unburned area (Figures~\ref{fig:ts_ward_pop} and~\ref{fig:ts_ward_hh}).

\subsubsection*{Summary}

The land readjustment reduced the area of residential lots by $15$\%.
Real rents rose throughout the city, and more so in the burned area.
When the reconstruction plan was completed in 1930, the total floor area almost returned to the pre-earthquake level, thanks to the increase in multi-story dwellings.
However, relative to the unburned area, the rent in the completely burned wards was about 28\% higher and the population about 11\% lower than before the earthquake.
On the other hand, the population in the unburned area increased moderately.

By the early 1930s, the total floor area in the burned area exceeded pre-earthquake levels due to the continued increase in the number of multi-story dwellings.
The population in the burned area had almost recovered to pre-earthquake levels by 1935, although the number of households remained lower.
The average unit rent decreased moderately, but it remained above its pre-earthquake level.

In short, even after suffering a major shock, the population in the city center almost recovered within about $10$ years.
This supports the evidence that the disaster did not fundamentally alter the pre-earthquake CBD, which was determined by geographical characteristics (Section~\ref{sec:sec_hb_cbd}).
However, it is also noteworthy that the average unit rent remained higher, and the number of households in the burned area remained below its pre-earthquake level as late as 1935.
In other words, the recovery in total floor area did not offset the reduction in residential land resulting from the land readjustment.
Therefore, the persistent decline in the number of households in the burned area did not simply reflect the physical reduction in land area; it reflected the market adjustment to this reduction, through which rents rose throughout the city and households relocated toward the suburbs.

\subsection{Block-level Evidence on the Role of Zoning}\label{sec:sec_ea_het}

The results in Section~\ref{sec:sec_ea_ward} indicate that the number of households and the population in the burned area declined as the market adjusted to the reduction in residential land.
This suggests that market-driven adjustments in rents influenced population distribution within the city.
On the other hand, did land-use regulations have an effect?
As discussed in Section~\ref{sec:sec_hb_cbd}, the zoning system had not regulated housing supply.
However, the introduction of the zoning system may have promoted population growth by increasing labor demand in the industrial zone.
In this section, I use block-level data to distinguish between burned blocks and industrial blocks, thereby identifying the effects through this alternative channel.

\subsubsection{Estimation Strategy}\label{sec:sec_ea_het1}

To obtain benchmark estimates, I consider an estimation strategy that leverages block-level information on fire destruction (Section~\ref{sec:sec_data_census}).
For block $i$ ($\in \{1,...,N\}$) at period $t$ ($\in \{1,...,T\}$), I specify the regression model as follows:
	\begin{align}\label{model_es_block}
	z_{i, t} = \phi + \sum_{\substack{j \in \{1,...,T\}\\ j \neq F-1}} \zeta_{j} D_{i, t}^{j} + \theta_{i} + \kappa_{t} + \epsilon_{i, t},
	\end{align}
where $z_{i, t}$ is either population or household, $\theta_{i}$ is a block fixed effect, $\kappa_{t}$ is a time fixed effect, and $\epsilon_{i,t}$ is a random error term.\footnote{For the block-level sample with a three-dimensional panel data structure used in Figure~\ref{fig:r_did_block_bin}, I consider a regression for each household-size bin. Results for the three-dimensional panel dataset pooling all household-size bins are materially similar to those from my baseline results. Despite this, I prefer to use a separate regression for each bin to relax the strong assumption of fixed-effects homogeneity across bins.}
The variable of my interest is the product terms $D_{i, t}^{j}= B_{i} \times I(t = j)$ for $j \in \{1,...,T\}$ and $j \neq F - 1$, where $B_{i}$ is an indicator variable for the burned blocks, and $F$ indicates the earthquake year.\footnote{The timing of the treatment is constant across cross-sectional units. This means that the heterogeneity in the disaster's impacts, arising from differences in treatment timing, is absent in this setting. In other words, the estimator that accounts for heterogeneity in the effects \citep[e.g.,][]{Sun2021-mt} yields the same estimates as those from the standard within-estimator for Equation~\ref{model_es_block}.}
As described in Section~\ref{sec:sec_ea_es}, the distribution of devastation was driven by both seismic intensity and wind patterns, making the assignments plausibly exogenous.
The land readjustment was carried out homogeneously in the burned blocks under a rigorous redistribution rule.
This historical fact supports the evidence for the systematic treatment on the burned blocks.

I also consider an expanded regression of Equation~\ref{model_es_block} to analyze potential treatment heterogeneity with respect to land-use regulation.
For block $i$ at period $t$, I consider the following regression:
	\begin{align}\label{model_es_het}
	z_{i, t} = \varphi + \sum_{\substack{j \in \{1,...,T\}\\ j \neq F-1}} \eta_{j}^{D} D_{i, t}^{j} 
	+ \sum_{\substack{j \in \{1,...,T\}\\ j \neq F-1}} \eta_{j}^{W} W_{i, t}^{j}
	+ \sum_{\substack{j \in \{1,...,T\}\\ j \neq F-1}} \eta_{j}^{X} X_{i, t}^{j}
	+ \vartheta_{i} + \varkappa_{t} + \varepsilon_{i, t},
	\end{align}
where $\vartheta_{i}$ is a block fixed effect, $\varkappa_{t}$ is a time fixed effect, and $\varepsilon_{i,t}$ is a random error term.
$W_{i, t}^{j}= Z_{i} \times I(t = j)$ for $j \in \{1,...,T\}$ and $j \neq F - 1$, where $Z_{i}$ is an indicator variable for the industrial blocks set in the zoning system.
Importantly, this zoning was based on the initial geographical distribution prior to the earthquake and was rarely altered by the devastation, so it showed no endogenous correlation with the burned blocks (Section~\ref{sec:sec_hb_cbd}).
The intersection between the burned and industrial blocks is considered in $X_{i, t}^{j}= B_{i} \times Z_{i} \times I(t = j)$.
This regression does not hypothesize a specific effect of zoning on population in the post-disaster burned area and thus, is not estimated within a triple-difference-in-differences framework.
I rather aim to evaluate whether the marginal effects of the disaster were associated with the land regulation in the burned blocks in each post-disaster census year by 1930.\footnote{Despite this, the identification assumption for the triple DID setting plausibly holds as the pre-treatment trend in the relative outcome of industrial and non-industrial blocks is similar across burned and unburned blocks. The parallel trends assumption required for interpreting the triple difference-in-differences estimator as the average treatment effect on the treated is not identical to that for the standard difference-in-differences estimator. \citet{Olden2022-qy} provides a summary of this difference.}

I employ a cluster-robust variance-covariance matrix estimator for statistical inference to address heteroskedasticity across blocks and serial correlation within each cluster \citep{Arellano1987-sc}.

\subsubsection{Results}\label{sec:sec_ea_het2}

\begin{figure}[htbp]
\centering
\captionsetup{justification=centering}
\subfloat[Impacts on Population\\ in Burned Blocks]{\label{fig:r_block_be_pop}\includegraphics[width=0.45\textwidth]{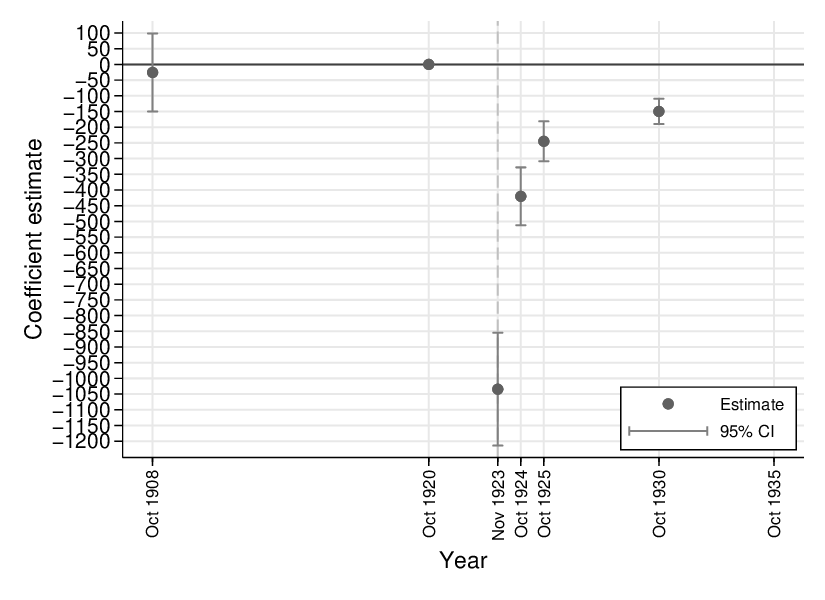}}
\subfloat[Impacts on Households\\ in Burned Blocks]{\label{fig:r_block_be_hh}\includegraphics[width=0.45\textwidth]{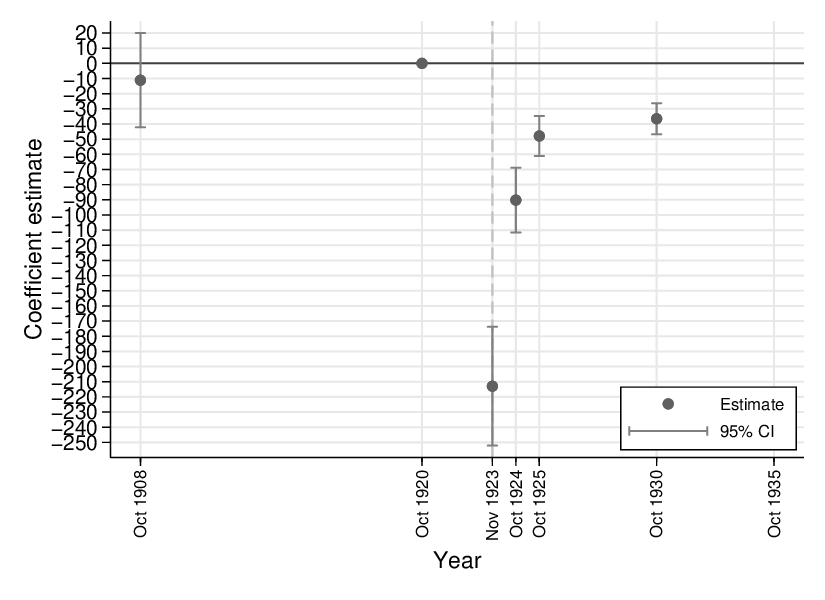}}\\
\subfloat[Impacts on Population\\ in Burned Blocks by Zones]{\label{fig:r_block_het_pop_burn}\includegraphics[width=0.45\textwidth]{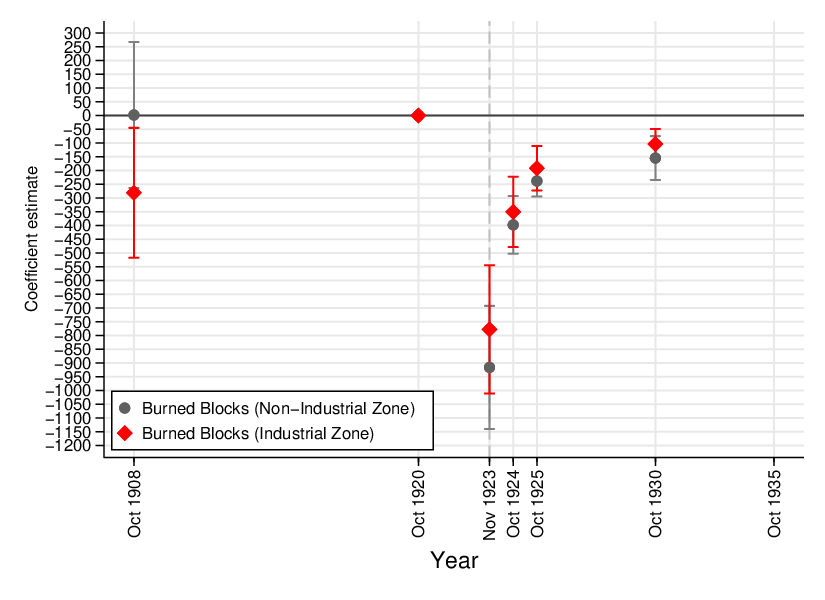}}
\subfloat[Impacts on Households\\ in Burned Blocks by Zones]{\label{fig:r_block_het_hh_burn}\includegraphics[width=0.45\textwidth]{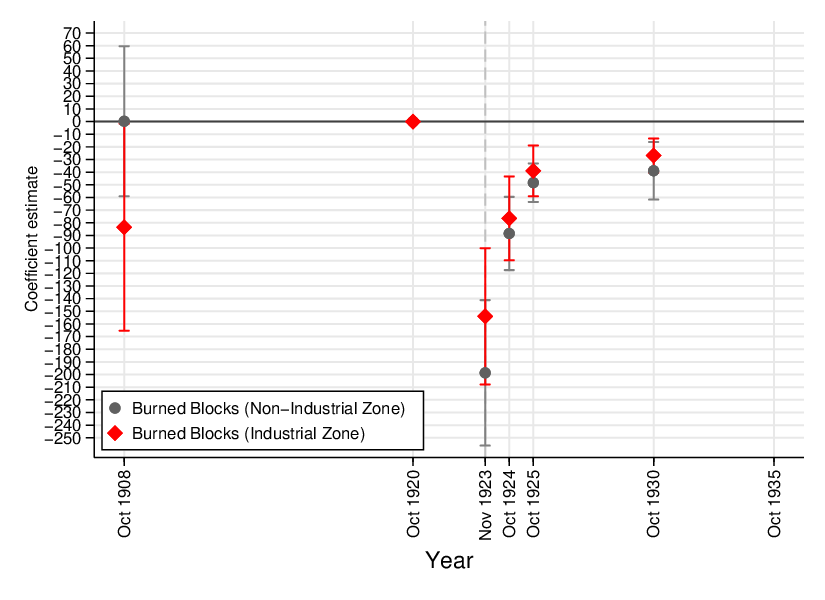}}\\
\caption{Results for the Population and Households: \\Evidence from the Block-level Census Statistics}
\label{fig:r_did_block_pop_hh}
\scriptsize{\begin{minipage}{450pt}
\setstretch{0.9}
Notes:
Figures~\ref{fig:r_block_be_pop} and~\ref{fig:r_block_be_hh} report the results for block-level population and households based on Equation~\ref{model_es_block}, respectively.
Both figures show the estimated coefficients of the interaction terms between the indicator variable for burned blocks and the year dummies.
The marginal effects of disaster calculated using the estimated coefficients for Equation~\ref{model_es_het} are shown in Figures~\ref{fig:r_block_het_pop_burn}--\ref{fig:r_block_het_hh_burn}.
The estimated effects on the non-industrial (industrial) blocks for each census year are illustrated in black (red) circles.
The industrial zone includes both commercial and manufacturing zones defined by the zoning system designed prior to the earthquake (Section~\ref{sec:sec_hb_cbd}).
All regressions include block and year fixed effects.
The $95$\% confidence intervals are shown in the black and red caps in each figure.
Standard errors in the regressions based on Equation~\ref{model_es_block} and~\ref{model_es_het} are clustered at the block-level.\\
Source: Created by the author using the block-level census statistics listed in Panel B-1 in Table~\ref{tab:sum}.
\end{minipage}}
\end{figure}

Figures~\ref{fig:r_block_be_pop} and~\ref{fig:r_block_be_hh} present the results for the population and households under Equation~\ref{model_es_block}, respectively.\footnote{Because block-level data covers small areas, it contains a small number of zero-censored observations. For example, the population data for 1908, 1920, and 1930 include 113, 43, and 57 censored blocks out of a total of $1,409$ blocks. Most of these are riverbank areas called (\textit{gashi}), which had few initial residential lands. These blocks are necessary to accurately depict population growth from 1908 to 1920 because the absolute population was relatively small in 1908. Despite this, the results are materially similar to my baseline results when these blocks are excluded (not reported).}
These results are similar to those obtained from ward-level data, supporting the robustness of my ward-level results.
The estimate for 1930 implies that the earthquake reduced the population of the burned blocks by approximately $150$ people.

The estimates in Figures~\ref{fig:r_block_be_pop} and~\ref{fig:r_block_be_hh} show the net effect of population decline resulting from land readjustment and the increase in potential labor demand resulting from land-use designation.
To disentangle both channels, Figures~\ref{fig:r_block_het_pop_burn} and ~\ref{fig:r_block_het_hh_burn} illustrate the suggested marginal effects of the disaster calculated from the estimates from Equation~\ref{model_es_het}.
Figure~\ref{fig:r_block_het_pop_burn} indicates that the estimated effects are nearly identical in industrial and non-industrial zones.
This holds even when the number of households is used as the dependent variable (Figure~\ref{fig:r_block_het_hh_burn}).
Although the marginal effect is slightly higher in the industrial zone than in the non-industrial zone, the difference is small and not statistically significant.\footnote{Online Appendix~\ref{sec:secd_r_block_het} further provides evidence that the marginal effect of zoning is statistically insignificant in most years, regardless of whether there was devastation or not.}

These results confirm that land regulation did not exert pressure on population growth through labor demand.
In this light, the population decline in the burned blocks can be attributed to the market adjustment to the reduction in residential land resulting from land readjustment, rather than to land-use regulation.
A similar conclusion is obtained when I use ward-level data on the total number of gainfully occupied persons from the 1920 and 1930 censuses instead of population.
The number of gainfully occupied persons decreased in the burned wards relative to the unburned wards, and this decrease did not differ significantly by the share of industrial blocks (Online Appendix~\ref{sec:secd_r_ward_het}).

\subsection{Margins of Adjustment}\label{sec:sec_ea_block}

The foregoing results show that the population in the burned area had almost recovered by 1935, whereas the number of households had not.
This section examines why.
The model treats each household as a single consumption unit.
It predicts the decline in the number of households and, through smaller housing per household, more households per unit of housing, but not changes in household size.
I first decompose the population change into these margins (Section~\ref{sec:sec_ea_decomp}).
I then examine each margin in turn: the number of households per dwelling through lodging (Section~\ref{sec:sec_ea_block2}) and average household size through the inclusion of non-kin members (Section~\ref{sec:sec_ea_block1}).

\subsubsection{Decomposing the Population Change}\label{sec:sec_ea_decomp}

The model treats each household as a single consumption unit and abstracts from multi-story dwellings (Section~\ref{sec:sec_tf}).
To gain an intuition for which margins of adjustment the model covers, I decompose the population density of each ward into three components:
\begin{align}\label{decomp}
\frac{\text{Pop}}{A}
= \frac{F}{A}
\times \frac{H}{F}
\times \frac{\text{Pop}}{H},
\end{align}
where $\text{Pop}$ is the population, $A$ is the ward area, $F$ is the residential floor area, and $H$ is the number of households.
The first component reflects both the reduction in residential land resulting from land readjustment, which corresponds to the decline in $\phi(m)$, and the construction of multi-story dwellings, which lies outside the model.
The second component is the empirical counterpart of the residential density in the model, $1/s(m, \hat{u}, y)$, measured per unit of floor area rather than land.
The third component, average household size, lies outside the model.
In terms of the extensive and intensive margins, the first two components together determine the number of households in each ward (the extensive margin), whereas the third component represents the intensive margin.
Because Equation~(\ref{decomp}) is an accounting identity, the log change in population density equals the sum of the log changes in the three components.
For the same reason, however, the decomposition is descriptive and does not identify the causal effects of each component.

\begin{table}[htbp]

\begin{center}
\captionsetup{justification=centering,margin=1.5cm}
\caption{Decomposition of Changes in Population Density: Evidence from the Ward-level Statistics 1920-1935}
\label{tab:decomp}
\scriptsize
\scalebox{0.95}[1]{
\setlength{\tabcolsep}{5pt}
\begin{tabular}{lccccccc}
\toprule[1pt]\midrule[0.3pt]
 & \multicolumn{3}{c}{1920--1930} & \multicolumn{3}{c}{1930--1935} & 1920--1935 \\
\cmidrule(lr){2-4}\cmidrule(lr){5-7}\cmidrule(l){8-8}
 & (1) Burned & (2) Unburned & (3) Difference & (4) Burned & (5) Unburned & (6) Difference & (7) Difference \\
\midrule
$\Delta \ln (\text{Pop}/A)$ & $-$0.076 & 0.021 & $-$0.097 & 0.112 & 0.000 & 0.112 & 0.015 \\
\midrule
$\Delta \ln (F/A)$          & 0.035    & 0.140 & $-$0.104 & 0.106 & 0.092 & 0.014 & $-$0.090 \\
$\Delta \ln (H/F)$          & $-$0.167 & $-$0.168 & 0.001 & $-$0.022 & $-$0.096 & 0.074 & 0.076 \\
$\Delta \ln (\text{Pop}/H)$ & 0.055    & 0.049 & 0.006    & 0.028 & 0.005 & 0.023 & 0.030 \\
\midrule[0.3pt]\bottomrule[1pt]
\end{tabular}
}
{\tiny
\begin{minipage}{440pt}   
\setstretch{0.85}
Notes: This table decomposes the log change in population density, $\text{Pop}/A$, into the log changes in floor area per ward area ($F/A$), the number of households per floor area ($H/F$), and average household size ($\text{Pop}/H$), so that the first row equals the sum of the remaining three rows (up to rounding).
All variables are summed over the wards in each group before computing the ratios.
Burned wards are those in which more than a quarter of the ward area was burned; the other wards are unburned wards.
Ward area excludes the Imperial Palace grounds and is fixed at its 1920 value.
``Difference'' is the change in the burned wards minus that in the unburned wards.
Column (7) equals the sum of Columns (3) and (6).
This decomposition is an accounting identity and does not identify causal effects.\\
Sources: Created by the author using the ward-level statistics listed in Panel A of Table~\ref{tab:sum}.
\end{minipage}
}
\end{center}
\end{table}

Table~\ref{tab:decomp} reports results based on the 1920, 1930, and 1935 ward-level census statistics.\footnote{The burned area in 1925 was still in the temporary housing period before the reconstruction plan was completed, which does not correspond to the residential equilibrium described by the model. Thus I exclude the 1925 data from the decomposition.}
Between 1920 and 1930, population density in the burned wards declined by $0.097$ log points relative to the unburned wards.
This relative decline is almost entirely accounted for by floor area: floor area per ward area grew by $0.104$ log points less in the burned wards, reflecting the reduction in residential land due to land readjustment.
This confirms that the construction of multi-story dwellings only partially offset the reduction in the floor area through the land readjustment.
The number of households per unit of floor area and average household size changed similarly in both groups.

Between 1930 and 1935, floor area changed little in relative terms ($0.014$), whereas population density recovered substantially ($0.112$).
This recovery reflected a more intensive use of the housing stock.
In fact, the number of households per floor area increased by $0.074$ log points relative to the unburned wards.
This is in line with the model's mechanism through which households consume less housing as rents rise.
Importantly, the average household size increased by $0.023$ log points, relative to the unburned wards.
This supports the evidence that the average household size in the areas affected by the fire has increased along with the rise in multi-story dwellings.

Over the entire period from 1920 to 1935 (Column 7 of Table~\ref{tab:decomp}), population density in the burned wards had almost returned to the level of the unburned wards ($0.015$), even though the relative shortfall in floor area persisted ($-0.090$).
This shortfall was more than offset by the increases in the number of households per floor area ($0.076$) and in average household size ($0.030$).
Of the combined contribution of these two components, about 70\% is accounted for by the number of households per floor area, which corresponds to the decrease in housing size in the model, and about 30\% by household size, which lies outside the model.

\subsubsection{Households per Dwelling: Lodging and Rent Sharing}\label{sec:sec_ea_block2}

\begin{figure}[htbp]
\centering
\resizebox{\textwidth}{!}{%
\begin{tikzpicture}[
  font=\footnotesize,
  hbox/.style={draw, rounded corners=1.5pt, inner sep=3pt, anchor=west, align=left},
  link/.style={draw, line width=0.6pt}
]
\def\ra{0}\def\rb{-1.35}\def\rc{-2.45}\def\rd{-3.85}\def\re{-4.95}\def\rf{-5.95}
\node[hbox] (H) at (0,\ra) {Ordinary\\household};
\node[hbox, right=4mm of H] (A) {[A] Kinship\\household};
\node[hbox, right=4mm of A] (A1) {[A-1] Without dependent cohabitants};
\node[hbox, right=4mm of A1] (A11) {[A-1-1] Non-shared dwelling};
\node[hbox,fill=exC!50] (A12) at (A11.west |- 0,\rb) {[A-1-2] Shared dwelling};
\node[hbox,fill=exC, right=4mm of A12] (A121) {[A-1-2-1] Landlord household\\(\textit{magashi setai})};
\node[hbox,fill=exC] (A122) at (A121.west |- 0,\rc) {[A-1-2-2] Lodger household\\(\textit{magari setai})};
\node[hbox,fill=inC] (B) at (A.west |- 0,\re) {[B] Kinship members\\+ servants/employees};
\node[hbox,fill=inC] (A2) at (A1.west |- 0,\rd) {[A-2] With dependent cohabitants (\textit{d\=okyonin})$^{\dagger}$};
\node[hbox,fill=inC, right=4mm of B] (B1) {[B-1] Without dependent cohabitants};
\node[hbox,fill=inC] (B2) at (B1.west |- 0,\rf) {[B-2] With dependent cohabitants (\textit{d\=okyonin})};
\foreach \p/\c in {H/A, A/A1, A1/A11, A12/A121, B/B1}
  \draw[link] (\p.east) -- (\c.west);
\foreach \p/\c in {H/B, A/A2, A1/A12, A12/A122, B/B2}
  \draw[link] ([xshift=6pt]\p.south west) |- (\c.west);
\node[draw, dashed, gray, rounded corners=3pt, inner sep=4pt, fit=(A121)(A122)] (surveybox) {};
\node[gray, anchor=north east, font=\scriptsize] at (surveybox.south east) {Covered by the SSDH (1930)};
\coordinate (mid) at (current bounding box.center);
\node[draw, rounded corners=2pt, inner sep=5pt, anchor=north]
  at ($(mid |- B2.south)+(0,-0.45)$) {\setlength{\fboxsep}{0pt}%
  \begin{tabular}{@{}c@{\hspace{1.5mm}}l@{\hspace{8mm}}c@{\hspace{1.5mm}}l@{}}
    \fcolorbox{black}{inC}{\rule{0pt}{3mm}\hspace{5mm}} &
    \begin{tabular}[c]{@{}l@{}}Non-kin members counted in the same household\\ ($\Rightarrow$ raises average household size)\end{tabular} &
    \fcolorbox{black}{exC}{\rule{0pt}{3mm}\hspace{5mm}} &
    \begin{tabular}[c]{@{}l@{}} Separate households sharing one dwelling\\ ($\Rightarrow$ raises the number of households per dwelling) \end{tabular}
  \end{tabular}};
\end{tikzpicture}}

\caption{Classification of the Households Measured in the Population Censuses}
\label{fig:type}

\scriptsize{\begin{minipage}{450pt}
\setstretch{0.85}
Notes: The census classifies households in two ways: ordinary households (futs\=u setai) and semi-households (jyun setai).
The former refers to the standard households; the latter includes individuals living in non-standard housing, such as jails, cheap lodging houses, and hospitals.
This study focuses on the former because the latter accounted for only about 1.3\% of total households in 1930 (Tokyo City Office 1932a, pp. 1--3).\\
Kinship households [A] (shinzoku setai) include [A-1] households without dependent cohabitants and [A-2] households with dependent cohabitants (d\=okyonin), who are financially dependent on the household and counted as its members.
Type [B] households include domestic servants and/or business employees (shiy\=onin), and [B-2] further includes dependent cohabitants.
[A-1] households are classified by living arrangement into [A-1-1] those in non-shared dwellings and [A-1-2] those in shared dwellings; the latter consist of [A-1-2-1] landlord households (magashi setai) and [A-1-2-2] lodger households (magari setai), which are financially independent and pay lodging rent for a portion of the landlord's dwelling.
Lodger households are counted as separate households.
Shaded boxes in blue indicate non-kin members counted within the same household; shaded boxes in orange indicate separate households sharing one dwelling.
The Survey of Shared-Dwelling Households (SSDH) used in Section~\ref{sec:sec_ea_block2} covers [A-1-2-1] and [A-1-2-2].
$\dagger$ Strictly, [A-2] households should not be classified as kinship households because dependent cohabitants are not necessarily kin; the 1930 census nonetheless classifies them as such to maintain consistency with the 1920 census classification.\\
Sources: Created by the author using Tokyo City Office (1932a; 1932b).
\end{minipage}}

\end{figure}
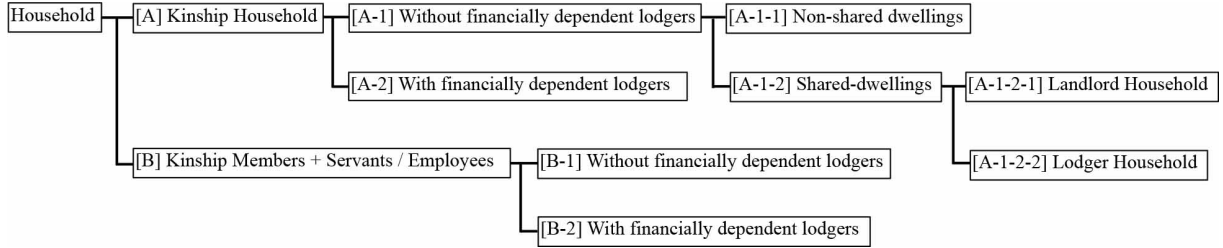

This section examines the first margin, the number of households per dwelling.
Lodger households (\textit{magari setai}) rented part of a landlord's dwelling and were counted as separate households ([A-1-2-2] in Figure~\ref{fig:type}).
An increase in lodging therefore raises the number of households per dwelling rather than average household size, and corresponds to the decrease in housing size per household in the model.
The SSDH introduced in Section~\ref{sec:sec_data_survey} covers all landlord and lodger households in 1930 ([A-1-2-1] and [A-1-2-2]) and records the rents paid by both, which allows me to examine the role of housing costs in this form of housing sharing.

\subsubsection*{Descriptive Analysis}

This survey identified $43,652$ landlord households (comprising $158,141$ individuals) and $55,665$ lodger households (comprising $98,784$ individuals), which corresponds to $24$\% ($/408,622$) of the ordinary households recorded in the 1930 census.\footnote{The descriptions in this section are from the Tokyo City Social Bureau (1930a).}
Since the average household size for these households was $2.6$ persons per household, the population sharing houses is smaller, accounting for $13$\% of the total population ($256,925$ out of $1,983,033$).\footnote{Note again that this survey does not focus on the households with servants/employees ([B] in Figure~\ref{fig:type}). The average household size is thus much smaller than that of the households with servants/employees.}

Of the $43,652$ landlord households, $35,768$ ($82$\%) were renters, while $7,884$ ($18.1$\%) owned their homes.
As for the type of buildings, $40,337$ ($92.4$\%) of the landlords' houses were two-story buildings.
The dominant arrangement was for the landlord to rent or own a two-story house and sublet the second floor to a lodger household.
$17,261$ ($39.5$\%) of the rental units were shop-house dwellings, which were relatively common in commercial districts such as Kanda, Shitaya, Asakusa, and Nihonbashi.
This likely represented a living arrangement in which workers lived with the shop owner (as an independent household).

The lodger's share of rent was 40--60\% of the total rent in most cases.
This implies that subletting was a practice of splitting the rent to cope with high rents in the city.

The sex ratio of landlord households was nearly balanced, with $78,669$ men and $79,472$ women.
Since it was rare for women to be the head of a household at that time, this indicates that landlord households were kinship households (consisting of a married couple or a married couple with children).
In contrast, the lodger households included $63,675$ men and $35,109$ women.
Regarding the composition of $55,665$ lodger households, $27,743$ ($49.8$\%) were single-person households, $18,069$ ($32.5$\%) were two-person households, $6,541$ ($11.8$\%) were three-person households, and $2,130$ ($3.8$\%) were four-person households.
Most single-person households were headed by men, while two-person households were predominantly married couples.
Therefore, it is reasonable to assume that most lodger households consisted of either single workers who had not yet formed families or married couples.

\subsubsection*{Estimation Strategy}

The descriptive analysis suggests that lodging was associated with workers' efforts to reduce their rent.
I assess this assumption using statistics on the landlord's and lodger's rents.
For block $i$, I specify a cross-sectional regression model as follows:
	\begin{align}\label{model_cs}
	l_{i} = \iota + \delta B_{i} + \psi_{g_{i}} + \upsilon_{i},
	\end{align}
where $l_{i}$ indicates dependent variable, $B_{i}$ is an indicator variable for the burned blocks introduced in Equation~\ref{model_es_block}, and $\upsilon_{i}$ is a random error term.
$l_{i}$ is either average monthly rent (or lodging rent) per unit space, average room space per lodger, average monthly lodging rent per lodger, or the ratio of rent and lodging rent.
I include ward fixed effect ($\psi_{g_{i}}$) to control for unobservable heterogeneity in ward-level responses to the disaster during the reconstruction period, which might be correlated with the exposure variable.
I employ the HC2 estimator to address heteroskedasticity across units \citep{Horn:1975tu}, albeit results are unchanged when I use the cluster-robust variance-covariance matrix estimator to allow for potential correlation within administrative wards.

\subsubsection*{Results}

\def\arraystretch{1.0}
\begin{table}[htbp]
\begin{center}
\captionsetup{justification=centering,margin=1.5cm}
\caption{Average Rents, Average Lodging Rents, and Average Living Space for Lodgers: Evidence from the Survey of Shared-Dwelling Households}
\label{tab:r_hs}
\scriptsize
\scalebox{0.88}[1]{
\begin{tabular}{lD{.}{.}{-2}D{.}{.}{-2}D{.}{.}{-2}D{.}{.}{-2}D{.}{.}{-2}}
\toprule[1pt]\midrule[0.3pt]
&\multicolumn{5}{c}{Dependent Variable}\\
\cmidrule(rrrrr){2-6}
&\multicolumn{2}{c}{Monthly Rent (yen / tatami)}
&\multicolumn{1}{c}{Tatami (mat)}
&\multicolumn{1}{c}{Monthly Rent (yen/ind.)}
&\multicolumn{1}{c}{(5) Rent Ratio}\\
\cmidrule(rr){2-3}
\cmidrule(r){4-4}
\cmidrule(r){5-5}
&\multicolumn{1}{c}{(1) Landlord}
&\multicolumn{1}{c}{(2) Lodger}
&\multicolumn{1}{c}{(3) Lodger}
&\multicolumn{1}{c}{(4) Lodger}
&\multicolumn{1}{c}{(Lodger / Landlord)}\\
\cmidrule(rrrrrr){1-6}
Burned Blocks				&0.334$***$	&0.184$***$	&-0.029	&0.767$***$	&-0.082$***$	\\
						&[0.030]		&[0.024]		&[0.074]	&[0.188]		&[0.015]		\\
						&(0.074)		&(0.056)		&(0.074)	&(0.188)		&(0.017)		\\\hline
Sample mean of the DV		&2.057		&2.131		&4.531	&9.670		&1.054		\\
Standard deviation of the DV	&0.390		&0.323		&1.021	&2.598		&0.160		\\
Ward FE 					&\multicolumn{1}{c}{Yes}		&\multicolumn{1}{c}{Yes}	&\multicolumn{1}{c}{Yes}	&\multicolumn{1}{c}{Yes}	&\multicolumn{1}{c}{Yes}\\
$F$-statistic $p$-value for	&\multicolumn{1}{c}{0.000}	&\multicolumn{1}{c}{0.000}&\multicolumn{1}{c}{0.000}&\multicolumn{1}{c}{0.000}&\multicolumn{1}{c}{0.000}\\
the Ward FE (zero slope)	&\multicolumn{1}{c}{}	&\multicolumn{1}{c}{}&\multicolumn{1}{c}{}&\multicolumn{1}{c}{}&\multicolumn{1}{c}{}\\
Observations				&\multicolumn{1}{c}{1096}		&\multicolumn{1}{c}{1096}	&\multicolumn{1}{c}{1092}	&\multicolumn{1}{c}{1092}&\multicolumn{1}{c}{1096}\\\midrule[0.3pt]\bottomrule[1pt]
\end{tabular}
}
{\scriptsize
\begin{minipage}{445pt}   
\setstretch{0.85}
***, **, and * denote statistical significance at the 1\%, 5\%, and 10\% levels, respectively.
The standard errors listed in the brackets are based on the HC2 estimator proposed by \citet{Horn:1975tu}.
The standard errors listed in the parentheses are based on the cluster-robust estimator proposed by \citet{Arellano1987-sc}.
$F$-statistics $p$-values for the null of the zero-slope hypothesis for the ward fixed effects are reported in each column.\\
Notes:
This table shows the results for the average unit rent paid by the landlord (Column (1)), average lodging unit rent paid by the lodgers (Column (2)), average number of \textit{tatami} mats per lodger (Column (3)), average monthly lodging rent per lodger (Column (4)), and the ratio between the rent and lodging rent (Column (5)).
The average rent and lodging rent are the price per unit area, \textit{tatami} mat (yen per $1.54 \text{m}^{2}$).\\
Source: Created by the author using the block-level complete survey statistics listed in Panel B-3 of Table~\ref{tab:sum}.
\end{minipage}
}
\end{center}
\end{table}

Column (1) shows that the average monthly unit rent charged by landlords in the burned blocks is approximately $0.3$ yen higher than that in the unburned blocks.
This corresponds to $86\%$ ($0.334/0.390$) of the standard deviation of this variable.
Since differences in average rent across wards are held constant by fixed effects, rents in the burned blocks were higher than those in the unburned blocks within the same ward.
This cross-sectional difference reflects both the pre-earthquake rent gradient (Prediction 1) and the post-earthquake rise in rents documented in Section~\ref{sec:sec_ea_ward}.
Column (2) shows the results for the average monthly unit lodging rents.
As with rent, the per-unit rent in burned blocks is $0.182$ yen higher than in unburned blocks.
This estimate corresponds to $57\%$ of the standard deviation ($0.184/0.323$).
Column (3) presents the result for the average number of \textit{tatami} mats per lodger.
The estimated coefficient is close to zero, indicating no significant difference in per-capita living area across lodger households in burned versus unburned blocks.
Consequently, the average lodging rents per lodger in the burned blocks are significantly higher than those in the unburned blocks.
The estimate in Column (4) is $0.767$, which is approximately $30\%$ of the standard deviation ($0.767/2.598$).

\begin{figure}[htbp]
\centering
\captionsetup{justification=centering}
\subfloat[Burned Blocks]{\label{fig:kden_rent_b}\includegraphics[width=0.45\textwidth]{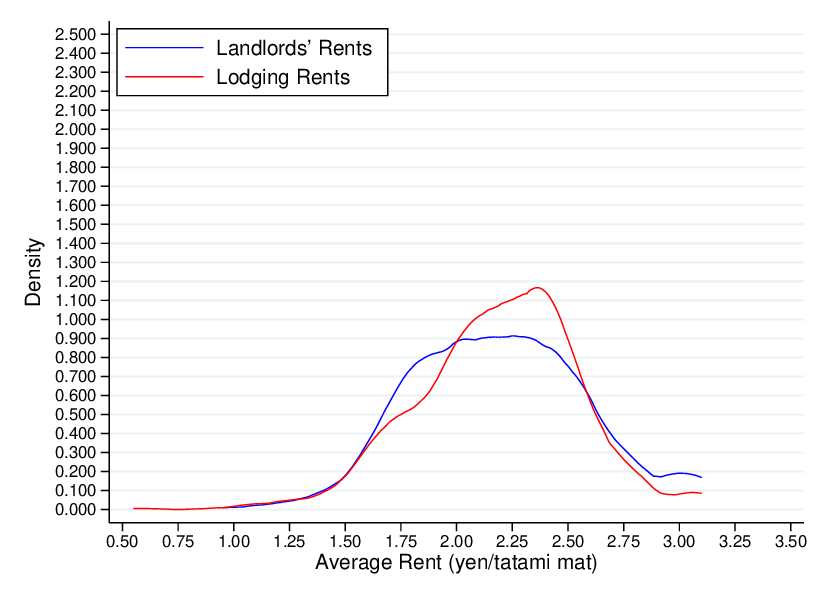}}
\subfloat[Unburned Blocks]{\label{fig:kden_rent_ub}\includegraphics[width=0.45\textwidth]{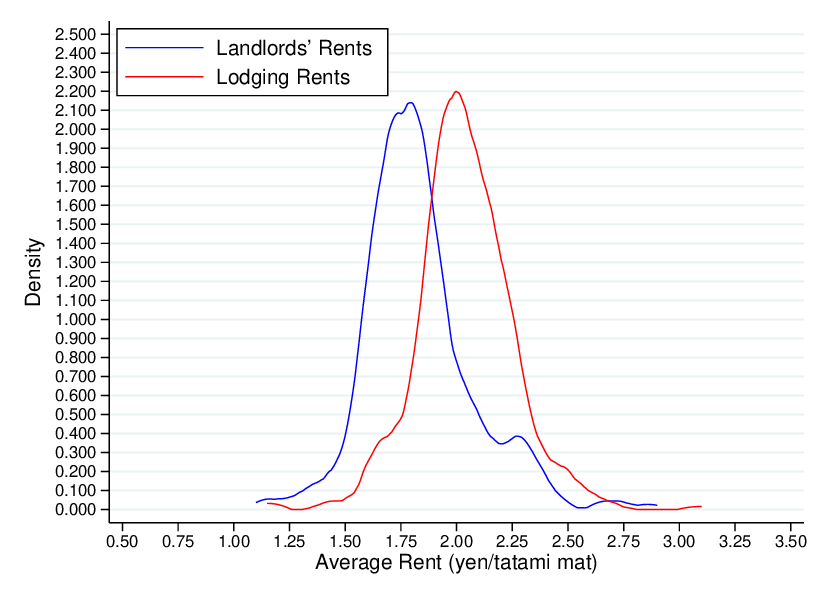}}\\
\caption{Density Estimates for the Average Unit Rents and Unit Rental Fees by Burned and Unburned Blocks}
\label{fig:kden_rent}
\scriptsize{\begin{minipage}{450pt}
\setstretch{0.9}
Notes:
Figures~\ref{fig:kden_rent_b} and~\ref{fig:kden_rent_ub} illustrate the estimated density of the average unit rents for landlords (blue) and lodging unit rents (red), respectively.
The average unit rents and lodging unit rents are in yen per unit area, \textit{tatami} mat ($1.54 \text{m}^{2}$).\\
Source: Created by the author using the block-level complete survey statistics listed in Panel B-3 of Table~\ref{tab:sum}.
\end{minipage}}
\end{figure}

Column (5) presents the results for the ratio of unit rents to lodging rents.
The estimate is $-0.082$ and statistically significant, indicating that unit payments are similar in both households in the burned blocks.
In fact, the ratio in the burned blocks is $1.01$, whereas it is higher at $1.13$ in the unburned blocks (Panel B-3 in Table~\ref{tab:sum}).
Figure~\ref{fig:kden_rent} illustrates the estimated density of unit rent in burned and unburned blocks.
While lodgers in unburned blocks pay relatively higher rents, the distributions of rent and lodging rent in burned blocks are fairly similar.
This means that landlords in the burned blocks bore a relatively larger share of the costs than those in the unburned area.

Although investigating the exact reason for this balanced rent payment in the burned blocks is beyond the scope of this study, it may reflect a cap on affordable rents among lodgers in the city.
My results imply that lodger households could live in the CBD without bearing a larger share of the rent than landlords, and paid roughly the same rent per lodger as in the unburned area.
Online Appendix Table~\ref{tab:r_rob_hs} finds that the average size of lodger households in the burned area was slightly larger than in the unburned area, but there was no statistically significant difference.
This implies that the lodger households did not share a single room with many people from other families, but used it with their partner.
These findings offer evidence on the potential mechanism behind the historical interpretation that urban low-income working-class people were able to form households during this period \citep{Nakagawa1985}.

The results are in line with the increase in the number of households per floor area in the burned wards.
Because the SSDH is a cross-section in 1930, however, they show differences in levels between burned and unburned blocks rather than changes over time.

\subsubsection{Household Size: Non-kin Members}\label{sec:sec_ea_block1}

This section examines the second margin, average household size.
Unlike lodgers, domestic servants, business employees, and dependent cohabitants were counted as members of the household with which they lived ([A-2] and [B] in Figure~\ref{fig:type}).
Their inclusion therefore raises average household size, a margin that lies outside the model.
Because servants and employees received housing and meals in kind as part of their employment, the cost of this form of housing sharing is not directly observable, unlike lodging rents.

\begin{figure}[htbp]
\centering
\captionsetup{justification=centering}
\subfloat[Burned Wards]{\label{fig:bar_pop_b}\includegraphics[width=0.5\textwidth]{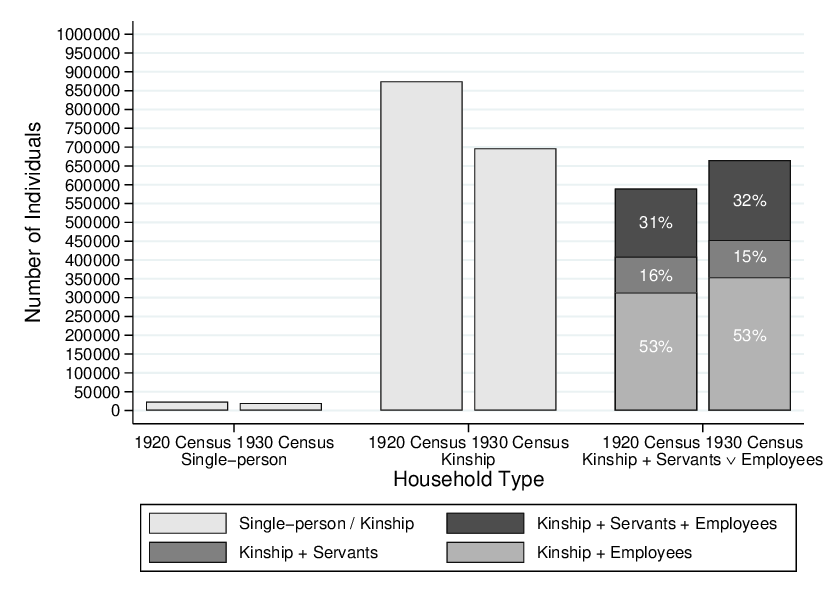}}
\subfloat[Unburned Wards]{\label{fig:bar_pop_ub}\includegraphics[width=0.5\textwidth]{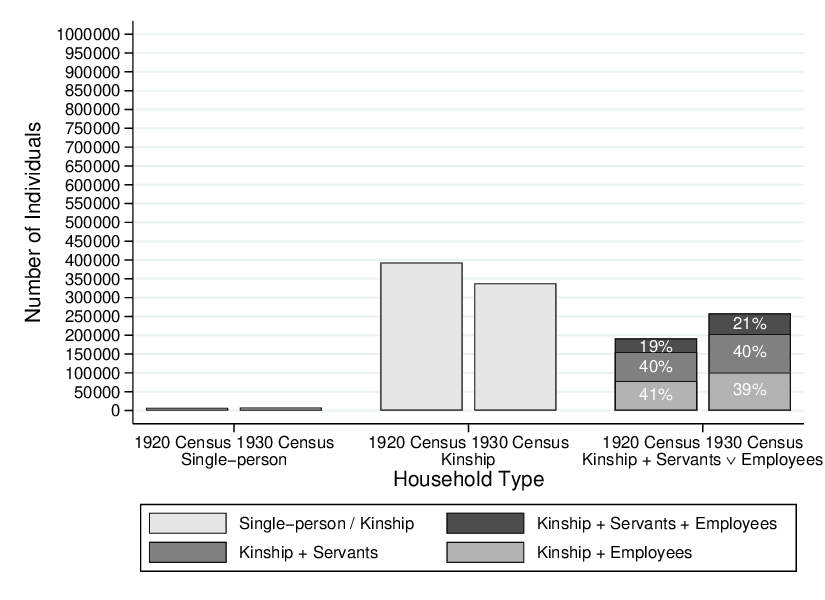}}\\
\caption{Number of Individuals by Type of Households and Burned/Unburned Wards: Evidence from the 1920 and 1930 Population Censuses}
\label{fig:bar_census}
\scriptsize{\begin{minipage}{450pt}
\setstretch{0.9}
Notes:
Figures~\ref{fig:bar_pop_b} and~\ref{fig:bar_pop_ub} illustrate the number of individuals by type of household in burned and unburned wards measured in the 1920 and 1930 censuses, respectively.
`Single-person' and `Kinship' indicate households with one person and with kin, respectively.
Both categories are included in the [A] household type in Figure~\ref{fig:type}.
`Kinship + Servants v Employees' indicates households with kin and domestic servants (\textit{kaji shiy\=onin}) and/or business employees (\textit{shokugy\=o shiy\=onin}).
This category corresponds to the [B] household type in Figure~\ref{fig:type}.
The share of each of the three types (Kinship + Servants + Employees; Kinship + Servants; Kinship + Employees) is shown as a percentage point.\\
Sources: Created by the author using Tokyo City Office (1922a; 1932a).
\end{minipage}}
\end{figure}

Figure~\ref{fig:bar_census} summarizes the changes in the household types by burned and unburned wards.
First, the number of individuals in kin-based households decreased in both wards.\footnote{Single-person households (included in [A]) decreased in the burned area and increased slightly in the unburned wards. This may reflect the rapid rise in rents in the city center, suggesting that some single-person households have moved to the outskirts, where rents are relatively lower.}
Second, the number of individuals in households with servants and/or employees increased in both wards.
In the burned wards, the proportion of individuals in households with business employees is high because the area contains representative industrial zones (Figure~\ref{fig:bar_pop_b}).
In the unburned wards, the proportion of individuals in households with domestic servants is high, as it includes many residential zones (Figure~\ref{fig:bar_pop_ub}).

\begin{figure}[htbp]
\centering
\captionsetup{justification=centering}
\subfloat[Burned Wards]{\label{fig:bar_pop_b_1930}\includegraphics[width=0.5\textwidth]{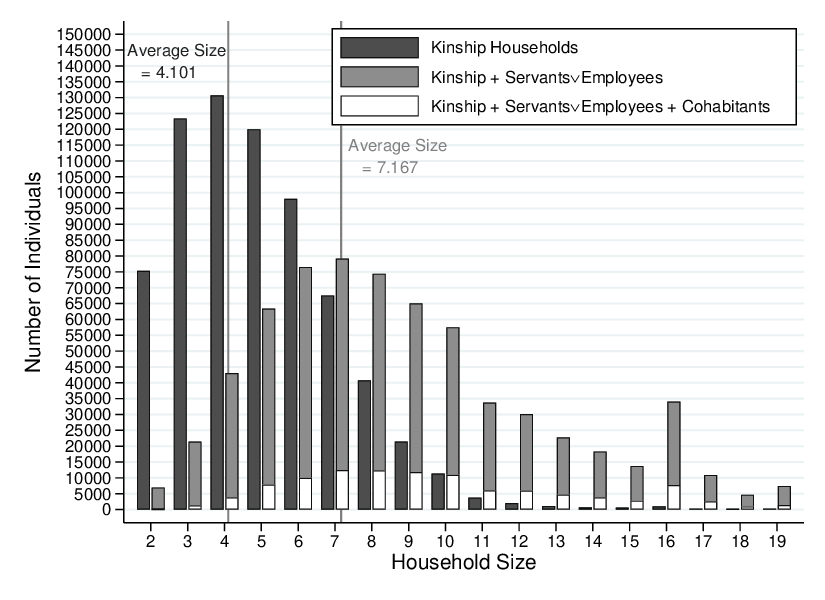}}
\subfloat[Unburned Wards]{\label{fig:bar_pop_ub_1930}\includegraphics[width=0.5\textwidth]{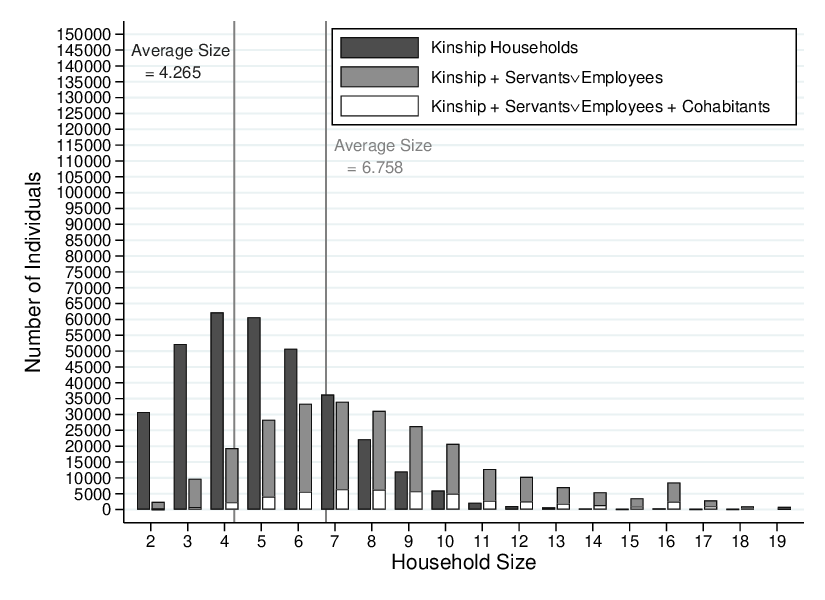}}\\
\caption{Number of Individuals by Type of Households and Burned/Unburned Wards: Evidence from the 1930 Population Census}
\label{fig:bar_census_1930}
\scriptsize{\begin{minipage}{450pt}
\setstretch{0.9}
Notes:
Figures~\ref{fig:bar_pop_b_1930} and~\ref{fig:bar_pop_ub_1930} illustrate the number of individuals by type of household and household size in burned and unburned wards in 1930, respectively.
`Kinship Households' (black) indicates that households include only kin.
This category corresponds to the [A] household type in Figure~\ref{fig:type}.
`Kinship + Servants v Employees' (gray) includes households that include kin, and domestic servants (\textit{kaji shiy\=onin}) and/or business employees (\textit{shokugy\=o shiy\=onin}).
This indicates the [B-1] household type shown in Figure~\ref{fig:type}.
`Kinship + Servants v Employees + Cohabitants' (white) includes households that include kin, domestic servants and/or business employees, and dependent cohabitants.
This indicates the [B-2] household type shown in Figure~\ref{fig:type}.
The average household size for the family households category is shown in dark gray, whereas that for the other category is shown in light gray.\\
Source: Created by the author using the Tokyo City Office (1932b).
\end{minipage}}
\end{figure}

The 1930 census report includes statistics on the number of individuals by household types and sizes.
Figure~\ref{fig:bar_pop_b_1930} illustrates that the distribution of kinship households ([A]) is skewed further to the right than that of households including servants/employees ([B]).
The average size for households including servants is approximately $7.2$ persons, which is about $3.1$ persons ($7.167$--$4.101$) more than that of kinship households.
A similar distribution is observed in the unburned wards (Figure~\ref{fig:bar_pop_ub_1930}).
This figure also highlights that households including dependent cohabitants ([B-2]) account for roughly $15\%$ of all households with servants/employees ($13.9\%$ and $16.8\%$ in the burned and unburned wards, respectively).

Against the backdrop of rising rents, while the number of independent kinship households decreased, the number of households with servants/employees and/or dependent cohabitants increased.
This complexification of household composition may be associated with the average household size.\footnote{
The Statistics Bureau offers an interesting observation: ``It should be recognized that the deepening economic and social recession over the past decade or so has made it difficult to establish and maintain independent households; on the one hand, this has prevented households that should have become independent from separating, and on the other hand, it has encouraged the merging of households that had already become independent... With regard to the qualitative structure of households, this is manifested in the widespread prevalence of mixed households'' (Tokyo City Office 1932a, pp.~10--11). The term ``mixed households'' here refers to households in which nonrelatives live with a kinship household. The increase in cases where people chose to live together for economic reasons is consistent with the rise in the (deflated) rent series following the disaster (Figure~\ref{fig:ts_ward_mrent}).}

\begin{figure}[htbp]
\centering
\captionsetup{justification=centering}
\subfloat[Population\\ in Burned and Unburned Blocks]{\label{fig:ts_block_pop_3b}\includegraphics[width=0.35\textwidth]{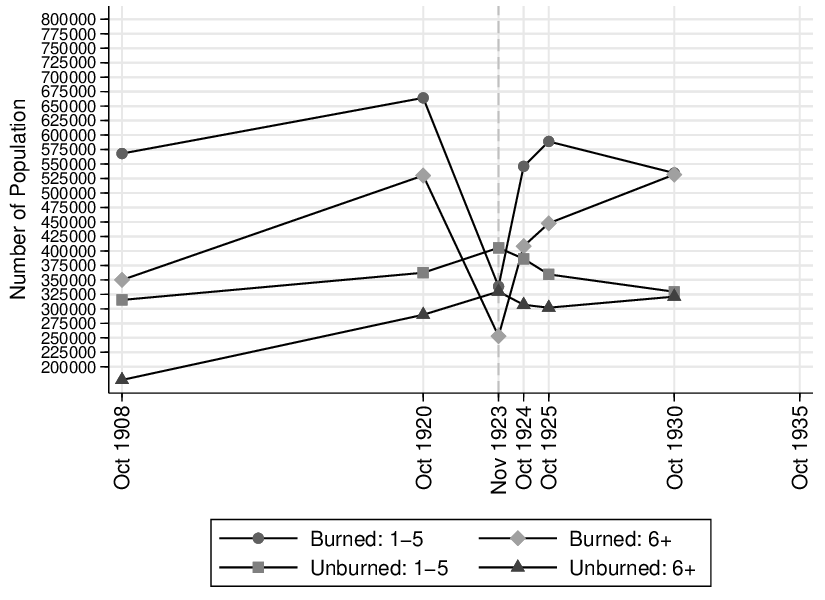}}
\subfloat[Households\\ in Burned and Unburned Blocks]{\label{fig:ts_block_hh_3b}\includegraphics[width=0.35\textwidth]{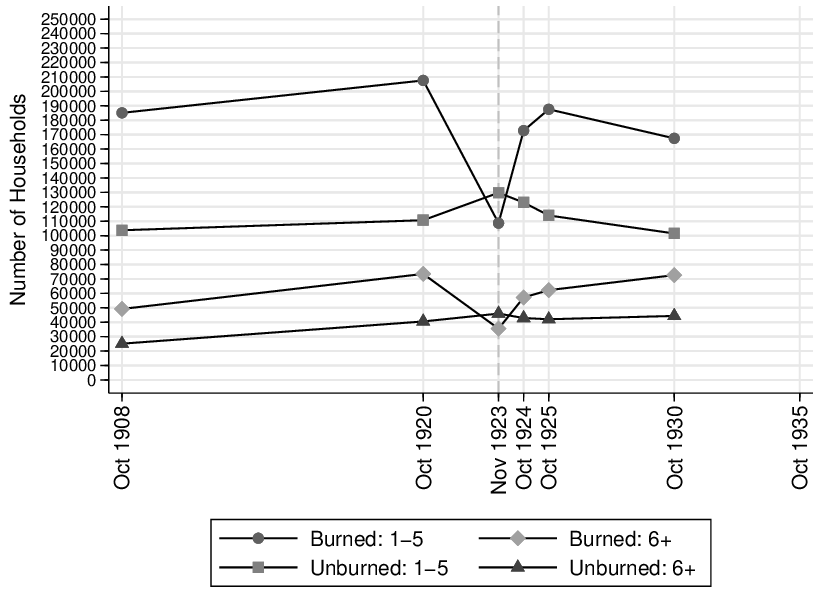}}
\subfloat[Household Size\\ in Burned and Unburned Blocks]{\label{fig:ts_block_size}\includegraphics[width=0.35\textwidth]{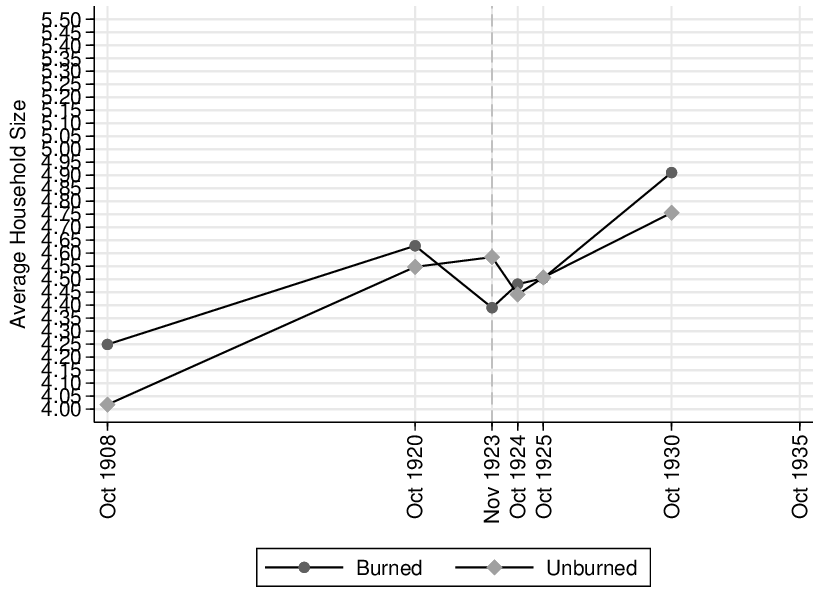}}
\caption{Population, Households, and Household Size\\ in the Burned and Unburned Blocks}
\label{fig:ts_block}
\scriptsize{\begin{minipage}{450pt}
\setstretch{0.9}
Notes:
Figures~\ref{fig:ts_block_pop_3b} and~\ref{fig:ts_block_hh_3b} depict the number of people and households by household size in the burned and unburned blocks between 1920 and 1930.
The household size is categorized into two groups using the 1920 mean household size, approximately $4.8$, as the threshold.
`1--5' and `6+' indicate the categories for the households with 1-5 persons and 6 or more persons, respectively.
Figure~\ref{fig:ts_block_size} shows the average household size in the burned and unburned blocks.
The dashed gray line indicates the earthquake year, 1923.\\
Sources: Created by the author using the block-level and block-size level census statistics listed in Panels B-1 and B-2 in Table~\ref{tab:sum}
\end{minipage}}
\end{figure}

Figures~\ref{fig:ts_block_pop_3b} and~\ref{fig:ts_block_hh_3b} show the changes in population and the number of households, broken down into small households (1--5 members) and large households (6+ members).\footnote{To avoid complexity, I have grouped household size into two categories. The interpretation would remain unchanged, even if households were divided into more detailed categories. A detailed figure showing the distribution of households across 10 categories is presented in Online Appendix Figure~\ref{fig:ts_block_bin}.}
Between 1925 and 1930, the number of small households decreased in the burned blocks, while the number of large households increased.
Figure ~\ref{fig:ts_block_size} shows that this led to an increase in the average household size in the burned blocks.
In 1920, the average household size was $4.5$ people in both burned and unburned blocks.
However, it had increased to about $5$ people per household in the burned blocks, exceeding that in the unburned blocks in 1930.

\begin{figure}[htbp]
\centering
\captionsetup{justification=centering}
\subfloat[Impacts on Household Size\\ in Burned Blocks]{\label{fig:r_block_size}\includegraphics[width=0.45\textwidth]{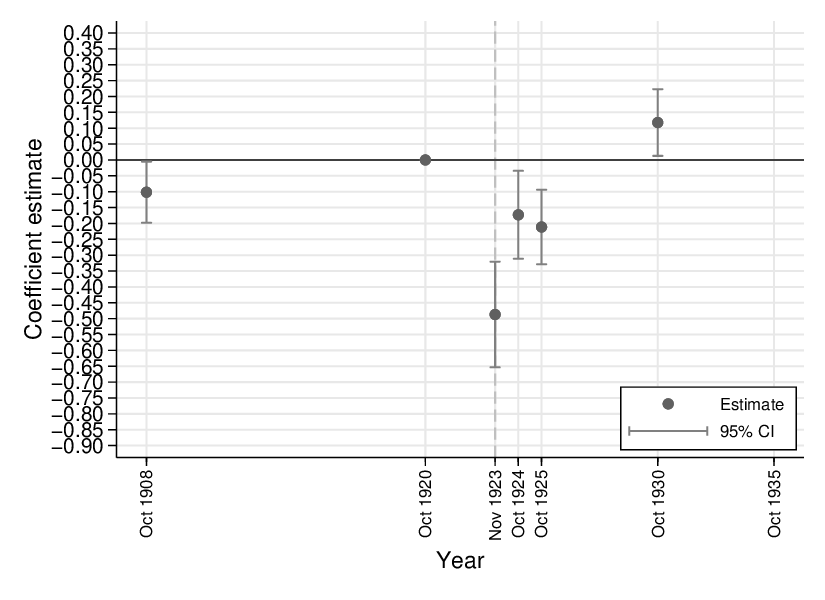}}
\subfloat[Impacts on Household Size\\ in Industrial and Non-Industrial Zones]{\label{fig:r_block_het_size_burn}\includegraphics[width=0.45\textwidth]{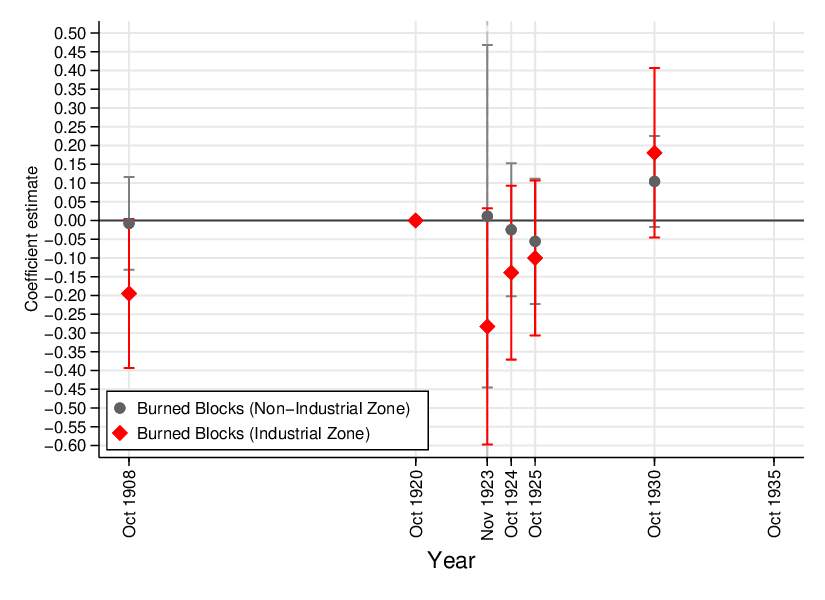}}
\caption{Results for the Average Household Size:\\ Evidence from the Block-level Census Statistics}
\label{fig:r_did_block}
\scriptsize{\begin{minipage}{450pt}
\setstretch{0.9}
Notes:
Figure~\ref{fig:r_block_size} shows the result for the average household size based on the regression of Equation~\ref{model_es_block}.
The average household size is the number of individuals per household.
Figure~\ref{fig:r_block_het_size_burn} shows the estimated marginal effects of disaster for the average household size under Equation~\ref{model_es_het}.
The estimated effects on the non-industrial (industrial) zone are illustrated in black (red) circles.
The `Industrial Zone' includes both commercial and manufacturing zones defined by the zoning system designed prior to the earthquake (Section~\ref{sec:sec_hb_cbd}).
The 95\% confidence intervals are shown in the black and red caps in each figure.
Standard errors in the regressions based on Equation~\ref{model_es_block} and~\ref{model_es_het} are clustered at the block-level.\\
Sources: Created by the author using the block-level census statistics listed in Panel B-1 of Table~\ref{tab:sum}.
\end{minipage}}
\end{figure}

Figure~\ref{fig:r_block_size} shows the results for the average household size under Equation~\ref{model_es_block}.
The estimate is $-0.5$ persons for 1923 and $-0.2$ persons for 1924--1925.
The estimate becomes positive in 1930, when the land area decreased and multi-story dwellings increased.
This implies that there were more cases in which many individuals lived in a single multi-story dwelling.

Figure~\ref{fig:r_block_het_size_burn} shows the marginal effects of disaster calculated for industrial and non-industrial zones based on the estimates from Equation~\ref{model_es_het}.
While regional differences in the estimated effects are not statistically significant, a trend toward an increase in average household size is observed in both industrial and non-industrial zones in 1930.
This is consistent with my baseline results for average household size (Figure~\ref{fig:r_block_size}), meaning that the increase in average household size in the burned blocks occurred independently of the zoning system.

\begin{figure}[htbp]
\centering
\captionsetup{justification=centering}
\subfloat[Burned Blocks]{\label{fig:kden_size_b}\includegraphics[width=0.45\textwidth]{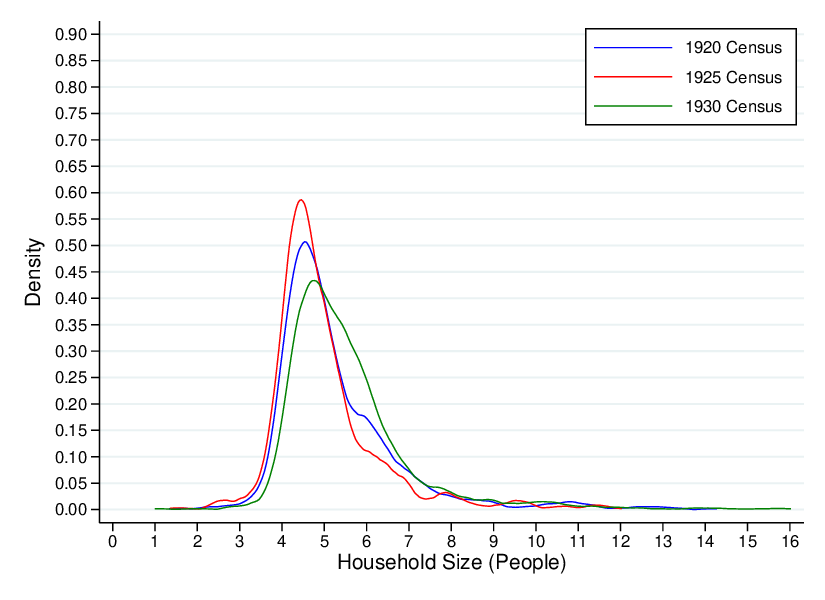}}
\subfloat[Unburned Blocks]{\label{fig:kden_size_ub}\includegraphics[width=0.45\textwidth]{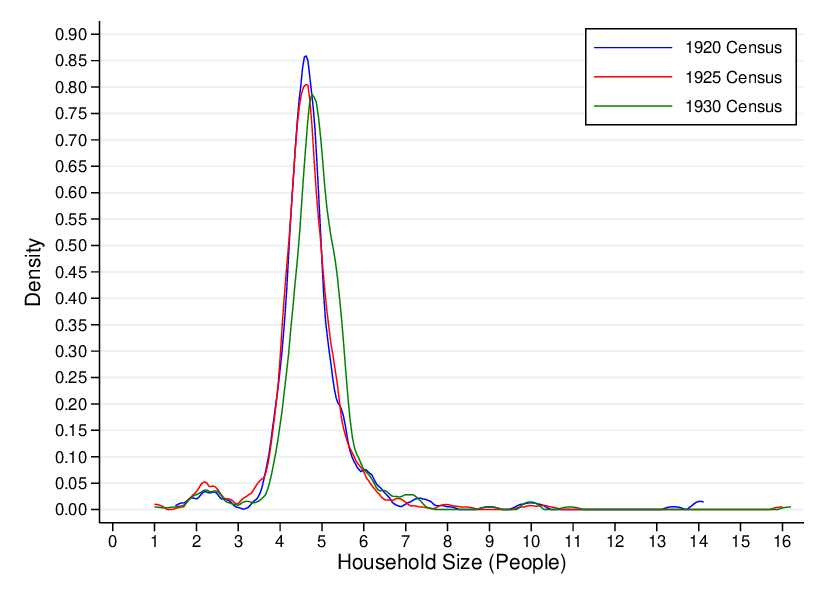}}\\
\caption{Density Estimates for the Average Household Size\\ by Burned and Unburned Blocks between 1920 and 1930}
\label{fig:kden_size}
\scriptsize{\begin{minipage}{450pt}
\setstretch{0.9}
Notes:
Figures~\ref{fig:kden_size_b} and~\ref{fig:kden_size_ub} illustrate the estimated density of the average household size measured in 1920 (blue), 1925 (red), and 1930 (green) censuses in the burned and unburned blocks, respectively.
To unify the range of the domain of definition, $16$ blocks with an average household size exceeding $17$ people are excluded from Figure~\ref{fig:kden_size_b}.\\
Sources: Created by the author using the block-level census statistics listed in Panel B-1 of Table~\ref{tab:sum}.
\end{minipage}}
\end{figure}

Figure~\ref{fig:kden_size} presents density estimates of the average household size.
In the burned blocks, the number of small households with five or fewer members declined steadily following the earthquake.
By 1930, as floor area increased thanks to the rise in multi-story dwellings, there was a marked increase in households with five to eight members (Figure~\ref{fig:kden_size_b}).
In contrast, average household size in unburned blocks showed little change from 1920 to 1925 (Figure~\ref{fig:kden_size_ub}).
From 1925 to 1930, the overall distribution shifted slightly to the right as multi-story dwellings increased even in unburned areas.
The increase in households with 5--8 members is consistent with the rise in households with servants/employees.

\begin{figure}[htbp]
\centering
\captionsetup{justification=centering}
\subfloat[Population: Raw Estimate for the 15 Household Size Bins]{\label{fig:r_block_pop}\includegraphics[width=0.85\textwidth]{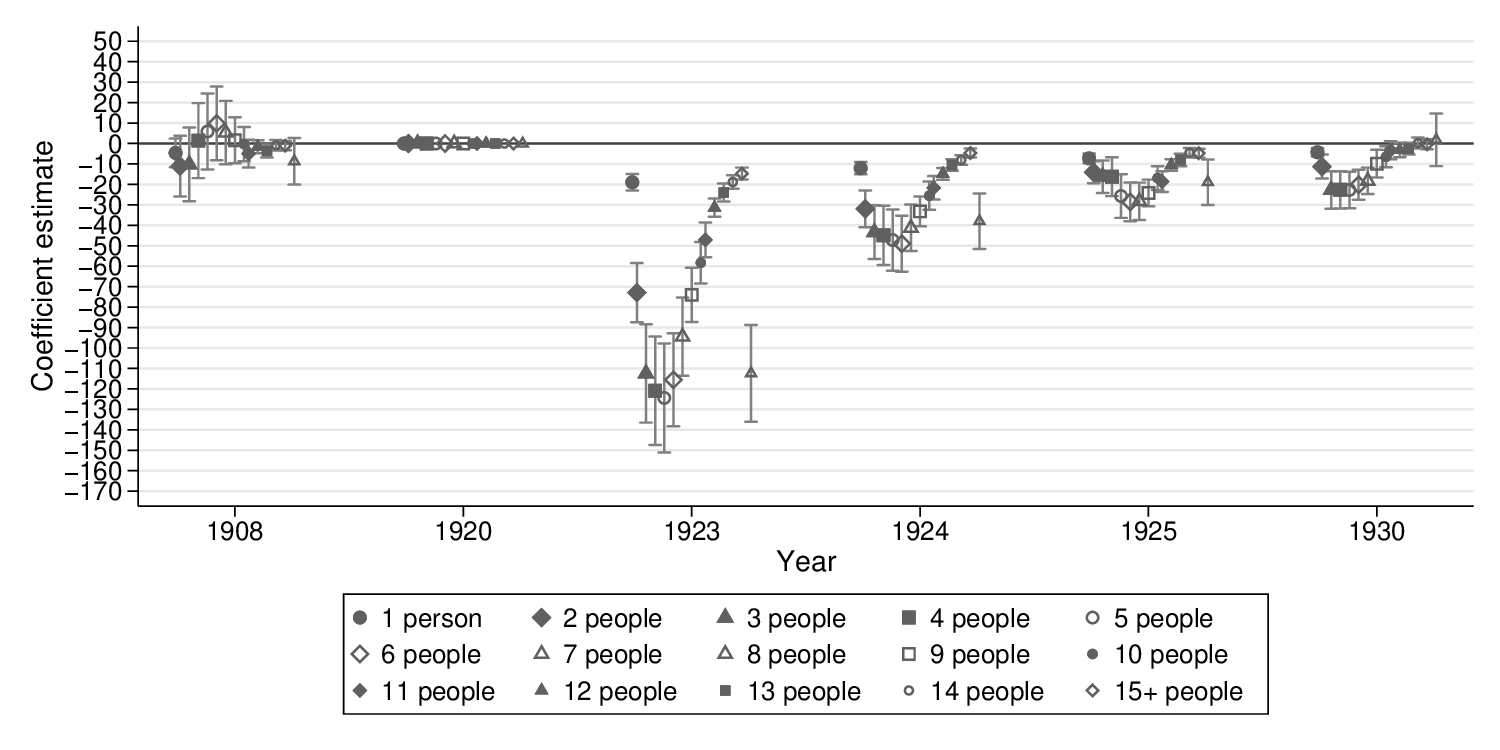}}\\
\subfloat[Population: Estimate / $\text{SD}_{g, 1920}$]{\label{fig:r_block_pop_mag}\includegraphics[width=0.85\textwidth]{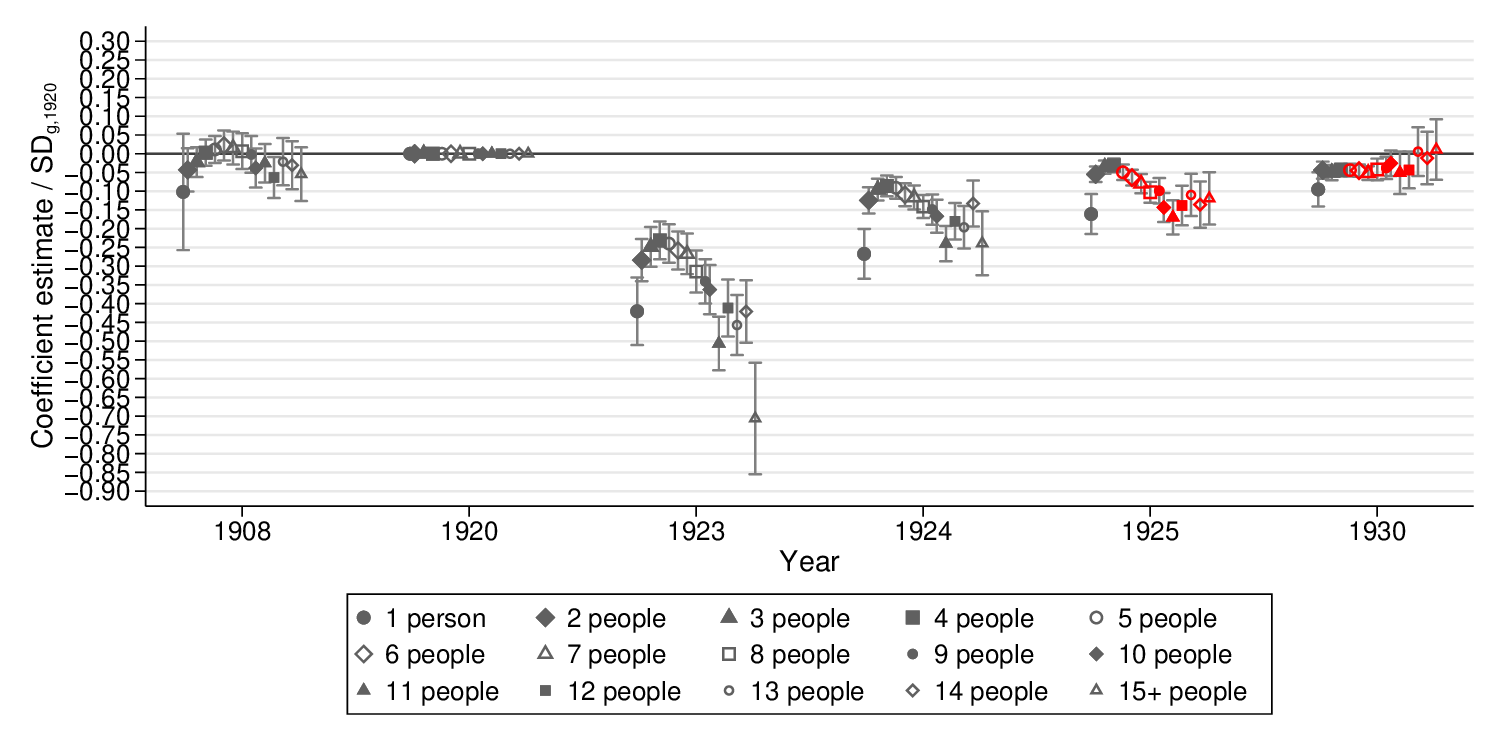}}\\
\caption{Results for the Number of Individuals by Household Size Bin:\\ Evidence from the Block-level Census Statistics}
\label{fig:r_did_block_bin}
\scriptsize{\begin{minipage}{450pt}
\setstretch{0.9}
Notes:
This figure summarizes the estimated coefficients on the interaction terms between the indicator variable for burned blocks and year dummies, conditioning on block- and year-fixed effects (Equation~\ref{model_es_block}).
Figure~\ref{fig:r_block_pop} summarizes the results for the number of individuals in the block-year panel data for each household-size bin (from $1$ to $15+$ individuals).
Figure~\ref{fig:r_block_pop_mag} shows the suggested magnitude of the disaster: the estimates shown in Figure~\ref{fig:r_block_pop} are weighted by the standard deviation of individuals within each household-size bin measured in the 1920 census ($\text{SD}_{g, 1920}$).
The estimates for the $5+$ people bins in the 1925 and 1930 subsamples are colored in red to highlight the relatively large increments in this period (Figure~\ref{fig:r_block_pop_mag}).
The $95$\% confidence intervals are obtained from the cluster-robust variance-covariance matrix estimator using the block as the cluster unit.\\
Sources: Created by the author using the block-size level census statistics listed in Panel B-2 of Table~\ref{tab:sum}.
\end{minipage}}
\end{figure}

Figure~\ref{fig:r_block_pop} shows the estimates from the block-year level regressions for 15 household size bins under Equation~\ref{model_es_block}.\footnote{I run 15 regressions across the bins in total. Since the distribution of households by category mirrors that of the population, the result presented here is essentially the same if the number of households were used as the dependent variable. Online Appendix Figure~\ref{fig:r_did_block_hh} presents the results for the households.}
The estimates are negative across all classes, confirming the results from the ward- and block-level datasets.
To examine the heterogeneity in magnitude across classes, Figure \ref{fig:r_block_pop_mag} summarizes the estimates weighted by the standard deviation in the 1920 sample for each bin.\footnote{The estimates in Figure \ref{fig:r_block_pop} are scaled according to the average number of people per bin, making it difficult to compare across classes. Generally, the standard deviation tends to be larger in classes with higher average population sizes. Thus, using the average population per bin as a weight yields materially similar results.}
It shows that the decline in magnitude from 1925 to 1930 is more evident in the $5+$ class than in single-person households (estimates highlighted in red).
In particular, while a clear trend could not be discerned from the density comparison regarding changes in the $11+$ person category (Figure~\ref{fig:kden_size_b}), this figure confirms that some categories had recovered to pre-earthquake levels by around 1930.
This result supports the evidence that, with the increase in multi-story dwellings from the late 1920s, the number of larger households increased proportionally.

\subsection{Robustness}\label{sec:sec_rob}

I have assessed the robustness of my findings from several perspectives.
These include the test for potential measurement error in the household statistics measured in the censuses, validation checks using alternative complete survey statistics, and assessment of the potential time-varying confounding factor.

\subsubsection*{Measurement Error}

I considered whether lodger households were counted as separate households in the census.
If the number of lodger households is not included in the census household count, the population figure shall be measured accurately, but the household count is under-reported, which would bias both the number of households per dwelling and average household size.
I conducted a set of statistical tests and found no evidence that the census undercounts lodger households.
Online Appendix~\ref{sec:secd_rob_me} summarizes the results.

\subsubsection*{Cross-sectional Evidence}

I used the SSDH (Section~\ref{sec:sec_data_survey}) to assess the validity of the panel-data results.
I confirmed that the average household size in landlord and lodger households in the burned area was statistically significantly larger than in the unburned area.
The sex ratio of landlord households in the burned blocks was significantly lower than that in the unburned area.
On the other hand, the sex ratio of lodger households was quite low on average, and a trend toward even lower ratios was observed in industrial blocks within the burned area.
These results are consistent with the panel data results.\footnote{The results for the sex ratio are summarized in Online Appendix~\ref{sec:secd_r_block_sr}.}
Online Appendix~\ref{sec:secd_rob_size} summarizes the results.

\subsubsection*{Transportation Cost}

When commuting costs to the CBD decrease, households can afford to live farther from the city center, which could reduce the population in the city center and promote suburbanization.
As part of the reconstruction plan, some reinforced concrete bridges were constructed (Online Appendix~\ref{sec:seca_rec_bridge}), which may have altered commuting costs in terms of both time and money \citep{Alvarez-Palau2025-ld}.
The monocentric model, however, implies that this channel differs from a reduction in residential land in its implications for rents and households.
A citywide decline in commuting costs flattens the bid rent function, lowering rents near the city center \citep{Wheaton1974-tx, BaumSnow2007}, whereas a reduction in residential land raises rents throughout the city (Section~\ref{sec:sec_tf}).
A localized improvement in accessibility in the burned area, in turn, would raise both rents and the number of households there.
The rise in real rents in both the burned and unburned areas, together with the decline in the number of households in the burned area, is therefore consistent with the reduction in residential land rather than with changes in commuting costs.

In my panel data analysis, changes in commuting costs that are common across the entire city are absorbed by the time fixed effect.
Additionally, as long as changes in commuting costs are not correlated with the spatial distribution of burned blocks, endogeneity issues do not arise in the regression.
Given this, I first examine whether changes in transportation costs exhibited spatial heterogeneity in the city.
I then use accessibility to hub stations as a proxy for commuting costs to quantitatively verify the robustness of the baseline results.

First, the primary modes of transportation within Tokyo as of 1930 were the national railway (\textit{sh\=oden}) and the city tram (\textit{shiden}).\footnote{Although the subway line between Asakusa and Shimbashi opened in 1927, it accounted for only $0.9$\% of the city's total passenger volume in 1930 (Tokyo City Electric Bureau 1936, p. 40). It was not until after the war that the subway became a major mode of transportation. The history of pre-war transportation in Tokyo is yet understudied. A few exceptional works include \citet{Suzuki2004} on the Metropolitan Electric Railway, \citet{Ono2010ib} on streetcars, and \citet{Yeo2010ib} on automobile transportation.}
These two modes accounted for $62.5$\% of the total number of passengers in the city.
Evidence from a 1931 survey conducted by Tokyo City among workers employed within the city indicates that $79.1$\% of commuters used the national railway or city trams, while $1.0$\% used buses, $13.5$\% used bicycles, and $6.4$\% used other means.\footnote{This survey covered $16,056$ workers who used a single mode of transportation, with no overlap. There were $10,461$ workers who used two or more modes of transportation; of these, $64$\% used the national railway and city trams, $28.4$\% used suburban trains, $5.4$\% used buses, and $2.2$\% used other modes. Since workers using two or more modes could be commuters from the suburbs, the number of users of private railways and buses, operating routes in the suburbs, was high. A similar survey conducted in 1934 confirmed the same trend. For details on these surveys, see Tokyo City Electric Bureau (1936, pp. 24--34).}
When considering commuting from the suburbs, the national railway was particularly important, as it operated lines connecting the outskirts to major city stations.
While the number of national railway stations increased slightly, there was almost no extension of tracks within the city.
Fares outside the city varied depending on distance, whereas they were uniform within the city.
The city tram system operated tracks that covered the entire city, and workers used the tracks to travel from the city limits toward the central district.
While the rates differed for one-way and round-trip travel, they remained flat throughout the 1920s.\footnote{The tracks originated as horse-drawn railways, and electrification began in the 1900s, leading to the expansion of the network. In the 1910s, fares were flat-rate: 5 \textit{sen} one way and 9 \textit{sen} round trip. In 1920, fares were raised slightly to a flat rate of 8 \textit{sen} one way and 15 \textit{sen} round trip. This rate was maintained throughout the 1920s. For details on the management of the Tokyo City Tram, see \citet{Ono2010ib}.}

Next, the public bus service began in 1919.
From the mid-1920s onward, passenger numbers increased as buses supplemented the tram system that was severely damaged by the earthquake.
Similarly, Taxi Automobile Co., Ltd. began operations in 1912.
Until the early 1920s, however, the number of vehicles was quite small, and the service did not develop into a significant mode of transportation.
After the earthquake, ridership gradually increased to compensate for the damage to the tram system.
Despite this, even as of 1930, bus passengers and taxis accounted for only $10.8$\% and $8.6$\% of total public transportation ridership, respectively (Tokyo City Electric Bureau 1936, p.~40).
The taxi fares were essentially constant within the city after the earthquake and unchanged throughout the 1920s.\footnote{The initial rate for taxis was $60$~\textit{sen} per mile, and the citywide fare was set at a flat rate of $1$~\textit{yen} in 1924. Although the fare was set at $50$~\textit{sen} per $2$ miles in June 1930, this rate was not actually enforced in practice \citep[82]{Yeo2010ib}. Consequently, the census data collected in October 1930 are unlikely to influence this revision, especially since taxis accounted for less than 10\% of total passengers at that time.}

To summarize, the cost of travel on the national railway and city tram lines was constant across the city, and rarely changed throughout the sample period.
This means that the potential decline in commuting costs would not lead to a population decline in the city center.
Moreover, automobile transport still had few users.
Because streetcars, buses, and taxis operated throughout the city at a uniform fare, commuting costs did not vary by area.
Thus, even if the decline of streetcars and the development of automobile transport altered average transportation costs, the effects can be captured by time-fixed effects.

To verify the robustness of this view, I conducted a regression analysis using an accessibility measure of the national railway as a control variable.
Generally, statistics on public transportation are compiled at the city-wide level or by operating entity, and statistics broken down by administrative unit are unavailable.
Given this, I calculated the distance from the ward centroid to the nearest railway station using the location information of each station provided by the Ministry of Land, Infrastructure, Transport and Tourism (Online Appendix~\ref{sec:secc_ward}).\footnote{I assume that commuting costs in districts located closer to the station are lower than those in districts farther away. The commuting costs include both time and financial expenses. For example, in districts closer to the station, not only are time costs lower, but financial costs are also likely to decrease because there are fewer transfers to streetcars or cars.}
The results incorporating this accessibility variable are nearly identical to my baseline results.
This confirms that there is no correlation between transportation costs within the city and the key treatment variables.
Online Appendix ~\ref{sec:secd_rob_trans} summarizes the results.

\section{Conclusion}\label{sec:sec_con}

The 1923 earthquake reduced most of central Tokyo to a wasteland.
The reconstruction plan increased the extent of public spaces, such as roads and parks, in the burned area, thereby reducing the land available for residential use.
Consistent with a closed-city monocentric model, real rents rose throughout the city, and the number of households in the burned area declined relative to its pre-earthquake level.
By the mid-1930s, the number of multi-story dwellings and total floor area in the burned area exceeded their pre-earthquake levels, and the average unit rent there declined moderately, although it remained above its pre-earthquake level.
The population in the burned area had returned to roughly its pre-earthquake level by 1935, whereas the number of households remained lower.
The zoning system designed before the earthquake had little effect on these changes.

The population recovered faster than the number of households because the burned area used its housing stock more intensively.
About $70\%$ of this adjustment is accounted for by more households per unit of floor area and about $30\%$ by larger households.
The first margin operated through lodging: lodger households rented rooms in other people's dwellings and were counted as separate households, and the unit rent paid by lodgers relative to that paid by landlords was lower in the burned area than in the unburned area.
The second margin, which lies outside the model, operated through the inclusion of non-kin members: as rents rose, the number of independent kinship households declined, while domestic servants and business employees increasingly lived in their employers' houses.

To summarize, the population in the burned area reflected the net effect of two opposing forces: fewer households due to the loss of residential land, partly offset by lodging, and larger households through the inclusion of non-kin members.
These findings show that, even when the population of a devastated city center recovers, a disaster can leave lasting changes in how its residents share housing.

\bibliographystyle{plainnat}
\bibliography{reference.bib}

\renewcommand{\refname}{Documents and Database}

\clearpage
\thispagestyle{empty}

\begin{center}
\qquad

\qquad

\qquad

\qquad

\qquad

\qquad

{\LARGE \textbf{
Appendices
}}
\end{center}
\clearpage
\appendix
\def\thesection{Appendix~\Alph{section}}
\def\thesubsection{\Alph{section}.\arabic{subsection}}
\setcounter{page}{1}

\section{Background Appendix}\label{sec:seca}
\setcounter{figure}{0} \renewcommand{\thefigure}{A.\arabic{figure}}
\setcounter{table}{0} \renewcommand{\thetable}{A.\arabic{table}}

\subsection{Zoning System}\label{sec:seca_zoning}

\begin{figure}[h]
\centering
\captionsetup{justification=centering,margin=1.5cm}
\includegraphics[width=0.55\textwidth]{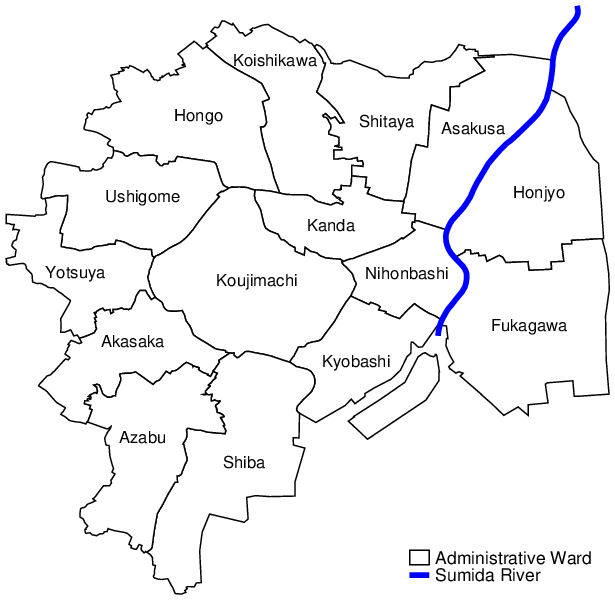}
\caption{Administrative Wards and Sumida River in Tokyo City}
\label{fig:map_tokyo}
\scriptsize{\begin{minipage}{400pt}
\setstretch{0.85}
Note:
This figure shows the 15 administrative wards in Tokyo City.
The Sumida River is illustrated on the border, surrounded by the Asakusa, Honjo, Nihonbashi, and Fukagawa Wards.
Kyobashi Ward encompasses Tsukishima Island in its southeastern section.
The administrative district borders from 1920 are used on this map, which were rarely changed throughout the 1920s and 1930s.
\\
Sources:
Created by the author using the official shapefile (Ministry of Land, Infrastructure, Transport and Tourism, database).
The location data of the Sumida River are obtained from the Ministry of Land, Infrastructure, Transport, and Tourism (website).
\end{minipage}}
\end{figure}

This section examines the spatiotemporal distribution of the commercial, manufacturing, and residential areas in Tokyo City.
Through this, I demonstrate that the zoning system introduced after the earthquake was based on the spatial distribution of initial land use.
Figure~\ref{fig:map_tokyo} shows the spatial distribution of the 15 administrative wards of the city and the Sumida River.

\subsubsection{Before the Introduction of the Zoning System: Initial Distribution}

\subsubsection*{Commercial Districs}

Prior to the Great Kanto Earthquake of 1923, commercial districts were aligned along an axis connecting Asakusa, Ueno (in Shitaya Ward), Kanda, Nihonbashi, Ningyocho (in Nihonbashi Ward), and Ginza (in Kyobashi Ward).
The districts of Asakusa, Kanda, and Nihonbashi, which encompassed these urban areas, had developed into wholesale centers.
This was due to the availability of water transport on the Sumida River and its tributaries (Horiuchi 1978, p.~44).
In other words, the initial spatial distribution of commercial districts was largely determined by the locational fundamentals provided by the sea and rivers.

\subsubsection*{Manufacturing Districs}

Manufacturing areas were concentrated in two regions.
One was the eastern part of the city, including the Honjo and Fukagawa Wards, and the other was the Kyobashi area, which included Tsukiji and Tsukishima.

The Honjo and Fukagawa Wards encompassed the Sumida River and the Onagi River, a tributary of the Sumida River.
Since a timber wharf (\textit{kiba}) was established in Fukagawa, it became possible to transport raw materials from Yokohama City in Kanagawa Prefecture.
A diverse range of factories developed here, including those in the textile, machinery, and chemical industries.
The Kyobashi district faced Tokyo Bay, offering convenient water transport.
It was also close to the Nihonbashi wholesale district, which had been a center of commercial capital since the Edo era.
Factories producing machinery, printing, and bookbinding flourished there.
Some factories in the Kyobashi area later moved to Shibaura in Shiba Ward and to Osaki-Shinagawa (along the Meguro River), adjacent to Shiba Ward but outside the city.
This led to the creation of four major industrial zones: Honjo-Fukagawa, Kyobashi, Shibaura, and Osaki-Shinagawa (outside the city).

Clearly, geographical features such as rivers and harbors played a major role in the formation of the manufacturing districts, as they did in the commercial districts.

\subsubsection*{Residencial Districs}

Residential districts in the city encompass all areas other than commercial and manufacturing zones.
This means that the western part of the city had relatively large residential areas.
Starting in the 1920s, the residential districts expanded further westward beyond the city limits, driven by the construction of private railway lines.
These areas are known as Shibuya and Meguro.

\subsubsection{After the Introduction of the Zoning System: Path Dependence}

\begin{figure}[h]
\centering
\captionsetup{justification=centering,margin=1.5cm}
\includegraphics[width=0.6\textwidth]{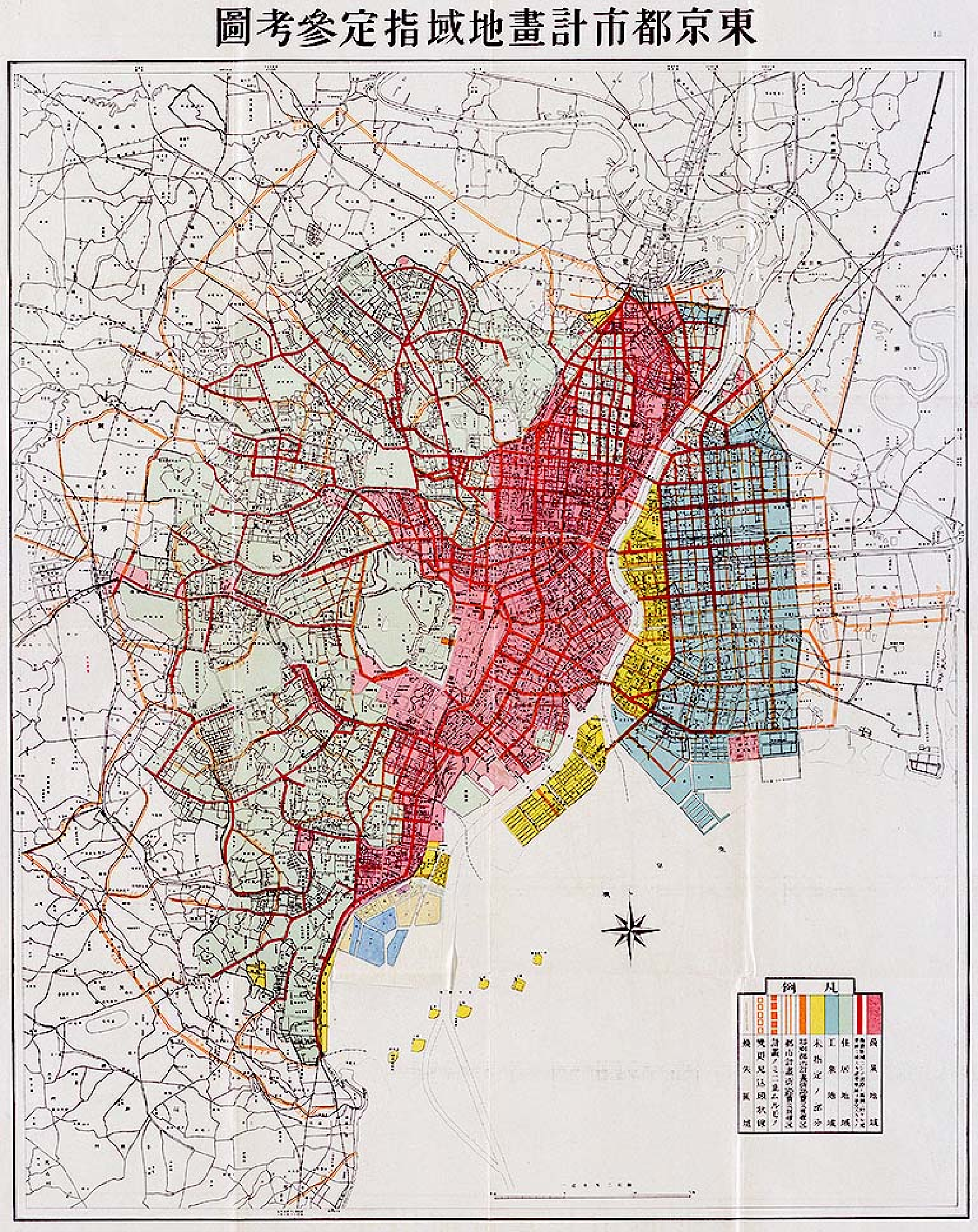}
\caption{Land-use Zones in Tokyo City}
\label{fig:zoning_1925}
\scriptsize{\begin{minipage}{400pt}
\setstretch{0.85}
Note:
This figure shows the land-use zones (\textit{y\=oto chiiki}) in Tokyo City based on the Tokyo City Planning Area (\textit{T\=oky\=o Toshi Keikaku Kuiki}), designed before the 1923 earthquake and enacted in 1925.
The red, blue, and green zones indicate the commercial, manufacturing, and residential zones, respectively.
The area colored in yellow indicates the undesignated area, which was designed as a buffer between the commercial zone (red) and the manufacturing zone (blue). 
Despite lacking an official designation, this area encompassed several principal manufacturing districts and thus constituted a de facto manufacturing zone.
\\
Source: National Archive of Japan (Fu B02002100), website.
Scale and tone were adjusted by the author using Adobe Photoshop 2024.
\end{minipage}}
\end{figure}

Tokyo City implemented urban planning projects under the Tokyo City Improvement Ordinance (\textit{T\=oky\=o Shiku Kaisei Jy\=orei}).\footnote{See Section~\ref{sec:seca_plan} for details on the city planning projects.}
The project under this ordinance was completed in 1914.
It covered only road improvements in Marunouchi and Hibiya on the east side of the Imperial Palace, as well as improvements in the Nihonbashi area.

In response to the nationwide trend of urbanization that emerged during World War I, the government then recognized the need for urban planning.
In 1918, the City Planning Division (\textit{Toshi Keikaku-ka}) was established within the Minister's Secretariat, Ministry of the Interior.
The City Planning Council (\textit{Toshi Keikaku Ch\=osakai}) was established within the division, and deliberations on the City Planning Bill began.
In 1919, the City Planning Act (\textit{Toshi Keikaku H\=o}) and the City Buildings Act (\textit{Shigaichi Kenchikubutsu H\=o}) were promulgated.
The City Planning Act incorporated the provisions of the Tokyo City Improvement Ordinance while adding three new elements: (1) designation of zoning districts, (2) implementation of land readjustment projects, and (3) expropriation of buildings and other structures.
Throughout these efforts, Tokyo City sought to halt unregulated urban development and promote planned urbanization.

In March 1920, a special committee was established within the Tokyo Local City Planning Committee (\textit{Toky\=o Toshi Keikaku, Toky\=o Chih\=o Iinkai}) of the Ministry of the Interior to draft a zoning system under the City Planning Act.
This special committee subsequently held more than $20$ meetings to formulate the draft, which was submitted to and approved by the Tokyo Regional Committee on Urban Planning on August 10, 1923.
The Minister of the Interior then submitted an application for approval to the Prime Minister.
However, following the Great Kanto Earthquake on September 1, 1923, the Minister of the Interior withdrew the application to reexamine and revise the draft.
The revised plan was approved in December 1924 and promulgated in January 1925.
Importantly, however, its content was virtually identical to the original.
The differences were minor and localized, such as the reclassification of the Shibaura district from an undesignated area to an industrial area, and changes to the road plan (Horiuchi 1978, p.~50).

Figure~\ref{fig:zoning_1925} shows the spatial distribution of zoning districts within Tokyo City.\footnote{
The land-use zoning system was also applied to the city's suburbs because the city planning area extended into them. Roughly speaking, the western suburbs were designated as residential zones, while the eastern suburbs were designated as manufacturing zones. Commercial zones were established in a scattered manner within parts of the residential zones. 
Although this is beyond the scope of the analysis in this study, the spatial distribution of land use zones in these suburbs can be viewed on the National Archives of Japan website (\url{https://www.archives.go.jp/exhibition/digital/henbou/contents/photo.html?m=49&ps=2&pt=1&pm=1}).\label{oa_fn_zoning}}
Clearly, the existing commercial, manufacturing, and residential zones were simply designated as each zone in the system.
Thus, commercial and manufacturing zones--which had been distributed based on geographical transportation conveniences, such as rivers and the sea--were maintained as is under the system.\footnote{This interpretation is consistent with that of the National Archives of Japan. See \url{National Archives of Japan}{https://www.archives.go.jp/exhibition/digital/henbou/contents/49.html}.}
In this sense, the zoning system imposed path dependency on the initial distribution of these industrial zones, which had been determined by locational fundamentals.

In fact, the government's explanatory memorandum explicitly states that zoning districts are determined based on multiple exogenous factors.\footnote{For the content of this paragraph, see Horiuchi (1978, pp.~52--53). His book also contains the explanations provided in the explanatory memorandum regarding each zoning district.}
It states that land use zones are determined by:
\begin{quote}
``comprehending the current state and history of land development, topography, climate, the status of land and water transportation infrastructure, and other various natural and man-made conditions.''
\end{quote}
Furthermore, the policy of ``examining natural and social conditions to select land where the objectives of each land use zone can be achieved and then designating the zones'' remained unchanged even after the war.
Although Tokyo City was reduced to ashes by the air raids of World War II, the spatial distribution of these zoning districts remained unchanged.
In summary, zoning districts were established by inheriting the initial distribution of industries shaped by geographical conditions, and this approach came to define them over the long term in Tokyo.

In addition to commercial and manufacturing zones, ``undesignated areas'' were set (Figure~\ref{fig:zoning_1925}).
These undesignated areas were established as a buffer zone between the manufacturing zones east of the Sumida River and the commercial and residential areas to the west.
While large factories could not be established in this undesignated zone (unlike the manufacturing zone), small-scale factories with low environmental hazard levels could be located there.
For example, Tsukishima--an island built on the Sumida River delta--originally had many machine factories, making it a reasonable candidate for designation as a manufacturing zone.
However, it was considered that ``given prevailing winds, it would pose a threat to the residential and commercial areas behind it'' (Horiuchi 1978, pp. 53--54).
In this way, areas with uncertainty regarding future city development were designated as undesignated zones.

\subsubsection{Revisions and Additions to the Zoning System in the Suburbs of Tokyo City}

The zoning system, designated in January 1925, underwent minor revisions in September 1926.
These revisions involved expanding commercial zones to accommodate the growth of residential areas in Tokyo City's suburbs, a trend evident since the 1910s.
As stated in the government's explanatory memorandum, this amendment was a localized adjustment concerning designated areas outside the city (footnote~\ref{oa_fn_zoning}) and did not alter the zoning system within the city (Horiuchi 1978, p.~59).
For the same reason, zoning districts were also designated in $33$ towns and villages in the city's suburbs in 1929 (for details on the designations, see Horiuchi 1978, pp. 60--64).
In October 1932, in light of population growth in the suburbs, the city area was expanded from 15 wards (the former city area) to 35 wards by incorporating 20 surrounding new wards (\textit{shin shiiki}).
As a result, this Greater Tokyo City (called \textit{Dai Toky\=o-shi}) and the city planning area came to coincide almost exactly.
Finally, the zoning districts were designated for the new wards in the suburbs of the former city area in July 1935 (Horiuchi 1978, pp. 68--70).

As is clear from the above, the changes and additions to the zoning system following the earthquake were targeted at suburban areas.
Consequently, the zoning system in Tokyo City (the former city area) remained unchanged from its initial designation.\footnote{Under the wartime regime, there were minor changes and additions to the zoning system. For example, the designation of open space districts for air defense. However, all of these were again targeted at suburban areas (Horiuchi 1978, pp.~71--76).}

\subsection{Great Kanto Earthquake of 1923}\label{sec:seca_gke}

\subsubsection{Overall Impacts}\label{sec:seca_gke_overall}

\begin{figure}[htbp]
\centering
\captionsetup{justification=centering}
\includegraphics[width=0.6\textwidth]{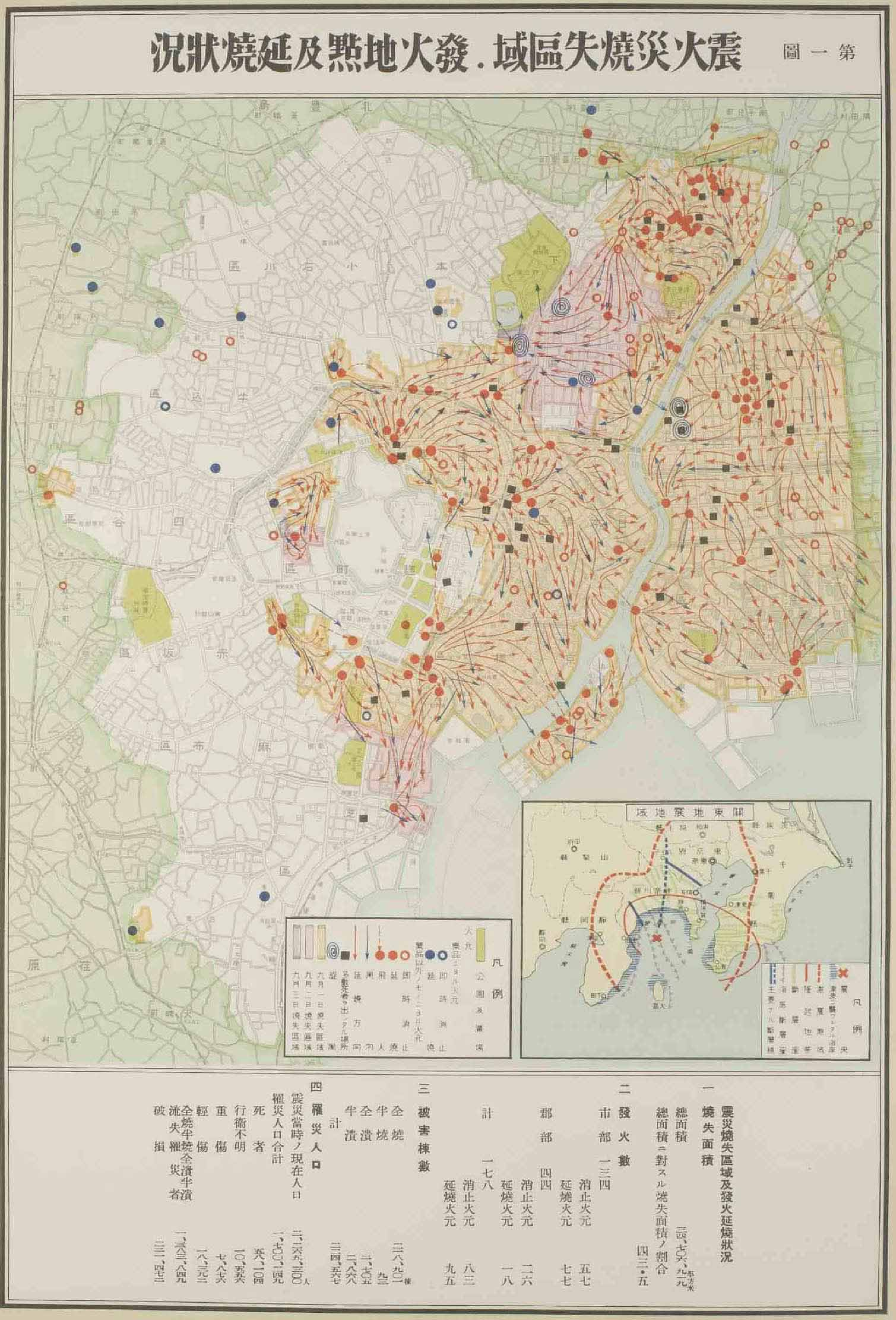}
\caption{Burned Area, Points of Ignitions, and Fire Spreading Routes}
\label{fig:fire}
\scriptsize{\begin{minipage}{400pt}
\setstretch{0.85}
Note:
This figure shows how fire spread after the earthquake in Tokyo City.
The yellow, pink, and light purple areas show the burned area (the light green surrounding area is not burned).
The yellow, pink, and purple areas are burned on 1st, 2nd, and 3rd September 1923, respectively.
The green lattices show the parks and plazas.
The points of ignition are shown in red circles, whereas the red arrows show the spreading routes of the fire.
The red-shaded circle without an arrow indicates the original ignition points (not caused by the chemicals).
The red-shaded circle with arrow(s) indicates the ignition points from spreading fire from other places.
The red-open circle indicates the ignition points, which were immediately extinguished.
The blue arrows show the wind direction.
The blue circle indicates the same meaning, but is ignited by the chemicals.
The black squares show the places where many deaths were recorded.
The spiral shows the places where the fire whirlwind (kasai senp\=u) occurred.
Source: Tokyo City Office (1930z, Figure 1). The author adjusted the scale and tone using Adobe Photoshop 2024.
\end{minipage}}
\end{figure}

The Great Kanto Earthquake occurred on September 1, 1923.
This earthquake caused the greatest material, human, and economic damage in Japan's history of natural disasters.
The epicenter was located in the northwestern part of Sagami Bay (Figure~\ref{fig:fire}), and the seismic intensity is estimated to have been M$8.1 \pm 0.2$ (Takemura 2003, p.~44).
According to the Cabinet Office (\textit{Naikakufu}), the economic damage amounted to approximately $5.5$ billion yen, representing a loss equivalent to about $37$\% of the GDP at the time, which was approximately $14.9$ billion yen.
Considering that the national budget was $1.4$ billion yen at the time, the magnitude of the loss becomes clear.\footnote{For details on these damages and losses, refer to the Cabinet Office's official website (\url{https://www.bousai.go.jp/kantou100/}, accessed on December 31, 2024 [in Japanese]).}

According to a survey by the Minister of the Interior, $694,621$ households were affected, with $91,544$ fatalities and $15,275$ people reported missing.
Table~\ref{tab:damage} shows a breakdown of the damage in Tokyo City.
As shown, $354,453$ households, which were approximately $73$\% ($354,453/483,000$) of the total number of households at the time of the disaster, were affected.
$3$\% ($68,660$ people) of the total city population were killed or went missing.

\def\arraystretch{1.00}
\begin{table}[htb]
\begin{center}
\caption{Households and Population Affected by the Great Kanto Earthquake}
\label{tab:damage}
\footnotesize
\scalebox{1.0}[1]{
\begin{tabular}{lcccc}
\toprule[1pt]\midrule[0.3pt]
&In Seven Affected&\multicolumn{3}{c}{Tokyo Prefecture}\\
\cmidrule(ll){3-5}
	&Prefectures&Total&City&Other than city\\\hline
Panel A: Affected Households	&&&&\\
Completely burned			&381,090	&311,962&300,924&11,038\\
Partially burned				&517		&366&239&127\\
Completely collapsed		&83,819	&16,684&4,222&12,462\\
Partially collapsed			&91,233	&20,122&6,336&13,786\\
Washed away				&1,390	&0&0&0\\
Damaged					&136,572	&47,985&42,732&5,253\\
Total						&694,621	&397,119&354,453&42,666\\
&&&&\\
Panel B: Affected Population	&&&&\\
Deaths					&91,344&59,593&58,104&1,489\\
Serious injury				&16,514&8,773&7,876&897\\
Missing					&13,275&10,904&10,556&348\\
Minor injuries				&35,560&20,199&18,392&1,807\\
\midrule[0.3pt]\bottomrule[1pt]
\end{tabular}
}
{\scriptsize
\begin{minipage}{400pt}
Notes: 
The affected areas include seven prefectures: Tokyo, Kanagawa, Chiba, Saitama, Shizuoka, Yamanashi, and Ibaraki.
Sources: 
The number of affected households and people is obtained from Tokyo City Office (1932aa, pp.~15--16) and Tokyo City Office (1932aa, pp.~17--18), respectively.
\end{minipage}
}
\end{center}
\end{table}

\subsubsection{Distribution of Devastated Area: Big Fire after the Hit}\label{sec:seca_gke_fire}

A key characteristic of earthquake damage is not only the shaking itself but also the fires.
More than $80$\% of the fatalities mentioned above were caused by homes being completely destroyed by fire (Tokyo City Office 1932aa, p.~17).\footnote{According to a report by the Ministry of the Interior, the area destroyed by fire (total damage) was 20 times (50 times) that of the Great Fire of London in 1666, four times (17 times) that of the Great Chicago Fire of 1871, and three times (seven times) that of the Great San Francisco Fire of 1906 (Tokyo City Office 1932aa, p. 20).}
The earthquake struck at 11:58 a.m., a time when many households were using open flames to prepare lunch.
Furthermore, due to a typhoon (central pressure of 997 hPa) near the Noto Peninsula on that day, quite strong winds were blowing (Takemura 2003, p.~13).
Since most buildings were wooden structures, these two conditions caused embers to be carried by the wind, thereby spreading the fire.

Figure~\ref{fig:fire} is a valuable resource that details the path of the fire's spread in Tokyo City.
The fire spread through flying embers, and the direction of the spread depended on the wind direction.
While most areas were engulfed by the fire on the day of the disaster, it is also clear that the fire continued to spread into the following day.
Furthermore, fire whirlwinds (\textit{kasai senp\=u}) occurred at various locations, and in some areas they caused numerous deaths.
Records also indicate that the earthquake's shaking caused water mains to burst, making firefighting difficult (Tokyo City Office 1932aa, p. 28).

\def\arraystretch{1.00}
\begin{table}[htb]
\begin{center}
\captionsetup{justification=centering,margin=1.5cm}
\caption{burned area and Number of Households Affected by the Great Kanto Earthquake: Statistics by Ward in Tokyo City}
\label{tab:damage_ward}
\footnotesize
\scalebox{1.0}[1]{
\begin{tabular}{lcccc}
\toprule[1pt]\midrule[0.3pt]
&&\multicolumn{3}{c}{Burned Area in Tokyo City}\\
\cmidrule(ll){3-5}
Ward			&Ward Area (ha)	&Number of town blocks&Area Burned (ha)&\% of Burned Area \\
				&				& destroyed by fire&&\\\hline
Kojimachi				&667			&40		&180		&27.0\\
Kanda				&382			&130		&297		&77.8\\
Nihonbashi			&292			&140		&292		&100.0\\
Kyobashi				&397			&207		&374		&94.2\\
Shiba				&867			&62		&230		&26.5\\
Azabu				&373			&1		&2		&0.5\\
Akasaka				&383			&15		&29		&7.6\\
Yotsuya				&270			&3		&7		&2.7\\
Ushigome				&535			&0		&0		&0.0\\
Koishikawa			&617			&5		&40		&6.5\\
Hongo				&518			&26		&89		&17.1\\
Shitaya				&477			&49		&238		&49.9\\
Asakusa				&482			&115		&474		&98.2\\
Honjyo				&583			&80		&545		&93.5\\
Fukagawa				&768			&132		&670		&87.1\\
Total					&7,611		&1005	&3466	&45.5\\
\midrule[0.3pt]\bottomrule[1pt]
\end{tabular}
}
{\scriptsize
\begin{minipage}{430pt}
Notes: 
This table shows the burned area across the 15 administrative wards of Tokyo City, as measured in the official reports compiled by the Tokyo City Office.
The original area is reported in Japanese units (\textit{tsubo}), which I converted into hectares.
The administrative ward area excludes the Imperial Palace grounds, where citizens are unable to live.
Sources: Tokyo City Office (1932aa, pp.~33--34).
The area for the Imperial Palace is from Tokyo City Office (1922, p.~20)
\end{minipage}
}
\end{center}
\end{table}

\begin{figure}[h]
\centering
\captionsetup{justification=centering,margin=1.5cm}
\includegraphics[width=0.55\textwidth]{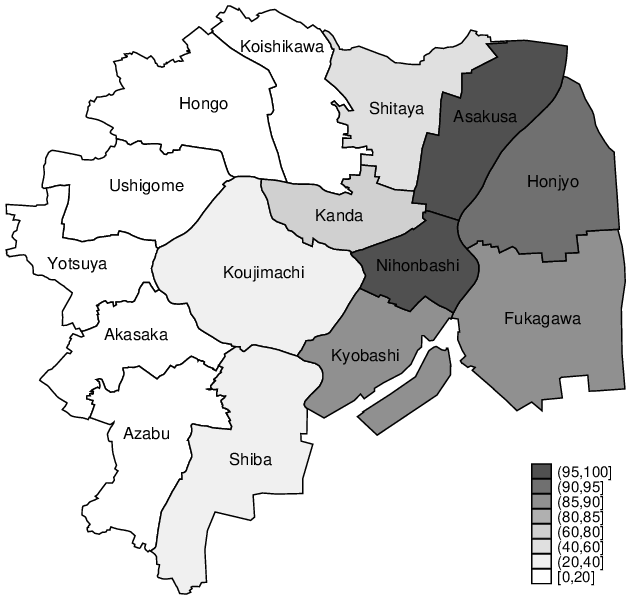}
\caption{Percentage of Burned Area in Administrative Ward}
\label{fig:map_ward_burned}
\scriptsize{\begin{minipage}{400pt}
\setstretch{0.85}
Notes:
This figure shows the share of burned area relative to total area for each administrative ward in Tokyo.
The original figures are listed in Table~\ref{tab:damage_ward}.
The administrative district borders from 1920 are used on this map.
Sources: Created by the author.
The data on the burned and total areas are obtained from the Tokyo City Office (1932aa, pp.~33--34).
The official shapefile is from the Ministry of Land, Infrastructure, Transport and Tourism (database).
\end{minipage}}
\end{figure}

Table~\ref{tab:damage_ward} summarizes the burned area by administrative ward in Tokyo City.
More than $40$\% of the 15 wards were engulfed by the fire, with the total burned area reaching $3,466$ ha.
Figure~\ref{fig:map_ward_burned} illustrates the proportion of the burned area relative to the total area of each ward.
What geographical characteristics influenced the spatial distribution of the burned area?
Importantly, the burned area did not depend solely on the ground's softness (i.e., its susceptibility to shaking).
Below, I summarize the spatial distribution of damage using a book by a seismologist who conducted a detailed analysis of shaking during the Great Kanto Earthquake (Takemura 2003, Chapter 4).

The first case involves areas where soft ground caused houses to collapse, which in turn led to fires.
West of the Sumida River, this applies to parts of Asakusa, Kanda, and Shitaya Wards.
East of the Sumida River, it applies to Honjo and Fukagawa Wards.
The ground in these areas consists largely of alluvial deposits and is prone to shaking during earthquakes.
The estimated average seismic intensity (0--7) in these areas ranged from $6-$ to $6+$, showing the strongest shaking within the city.
The estimated total collapse rates for residential buildings were $7.0$\% in Asakusa, $7.6$\% in Kanda, $2.4$\% in Shitaya, $15.6$\% in Honjo, and $8.9$\% in Fukagawa Wards.
As shown in Figure~\ref{fig:map_ward_burned}, the fire destruction rate was also high in these areas.
With the exception of Shitaya, the fire destruction rate was $80$\% or higher.

The second case involves areas with ground that is relatively resistant to seismic shaking and where few buildings collapsed.
This applies to the wards of Kojimachi, Shiba, Azabu, Akasaka, Yotsuya, Ushigome, Koishikawa, and Hongo, located west of the Sumida River.
The ground in these areas consists largely of Pleistocene terraces, which are more resistant to shaking than alluvial deposits.
The estimated average seismic intensity ranged from $5+$ to $6-$, indicating relatively mild shaking within the city.
The estimated total residential building collapse rate ranged from $0.4$\% in Yotsuya to $3.1$\% in Akasaka.
As shown in Figure~\ref{fig:map_ward_burned}, the fire description rate was relatively low in these areas.

The third case involves areas that, despite having relatively resistant ground, experienced widespread fire spread due to sparks.
The Nihonbashi and Kyobashi Wards fall into this category.
Because these areas feature a buried wave-cut platform, called the Nihonbashi Plateau, the base of the alluvial deposits is quite shallow.
Consequently, the estimated average seismic intensity for these two wards was relatively low at $5+$, and the estimated residential building collapse rates were the lowest in the city at $0.3$\% in Nihonbashi and $0.4$\% in Kyobashi Wards.
Accordingly, Figure~\ref{fig:fire} shows that there were few ignition points in either ward.
However, due to widespread flying embers carried by the prevailing wind on that day, combined with the spread of fire from other areas, these two central wards were almost entirely destroyed (Takemura 2023, pp. 27--29).

To summarize, the distribution of fire damage exhibits exogenous variation due to the ground and weather conditions.

\subsection{Governmental Responses and Population Movements in the Immediate Aftermath}\label{sec:seca_gov}

\subsubsection{Immediate Responses}\label{sec:seca_response}

I use an official report titled the History of the Taisho Earthquake (\textit{Taish\=o Shinsaishi}), published by the Social Bureau, Ministry of the Interior (1926aa), to summarize policy responses from the time of the earthquake through several months afterward.

\subsubsection*{Operations by Governement}

In response to the extensive damage, the government established the Temporary Bureau for Earthquake Disaster Relief (\textit{Rinji Shinsai Ky\=ugo Jimukyoku}) on the day of the disaster.
With the Prime Minister serving as its head, the headquarters comprised $11$ departments: General Affairs, Food, Shelter, Materials, Transportation, Drinking Water, Health and Medical Care, Security, Information, Donations, and Accounting and Finance.

On September 2, the government issued an emergency requisition order (\textit{Hijy\=o Ch\=ohatsurei}) and a state of martial law (\textit{Kaigenrei}).
This enabled the government to maintain public order and provide relief to disaster victims.
These measures continued until September 5, when government functions were restored.

On September 7, three emergency imperial decrees (\textit{Kinky\=u Chokurei}) were promulgated and enforced.
These were the Public Order Maintenance Ordinance (\textit{Chian Iji Rei}, Imperial Ordinance No. 403), the Payment Deferral Ordinance (\textit{Shiharai Enki Rei}, Imperial Ordinance No. 404), and the Anti-Profiteering Ordinance (\textit{B\=ori Torishimari Rei}, Imperial Ordinance Nos. 407--408 and 411).
Through these measures, the government sought to maintain economic activity while ensuring public safety.
In particular, the Anti-Profiteering Ordinance imposed price controls on essential goods such as food, building materials, and fuel.
It was crucial for curbing post-earthquake inflation while awaiting the restoration of the supply of goods.

On September 11 and 12, an ordinance exempting import duties (Imperial Ordinance No. 420) was issued to increase the supply of goods.\footnote{For this paragraph, I also referred to the Record of the Temporary Commodity Supply Program (\textit{Rinji Busshi K\=oky\=u Jigy\=oshi}), an official record published by the Accounting Department, Reconstruction Bureau (1926, pp.~8--9).}
In addition, recognizing the limitations of securing supplies through the market, the Temporary Commodity Supply Ordinance (\textit{Rinji Busshi Ky\=oky\=u Rei}, Imperial Ordinance No. 420) was issued on September 22.
This decree was intended to ensure the smooth supply of daily necessities.
It stipulated that the government take the lead in purchasing and stockpiling various goods and could restrict or prohibit their export.
Furthermore, the Special Account Ordinance for the Temporary Commodity Supply Program (\textit{Rinji Busshi Ky\=oky\=u Tokubetsu Kaikei Rei}, Imperial Ordinance No. 421) was issued simultaneously.
It allowed these matters to be handled through a special account, separate from the general account.
Section~\ref{sec:seca_wood} summarizes examples of timber procurement by the Materials Department (\textit{Zairy\=o-bu}) under these decrees.

For disaster victims, an imperial ordinance regarding tax relief measures (Imperial Ordinance No. 410) was issued on September 12.
This waived income tax and business tax for the 1923 fiscal year for disaster victims.
In addition, building regulations were relaxed: Ordinance Concerning Temporary Buildings within Areas Subject to the City Building Act (\textit{Shigaichi Kenchikubutsu H\=o Tekiy\=o Kuiki-nai ni Okeru Kari Kenchikubutsu ni Kansuru Ken}, Imperial Ordinance No.414) was issued on September 16.
This stipulates that structures for which construction began by the end of February 1924 and were to be dismantled by August 1928\footnote{This period was amended and extended on February 18, 1924, later.} would be exempt from the standard provisions of the City Building Act.
This allowed residents to construct simple temporary housing.

In September of 1923, various regulations aimed at sustaining daily life were implemented through the government's emergency ordinances.

The government also had to begin discussing long-term reconstruction plans for Tokyo.
On September 19, the Imperial Capital Reconstruction Council was established.
The Imperial Capital Reconstruction Agency was then set up on the 27th as its executive body.
Details regarding this Imperial Capital Reconstruction Plan carried out through the 1920s are summarized in Section~\ref{sec:seca_rec}.

\subsubsection*{Operations by Tokyo City}

On 12th October 1921, Tokyo City established five departments--General Affairs, Relief, Public Works, Finance, and Electricity--in accordance with the Emergency Disaster Operations Regulations (\textit{Hijy\=o saigai Shomu Kitei}, Tokyo City Directive A No. 41).
Until September 22, 1923, disaster relief operations were carried out under a provisional system.\footnote{Tokyo City Office (1932aa, pp. 63--65; 517--518; Social Bureau, Ministry of the Interior (1926aa, p. 506).}
Relief operations covered a wide range of activities, but the following are some of the most representative examples.

Immediately after the earthquake, evacuees gathered in parks and public squares.
Although the number of evacuees gradually decreased, more than $25,000$ people were still taking shelter in the vicinity of Tokyo Station alone, as of September 7.
In response, the city set up military tents starting on September 17 and was housing over $4,000$ people by October 10.
A maternity ward was established using these tents.
Meal services and food distributions were provided until September 30.
Thereafter, daily necessities were distributed exclusively to disaster victims.

$76$\% percent ($162$ out of $213$) of hospitals in the city were destroyed by fire (Tokyo City Office 1925s, p.~59).
The medical teams were set up in open spaces such as parks to treat the injured and sick.
More than $180$ medical facilities were established within the city, treating $1,575,297$ people between September 1 and November 30.
However, the number of patients with infectious diseases during those three months was $3,705$ ($1,645$ with dysentery; $1,780$ with typhoid fever; and $280$ with other diseases), which is reported to be nearly double the figure for the same period the previous year (Tokyo City Office 1932aa, pp~508--516).
Relatedly, of the $971$ bathhouses in the city, $631$ were destroyed by fire.
The temporary bathhouses were set up in municipal barracks, primarily in the burned area (Tokyo City Office 1932aa, pp. 518--519).

Some disaster victims lost their jobs.
The report indicates this led to a temporary surplus of labor in the labor market.
The city hired them to clear the burned-out sites and to construct temporary residences.
Wages were determined in consultation with the Chamber of Commerce and publicly announced to the workers.

\subsubsection{Temporary Housing}\label{sec:seca_housing}

The population survey on November 15 was the most comprehensive survey immediately after the disaster.\footnote{This survey was designed by the government to distribute monetary grants from the Emperor to all disaster victims. As a result, a complete census was conducted in Tokyo Prefecture and the six surrounding prefectures of the Kanto region (Takemura 2023, p. 35).}
Using this survey, Takemura (2003, pp. 29--31) provided estimates of the number of affected people as follows.
The population of Tokyo City prior to the earthquake was roughly $2.27$ million, of whom $1.7$ million were affected by the disaster.
Of the affected population, approximately $70,000$ were killed or went missing, leaving approximately $1.63$ million survivors.
Of the survivors, $1.36$ million lost their homes, and $670,000$ of them evacuated outside Tokyo.
As a result, the population had decreased to approximately $1.53$ million ($2.27$ million minus $700,000$ minus $670,000$) as of November 15.
In other words, Tokyo's population decreased by $740,000$ (of whom approximately $70,000$ were dead or missing) in the two and a half months following the earthquake.

According to Takemura (2023, p.~35), $690,000$ people ($1.36-1.67$ million) who had lost their homes remained in the city.
Those who lost their homes fell into one of the following categories: (1) those using publicly provided temporary housing, (2) those staying with acquaintances or in private residences, or (3) those erecting temporary structures on their own.
As of November 15, $86,000$ people were living in public temporary housing (1), accommodating just over $12$\% ($86,000/690,000$) of those who had lost their homes.
In other words, the remaining $600,000$-plus people fell into categories (2) or (3) as of mid-November.
Regarding case (2), Takemura (2023, pp. 31--34) has compiled several examples: immediately after the disaster, not only the homes of acquaintances but also the mansions of corporate executives were temporarily opened to the public.

As for case (3), I attempt to infer the intentions of the disaster victims who built temporary housing on their own based on the available sources.
In the Tokyo City Office (1932aa, p. 66), a section titled `return of residents to the affected area' states as follows:

\begin{quote}
``...The number of disaster victims returning to the burned area of their former residences is steadily increasing. However, most of these returnees are doing so in conjunction with the construction of temporary shops and residences along the so-called `main thoroughfares' facing the streets, while many of those who lived in rented housing before the earthquake are waiting for new rental properties to be built. Nevertheless, since landlords consider the construction of rental housing to be highly disadvantageous at this time, when city planning and land readjustment have not yet been finalized, and are therefore not rushing to begin construction, the backstreets--which previously had many tenants--generally have few returnees.''
\end{quote}

It appears that, immediately after the hit, homeowners rushed to build temporary housing, while disaster victims who had been renting were forced to rely on connections or other means of support.

\begin{figure}[h]
\centering
\captionsetup{justification=centering,margin=1.5cm}
\includegraphics[width=0.55\textwidth]{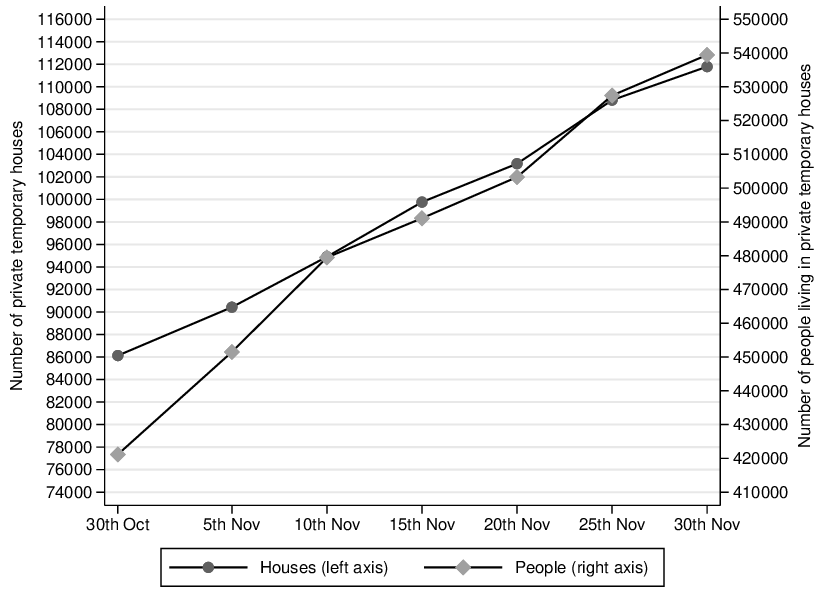}
\caption{Temporary Houses and People between 30th October and 30th November 1923}
\label{fig:private_temporary_houses}
\scriptsize{\begin{minipage}{400pt}
\setstretch{0.85}
Notes:
This figure shows time-series plots of the number of private temporary houses and the number of people living in them in Tokyo City between 30th October and 30th November 1923.
Source: Created by the author using Tokyo City Office (1932aa, p.~66).
\end{minipage}}
\end{figure}

Despite this hardship, the number of privately built temporary housing units rose steadily.
Figure~\ref{fig:private_temporary_houses} illustrates the trends in the number of privately built temporary housing units and the population residing in them.\footnote{In the source material, the statistics are compiled under the title ``Number of Households and Population Returning to Burned Sites'' (Tokyo City Office 1932aa, p.~66). Since this is clearly distinct from publicly built temporary housing, I defined it as privately built housing.}
According to this data, there were already $99,760$ housing units and $491,081$ people living as of November 15.
This represents approximately $80$\% of the slightly over $600,000$ people who had lost their homes, excluding residents of public temporary housing ($86,000$ people).
Two and a half months after the earthquake, one can imagine that a considerable number of privately built temporary housing units had been erected throughout the city.

Immediately after the earthquake, the supply of construction materials decreased, and soaring material prices made it difficult to proceed with construction.
There are two possible reasons why the number of privately built temporary housing units increased so rapidly in that situation.
First, this relates to the definition of ``temporary housing.''
Immediately after the earthquake, people gathered wood on their own to build dwellings; these were often crude structures with walls made of a few pieces of wood, covered with corrugated iron roofs or cloth.
These simple temporary dwellings must be included in the statistics above.
Second, the effectiveness of the government's housing supply policies.
Through the Ordinance Concerning Temporary Buildings within Areas Subject to the City Building Act (Imperial Ordinance No.414) mentioned above, the construction of simple dwellings became possible.
In addition, the government's price controls on construction materials and timber procurement immediately functioned reasonably well (Tokyo City Office 1932aa, p.~68).
This is consistent with the government's domestic timber procurement efforts being successful, leading to a dramatic increase in timber supply starting the month after the earthquake (Section~\ref{sec:seca_wood}).

\begin{figure}[h]
\centering
\captionsetup{justification=centering,margin=1.5cm}
\includegraphics[width=0.55\textwidth]{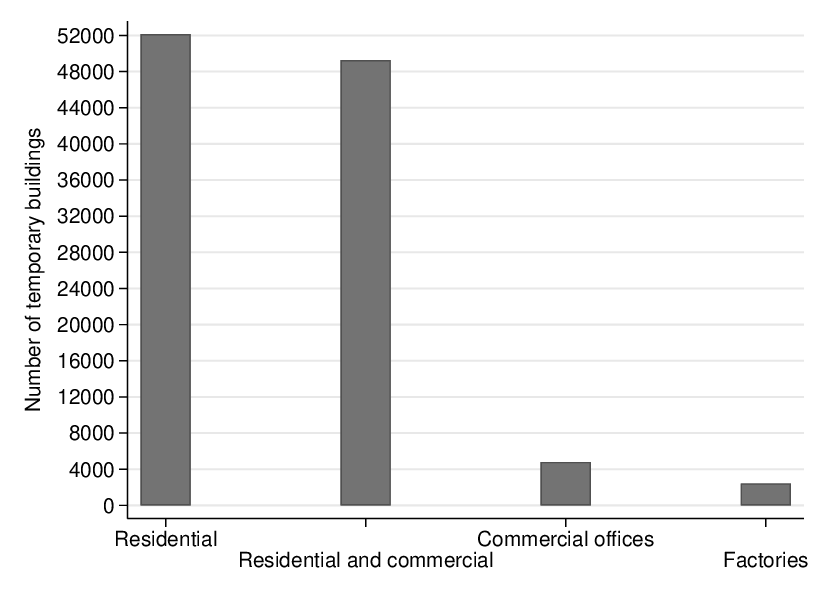}
\caption{Temporary Buildings in Tokyo City on 23rd November 1923}
\label{fig:hist_buildings}
\scriptsize{\begin{minipage}{400pt}
\setstretch{0.85}
Note:
This figure shows the number of temporary buildings categorized by type.
The number of residential buildings is $52,133$.
The number of buildings used for residential and commercial purposes is $49,261$.
The number of buildings used as commercial offices is $4,796$, whereas those used as factories is $2,418$.
The statistics are based on the survey on 23rd November 1923.\\
Source: Created by the author using Tokyo City Office (1932aa, p.~69--70).
\end{minipage}}
\end{figure}

Finally, I examine the types of temporary structures.
Figure~\ref{fig:hist_buildings} shows a histogram calculated using the statistics on temporary structures as of November 23, 1923.
The total number of buildings was $108,000$, of which $101,000$ were residences or residential-commercial buildings.
This figure generally matches the data in Figure~\ref{fig:private_temporary_houses}.
The temporary structures also include commercial offices and factories (Figure~\ref{fig:hist_buildings}).
In other words, a number of companies and factories had resumed economic activity by at least mid-November.

In summary, various temporary structures were scattered throughout the city immediately after the earthquake.
While some privately built temporary housing resembled crude shacks, others, such as the publicly built temporary housing, were relatively functional as dwellings.
Starting at the end of the year, timber began arriving from overseas, and the supply of timber stabilized (Section~\ref{sec:seca_wood}).
As a result, the quality of the temporary structures must have been gradually improved.
From 1924 onward, existing temporary structures were gradually replaced with permanent buildings, following the land adjustments associated with the reconstruction plan (Section~\ref{sec:seca_rec}).
The new layout of the burned area had then been largely finalized around 1930, when the reconstruction plan was completed.

\subsection{History of City Planning in Tokyo}\label{sec:seca_plan}

\subsubsection{City Planning before the Earthquake}\label{sec:seca_before}

In this section, I review the city planning that took place in the city prior to the disaster.
Through this review, I show that the reconstruction plans following the earthquake were fundamentally different from those implemented before the disaster.

The most comprehensive account of Tokyo's history is the Centennial History of Tokyo (\textit{T\=oky\=o Hyakunenshi}), compiled by the Tokyo Metropolitan Government.
This six-volume work (plus one appendix), with each volume exceeding $1,000$ pages, provides a comprehensive record of history from prehistoric times to the present.
I use this document as the primary source, supplemented by other important works such as Fiscal History of the Tokyo Metropolitan Government (\textit{T\=oky\=oto Zaiseishi}).

Following the Meiji Restoration, the name was changed from Edo to Tokyo in 1868.
This marked the establishment of the early Tokyo Prefecture.
The Meiji government, seeking to stabilize the unified local government system, enacted the Three New Laws in 1878.
This includes the Act for the Organization of Counties, Urban Wards, Towns, and Villages (\textit{Gun-ku-ch\=o-son Henesei H\=o}); Regulations for Prefectural Assemblies (\textit{Fuken-kai Kisoku}); Regulations for Local Taxation (\textit{Chih\=o-zei Kisoku}).
Under the Act for the Organization of Counties, Urban Wards, Towns, and Villages, Tokyo Prefecture was divided into 15 wards (\textit{ku}) and 6 districts (\textit{gun}).
The Imperial Constitution was promulgated in 1889, whereas the City and Town and Village Codes (\textit{Shisei-Ch\=osonsei}) and Prefectural and District Codes (\textit{Fukensei-Gunsei}) were enacted in 1888 and 1890, respectively.
Based on this City Code (\textit{Shisei}), the 15 wards within Tokyo Prefecture were reorganized into Tokyo City.

Under the City Code (\textit{Shisei}), the City Council (\textit{Shi Sanjikai})--composed of the mayor (\textit{Shich\=o}), the deputy mayor (\textit{Jyoyaku}), and honorary council members (\textit{Meiyoshoku Sanjikaiin})--held executive authority.
Because Special Provisions of the City Code were applied to Tokyo City, the governor (\textit{Chiji}) and the prefecture secretary (\textit{Fu Shokikan}) concurrently served as the mayor and deputy mayor.
Ward chiefs (\textit{Kuch\=o}) were appointed by the City Council and were designated to assist the prefectural governor, a national bureaucrat, in the city's administrative affairs.
In this sense, Tokyo City's autonomy was initially subject to the supervision of the prefectural government.
With the abolition of the Special Provisions in 1898, restrictions on the city's administrative authority were lifted, and Tokyo City became able to elect its own mayor.\footnote{Furthermore, the Act on the Supervision of Administration in the Six Major Cities was enacted in 1922, eliminating the requirement for the prefectural governor's permission regarding national affairs under the mayor's jurisdiction. As a result, Tokyo City strengthened its autonomy.}
In October 1898, the Tokyo City Office was established within the Tokyo Prefectural Office in Yurakucho.

In the 1860s and 1870s, following the Meiji Restoration, a movement emerged to create Western-style streetscapes in parts of Tokyo.
For example, a foreign settlement was established in Tsukiji in 1869, and the national railway line between Shimbashi and Yokohama opened in 1872.
In Ginza, near Shimbashi, roads were laid out in the 1870s based on a design by the Irish architect Thomas James Waters (1842--1898), and a brick-lined district was created.\footnote{This was a project undertaken by the Ministry of Finance, the Ministry of Public Works, and the Tokyo Prefectural Government following the Great Fire of Ginza in February 1872. The intention was to improve Tokyo's appearance in preparation for treaty revision negotiations with Western nations. The model for Ginza Brick Street is said to have been Regent Street in the UK (Ishizuka and Narita 1986, pp. 23--27).}
Thus, early urban planning in Tokyo was implemented in only a very limited area of the city center. \footnote{Ishizuka and Narita (1986, pp.~24--36) summarize the development of urban areas outside of urban planning during this period. For example, the Nihonbashi area, located near Ginza, was originally a hub for Edo wholesalers and became the center of Japan's economy and finance after the Meiji era. The merchants of Nihonbashi were wholesalers who handled a wide variety of goods and employed numerous servants; they also owned the row houses where their servants lived. They were large-scale landowners and merchants--such as Mitsui, Mitsubishi, Okura, and Yasuda--who would later grow into \textit{zaibatsu}, diversified business conglomerates. For instance, in Kabuto-ch\=o, within the Nihonbashi district, the Mitsui Group established its business headquarters; later, Eiichi Shibusawa opened the First National Bank there, and a stock exchange was also established. Nihonbashi Kabuto-ch\=o remains a financial hub today, home to the Tokyo Stock Exchange (Ishizuka and Narita 1986, pp.~30--32).}

City planning on a larger scale began to take shape in the 1880s.
Between 1880 and 1882, a series of major fires broke out in the central Kanda Ward.
In response, the Tokyo Prefecture implemented several fire prevention projects.
These included slum clearance in Kanda Hashimoto-ch\=o, the improvement of roads and rivers to prevent fires in burned area, and the conversion of building roofs in the four central wards (Nihonbashi, Kyobashi, Kanda, and Kojimachi) to non-combustible materials (Ishida 1987, pp. 53--58).
Although these projects were larger in scale than the city development of the 1870s, they were still limited to only a small portion of central Tokyo.\footnote{Ishida (1987, pp. 69--126) details the slum clearance in Kanda Hashimoto-ch\=o.}
Yoshikawa Masaaki, Governor of the Prefecture, submitted a memorandum on urban reform to the Minister of the Interior in 1884.
In response, the Ministry of the Interior established a committee to review urban reform.
Following deliberations, the Tokyo City Improvement Ordinance (\textit{T\=oky\=o Shiku Kaisei Jy\=orei}) was enacted in 1888 and came into effect in 1889.
The initial plan called for concentrating political and economic functions around the Imperial Palace and included road construction, railway development, canal excavation, and the establishment of parks, markets, crematoriums, cemeteries, and water supply infrastructure, as well as the construction of parks, markets, crematoriums, and cemeteries.
However, due to limited financial resources, the plan itself was scaled back in 1903.
As a result, road improvements were carried out only in Marunouchi and Hibiya on the east side of the Imperial Palace, as well as in the Nihonbashi area, and the urban renewal project was completed in 1914.
Although this plan was larger in scale than the urban planning carried out in the 1860s and 1870s, it ultimately amounted to little more than improvements to the area surrounding the Imperial Palace.

Ishida (1987, pp. 13; 235) notes that, while efforts were made during this period to advance urban infrastructure development through legal frameworks and planning, funding for urban planning--deemed non-essential and non-urgent compared with the policy of enriching the nation and strengthening the military--was severely constrained.
Projects based on the Tokyo City Improvement Ordinance ``lacked a vision to anticipate and plan for urban expansion''.\footnote{The Tokyo City Improvement Ordinance is considered the foundation of Japan's urban planning system, and a body of research has accumulated in the field of architecture. Ishida (1987, p.~33) provides a review of these studies.}

\subsubsection{Imperial Capital Reconstruction Plan}\label{sec:seca_rec}

By the 1910s, the expansion of the city had become pronounced due to industrialization, and the City Planning Act (\textit{Toshi Keikaku H\=o}) and Urban Building Act (\textit{Shigaichi Kenchikubutsu H\=o}) were enacted in 1919.
This introduced planning techniques to regulate urban development, including land readjustment, building lines, and zoning.
However, it was not until the 1930s that measures to control urban expansion were actually implemented.\footnote{An overview of urban planning from the 1930s onward is summarized in Ishida (1987, pp. 235--242).}
This was because the Great Kanto Earthquake of 1923 led to the 1920s urban planning being superseded by reconstruction efforts (Ishida 1987, p. 235).
In other words, modern urban planning in Tokyo City was implemented following the Great Kanto Earthquake.

The post-earthquake Imperial Capital Reconstruction Plan (\textit{Teito Fukk\=o Keikaku}) was a large-scale project aimed at rebuilding the burned area. In fact, the Imperial Capital Reconstruction Plan boldly rebuilt the areas destroyed by the earthquake. Ishida (1987, p.~14) states:

\begin{quote}
``The urban planning for the reconstruction following the Great Kanto Earthquake of 1923--1930 was a major undertaking in the history of modern Japanese urban planning. However, from the perspectives of the development of land readjustment techniques suitable for regulated urban areas and the expansion and strengthening of the urban planning technocratic class, it can be viewed as having powerfully advanced the characteristic of this period: the establishment of the urban planning system.'' 
\end{quote}

Below, I summarize the outline of the reconstruction plan using the History of Land Readjustment in the Imperial Capital Reconstruction Project (\textit{Teito Fukk\=o Kukaku Seiri Shi}), published by the Tokyo City.
This six-volume work is a valuable resource that provides detailed information on the Imperial Capital Reconstruction Plan.

The draft of the Imperial Capital Reconstruction Project, prepared immediately after the earthquake, consisted of four major components.
The first item proposed establishing the Imperial Capital Reconstruction Council (\textit{Teito Fukk\=o Shingikai}) as the body responsible for deliberating on and determining the highest-level policies for the Imperial Capital's reconstruction.
It was promulgated by imperial ordinance on September 19, 1923 (Tokyo City Office 1932aa, p.~84).\footnote{For the composition of the Imperial Capital Reconstruction Council, see Tokyo City Office (1932aa, pp.~109--110). The President was to be the Prime Minister, and the members were to be Ministers of State, appointed officials, and experts.}
In the second item, the establishment of the Imperial Capital Reconstruction Agency (\textit{Teito Fukk\=o-in}) was proposed as the body responsible for planning and executing reconstruction projects, and it was enacted on September 27, 1923 (Tokyo City Office 1932aa, p.~90).
The third item designated long-term domestic and foreign bonds as the financial resources for the reconstruction plan, and the fourth item outlined the general framework for the readjustment of the disaster-stricken areas.
Specifically, a plan was devised for the government to issue public bonds to purchase the entire area and carry out land readjustment, selling and leasing land as necessary (Tokyo City Office 1932aa, p.~82).

However, because the budget was significantly reduced at the 47th Extraordinary Council meeting in December 1923, it became impossible for the national government to implement the entire plan.
As a result, the final reconstruction plan was to be implemented jointly by the national government, Tokyo Prefecture, and Tokyo City (Tokyo City Office 1932aa, p. 379).
The final budget totaled $690,316,115$ yen, comprising $324,987,465$ yen for projects to be implemented by the Minister of the Interior (national government), $22,004,036$ yen for projects to be implemented by the Governor of Tokyo Prefecture, and $343,324,614$ yen for projects to be implemented by the Mayor of Tokyo.\footnote{In addition, the Mayor of Tokyo was responsible for the construction of a branch of the Central Wholesale Market, emergency sewer improvements, iron water pipe construction, and the second phase of reconstruction facilities for the railway transfer project, totaling $53,840,000$ yen; adding this brings the grand total to $744,156,115$ yen (Tokyo City Office 1932aa, p.~404).}
This amounted to approximately $50$\% of the national budget at the time ($700$ million yen out of $1.4$ billion yen).
This illustrates just how large-scale the reconstruction plan was.

The reconstruction plan for Tokyo was broadly divided into `land readjustment,' `improvements to streets, bridges, canals, and parks,' and `other facilities (water and sewer systems, markets, educational and health facilities, social services, and electric utilities)' (Tokyo City Office 1932aa, p. 403).

\begin{figure}[htbp]
\centering
\captionsetup{justification=centering,margin=1.5cm}
\includegraphics[width=0.6\textwidth]{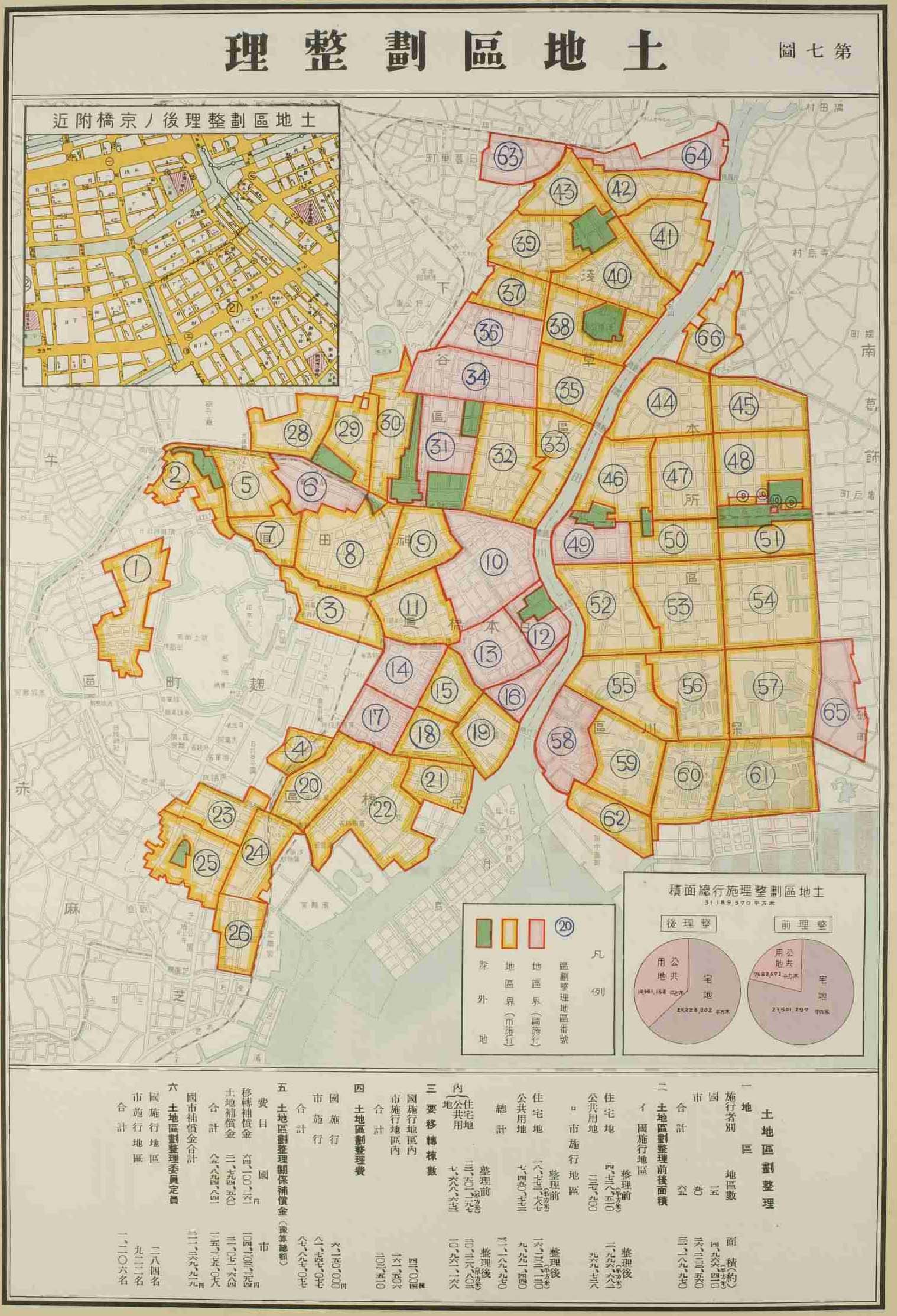}
\caption{66 Districts for Land Readjustment Plan in Tokyo City}
\label{fig:readj_plan}
\scriptsize{\begin{minipage}{400pt}
\setstretch{0.85}
Notes:
This figure shows the districts for the land readjustment in Tokyo City.
The districts colored in pink (yellow) show the districts readjusted by the government (Tokyo City).
The green area indicates the area excluded from the land readjustment plan.\\
Source:  Tokyo City Office (1930z, Figure 7). Tone adjusted by the author using Adobe Photoshop 2024.
\end{minipage}}
\end{figure}

The area destroyed by the fire following the earthquake totaled $10,487,470$ tsubo.
Land readjustment was carried out on $9,493,806$ tsubo, accounting for approximately $91$\% of this total.
The land readjustment project first divided the burned area into $66$ districts, with $15$ districts administered by the Imperial Capital Reconstruction Bureau\footnote{In February 1924, the Imperial Capital Reconstruction Agency (\textit{Teito Fukk\=o-in}) was abolished, and its functions were transferred to the Imperial Capital Reconstruction Bureau (\textit{Teito Fukk\=o-kyoku}), an affiliated bureau of the Minister of the Interior.} and $51$ districts by the Mayor of Tokyo (Tokyo City Land Readjustment Bureau--later the Reconstruction Projects Bureau).
Of the $51$ districts under the jurisdiction of Tokyo City, the Akasaka-Tameike area ($27$ districts: $66,013$ tsubo) was isolated from the other districts and had a generally orderly layout, so it was not included in the project; thus, there were effectively $50$ districts.
The total area of these $50$ districts was $7,533,715$ tsubo (approximately $80$\% of the total land readjustment area) (Tokyo City Office 1932aa, p. 407).
Figure~\ref{fig:readj_plan} shows these $66$ land readjustment districts.

\subsubsubsection{Land Readjustment Project}\label{sec:seca_rec_land}

This section provides an overview of the land readjustment project, a main component of the reconstruction plan.
The key steps in the land readjustment process are summarized as follows.

First, the current status of land and buildings is organized for each district to create a land readjustment plan.
In formulating the land readjustment plan, the Land Readjustment Committee (\textit{Tochi Kukaku Seiri Iin})--composed of members selected from landowners and others in the district--served as the advisory body.
Next, if the total area of residential land after implementation decreases by $10$\% or more compared to the total area before implementation, land compensation is paid for the area (exceeding that $10$\%) (Tokyo City Office 1932ab, pp. 300--301).
In other words, it can be interpreted that up to $10$\% of the land owned was provided free of charge.
The area and price of the land owned were determined based on the land register.
The compensation amount was determined by multiplying the area (exceeding that $10$\%) by the average price of residential land within the district prior to the readjustment.\footnote{Tokyo City Office (1932ab, p. 302) presents the specific calculation formula.}
In the compensation decision-making process, a Compensation Review Board (\textit{Hosh\=o Shinsakai}) functioned alongside the Land Readjustment Committee.
Compensation costs were jointly borne by the national government and the city.\footnote{Tokyo City Office (1932ab, pp. 324--325) presents the specific calculation formula.}

Next, I examine the land readjustment process point by point.\footnote{For this paragraph, I referred to the Tokyo City Office (1932ab, pp.~84--85; 88--96).}
First, the survey of land and land-related rights served as the basis for land exchange planning and land exchange dispositions.
It generally began between April and September 1924.
Surveys of land and landowners were conducted at tax offices and land registry offices.
While land area determinations were based on land register areas and declared leasehold areas, numerous discrepancies between actual measured areas and the land register and declared leasehold areas were noted, leading to corrections of the land register areas at the tax offices.
Consequently, deadlines for correcting land register areas were set for each district.
For applications submitted by the deadline, actual measurements and corrections were carried out in the landowners' presence.
The deadlines for determining land registry areas were generally set between July 1925 and September 1926.
It is noted that adjusting land registry areas took time and hindered the progress of the land readjustment project.

Second, I summarize the process from the creation of the readjustment plan to the decision on readjustment.\footnote{I used Tokyo City Office (1932ab, pp.~126--129; 130--131; 134--136; 202--203; 270--274) for describing this paragraph.}
A large number of personnel were mobilized for the surveying work required to formulate the land readjustment plan (including surveys of existing buildings).
In the mayor-administered area alone, the number of temporary laborers is reported to be $ 676,881$, and the survey was completed in 1924.
The procedures for land readjustment design were carried out in accordance with the circular issued by the Director of the Land Readjustment Department (\textit{Seiri Buch\=o}) of the Imperial Capital Reconstruction Bureau (\textit{Teito Fukk\=o-kyoku}) in June 1924.
Final boundary surveying (\textit{kakutei sokury\=o}) involved driving stakes into the ground and conducting measurements based on the results of the land readjustment design, and it was completed between January 1925 and November 1929.
Concurrently, decisions on land exchange dispositions by Land Readjustment Committees were generally made between April 1927 and February 1930.
Finally, public notices regarding land exchange dispositions in each region were issued between October 1926 and March 1930.

Third, I describe the process from the start of construction to its completion.\footnote{The descriptions in this paragraph are from Tokyo City Office (1932ab, pp. 375--377; 397).}
Construction associated with the land readjustment project was carried out between June 1925 and June 1930, but in most cases, work began around 1927--28 and was completed by the end of 1929.
Records indicate that the relocation of existing buildings also began in 1925 and was completed by the end of 1928.
Therefore, it is reasonable to assume that most land readjustment projects had been completed across all districts by the end of the 1920s.

\begin{figure}[]
\centering
\captionsetup{justification=centering}
\subfloat[Relocation Plan]{\label{fig:relocation1}\includegraphics[width=0.40\textwidth]{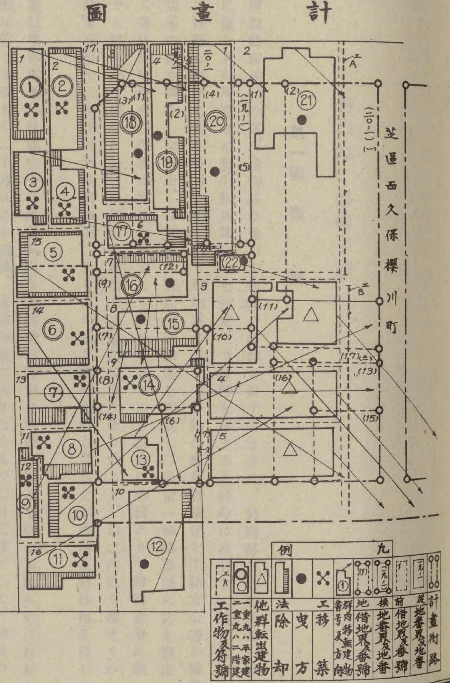}}
\hspace{5pt}
\subfloat[Relocated Plot]{\label{fig:relocation2}\includegraphics[width=0.323\textwidth]{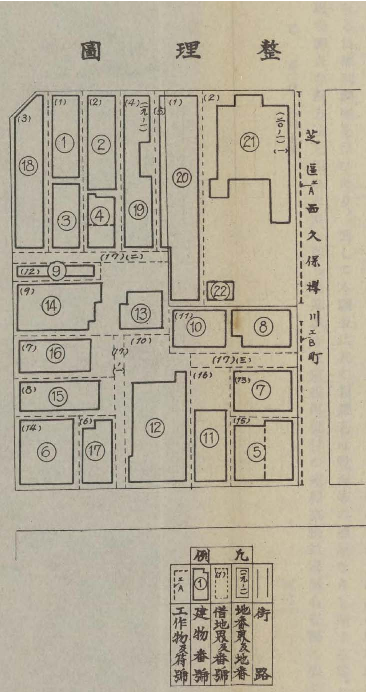}}
\caption{An Example of Relocation of Housing (Group \#66 in Area \#23)}
\label{fig:relocation}
\scriptsize{\begin{minipage}{450pt}
\setstretch{0.9}
Notes:
The left photo shows the relocation plan for Mobile Group \#66 in Area \#23.
The area surrounded by the lines with small white circles indicates the new plots created by land readjustment.
Each small block indicates the house that needs to be relocated.
Each arrow starting from each block indicates the new location.
The sheds in each block show the part of the house that needs to be cut.
The house with a small black circle indicates the building that can be moved without demolition, though with some adjustments.
The house marked with a cross indicates the building that requires demolition to relocate.
The right photo depicts the relocated plot (note that the circled numbers in the houses are identical in both figures).\\
Source: Tokyo City Office (1932ab, p.~416).
Tone adjusted by the author using Adobe Photoshop 2024.
\end{minipage}}
\end{figure}

During the land readjustment process, it became necessary to relocate buildings.\footnote{Tokyo City Office (1932ab, pp.~397--399; 404--410; 441) is used to describe this paragraph.}
The total number of existing buildings was $168,030$, of which approximately $96$\% (161,515 buildings) were deemed necessary to relocate.\footnote{This figure includes the temporary housing (Section~\ref{sec:seca_response}). According to documents summarizing applications for temporary structures within the mayor's enforcement area, $33,562$ new construction applications were filed in the land readjustment districts, of which $27,162$ were approved. For reconstruction, $2,380$ applications were filed, of which $1,734$ were approved. For additions, $11,110$ applications were filed, of which $8,058$ were approved. For repairs and other work, $806$ applications were filed, of which $488$ were approved. Permits were denied in cases where the construction would interfere with roads, canals, parks, or land readjustment, or when the floor area exceeded the permitted limit (Tokyo City Office 1932ab, pp. 639--647).}
The total area subject to relocation is reported to be $2,737,062$ tsubo.
Most of these were wooden structures used as general residences.
Since land readjustment reduced the available residential land, buildings were often moved without being completely dismantled -- with only some parts removed -- when the distance to be relocated was short.
Conversely, when the distance was long, they were typically dismantled, with some parts removed, before being relocated.
The total floor area of the $159,905$ wooden buildings scheduled for relocation was $2,721,666$ tsubo before relocation.
After relocation, the number of buildings was $159,891$, with a total floor area of $2,270,086$ tsubo.
Relocation costs for cases not involving new construction or renovation were borne by the national government and Tokyo City.
The determination of compensation payments was completed by December 1929 at the latest.
Figure~\ref{fig:relocation} shows an example of relocation in Area \#23.

\begin{figure}[]
\centering
\captionsetup{justification=centering}
\subfloat[Dwellings in Shiba Park]{\label{fig:tmp_house1}\includegraphics[width=0.40\textwidth]{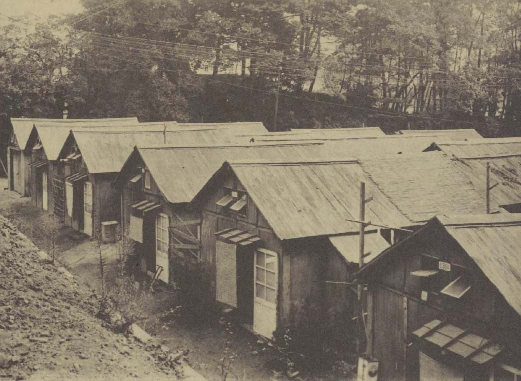}}
\hspace{5pt}
\subfloat[Dwellings in Oshiage]{\label{fig:tmp_house2}\includegraphics[width=0.40\textwidth]{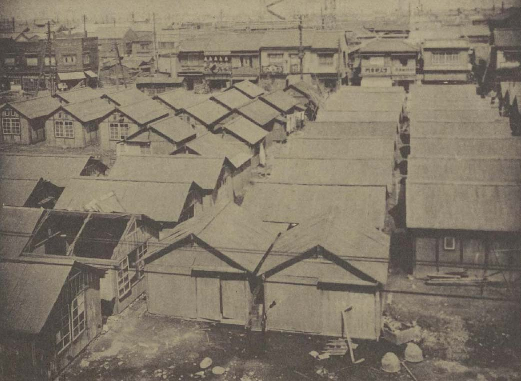}}
\caption{Temporary Dwelling for Relocation}
\label{fig:tmp_house}
\scriptsize{\begin{minipage}{450pt}
\setstretch{0.9}
Notes:
This figure shows example photos of the temporary housing for relocation.
The left photo depicts the temporary housing built in Shiba Park in Shiba Ward.
The right photo depicts the temporary housing in Oshiage Town in Honjyo Ward.
Source: Tokyo City Office (1932ab, p.~photo).
Tone adjusted by the author using Adobe Photoshop 2024.
\end{minipage}}
\end{figure}

In cases where the relocation of existing buildings was required, residents lived in temporary shelters (\textit{rinji sh\=uy\=o naya}).\footnote{For this paragraph, I used Tokyo City Office (1932ab, pp.~591; 592).}
The temporary shelters included both mobile dwellings (i.e., those that were easy to assemble, disassemble, and move) and fixed dwellings.
The city provided 49 fixed dwellings, $2,179$ mobile dwellings, and $570$ mobile storage sheds. 
These temporary dwellings were provided free of charge.
Figure~\ref{fig:tmp_house} shows an example of a mobile dwelling.

\begin{figure}[]
\centering
\captionsetup{justification=centering}
\subfloat[Before Readjustment]{\label{fig:Nihonbashi1}\includegraphics[width=0.450\textwidth]{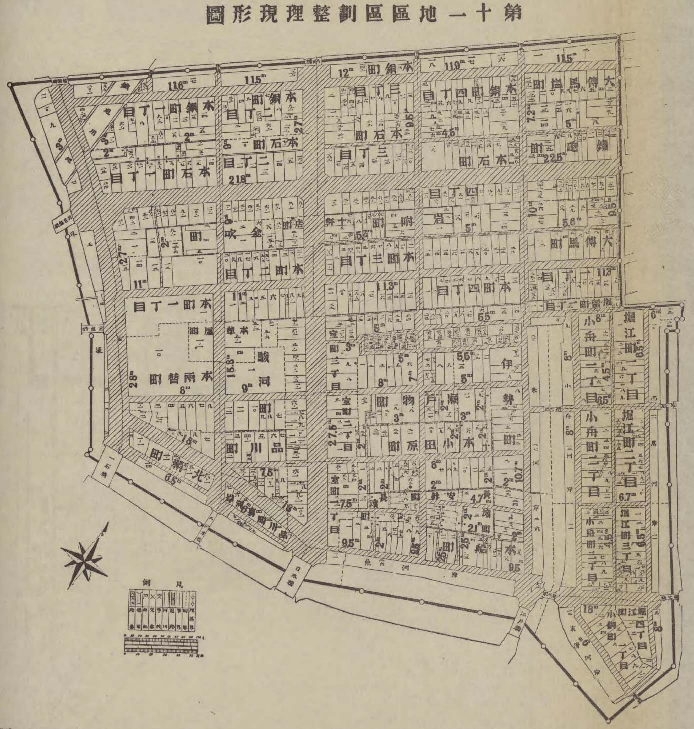}}
\hspace{5pt}
\subfloat[After Readjustment]{\label{fig:Nihonbashi2}\includegraphics[width=0.445\textwidth]{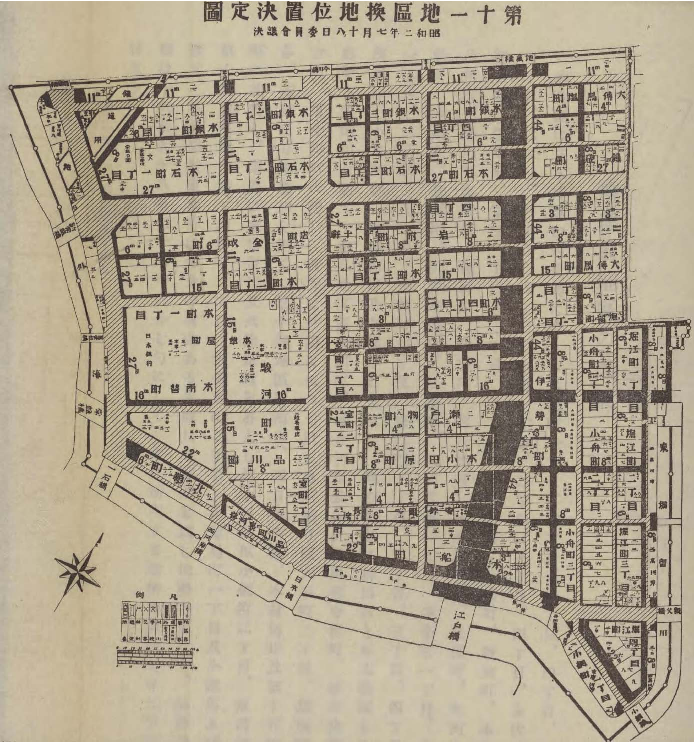}}
\caption{An Example of Readjustment: Area \#12 in Nihonbashi Ward}
\label{fig:readjustment}
\scriptsize{\begin{minipage}{450pt}
\setstretch{0.9}
Notes:
This figure depicts the land readjustment in Area\# 12 in Nihonbashi Ward, which includes the Bank of Japan, Mitsukoshi Department Store, and Nihonbashi Bridge.
Figure~\ref{fig:Nihonbashi1} shows the location of buildings, roads, river, and bridges before the readjustment.
Figure~\ref{fig:Nihonbashi2} shows those after the readjustment.
In Figure~\ref{fig:Nihonbashi2}, the black areas indicate roads newly established or expanded as part of the readjustment process.
Similarly, the shaded area indicates the location of the road that existed before the readjustment.
The most prominent road in Figure~\ref{fig:Nihonbashi2} is the \textit{Sh\=owa-d\=ori} Avenue, the most representative road established as part of the readjustment plan (Figure~\ref{fig:showa_dori}).
\textit{Edo-bashi} Bridge was newly built at the point where the \textit{Sh\=owa-d\=ori} Avenue crosses the river.
The bridge to the left of \textit{Edo-bashi} Bridge is the Nihonbashi Bridge.
The large blocks without houses to the left of the map show the area with the Bank of Japan and Mitsukoshi Department Store.\\
Source: Tokyo City Office (1932ad, p.~805).
The author adjusted the scale and tone using Adobe Photoshop 2024.
\end{minipage}}
\end{figure}

Finally, I summarize the changes in residential lot sizes resulting from the land readjustment.
Since the reconstruction plan included street widening and park development, the size of residential lots generally decreased.
Figure~\ref{fig:readjustment} shows a representative example of land readjustment in Area \#12, which includes a couple of representative buildings, roads, and bridges of Tokyo, such as the Bank of Japan, Showa-dori Avenue, and Nihonbashi Bridge.
Figures~\ref{fig:Nihonbashi1} and~\ref{fig:Nihonbashi2} show the area before and after the readjustment, respectively.
The expanded and newly established roads are painted in black in Figure~\ref{fig:Nihonbashi2}, whereas the old existing part of the roads is shaded in both figures.
Clearly, many roads were added and expanded.
The most obvious one is the broad road called \textit{Sh\=owa-d\=ori} Avenue with \textit{Edo-bashi} Bridge illustrated on the right of the map.

On average, the area of residential land in the $65$ land readjustment zones decreased by $15.3$\% compared to before land grading (Table~\ref{tab:res_area}).
The proportion of residential land within the total area of these zones fell from $75.3$\% to $63.8$\% (Tokyo City Office 1932ab, pp. 204--208).
On the other hand, the area of public land in these zones increased by $47.4$\% compared to before the land was graded.
As a result, the proportion of public land within the total area of these zones rose from $24.6$\% to $36.2$\%.
In short, the land readjustment project generally aimed to improve urban functions by converting part of the residential land into public land and expanding streets, thereby improving traffic conditions and enhancing firefighting capabilities.
In turn, the reduction in residential land area in the burned area meant that the number of dwellings that could be built was physically constrained.

\def\arraystretch{1.00}
\begin{table}[htb]
\begin{center}
\captionsetup{justification=centering}
\caption{Changes in the Area of Residential and Public Land within Land Readjustment Zones: Before and After the Imperial Capital Reconstruction Plan}
\label{tab:res_area}
\footnotesize
\scalebox{0.90}[1]{
\begin{tabular}{lcccccc}
\toprule[1pt]\midrule[0.3pt]
Land&\multicolumn{2}{c}{Before readjustment}&\multicolumn{2}{c}{After readjustment}&\multicolumn{2}{c}{Difference}\\
\cmidrule(ll){2-3}\cmidrule(ll){4-5}\cmidrule(ll){6-7}
&Area&\% of total land&Area&\% of total land&Area&Rate of change (\%)\\\hline
Residential land	&7,101,686&75.3&6,014,824&63.8&-1,086,862&-15.3\\
Public land		&2,316,208&24.6&3,414,789&36.2&1,098,582&47.4\\
\midrule[0.3pt]\bottomrule[1pt]
\end{tabular}
}
{\scriptsize
\begin{minipage}{440pt}
Note: 
The total area within the reorganization zone is $9,427,793$ \textit{tsubo}.
Sources:
The data on the residential land area is obtained from Tokyo City Office (1932ab, pp.204--208).
The data on the area of public land is obtained from Tokyo City Office (1932ab, pp.208--212).
\end{minipage}
}
\end{center}
\end{table}

\subsubsubsection{Street Expansion}\label{sec:seca_rec_land}

Before the Great Kanto Earthquake, many streets in Tokyo City were said to have been ``constructed around Edo Castle during the Edo period to prepare for urban warfare''; consequently, they were ``extremely winding and narrow, with generally narrow right-of-ways'' (Tokyo City Office 1932aa, p.~413).
The Nihonbashi area had been a commercial center since the Edo period and thus has many relatively wide, straight roads.
However, Figure~\ref{fig:Nihonbashi1} shows that there are still numerous back streets that are quite narrow.
To improve this situation, the reconstruction plan called for straightening and widening the streets.
As a result, ``roads prior to the implementation of the reconstruction project... accounted for only $11.7$\% of the city's total area, but after the completion of the reconstruction project… they accounted for $18$\% of the city's total area'', indicating a significant increase in road area (Tokyo City Office 1932aa, p. 414).
Table~\ref{tab:road} shows the average width, length, and area of streets before and after the street reorganization, as determined by a survey.
Road paving was also incorporated into the plan.

\begin{figure}[htbp]
\centering
\captionsetup{justification=centering,margin=1.5cm}
\includegraphics[width=0.6\textwidth]{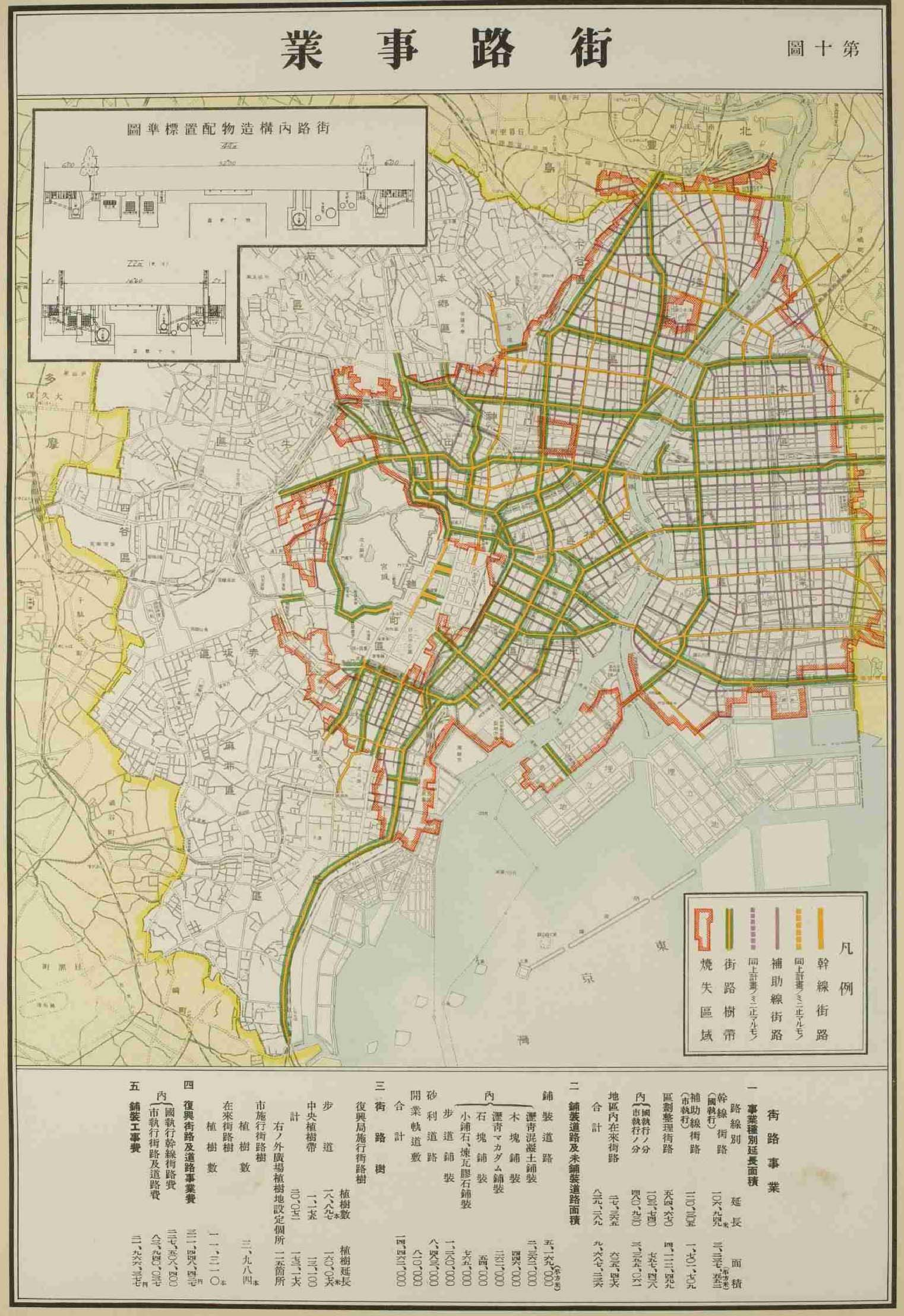}
\caption{Road Expansion Project in Tokyo City}
\label{fig:roads}
\scriptsize{\begin{minipage}{400pt}
\setstretch{0.85}
Note:
This figure shows the roads considered in the road expansion project in the reconstruction plan.
The area surrounded by the red line shows the burned area.
The yellow roads show the main roads.
The yellow and green roads show the main roads with trees.
The thinner roads colored in purple show the supplementary streets.
The dashed lines indicate the planned streets, but are not included in the project.\\
Source:  Tokyo City Office (1930z, Figure 10).
Scale adjusted by the author using Adobe Photoshop 2024.
\end{minipage}}
\end{figure}

\begin{figure}[]
\centering
\captionsetup{justification=centering,margin=1.5cm}
\subfloat[Sh\=owa-d\=ori Avenue]{\label{fig:showa_dori}\includegraphics[width=0.46\textwidth]{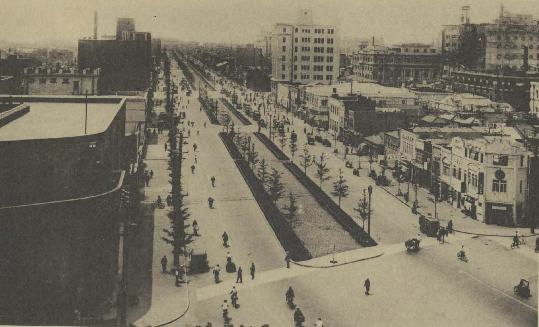}}
\hspace{5pt}
\subfloat[Kiyosubashi Bridge]{\label{fig:kiyosubashi}\includegraphics[width=0.455\textwidth]{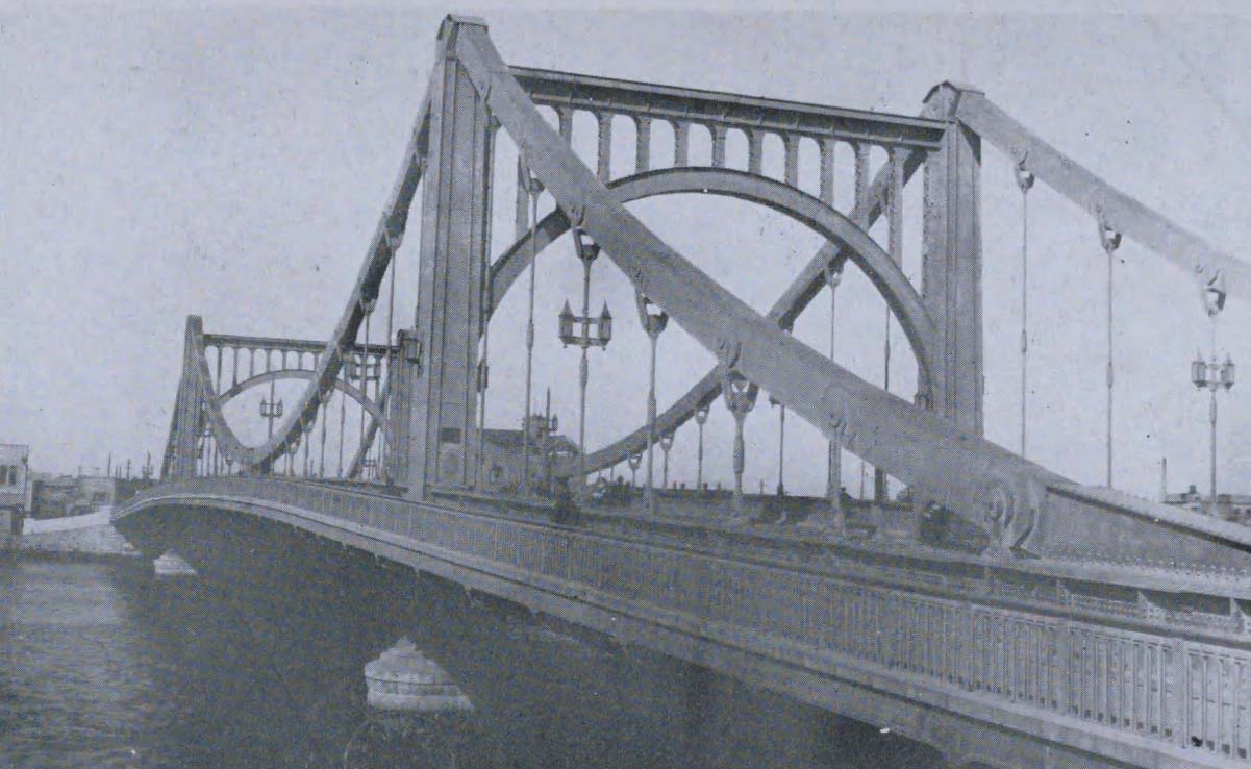}}
\caption{Sh\=owa-d\=ori Avenue and Kiyosubashi Bridge}
\label{fig:road_bridge}
\scriptsize{\begin{minipage}{450pt}
\setstretch{0.9}
Notes:
This figure shows example photos of the road and bridge established as part of the reconstruction plan in Tokyo City.
The left photo depicts the \textit{Sh\=owa-d\=ori} Avenue, labeled as the Arterial Road \#1.
The right photo depicts the \textit{Kiyosubashi} Bridge over the Sumida River (Figure~\ref{fig:bridges}).
Both infrastructures exist today.\\
Sources:
\textit{Sh\=owa-d\=ori}: Tokyo City Office (1932aa, p.~photos)
\textit{Kiyosubashi} Bridge: Association for Reconstruction Research (1930, p.~photos).
The author adjusted the scale using Adobe Photoshop 2024.
\end{minipage}}
\end{figure}

Figure~\ref{fig:roads} shows the spatial distribution of roads that were planned as part of road construction projects.
Despite the air raids during World War II, many of these roads are still in use today.
Figure~\ref{fig:showa_dori} is a photograph of \textit{Sh\=owa-d\=ori} Avenue, a representative arterial road (Arterial Road No. 1), which remains a major north-south thoroughfare running through the center of Tokyo to this day.
In this light, the land readjustment projects created a path dependency in road construction that persists to today.

\def\arraystretch{1.00}
\begin{table}[htb]
\begin{center}
\captionsetup{justification=centering,margin=1.5cm}
\caption{Street Expansion in Tokyo City Before and After the Imperial Capital Reconstruction Plan}
\label{tab:road}
\scriptsize
\scalebox{0.90}[1]{
\begin{tabular}{lcccccc}
\toprule[1pt]\midrule[0.3pt]
&\multicolumn{3}{c}{Before Reconstruction Plan}&\multicolumn{3}{c}{After Reconstruction Plan}\\
\cmidrule(lll){2-4}\cmidrule(lll){5-7}
&Avg. road width ($m$)	&Expansion ($m$)	&Area ($m^{2}$)
&Avg. road width ($m$)	&Expansion ($m$)	&Area ($m^{2}$)\\\hline
Within target area		&9.7&606,100&5,873,259&13.7&729,029&9,992,058\\
Outside target area		&8.4&405,250&3,410,608&8.5&418,528&3,559,874\\
\midrule[0.3pt]\bottomrule[1pt]
\end{tabular}
}
{\scriptsize
\begin{minipage}{440pt}
Notes: 
This survey was conducted by the Tokyo City Public Works Bureau in October 1930.
The term `target area' refers to the area covered by a land readjustment project.
Source: Tokyo City Office (1932aa, p.~414).
\end{minipage}
}
\end{center}
\end{table}

\subsubsubsection{Bridge Improvement}\label{sec:seca_rec_bridge}

A total of $359$ bridges were affected by the earthquake: $289$ were destroyed and $70$ were damaged.\footnote{I used the Tokyo City Office (1932aa, pp.~415--422; 423--454; 458--459) to describe this section.}
Regarding new construction, the city built $57$ bridges and the national government built $1$ bridge.
As for the bridges associated with major arterial roads, the city constructed $129$ bridges associated with medium-width roads, and the national government constructed $96$ bridges.
The plan also included reconstruction work for bridges that had survived the fire but sustained damage.

\begin{figure}[htbp]
\centering
\captionsetup{justification=centering}
\subfloat[Location of Bridges]{\label{fig:bridges_map}\includegraphics[width=0.45\textwidth]{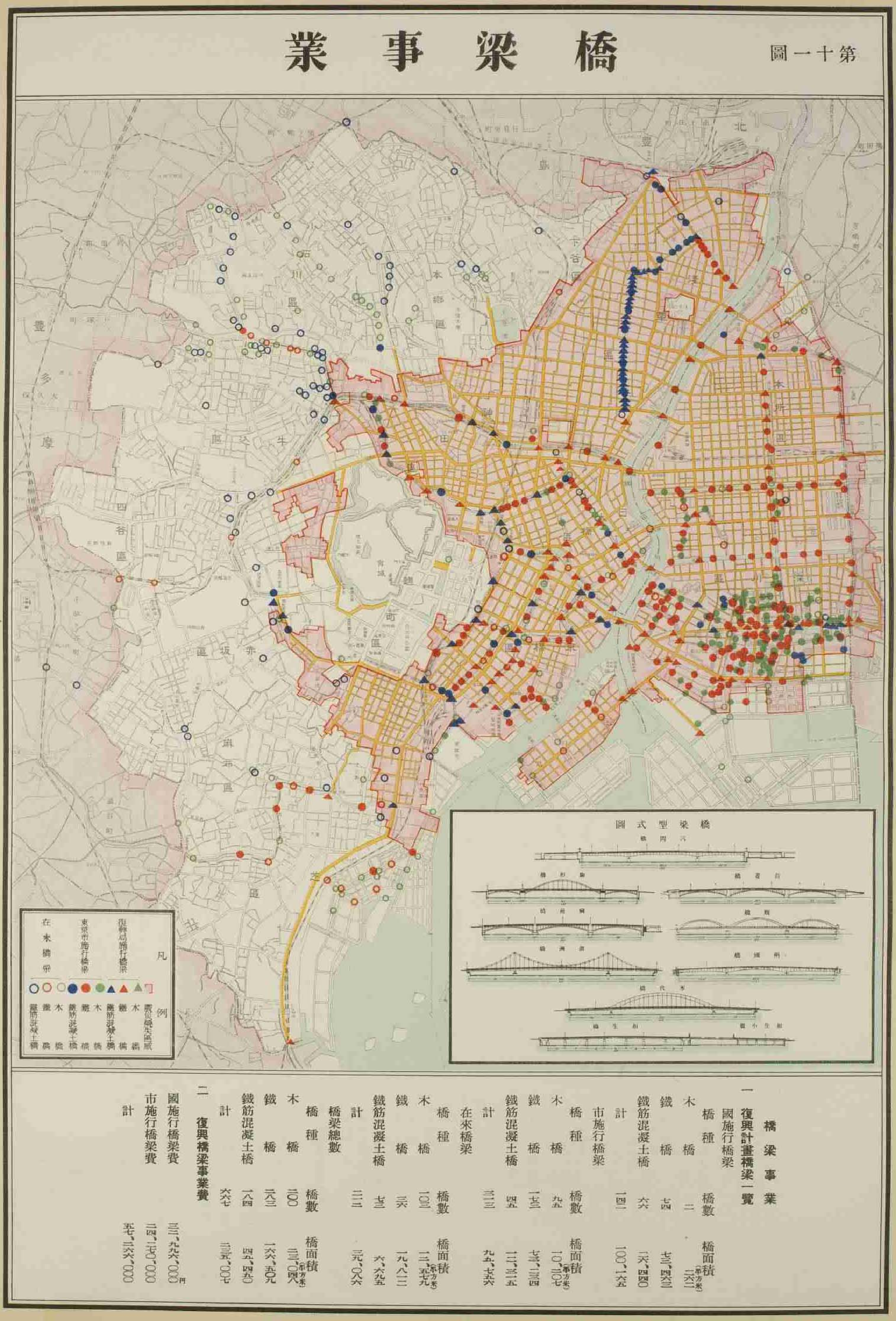}}
\hspace{5pt}
\subfloat[Main Bridges on Sumida River]{\label{fig:bridges_ex}\includegraphics[width=0.45\textwidth]{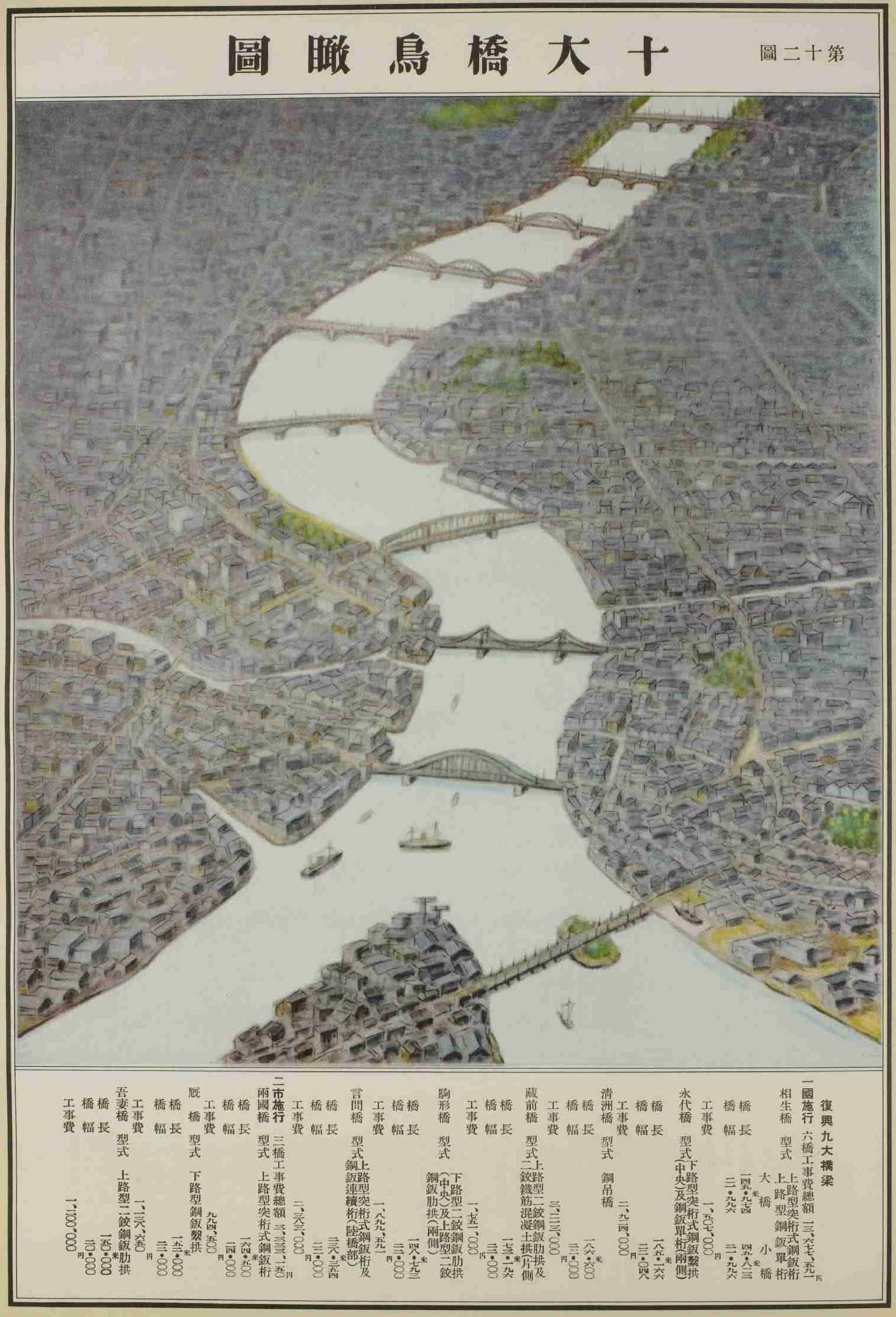}}
\caption{Bridge Improvement Project in Tokyo City}
\label{fig:bridges}
\scriptsize{\begin{minipage}{450pt}
\setstretch{0.9}
Notes:
Figure~\ref{fig:bridges_map} shows the location of bridges considered in the reconstruction project.
The wooden bridges are shown in green, whereas the iron bridges are illustrated in red.
The reinforced concrete bridges are colored in blue.
The bridges constructed and/or reconstructed by the government (Tokyo City) are shown as triangles (shaded circles).
The bridges that existed before the earthquake are shown in a circle.\\
Figure~\ref{fig:bridges_ex} shows an example of the large bridges crossing the Sumida River.
From the bottom to the top, \textit{Aioibashi} Bridge, \textit{Eitaibashi} Bridge, \textit{Kiyosubashi} Bridge, \textit{Ry\=ogokubashi} Bridge, \textit{Kuramaebashi} Bridge, \textit{Umayabashi} Bridge, \textit{Komagatabashi} Bridge, \textit{Azumabashi} Bridge, and \textit{Kototoibashi} Bridge.
All these bridges still exist today.
Sources: Tokyo City Office (1930z, Figures 11--12), website. The author adjusted the scale using Adobe Photoshop 2024.
\end{minipage}}
\end{figure}

Many of these bridges are still in use today, despite the air raids of World War II.
Figure~\ref{fig:bridges_map} shows the spatial distribution of bridges compiled by the city during the reconstruction planning phase.
Most of the bridges were made of steel or reinforced concrete, while wooden bridges were relatively few.
This demonstrates that many bridges were designed to be robust in preparation for disasters.
Figure~\ref{fig:bridges_ex} depicts the large iron bridges spanning the Sumida River.
All of these bridges have been renovated but are still visible today.
Figure~\ref{fig:kiyosubashi} is a photograph of the particularly representative \textit{Kiyosubashi} Bridge, which still retains its original form from the time of its completion.

\subsubsubsection{Other Projects}\label{sec:seca_rec_other}

There were 11 canal renovations, 1 new canal, and 1 land reclamation project, for a total of 13.
There were 18 bridges built over these canals, all of which were constructed by the national government.
Other reconstruction projects included water supply systems, small parks, elementary schools, a municipal hospital, and social welfare facilities.

\subsubsubsection{Opposition Movements}\label{sec:seca_rec_opp}

The land readjustment project began in March 1924.
Initially, opposition to the land readjustment arose.\footnote{About the opposition movements, I referred to Tokyo City Office (1932ab, pp.~2--3; 4). Regarding the land readjustment commissioners, I use Tokyo City Office (1932ab, pp.~35; 36--39; 46--52; 57--60).}
In response, the government had widely publicized the relationship between land readjustment and reconstruction through newspapers and other promotional materials, and held lectures in various locations.
It is reported that as enthusiasm for the Imperial Capital Reconstruction grew among residents, the opposition movement gradually subsided.
As mentioned earlier, most of the land readjustment designs were completed by December 1924.
Despite being a large-scale land readjustment project, the final tally was limited to 65 appeals, 31 administrative lawsuits, and 50 civil lawsuits.

The land readjustment committee played an important role in implementing the land readjustment project.
Article 5 of the Special City Planning Act stipulated that the government and the city were required to seek the opinion of the Land Readjustment Committee when carrying out land exchange dispositions or distributing compensation payments.
As examined in Section~\ref{sec:seca_rec_land}, the Land Readjustment Committee was composed of ``members elected from among landowners and leaseholders within the district (with an equal number of members representing landowners and leaseholders).''

Land Readjustment Commissioners were elected by secret ballot, with elections generally held between May 1924 and February 1925.
When vacancies arose, an equal number of commissioners were elected to fill them.
The total number of committee members was $938$ in the Tokyo City implementation area and $284$ in the Minister of the Interior implementation area.
In contrast, the number of landowners and leaseholders registered on the electoral roll was $39,626$ in the Tokyo City implementation area and $11,246$ in the Minister of the Interior's implementation area.
Therefore, approximately $2.5$\% of all landowners and leaseholders served as committee members.
It is clearly stated that the committee's activities were very frequent.
The committees generally met between June 1924 and February 1930, with each district holding approximately $20$ to $70$ sessions, including both committee and consultative meetings.

These historical facts suggest that the land readjustment project was made possible by the Land Readjustment Committee, which served as a liaison between area residents and the local government.

\subsection{Demand and Supply of Woods after the Earthquake}\label{sec:seca_wood}

In this section, I examine post-earthquake demand and supply for timber--particularly trends in timber used for construction--based on an official report named the Great Kanto Earthquake and Firewood and Charcoal (\textit{Kant\=odaisinsai to Mokuzai Oyobi Shintan} ) published by the Forestry Bureau, Ministry of Agriculture and Commerce in 1924.

The earthquake destroyed not only houses and furniture but also public infrastructure, including schools, bridges, and utility poles.
Many wholesalers who held large inventories of lumber were located in the eastern part of the city, such as Honjo and Fukagawa Wards.
The city's lumber supply was virtually wiped out by the disaster, as the fire burned the main lumber yard in both wards.
The earthquake struck at a time when large quantities of materials had been imported from the United States and Russia and were being stored in preparation for the surge in construction demand expected in early autumn.
The fire destroyed nearly $100$\% of the lumber inventory and more than $60$\% of the building materials inventory in the city.
Only timber stored underwater escaped damage (Forestry Bureau, Ministry of Agriculture and Commerce 1924, pp. 10--13).
Consequently, demand for timber increased dramatically after the disaster.
The most urgent need was to provide temporary housing for disaster victims and affected businesses who had lost their homes (Figure~\ref{fig:tmp_house}).
In response, the national government and Tokyo City collected timber from across the country to meet the surging demand (Forestry Bureau, Ministry of Agriculture and Commerce 1924, pp.~1--3).

\subsubsection{Procurements of Woods}\label{sec:seca_wood_a}

\subsubsection*{Temporary Earthquake Relief Office and Donations}

On September 2, 1923, the day after the earthquake, the government established a Temporary Earthquake Relief Office (\textit{Rinji Shinsai Ky\=ugo Jimukyoku}) (see Section~\ref{sec:seca_response}).
The Materials Division was tasked with procuring lumber and decided to collect building materials for $120,000$ temporary housing units ($3$ tsubo per unit) to serve as shelters for disaster victims.
On September 3, officials were dispatched to various locations throughout the country to procure timber.
The destinations included Akita and Aomori Prefectures, major timber-producing regions; Osaka, Kobe, Nagoya, and Shizuoka, where timber distribution markets were located; and the prefectures adjacent to Tokyo Prefecture, which served as a key transportation hub.

Table~\ref{tab:wood} lists the quantities of lumber procured by region.
It shows that lumber from distant areas, such as Osaka and Aichi Prefectures, was transported by sea, while lumber from prefectures in the Kant\=o region was transported by rail.
As of October 18, $94$\% ($258,000$ \textit{koku}) had arrived, and by mid-December, procurement of nearly all the lumber was complete.
This means that most of the total had arrived within a month and a half after the earthquake.
This is consistent with the fact that the construction of temporary housing in Tokyo City was nearly complete by the end of 1923 (Section~\ref{sec:seca_wood_d}).

\def\arraystretch{1.00}
\begin{table}[htb]
\begin{center}
\captionsetup{justification=centering,margin=1.5cm}
\caption{
Procurement of Lumber by the Temporary Earthquake Relief Office and Tokyo City
}
\label{tab:wood}
\scriptsize
\scalebox{0.9}[1]{
\begin{tabular}{lclclc}
\toprule[1pt]\midrule[0.3pt]
\multicolumn{4}{c}{Provisional Earthquake Relief Bureau}&\multicolumn{2}{c}{Tokyo City}\\
\cmidrule(llll){1-4}\cmidrule(ll){5-6}
\multicolumn{2}{c}{Procurement bu sea}&\multicolumn{2}{c}{Procurement via land transport}&&\\
\cmidrule(ll){1-2}\cmidrule(ll){3-4}
			&Procurement	&			&Procurement		&&Procurement\\
Prefecture		&volume (\textit{koku})	&Prefecture	&volume (\textit{koku}) 	&Source of supply&volume (\textit{koku})\\\hline
Osaka		&58,000		&Tochigi		&37,000		&Aichi		&43,126\\
Aichi			&20,000		&Akita		&54,600		&Shizuoka	&23,612\\
Shizuoka		&60,000		&Chiba		&200			&Osaka		&6,543\\
Aomori		&15,300		&Tokyo Obayashi Forestry Office	&5,000	&Toyama&3,648\\
			&			&(Government-supplied materials)	&		&		&\\
Hyogo		&10,000		&Toyama		&500			&Fukui		&2,247\\
Hokkaido		&9,000		&Nagano		&150			&Nagano		&2,403\\
			&			&Gunmna		&4,000		&Hokkaido	&75,449\\
			&			&			&			&Korea and Andong&84,054\\
			&			&			&			&County		&\\\hline
Total			&172,300		&Total		&101,450		&Total		&241,082\\
\midrule[0.3pt]\bottomrule[1pt]
\end{tabular}
}
{\scriptsize
\begin{minipage}{410pt}
Notes:
Columns 1--4 show the volume of lumber procured reported by the Temporary Earthquake Relief Office.
Although the document originally stated that the total volume procured by sea was $182,300$ \textit{koku}, this has been corrected to $172,300$ \textit{koku} to keep consistency in the descriptions.
Consequently, the total volume procured by sea and land is $273,750$ \textit{koku}.
This lumber was primarily used to construct temporary housing.
Columns 5--6 show the status of lumber procurement by Tokyo City.
While the lumber procured by Tokyo City was primarily used for city-owned buildings, some of that was sold at distribution centers within the city.
Source: Forestry Bureau, Ministry of Agriculture and Commerce (1924, pp.~48--49; 64--65).
\end{minipage}
}
\end{center}
\end{table}

In addition to supplies procured from various parts of the country, donations of lumber were also received from abroad.
Approximately $110,000$ \textit{koku} were donated by the United States, the Philippines, Canada, and others.
Of which, $68,000$ \textit{koku} were donated to Tokyo Prefecture.
Donations from other prefectures within Japan and from private individuals totaled approximately $77,000$ \textit{koku} (Forestry Bureau, Ministry of Agriculture and Commerce 1924, pp.~49--50).

\subsubsection*{Imperial Capital Reconstruction Agency}

The Imperial Capital Reconstruction Agency (\textit{Teito Fukk\=o-in}) was a government body established on September 27 under the Imperial Capital Reconstruction Agency Ordinance.\footnote{Again, the Imperial Capital Reconstruction Agency (\textit{Teito Fukk\=o-in}) was abolished in February 1924. The Imperial Capital Reconstruction Bureau (\textit{Teito Fukk\=o-kyoku}) then assumed its functions.}
All timber purchased by the Reconstruction Agency was held and managed by its Bureau of Material Supply.
The Reconstruction Agency contracted to purchase $95,000$ \textit{koku} of timber domestically and $1,040,000$ \textit{koku} from overseas, primarily from the United States and Canada.
Deliveries of the procured timber began in December from domestic sources and in January 1924 from overseas sources.
It is reported that all timber from North America had arrived at the Port of Yokohama by June 1924.
The construction of temporary housing in Tokyo City was nearly complete by the end of 1923 (Section~\ref{sec:seca_wood_d}).
Therefore, the timber procured by the Reconstruction Agency must be used to implement the reconstruction plan rather than to construct temporary housing.

\subsubsection*{Tokyo City}

Tokyo City handled the consignment sale of lumber procured by the Temporary Earthquake Relief Office.
In addition, the city procured approximately $240,000$ \textit{koku} of lumber from markets across the region for use in constructing public buildings (such as schools and government offices) and for distribution to the general public.
According to statistics measured in April 1924, approximately $60,000$ \textit{koku} of the procured timber was used directly by the city for the construction of city-owned buildings, approximately $97,000$ \textit{koku} was sent to distribution centers and sold to citizens, and the remainder remained unprocessed (Forestry Bureau, Ministry of Agriculture and Commerce 1924, pp. 65).
It is interesting that the city procured timber for sale to citizens.
This suggests the need to cope independently with soaring market prices.

\subsubsection*{Procurement Through the Market}

As already summarized, the lumber procured by the government was used to construct temporary housing and to implement reconstruction policies.
Meanwhile, the lumber procured by Tokyo City was used to rebuild city-owned public facilities, and a portion was sold to citizens.
Combining the timber procured domestically by the Temporary Earthquake Relief Bureau ($270,000$ koku) with donated timber ($190,000$ koku) brings the total to $460,000$ \textit{koku}.
In addition, the Imperial Capital Reconstruction Agency procured approximately $1.15$ million koku, and Tokyo City procured $240,000$ \textit{koku}.

According to government estimates, however, the value of houses lost in Tokyo City due to the earthquake, when converted to timber, amounted to over $13$ million koku (Forestry Bureau, Ministry of Agriculture and Commerce 1924, pp. 8--9).
In other words, the timber procured through the policies amounted to less than $15$\% ($185/1300$) of the timber lost in the earthquake.
Since the number of dwellings decreased after the earthquake due to population outflow from the city, it was likely unnecessary to compensate for all these losses.
Furthermore, the statistics cited in this report may not be exhaustive.
However, it may be clear that much of the timber in the disaster-stricken area was likely supplied through the market.
Since the official report does not provide a comprehensive description of the lumber market, I support this point by examining the volume of timber imported into Tokyo City in Section~\ref{sec:seca_wood_trans}.

\subsubsection{Transportation of Woods}\label{sec:seca_wood_trans}

\begin{figure}[]
\centering
\captionsetup{justification=centering,margin=1.5cm}
\includegraphics[width=0.6\textwidth]{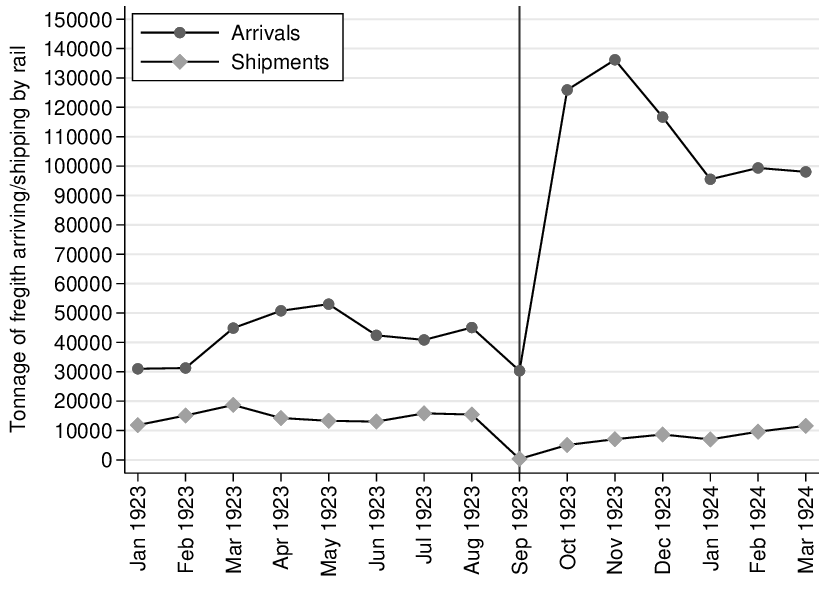}
\caption{Tonnage of Wood Freight Arriving/Shipping at Stations in Tokyo from January 1923 to March 1924}
\label{fig:wood_transport_rail}
\scriptsize{\begin{minipage}{350pt}
\setstretch{0.85}
Notes:
This figure shows the tonnage of wood freight arriving at Shibaura Port in Tokyo from January to December 1923.
Source: Created by the author using Forestry Bureau, Ministry of Agriculture and Commerce (1924, p.~85).
\end{minipage}}
\end{figure}

Figure~\ref{fig:wood_transport_rail} shows the volume of timber arriving at and departing from $23$ railway stations in Tokyo and its suburbs.
The volume of arrivals decreased in September, immediately following the earthquake.
By the following month of October, however, the volume of timber arriving had begun to exceed three times the pre-earthquake level.\footnote{Immediately after the earthquake, as the railways were temporarily out of service, approximately $40,000$ koku of timber was delivered to Tokyo via small boats traveling down rivers such as the Sumida River (Forestry Bureau, Ministry of Agriculture and Commerce 1924, p.~82). Even on the railways that were operational, the transport of relief supplies took precedence until mid-October, and almost no timber arrived until then (Forestry Bureau, Ministry of Agriculture and Commerce 1924, pp. 132--133).}
Although the tonnage of arrivals had settled in 1924, it remained high compared to before the earthquake.
In particular, the high tonnage of arrivals through the end of 1923 is consistent with most temporary housing construction taking place that year.\footnote{The timber procured by the Temporary Earthquake Relief Bureau was transported free of charge. It is unclear whether this timber is reflected in Figure~\ref{fig:wood_transport_rail}. However, the trend shown in this figure is consistent with statistics on the number of freight cars, excluding those used to transport timber procured by the Temporary Earthquake Relief Bureau (Forestry Bureau, Ministry of Agriculture and Commerce 1924, pp. 135--136).}
Shipment volumes also returned to pre-earthquake levels by around March 1924.

\begin{figure}[]
\centering
\captionsetup{justification=centering,margin=1.5cm}
\includegraphics[width=0.6\textwidth]{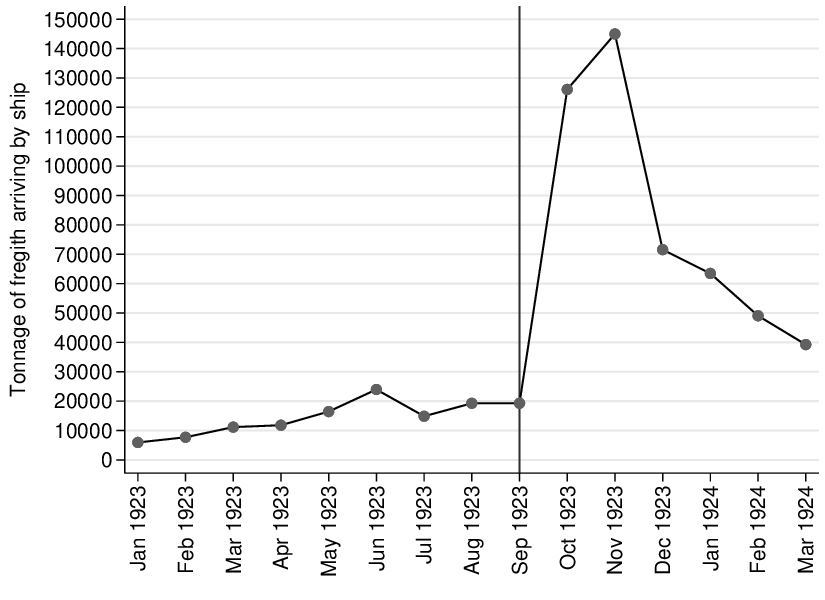}
\caption{Tonnage of Wood Freight Arriving/Shipping at Shibaura Port in Tokyo from January 1923 to March 1924}
\label{fig:wood_transport_ship}
\scriptsize{\begin{minipage}{350pt}
\setstretch{0.85}
Note:
This figure shows the tonnage of wood freight arriving at Shibaura Port in Tokyo from January 1923 to March 1924.
Source: Created by the author using Forestry Bureau, Ministry of Agriculture and Commerce (1924, p.~85).
\end{minipage}}
\end{figure}

Figure~\ref{fig:wood_transport_ship} shows the status of timber imports at Shibaura Port in Shiba Ward in Tokyo.\footnote{The ports of Tokyo and Yokohama had served as ports of entry for timber even before the earthquake, yet they exported almost no timber.}
Although imports did not increase immediately in September, the volume of imports rose to more than six times the pre-earthquake level in October and peaked in November 1923.
This is thought to reflect the arrival of timber procured domestically by the Temporary Earthquake Relief Bureau (Section~\ref{sec:seca_wood_a}).
Although the volume of imports decreased slightly in December 1923, it remained four times the pre-earthquake level.
As noted in Section~\ref{sec:seca_wood_a}, this reflects the arrival of timber ordered by the Imperial Capital Reconstruction Agency from both domestic and international sources.
Shipping rates rose temporarily both domestically and internationally after the earthquake, but are said to have returned to normal levels by March 1924 (Forestry Bureau, Ministry of Agriculture and Commerce 1924, pp.~139--144).

According to the report, $8$ million koku of timber had been delivered to Tokyo City as of the end of March 1924.
Of these, $2.2$ million \textit{koku} were imported, and $5.8$ million koku were transported from within the country (Forestry Bureau, Ministry of Agriculture and Commerce 1924, pp. 90--91).
In other words, more than $60$\% ($800/1300$) of the timber lost in the disaster was procured within seven months of the event.
Given that $1.85$ million \textit{koku} of timber was procured through government policy, it appears that approximately $77$\% ($615/800$) of the total volume procured by the end of March 1924 was obtained through the market.

In summary, while the timber used for the construction of temporary housing and public facilities was procured through government policy, the majority was ultimately sourced on the market.

\subsubsection{Wholesalers and Retailers}\label{sec:seca_wood_b}

As mentioned earlier, timber wholesalers were concentrated in the burned area of Honjo and Fukagawa.
Of the $420$ wholesalers that existed before the earthquake, $418$ were affected by the disaster.
Although retail stores were more geographically dispersed than wholesalers, $222$ of the $446$ timber retail stores were destroyed.
Although most of them had resumed operations by the end of December 1923, the timber market remained inoperative for several months following the earthquake.
Furthermore, the timber yard in Fukagawa was destroyed, and rivers and canals were blocked by debris, including burned boats and household goods, making sea transport difficult.
Of the timber wholesalers affected by the disaster, $256$ (roughly $60$\%) had resumed operations by mid-November 1923, two and a half months after the earthquake.
Furthermore, by the end of March 1924, nearly all $398$ businesses had resumed operations (Forestry Bureau, Ministry of Agriculture and Commerce 1924, pp. 21--27).

Importantly, the timber market in Tokyo City was virtually non-functional for the first two to three months following the earthquake.
During this period, timber distribution relied entirely on supplies from the national government and Tokyo City.
It was not until around March 1924 that both the number of wholesalers and the number of retail stores returned to pre-earthquake levels.
By that time, market functions must have recovered to some extent.

\subsubsection{Wood Processing Factories}\label{sec:seca_wood_c}

Prior to World War I, it was common practice in Tokyo City to import lumber rather than raw timber.
From 1920 onward, as imports of timber from the United States and Russia increased, the number of sawmills in the city grew to meet the demand.
It is reported that there were $175$ sawmills in Tokyo prior to the earthquake.

Since these sawmills were located in Honjo and Fukagawa alongside timber wholesalers, they could not escape destruction in the fire.
Before the earthquake, there were $93$ major sawmills with a total horsepower of $6,900$,
$73$ of these were damaged in the disaster, reducing the total horsepower to $400$ (Forestry Bureau, Ministry of Agriculture and Commerce 1924, pp. 13--14).\footnote{No changes in horsepower are recorded for the $82$ small-scale mills other than the major ones. It is likely that these mills were workshops with low horsepower.}
The report states:
\begin{quote}
``The revival of the Tokyo timber market can only be achieved by restoring sawmills, processing the remaining materials from the fire into lumber, and using it for the construction of temporary sales offices and sales. Therefore, through sawmill operators and lumber merchants, efforts were immediately launched to simultaneously resume operations at existing sawmills and establish new ones (Forestry Bureau, Ministry of Agriculture and Commerce 1924, p. 27).'' 
\end{quote}
The government determined that the resumption of sawmill operations was essential for the construction of temporary structures.

The government facilitated the procurement of equipment and machinery, and the Industrial Bank of Japan (\textit{k\=ogy\=o gink\=o}) authorized low-interest loans.
Then, six temporary factories began operations in Kiba in Fukagawa Ward in mid-October 1923.
From November onward, the number of temporary factories gradually increased, and lumber production capacity is said to have ``significantly improved.''
By the end of December, the number of sawmills in Tokyo City (though many were temporary) had reached approximately twice the pre-earthquake level ($268$ mills), and total horsepower rose to $7,100$ horsepower.
By the end of 1923, levels had returned to pre-earthquake levels (Forestry Bureau, Ministry of Agriculture and Commerce 1924, p.~28).

As noted in Section~\ref{sec:seca_wood_b}, lumber procurement in 1923 was led by the national government and Tokyo City.
Similarly, the reconstruction of sawmills through the end of 1923 was led by the government.
In other words, the government took the lead in lumber procurement and processing during the first 3--4 months following the earthquake, driven by a strong sense of urgency.
By the end of 1924, the number of sawmills in Tokyo had reached $430$, with a total horsepower of $12,000$ (Forestry Bureau, Ministry of Agriculture and Commerce 1924, p.~28).

\subsubsection{Price of Wood in Tokyo}\label{sec:seca_wood_price}

Wholesale timber prices in November 1923 rose by an average of approximately $170$\% compared to those immediately prior to the disaster (Forestry Bureau, Ministry of Agriculture and Commerce 1924, p. 121).
It is striking that such a significant increase in wholesale prices was observed despite government price controls.
The prices peaked around November 1923, when imported timber had not yet arrived, and had almost returned to pre-earthquake levels by early May 1924.

In general, systematic statistics on retail prices are not available (Hunter and Ogasawara 2019).
Statistics on lumber, which does not fall into the category of everyday goods, are particularly hard to come by.
However, the report states that retail prices for lumber in Tokyo followed the same trends as wholesale market prices (Forestry Bureau, Ministry of Agriculture and Commerce 1924, p. 126).
Due to government price controls, retailers found it difficult to adjust prices freely.
In addition, retailers' profit margins were not large because timber shipments began only after the controls were lifted.

Of course, since this information is based on government reports, one cannot rule out the possibility that actual retail prices rose even higher.
However, the key point is that timber prices rose only temporarily immediately after the earthquake and began to decline after the start of the new year.
This confirms that the Tokyo timber market functioned normally, except for the immediate aftermath of the disaster.
The fact that the rise in timber prices did not persist over the long term would be crucial to Tokyo's recovery.

\subsubsection{Temporary House Construction Project}\label{sec:seca_wood_d}

Immediately after the fire, people gathered in public squares such as those in front of the Imperial Palace, Ueno, Asakusa, and Hibiya.
They subsequently moved to the homes of acquaintances and relatives, both within and outside the prefecture.
Since many had no place to live, however, they ended up staying temporarily in public facilities such as schools or large private residences.
The Temporary Earthquake Relief Office was compelled to construct and manage temporary housing (called barracks) to accommodate these evacuees (Forestry Bureau, Ministry of Agriculture and Commerce 1924, p. 39).
As examined in Section~\ref{sec:seca_response}, the government issued Imperial Ordinance No. 414 on September 16, 1923, allowing people to construct temporary housing that did not comply with conventional building codes.
These temporary housing units were generally intended to be built in burned area to benefit disaster victims (Forestry Bureau, Ministry of Agriculture and Commerce 1924, pp. 45--46).

Temporary housing was built in Shiba Riky\=u, Aoyama Gaien, and Ueno, Asakusa, and Shiba Parks.
Construction was rushed in cooperation with the Tokyo Prefectural Government and the Metropolitan Police Department.
They also constructed public- and privately funded temporary housing, and $138,756$ units had been provided in Tokyo by the end of November 1923 (Forestry Bureau, Ministry of Agriculture and Commerce 1924, p. 41).
Some of the disaster victims also rebuilt their homes on their own.
To assist these victims, the city expedited the reconstruction of housing by selling timber procured by the city directly to them (Forestry Bureau, Ministry of Agriculture and Commerce 1924, p. 40).

Most of the temporary housing construction was completed by the end of 1923.
Accordingly, the proportion of temporary housing units to the reported number of affected households was $47$\% at the end of November 1923, $54$\% at the end of December, $57$\% at the end of January 1924, and $61$\% at the end of February (Forestry Bureau, Ministry of Agriculture and Commerce 1924, p. 41).
Some disaster victims moved within or outside the prefecture, and others were rebuilding their homes after the earthquake.
Thus, it is reasonable that the number of affected households and the proportion of temporary housing supply do not align (Section~\ref{sec:seca_housing}).

The completion of most temporary housing by the end of 1923 is consistent with the descriptions in Section~\ref{sec:seca_wood_b} and Section~\ref{sec:seca_wood_c}.
The government and the city procured timber from domestic and international sources and rapidly rebuilt sawmills.
The city then constructed temporary housing with remarkable organization within just three to four months of the earthquake.
Importantly, these temporary housing units included housing for shopkeepers (Section~\ref{sec:seca_housing}; Forestry Bureau, Ministry of Agriculture and Commerce 1924, pp.~43--44).
Even though these were temporary housing units, residents had access to an environment where they could purchase items necessary for daily life.

\clearpage
\section{Theoretical Framework Appendix}\label{sec:secb}
\setcounter{figure}{0} \renewcommand{\thefigure}{B.\arabic{figure}}
\setcounter{table}{0} \renewcommand{\thetable}{B.\arabic{table}}

\subsection{Setting}\label{sec:secb1}

Theoretical research on residential choice within cities has accumulated in the field of urban economics.
Based on the theory of land use proposed by Alonso (1964), theories of land rental and residential choice in cities have been studied (e.g., Mills 1967; Beckmann 1969; Montesano 1972).
Straszheim (1987) and Fujita (1989) provide comprehensive reviews of this theory.
Below, I outline the theoretical background that determines population distribution in a monocentric model, drawing on the simplified model of Wheaton (1974).

Consider a closed monocentric city with its economic activity centered in the city center; that is, the number of households, $n$, is fixed, and the equilibrium utility level is determined endogenously.
The $n$ households (or consumers) have identical preferences and commute to the city center, where they earn income ($y$).
Given their income, households consume housing ($s$) and a composite good ($x$).
Both goods are normal goods, and the household's utility function is given by:
\begin{align}\label{utility}
u(x, s).
\end{align}
The utility function is assumed to be strictly increasing in each good, twice differentiable, and strictly quasi-concave.
Commuting from a residence at distance $m \geq 0$ to the city center incurs a commuting cost, $C(m)$, which is strictly increasing in distance ($\partial C/\partial m > 0$).
The price per unit area of housing, $R(m)$, depends on the residential location, and the composite good is the num\'eraire.
As in Wheaton (1974), I treat housing as land, so that $s$ represents the land area consumed by a household.
The household's budget constraint is then given by:
\begin{align}
sR(m) = y - C(m) - x.
\end{align}
This implies that the household faces a trade-off between the size of the home and its proximity to the city center where it works.
Under this trade-off, households choose a location that affects both rent and commuting costs to maximize utility:
\begin{equation}
\begin{aligned}
&\max_{x, s} &&u(x, s) \\
&\textrm{subject to} &&sR(m) = y - C(m) - x. \nonumber
\end{aligned}
\end{equation}
The first-order conditions imply the following relationship:
\begin{align}\label{mrg}
\frac{\partial u / \partial s}{\partial u / \partial x} = \frac{y-C(m)-x}{s}.
\end{align}
Since households have identical preferences and can move freely within the city, all households achieve the same level of utility, $\hat{u}$, regardless of their residential location in the residential equilibrium.

\subsection{Bid Rent and Housing Size Functions}\label{sec:secb2}

The maximum rent per unit area of housing that a household is willing to pay at distance $m$ while achieving a given utility level $u$ is called the bid rent.
The bid rent function (Alonso 1964) is defined as follows:
\begin{align}\label{brf_def}
R(m, u, y) = \max_{x, s} \left[\frac{y-C(m)-x}{s} \middle| u(x, s) = u \right].
\end{align}
That is, for a household consuming $(x, s)$ to achieve utility level $u$ at distance $m$, the amount payable per unit area is $(y-C(m)-x)/s$.
Let $(x^{*}, s^{*}) = (x(m, \hat{u}, y), s(m, \hat{u}, y))$ denote the consumption bundle that solves this problem at the equilibrium utility level $\hat{u}$, obtained from Equation~(\ref{mrg}) together with $u(x, s) = \hat{u}$.
The equilibrium bid rent is then expressed as:
\begin{align}\label{brf}
\hat{R}(m, \hat{u}, y) \equiv R(m, \hat{u}, y) = \frac{y-C(m)-x(m, \hat{u}, y)}{s(m, \hat{u}, y)}.
\end{align}

First, I analyze the relationship between market rental prices and residential locations.
Applying the envelope theorem to Equation~(\ref{brf_def}), one obtains the following relationship:
\begin{align}\label{brf_distance}
\frac{d\hat{R}(m, \hat{u}, y)}{dm} = - \frac{\partial C(m)/\partial m}{s(m, \hat{u}, y)} < 0.
\end{align}
Equation~(\ref{brf_distance}) suggests that market rental prices decrease as the distance from the city center increases: higher commuting costs in the suburbs are offset by lower rents.

Next, I analyze the relationship between housing size and the location of residence.
Because utility is fixed at $\hat{u}$ in equilibrium, the optimal housing size can be expressed as a Hicksian (compensated) demand function evaluated at the housing price $\hat{R}$ and utility level $\hat{u}$:
\begin{align}\label{hicks}
s(m, \hat{u}, y) = s^{H}\bigl(\hat{R}(m, \hat{u}, y), \hat{u}, y\bigr).
\end{align}
Note that housing size depends on distance $m$ only through the housing price $\hat{R}$.
By the compensated law of demand, Hicksian housing demand decreases in its own price, $\partial s^{H}/ \partial \hat{R} < 0$.
This property reflects the substitution effect alone and follows from the strict quasi-concavity of the utility function.
Differentiating Equation~(\ref{hicks}) with respect to $m$ and using Equation~(\ref{brf_distance}), one obtains the following relationship:
\begin{align}\label{size_distance}
\frac{d s(m, \hat{u}, y)}{d m}
= - \frac{\partial s^{H}(\hat{R}, \hat{u}, y)}{\partial \hat{R}} \, \frac{\partial C(m)/\partial m}{s(m, \hat{u}, y)} > 0.
\end{align}
Equation~(\ref{size_distance}) suggests that housing size increases as the distance from the city center increases.

\subsection{City Boundary and Household Density}\label{sec:secb3}

To analyze the relationship between distance from the city center and household density, I introduce the city boundary.
Suppose that land beyond the city boundary ($m^{*}$) is used for agricultural production that generates an agricultural rent ($R^{*} \geq 0$) per unit area.
The following condition then holds at the city boundary:
\begin{align}\label{boundary}
R(m^{*}, \hat{u}, y) = R^{*},
\end{align}
and the market rental price is given by:
\begin{equation*}
\hat{R}(m, \hat{u}, y) =
\begin{cases}
R(m, \hat{u}, y) & \text{if} \quad m \leq m^{*}, \\
R^{*} & \text{if} \quad m > m^{*}.
\end{cases}
\end{equation*}
Let $\phi(m) \in (0, 1]$ denote the share of land at distance $m$ available for residential use; the remainder is used for roads, parks, rivers, and other public purposes.
The standard model with a featureless circular plain corresponds to $\phi(m) = 1$, and $\phi(m)\,2\pi m$ is a general land distribution function (Fujita 1989).
Since all residential land within the city is consumed, the total number of households is expressed as follows:
\begin{align}\label{hhs}
\int_{0}^{m^{*}} \frac{\phi(m)\, 2 \pi m}{s(m, \hat{u}, y)}\,dm = n.
\end{align}
Note that $\phi(m)$ does not enter the household problem; it affects the equilibrium only through this land-market condition.
Thus, Equations~(\ref{brf_def})--(\ref{size_distance}) hold for any $\phi(m)$.
Equations~(\ref{boundary}) and (\ref{hhs}) jointly determine the equilibrium pair $(\hat{u}, m^{*})$.

As suggested by Equation~(\ref{size_distance}), housing size increases toward the city boundary.
Let $\Pi(m)$ denote the residential density per unit of residential land:
\begin{align}\label{rd}
\Pi(m)=\frac{1}{s(m, \hat{u}, y)}.
\end{align}
Differentiating Equation~(\ref{rd}) with respect to $m$ and using Equation~(\ref{size_distance}), one obtains:
\begin{align}\label{rd_distance}
\frac{d\Pi(m)}{dm} = -\frac{1}{s(m, \hat{u}, y)^{2}}\,\frac{ds(m, \hat{u}, y)}{dm} < 0.
\end{align}
Thus, the residential density per unit of residential land decreases toward the city boundary.
When the share of residential land is spatially uniform, $\phi(m)=\bar\phi$, which approximates the pre-earthquake state, the number of households per unit of land, $\bar\phi\,\Pi(m)$, is also higher near the city center and lower in the suburbs.
Population density follows the same pattern only if average household size does not vary systematically with distance (see Section~\ref{sec:sec_ea_block} of the main text).

\subsection{Influence of the Land Readjustment after the Earthquake}\label{sec:secb4}

\begin{figure}[htbp]
\centering
\captionsetup{justification=centering,margin=1.5cm}
\subfloat[Rent Function in the Market]{\label{fig:brf}\includegraphics[width=0.5\textwidth]{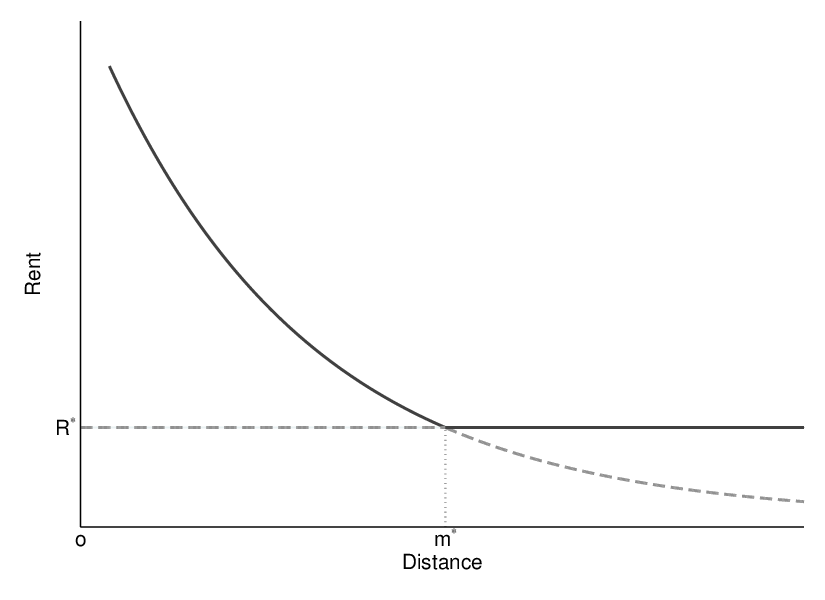}}
\subfloat[Shift of the Rent]{\label{fig:brf_shift}\includegraphics[width=0.5\textwidth]{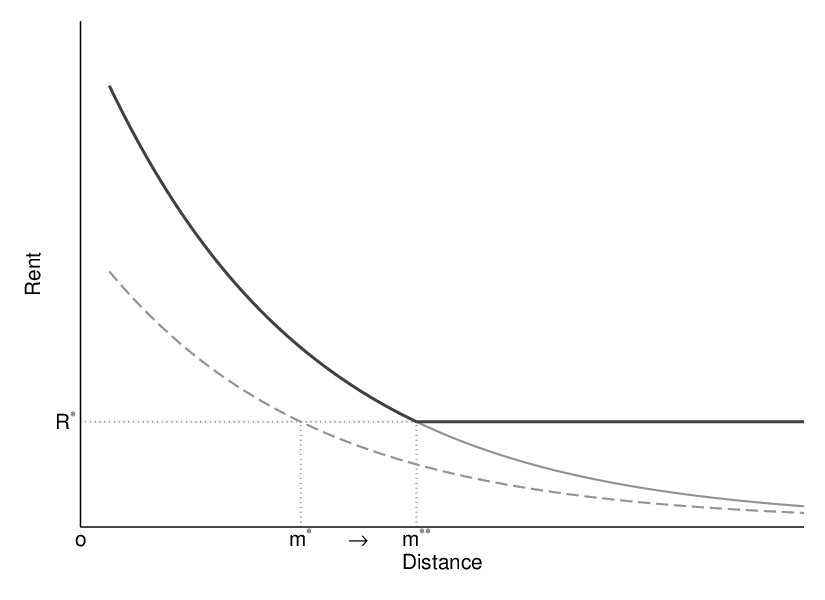}}
\caption{Rent and Distance to the City Center}
\label{fig:brf_all}
\scriptsize{\begin{minipage}{450pt}
\setstretch{0.9}
Notes:
Figure~\ref{fig:brf} shows market rents as a function of distance from the city center.
$m^{*}$ and $R^{*}$ denote the city boundary and the agricultural rent outside the city, respectively.
Figure~\ref{fig:brf_shift} illustrates the effect of an upward shift of the bid rent function on the city boundary; $m^{**}$ represents the city boundary under the shifted function.\\
Source: Created by the author.
\end{minipage}}
\end{figure}

As explained in the main text, the share of residential land in the land readjustment zones decreased from $75.3$\% before the earthquake to $63.8$\% after the post-earthquake land readjustment.
In the model, this corresponds to a reduction in $\phi(m)$ in the burned area.
Throughout, I hold the commuting cost function $C(m)$ fixed; although road widening could have lowered commuting costs, transportation fares within the city remained largely unchanged during the sample period (Section~\ref{sec:sec_rob}).

The effect of a reduction in $\phi(m)$ can be derived in three steps.
First, because $\partial \hat{R}/\partial \hat{u} < 0$ and $\partial \hat{R}/\partial m < 0$, Equation~(\ref{boundary}) implies that the city boundary is a decreasing function of the equilibrium utility:
\begin{align}\label{boundary_u}
\frac{d m^{*}}{d \hat{u}} = - \frac{\partial \hat{R}(m^{*}, \hat{u}, y)/\partial \hat{u}}{\partial \hat{R}(m^{*}, \hat{u}, y)/\partial m} < 0.
\end{align}
A lower utility level allows households to bid more for land, so that the city extends further outward.

Second, housing size increases in the utility level:
\begin{align}\label{size_u}
\frac{\partial s(m, \hat{u}, y)}{\partial \hat{u}}
= \frac{\partial s^{H}(\hat{R}, \hat{u}, y)}{\partial \hat{R}}\,\frac{\partial \hat{R}(m, \hat{u}, y)}{\partial \hat{u}}
+ \frac{\partial s^{H}(\hat{R}, \hat{u}, y)}{\partial \hat{u}} > 0,
\end{align}
where the first term is positive because $\partial s^{H}/\partial \hat{R} < 0$ by the compensated law of demand and $\partial \hat{R}/\partial \hat{u} < 0$, and the second term is positive because housing is a normal good.
Combining Equations~(\ref{boundary_u}) and (\ref{size_u}), the left-hand side of Equation~(\ref{hhs}) is a decreasing function of $\hat{u}$, so that the equilibrium utility is uniquely determined.

Third, a reduction in $\phi(m)$ in the burned area lowers the left-hand side of Equation~(\ref{hhs}) at the initial utility level; that is, residential land becomes insufficient to accommodate all households.
Because the number of households ($n$) is fixed in a closed city, the equilibrium utility must decline to $\hat{u}' < \hat{u}$ to restore Equation~(\ref{hhs}).
Consequently,
(i) the bid rent function shifts upward at every location, $R(m, \hat{u}', y) > R(m, \hat{u}, y)$;
(ii) the city boundary moves outward, $m^{**} > m^{*}$ (Figure~\ref{fig:brf_shift}); and
(iii) housing size decreases at every location, $s(m, \hat{u}', y) < s(m, \hat{u}, y)$.
In other words, the reduction in residential land resulting from land readjustment increases market rental prices within the city and leads to suburbanization.

These results have two implications for the spatial distribution of households.
First, the number of households within the pre-earthquake city boundary decreases, since some households now reside between $m^*$ and $m^{**}$:
\begin{align}\label{pop_inner}
\int_{0}^{m^{*}} \frac{\phi'(m)\, 2 \pi m}{s(m, \hat{u}', y)}\,dm
= n - \int_{m^{*}}^{m^{**}} \frac{\phi'(m)\, 2 \pi m}{s(m, \hat{u}', y)}\,dm < n,
\end{align}
where $\phi'(m)$ denotes the share of residential land after the land readjustment.
Second, the number of households in the burned area, $B$, also decreases.
The change can be decomposed as follows:
\begin{align}\label{decomp_b}
N_B' - N_B
= \int_B \frac{\bigl(\phi'(m)-\phi(m)\bigr)\,2\pi m}{s(m,\hat u',y)}\,dm
+ \int_B \phi(m)\,2\pi m\left[\frac{1}{s(m,\hat u',y)}-\frac{1}{s(m,\hat u,y)}\right]dm,
\end{align}
where $N_B=\int_B \phi(m)\,2\pi m/s(m,\hat u,y)\,dm$ and $N_B'=\int_B \phi'(m)\,2\pi m/s(m,\hat u',y)\,dm$ denote the number of households in the burned area before and after the land readjustment, respectively.
The first term is negative, reflecting the decline in residential land, whereas the second term is positive, reflecting the decrease in housing size.
The net effect is nevertheless negative.
Outside the burned area, $\phi(m)$ is unchanged, housing size decreases, and the city boundary expands, so the number of households outside the burned area increases.\footnote{This means that:
$
N_{B^c}' = \int_{B^c \cap [0,m^{**}]} \frac{\phi(m)\,2\pi m}{s(m,\hat u',y)}\,dm
\;>\;
\int_{B^c \cap [0,m^{*}]} \frac{\phi(m)\,2\pi m}{s(m,\hat u,y)}\,dm = N_{B^c}.
$}
Because the total number of households is fixed at $n$, the number of households in the burned area must decline: $N_B' = n - N_{B^c}' \;<\; n - N_{B^c} = N_B$.

The model abstracts from three features that are relevant to the empirical analysis.
First, the model treats each household as a single consumption unit.
The population in the burned area, however, equals the number of households multiplied by the average household size.
Thus, if households absorb higher rents by increasing household size, for example, by taking in dependent cohabitants (\textit{d\=okyonin}) or living with employees (\textit{shiy\=onin}), the decline in population can be smaller than the decline in the number of households.
I examine this channel empirically in Section~\ref{sec:sec_ea_block1}.

Second, the bid rent at each location depends on $\phi(m)$ only through the equilibrium utility level, so the model predicts a citywide increase in rents rather than a differential increase in the burned area.
In contrast, the model does predict a differential change in the number of households: it declines in the burned area and increases elsewhere.
Because the empirical analysis compares the burned and unburned areas, I interpret the results for rents as being consistent with, rather than a direct test of, the model's predictions (Online Appendix~\ref{sec:rent_response}).

Third, the model abstracts from the construction of multi-story dwellings, which would allow floor area to increase on a given plot.
This simplification is reasonable given the technical constraints on multi-story wooden houses during the study period (Section~\ref{sec:sec_ea_ward}).

\subsection{Rent Responses to the Land Readjustment}\label{sec:rent_response}

Section~\ref{sec:secb4} shows that a reduction in $\phi(m)$ in the burned area lowers the equilibrium utility from $\hat u$ to $\hat u'$ and thereby shifts the bid rent function upward at every location. Because $\phi(m)$ does not enter the household problem, the bid rent at each location depends on the land readjustment only through the equilibrium utility. This section characterizes how the size of the rent increase varies across locations, which informs the interpretation of the event-study results for rents in Section~\ref{sec:sec_ea_ward2}.

\paragraph*{Level change}
Applying the envelope theorem to Equation~(\ref{brf_def}) with respect to the utility level yields
\begin{equation}\label{eq:rent_level}
\frac{\partial R(m,u,y)}{\partial u} = -\frac{1}{s(m,u,y)\,\partial u/\partial x} < 0 .
\end{equation}
A decline in utility therefore raises rents at every location.
The absolute increase is larger where $s\cdot\partial u/\partial x$ is smaller.
Because housing size is smallest near the city center (Equation~\ref{size_distance}), the absolute increase in rents tends to be larger near the city center.
Locations between the original boundary $m^*$ and the new boundary $m^{**}$ also experience an increase in rents, from the agricultural rent $R^*$ to $R(m,\hat u',y)$.

\paragraph*{Proportional change}
Dividing Equation~\eqref{eq:rent_level} by $R$ and using the first-order condition~(\ref{mrg}), $\partial u/\partial s = R\,\partial u/\partial x$, one obtains
\begin{equation}\label{eq:rent_prop}
\frac{\partial \ln R(m,u,y)}{\partial \ln u} = -\frac{u}{s\,\partial u/\partial s} \equiv -\frac{1}{\varepsilon_s(m)} ,
\end{equation}
where $\varepsilon_s(m)$ denotes the elasticity of utility with respect to housing evaluated at the optimal bundle at distance $m$. The proportional increase in rents is larger where this elasticity is smaller.
Because the optimal bundle varies with distance, the sign of the difference in proportional rent changes between central and outlying locations is not determined in general.

\clearpage
\section{Data Appendix}\label{sec:secc}
\setcounter{figure}{0} \renewcommand{\thefigure}{C.\arabic{figure}}
\setcounter{table}{0} \renewcommand{\thetable}{C.\arabic{table}}

\subsection{Ward-level Statistics}\label{sec:secc_ward}

\subsubsection*{Residence and Rents}

Tokyo City compiled information on wooden residential buildings on December 31 each year.
Systematic statistics from the survey are documented in the \textit{T\=oky\=o-shi T\=okei Nenpy\=o} (Tokyo City Statistical Yearbook, TCSY).
I digitized statistics on the number of wooden residential buildings, the number of multi-story wooden residential buildings, and the floor area of wooden residential buildings for 1919--1935, using the 16th--33rd volumes of the ASCT.
The TCSY also reports average monthly unit rents.
This is described as: ``the average price of residential properties transacted over the course of a year.''
Given that recording all transactions must be practically difficult, these statistics probably reflect a compilation of prices from all contracts transacted in some areas within each ward.
I used the 17th--33rd volumes of the TCSY to obtain the unit rent price data for 1919--1935.
Regarding the statistics on floor area and rents, I converted the unit from \textit{tsubo} (a Japanese unit of land area) to square meters ($\text{m}^{2}$).

\subsubsection*{Ward Area}

The total ward area used for normalization is derived from volumes 17-33 of the TCSY.
The ward area statistics are measured on the 1st of January each year.
The area belonging to the imperial households, where ordinary households can not live, is excluded from the total ward area.
Given that the residential statistics are measured on the 31st of December, I assigned the ward area measured in the subsequent year.
For example, for the residential statistics measured on 31st December 1919, I used the ward area measured on 1st January 1920, meaning there is only one day gap in the assignment.

\subsubsection*{Population}

I digitized a series of the Original Tables of Tokyo City Statistics (\textit{T\=oky\=o-shi Shisei T\=okei Genpy\=o}, OTTCS) conducted in 1908, 1920, 1923, 1924, 1925, and 1930.\footnote{Precisely, the name of document for 1908 is \textit{T\=oky\=o-shi Shisei Ch\=osa Genpy\=o, Dai Ikkan, Meiji Yonjy\=uichi-nen} (Tokyo City Office 1909), for 1920 is \textit{T\=oky\=o-shi Shisei T\=okei Genpy\=o, Dai Ikkan, Taish\=o Ky\=u-nen} (Tokyo City Office 1922a), for 1923 is \textit{Shinsai Chokugo No Shisei T\=okei} (Tokyo City Office 1925a), for 1924 is \textit{T\=oky\=o-shi Shisei Ch\=osa T\=okei Genpy\=o, Dai Ikkan, Taish\=o Jy\=usan-nen} (Tokyo City Office 1926a), 1925 is \textit{T\=oky\=o-shi Shisei T\=okei Genpy\=o, Taish\=o Jy\=uyo-nen} (Tokyo City Office 1927a), 1930 is \textit{T\=oky\=o-shi Shisei T\=okei Genpy\=o, Sh\=owa Go-nen} (Tokyo City Office 1932a). The characteristics are very similar across these documents; I systematically refer to them as \textit{T\=oky\=o-shi Shisei T\=okei Genpy\=o}.}
The 1920, 1925, and 1930 statistics are compiled in the regular census years, and the 1908 survey is a pilot population census by Tokyo City.
These are the comprehensive censuses of all residents of Tokyo City conducted on October 1 each year, using the same methodology as the national population census.
The complete surveys to investigate the aftermath of the earthquake were also conducted on 15th November 1923 and 1st October 1924.
Both reports are quite useful for capturing the immediate responses of the city population after the disaster.
For the 1935 data, I used the prefectural section of the official population census report, titled \textit{Kokusei Ch\=osa H\=okoku} (Report of the Population Census), by the Statistics Bureau of the Cabinet (1937).

\subsubsection*{Railway Stations}

The location information of railway stations was obtained from the official shapefile file created by the Ministry of Land, Infrastructure, Transport and Tourism (\url{https://nlftp.mlit.go.jp/ksj/gml/datalist/KsjTmplt-N05-v1_3.html}, accessed 29th July 2022).
Figure~\ref{fig:map_station} illustrates the example figure for the spatial distribution of stations.
The shortest distance from the ward centroid to the nearest railway station is calculated using the shapefile provided by the Ministry of Land, Infrastructure, Transport and Tourism (Gy\=osei Kuiki Data).

\subsection{Block-level Statistics}\label{sec:secc_block}

\subsubsection{Population, Household, Household Size, and Sex Ratio}\label{sec:secc_block1}

The OTTCS provides the block-level statistics on the population by gender and the number of households.
This allows the creation of block-level data on the population and households for 1908, 1920, 1923, 1924, 1925, and 1930.
The sex ratio is calculated as the number of females divided by the number of males.
The number of households is recorded by distinguishing between ordinary households (\textit{futs\=u setai}) and quasi-households (\textit{jyun setai}).
While ordinary households are standard households that own their homes, quasi-households are those residing in boarding houses, hospitals, prisons, and similar facilities.
Consequently, nearly all statistical data is classified under ordinary households, and this paper excludes quasi-households from its analysis.

\subsubsection{Population by Household Size Bin}\label{sec:secc_block1_bin}

The OTTCS also contains statistics on the population by household size.
Household size is recorded in individual categories ranging from one-person households to 15-person households. 
For households of 16 or more people, it is divided into four categories: 16--20, 21--25, 26--30, and 31 or more.
Since the number of households in the $10+$ category is relatively small, I have aggregated them into a single category.

\subsubsection{Rents and Lodgers' Rents}\label{sec:secc_block2}

I obtained data on landlords' and lodgers' average monthly unit rents from the official report of the Survey of Shared-Dwelling Households (\textit{D\=okyo Setai ni Kansuru Ch\=osa}, SSDH), published by the Tokyo City Social Bureau (1930a).
This report documents the number of landlord (lodger) households by block and by average monthly unit rent bin for landlords (lodgers).
The unit used is the \textit{tatami} mat, a traditional Japanese floor mat made from rice straw (\textit{wara}) and rush (\textit{igusa}).
This is roughly $1.54$ square meters in prewar Tokyo (\textit{Edo-ma}).

The number of bins for the monthly unit rent is $20$, including an open-ended bin.
I rounded the end bin using the width of the adjacent closed bin (i.e., $0.2$ yen).
Similarly, the number of bins of the monthly unit rental fee is 22, including an open-ended bin.
The open-ended bin is rounded using the width of the adjacent closed bin (i.e., $0.2$ yen).
The average monthly unit rent and rental fee are then calculated as weighted averages across all bins.
For example, the average monthly unit rental fee for block $i$ is defined as follows:
	\begin{align}\label{ave_rent}
	\textit{Unit Rental Fee}_{i} = \frac{\sum_{j=1}^{22} b_{i, j}^{\text{fee}} \times h_{i, j}}{\sum_{j=1}^{22} h_{i,j}},
	\end{align}
where $b$ and $h$ show the class value and the number of households in $j$-th bin, respectively.

The report also documents the number of households by block and by 11 bins, with respect to the number of tatami mats per lodger.
The open-ended bin is rounded using the width of the adjacent closed bin (i.e., 1 tatami mat).
For each block $i$, the average number of tatami mats per lodger is then calculated as a weighted average across all bins as follows:
	\begin{align}\label{ave_mat}
	\textit{Mats}_{i} = \frac{\sum_{j=1}^{11} b_{i, j}^{\text{mat}} \times h_{i, j}}{\sum_{j=1}^{11} h_{i,j}},
	\end{align}
where $m$ shows the class value for the number of mats.

The average monthly rental fee per lodger is calculated as the product of the monthly unit rental fee in Equation~\ref{ave_rent} and the average number of tatami mats per lodger in Equation~\ref{ave_mat}.

\clearpage
\section{Empirical Analysis Appendix}\label{sec:secd}
\setcounter{figure}{0} \renewcommand{\thefigure}{D.\arabic{figure}}
\setcounter{table}{0} \renewcommand{\thetable}{D.\arabic{table}}

\subsection{Additional Descriptive Statistics}\label{sec:secd1}

\subsubsection{Residential Buildings}\label{sec:seca_housing}

\begin{figure}[htbp]
\centering
\captionsetup{justification=centering}
\subfloat[Buildings in Burned Wards]{\label{fig:ts_ward_bldg_ba}\includegraphics[width=0.5\textwidth]{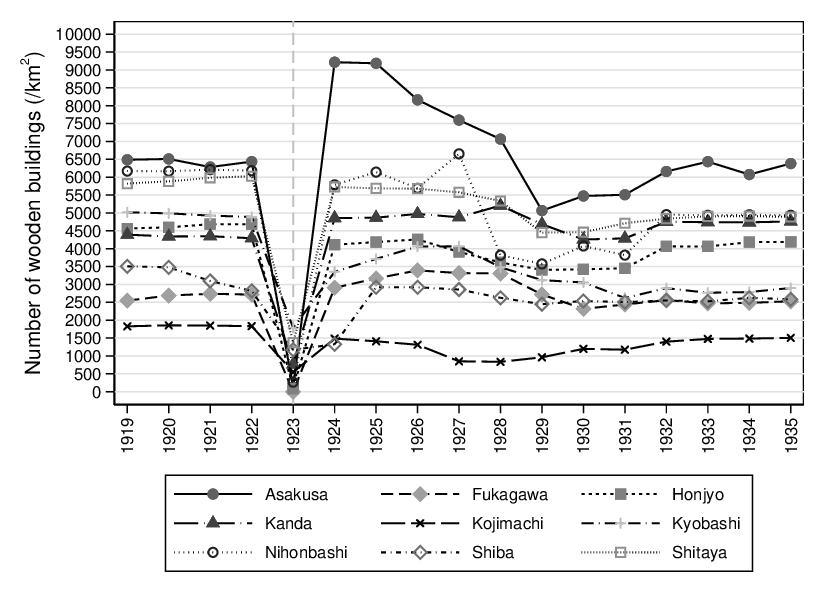}}
\subfloat[Buildings in Unburned Wards]{\label{fig:ts_ward_bldg_uba}\includegraphics[width=0.5\textwidth]{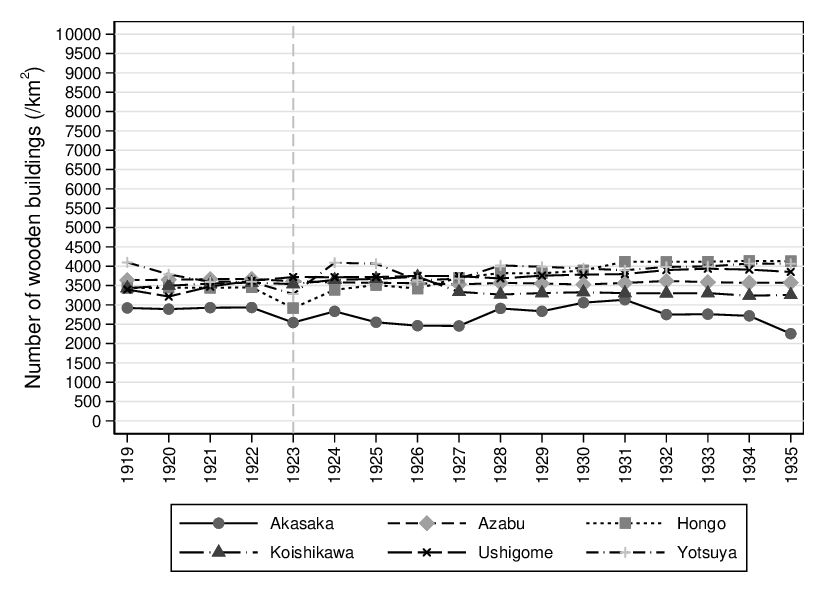}}\\
\subfloat[Multi-story Dwellings in Burned Wards]{\label{fig:ts_ward_bldg_mf_ba}\includegraphics[width=0.5\textwidth]{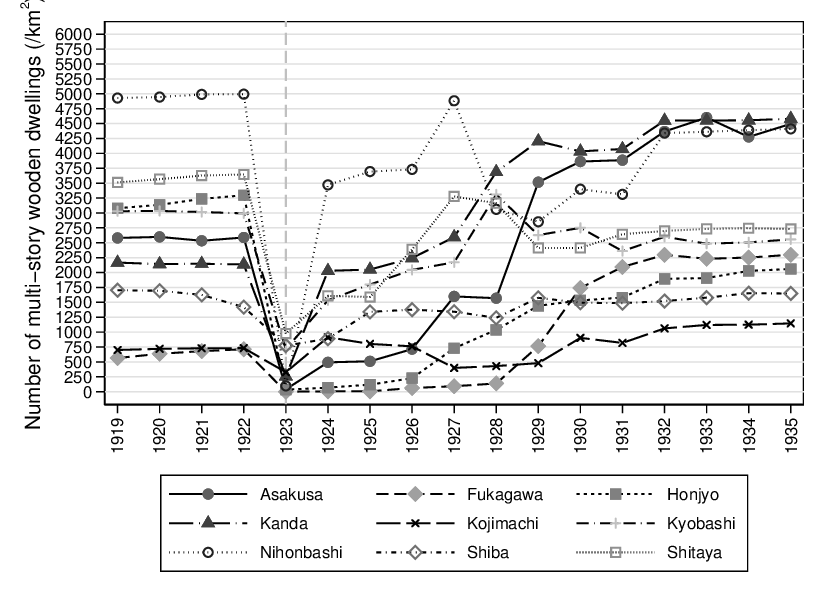}}
\subfloat[Multi-story Dwellings in Unburned Wards]{\label{fig:ts_ward_bldg_mf_uba}\includegraphics[width=0.5\textwidth]{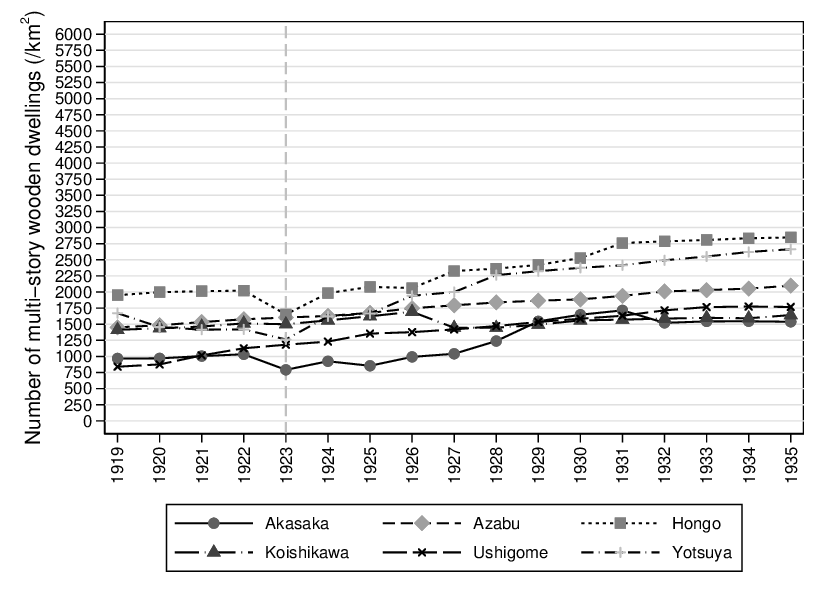}}\\
\subfloat[Floor Area in Burned Wards]{\label{fig:ts_ward_bldg_fs_ba}\includegraphics[width=0.5\textwidth]{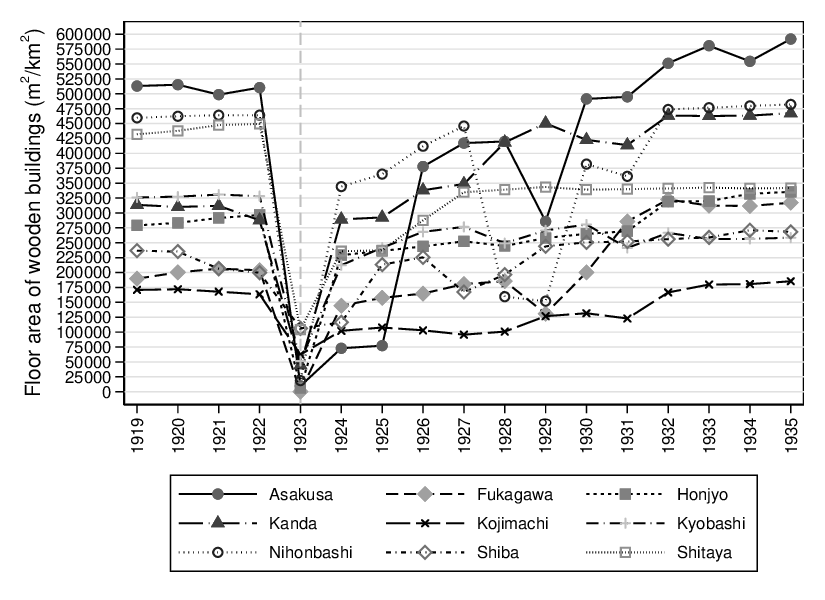}}
\subfloat[Floor Area in Unburned Wards]{\label{fig:ts_ward_bldg_fs_uba}\includegraphics[width=0.5\textwidth]{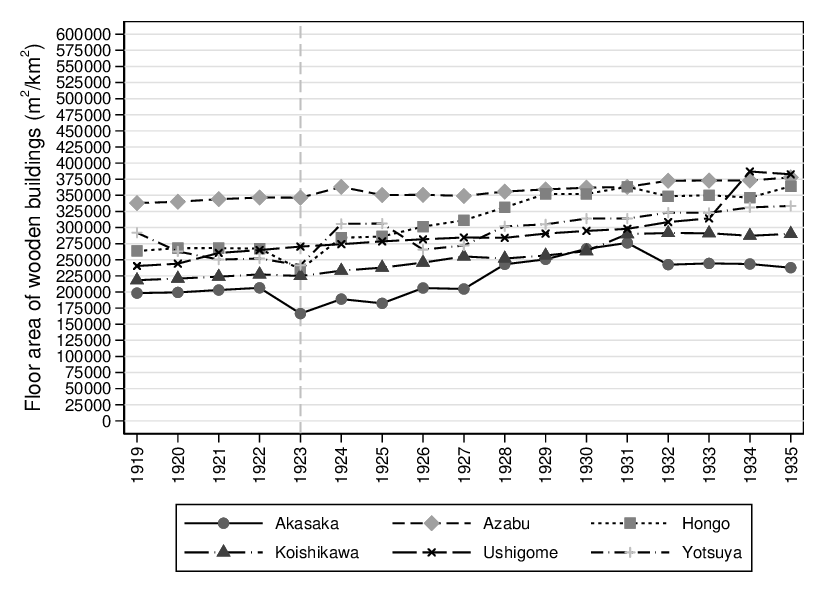}}
\caption{Wooden Residential Buildings, Multi-story Dwellings, and Floor Area\\ in Tokyo City}
\label{fig:ts_ward_building}
\scriptsize{\begin{minipage}{450pt}
\setstretch{0.9}
Notes:
Figures~\ref{fig:ts_ward_bldg_ba} and~\ref{fig:ts_ward_bldg_uba} illustrate the number of wooden residential buildings (per square kilometer) in burned and unburned wards, respectively.
Figures~\ref{fig:ts_ward_bldg_mf_ba} and~\ref{fig:ts_ward_bldg_mf_uba} illustrate the number of wooden multi-story dwellings (per square kilometer) in burned and unburned wards, respectively.
Figures~\ref{fig:ts_ward_bldg_fs_ba} and~\ref{fig:ts_ward_bldg_fs_uba} illustrate the floor area of the wooden buildings (per square kilometer) in burned and unburned wards, respectively.
Burned wards are those that lost more than a quarter of the ward area due to the fire, whereas the unburned wards are the remaining wards.
The administrative ward area excludes the Imperial Palace grounds, where citizens are unable to live.\\
Sources:
Created by the author using the ward-level census statistics listed in Panel A of Table~\ref{tab:sum}.
\end{minipage}}
\end{figure}

Since wooden structures were the predominant type of housing, I focused solely on wooden houses in this study.
Table~\ref{tab:housing} summarizes changes in the number of wooden houses and their total floor area.

Panel A of Table~\ref{tab:housing} shows that in the burned area, the number of residential buildings per square kilometer decreased by slightly more than $900$ between 1920 and 1930.
This figure is clearly higher than that in the unburned area.
Even by 1935, the number had decreased by more than $700$ compared to 1920, indicating that the number of residential buildings in the burned area had been declining over the long term.

Panel B shows changes in the floor area per wooden house.
As of 1920, the floor area per house in the burned area was about $2$ square meters smaller than in the unburned area.
In both the burned and unburned area, residential floor area per building increased from 1920 to 1935, but the increase in the burned area was more pronounced.
As a result, by 1935, the residential floor area per building in the burned area was approximately $2$ square meters larger than in the unburned area.
As discussed, this was due to an increase in the number of multi-story dwellings.

Figure~\ref{fig:ts_ward_building} summarizes the statistics on the wooden buildings by ward.

\def\arraystretch{1.00}
\begin{table}[htb]
\begin{center}
\captionsetup{justification=centering,margin=1.5cm}
\caption{Changes in the Number of Residential Buildings and Area for Residential Land (1920-1930)}
\label{tab:housing}
\footnotesize
\scalebox{1.0}[1]{
\begin{tabular}{lccccc}
\toprule[1pt]\midrule[0.3pt]
&\multicolumn{3}{c}{Survey Year}&\multicolumn{2}{c}{Difference}\\
\cmidrule(lll){2-4}\cmidrule(ll){5-6}
Panel A: Number of buildings/$\text{km}^{2}$	&1920	&1930	&1935	&1920-1930	&1920-1935\\\hline
Burned Wards							&3828	&2920	&3107	&-908		&-721\\
Unburned Wards						&3261	&3328	&3249	&67			&-11\\
Difference								&567		&-408	&-143	&-975		&-710\\
&&&&&\\
Panel B: Residential area $m^2$/building	&1920	&1930	&1935	&1920-1930	&1920-1935\\\hline
Burned Wards					&72.5	&89.4	&95.8	&16.9		&23.3\\
Unburned Wards				&74.3	&85.6	&94.3	&11.3		&20.1\\
Difference						&-1.8	&3.8		&1.5		&5.6			&3.3\\
\midrule[0.3pt]\bottomrule[1pt]
\end{tabular}
}
{\scriptsize
\begin{minipage}{380pt}
Notes:
Panel A shows the number of wooden buildings (count) per square kilometer.
Panel B summarizes the average residential area (square meters) per wooden building.
The burned wards are the wards where more than a quarter of the administrative area was burned, whereas the unburned wards are the other wards.
The administrative area does not include the land of the Imperial Palace.
Sources:
The data on the burned area are from Tokyo City Office (1932aa, pp.~33--34).
The data on the ward area, number of buildings, and residential areas are obtained from Tokyo City Office (1922; 1932; 1937). 
\end{minipage}
}
\end{center}
\end{table}

\subsubsection{Rent}\label{sec:seca_rent}

Rapid urbanization throughout World War I led to a shortage of housing supply in Tokyo City.
Although urbanization in nearby rural areas progressed gradually, housing supply in the city could not keep pace with the population growth driven by industrialization.
The main reason cited for the lack of housing supply was the soaring cost of construction (Ishizuka and Narita 1986, p. 136).
As a result, rental housing disputes (\textit{shakuya funs\=o}) arose, with residents opposing rent increases.

In 1921, the government enacted the Land Lease and House Lease Law (\textit{Shakuchi Shakka H\=o}) with the aim of alleviating the housing shortage.
Although the policy involved providing loans for housing construction, the number of units built remained small, and it failed to resolve the housing supply shortage (Ishizuka and Narita 1986, p.~139).
In practice, low-income households were excluded from eligibility for construction loans from the outset.

According to a 1922 survey report, the rate of rental housing among the new middle class exceeded $90$\%.
Since renting is also common among low-income groups, almost all households in Tokyo City live in rental housing (Ishizuka and Narita 1986, p.~137).
This underscores the importance of examining the earthquake's impact on rent.

For census years with available demographic data (1920, 1930, and 1935), statistics on average rent within the city can be obtained from the TCSY.
Table~\ref{tab:rent} shows the average unit rent for wooden residential properties.
As of 1920, the average rent in the burned area was $0.06$ yen higher than in the unburned area.
A trend of rising rent was observed in both areas.
However, toward the end of the reconstruction plan in 1930, the gap in rental prices between the two areas had widened significantly.
In addition to the citywide rise in rents, the average rental price there may have risen as a result of land readjustment-driven renewal of housing and public spaces in the burned area.

Although multiple factors likely contribute to the rise in rental prices, the primary issue may be on the supply side.
The number of residential buildings in the burned area physically decreased after the earthquake (Section~\ref{sec:seca_housing}).
On the other hand, given that there was no significant change in the number of residential buildings in the unburned area (Table~\ref{tab:housing}), rental prices in the burned area are likely to rise.

The price gap in 1935 had narrowed slightly from 1930.
This was due to a slight increase in rental prices in areas not destroyed by fire and a slight decrease in areas destroyed.
This is consistent with the improvement in housing supply in the areas that were destroyed (Table~\ref{tab:housing}).
However, the unit price difference remained larger than it was before the earthquake.

In summary, by around 1930, a clear gap had emerged between average rent in the burned and unburned areas.
By the mid-1930s, as housing supply in the burned area increased, the price gap narrowed, but it remained wider than before the earthquake.
Thus, rents rose in both areas, and the increase was larger in the burned area, which is consistent with the model's prediction of a citywide rise in rents (Online Appendix~\ref{sec:rent_response}).

\def\arraystretch{1.00}
\begin{table}[htb]
\begin{center}
\captionsetup{justification=centering,margin=1.5cm}
\caption{Changes in the Average Unit Rent (yen/$m^2$) (1920--1935)}
\label{tab:rent}
\footnotesize
\scalebox{1.0}[1]{
\begin{tabular}{lccccc}
\toprule[1pt]\midrule[0.3pt]
&\multicolumn{3}{c}{Survey Year}&\multicolumn{2}{c}{Difference}\\
\cmidrule(lll){2-4}\cmidrule(ll){5-6}
Average rent/$m^{2}$			&1920&1930&1935&1920-1930&1920-1935\\\hline
Burned Wards					&0.12&0.51&0.35&0.39&0.23\\
Unburned Wards				&0.06&0.15&0.16&0.09&0.09\\
Difference						&0.06&0.36&0.19&0.31&0.14\\
\midrule[0.3pt]\bottomrule[1pt]
\end{tabular}
}
{\scriptsize
\begin{minipage}{290pt}
Note: 
This table shows the average unit rent for wooden housing units traded each year (yen/$m^{2}$).
Rents are deflated by the urban consumer price index excluding rent.
Burned wards are wards where more than a quarter of the administrative area burned, while unburned wards are the others.
The administrative area does not include the land of the Imperial Palace.
Sources: 
The data on the burned area are from Tokyo City Office (1932aa, pp.~33--34).
The data on the ward area, residential areas, and rents are obtained from Tokyo City Office (1922; 1932; 1937).
The data on the consumer price index are from Ohkawa et al. (1967, pp~135--136).
\end{minipage}
}
\end{center}
\end{table}

\subsubsection{Population, Household, and Population Density}\label{sec:seca_hh_pop}

\begin{figure}[htbp]
\centering
\captionsetup{justification=centering}
\subfloat[Population in Burned Wards]{\label{fig:census_pop_ba}\includegraphics[width=0.5\textwidth]{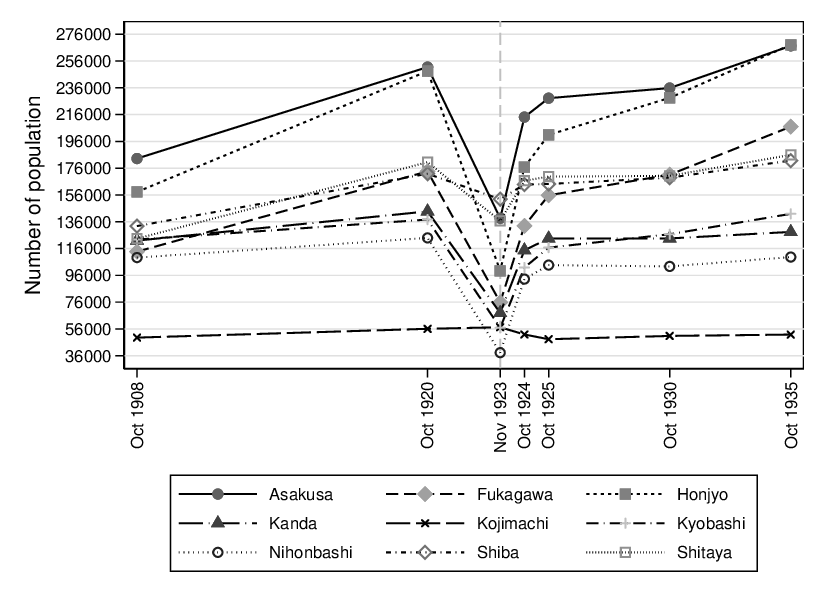}}
\subfloat[Population in Unburned Wards]{\label{fig:census_pop_uba}\includegraphics[width=0.5\textwidth]{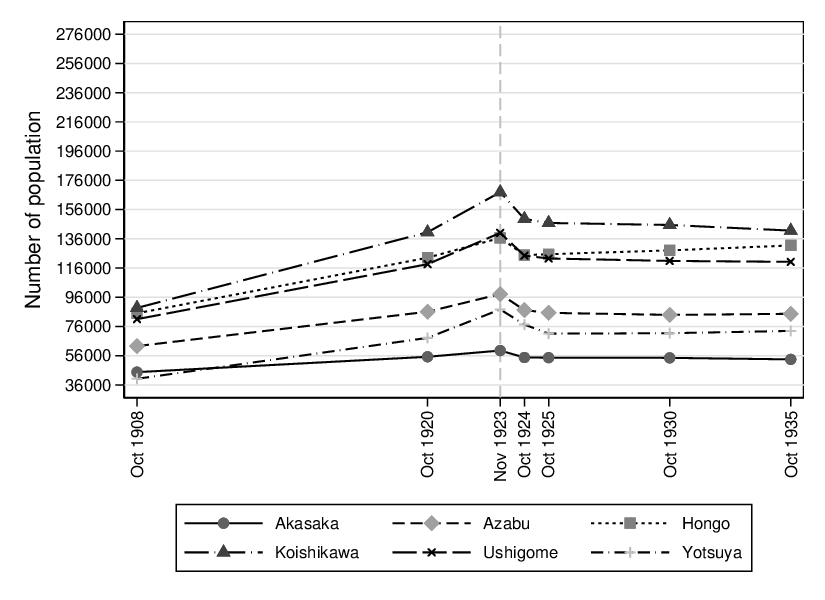}}\\
\subfloat[Households in Burned Wards]{\label{fig:census_hh_ba}\includegraphics[width=0.5\textwidth]{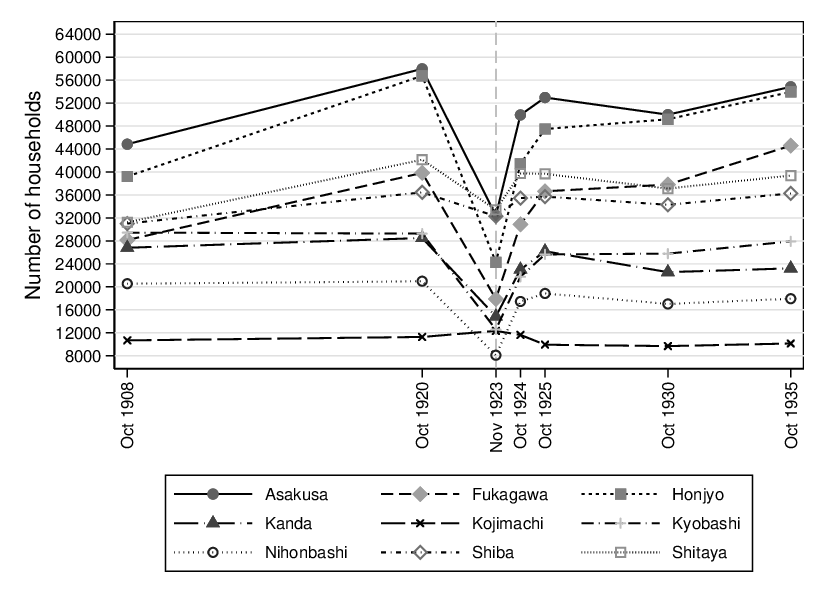}}
\subfloat[Households in Unburned Wards]{\label{fig:census_hh_uba}\includegraphics[width=0.5\textwidth]{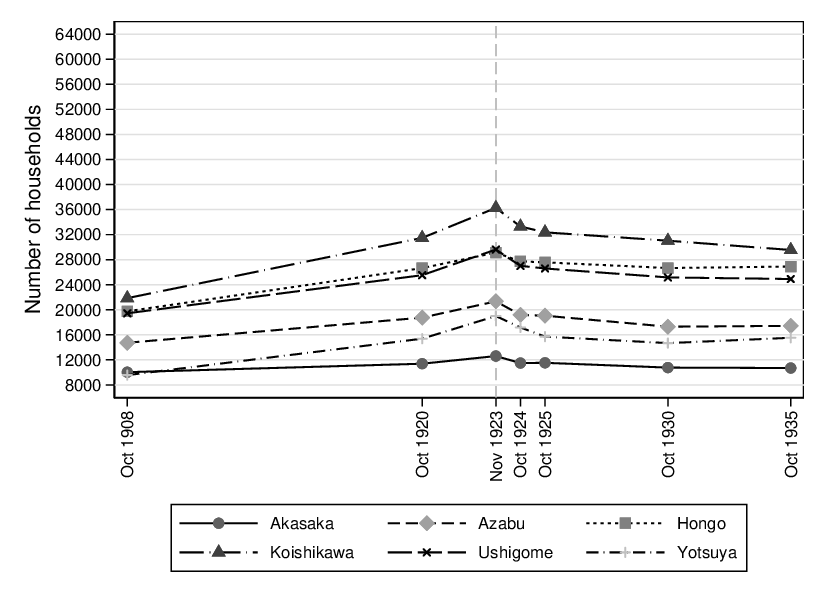}}\\
\subfloat[Population Density in Burned Wards]{\label{fig:ts_ward_pden_ba}\includegraphics[width=0.5\textwidth]{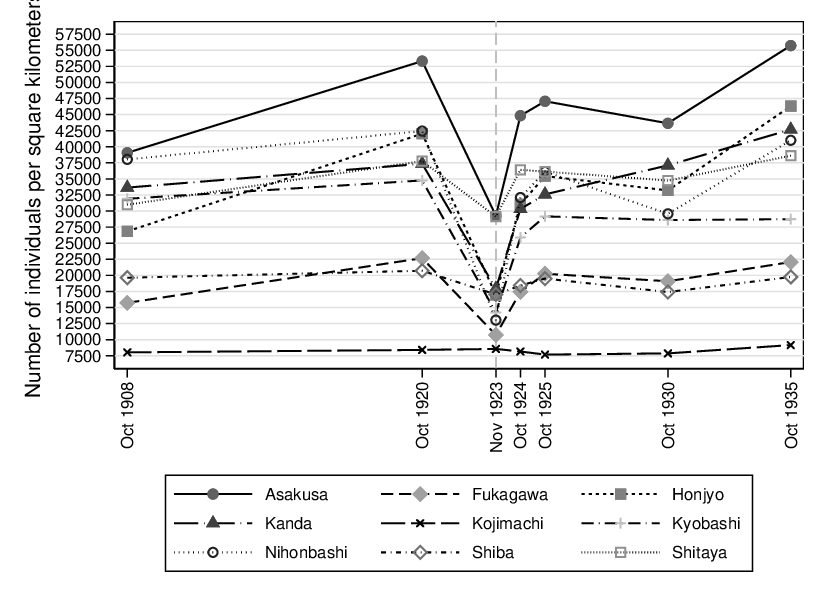}}
\subfloat[Population Density in Unburned Wards]{\label{fig:ts_ward_pden_uba}\includegraphics[width=0.5\textwidth]{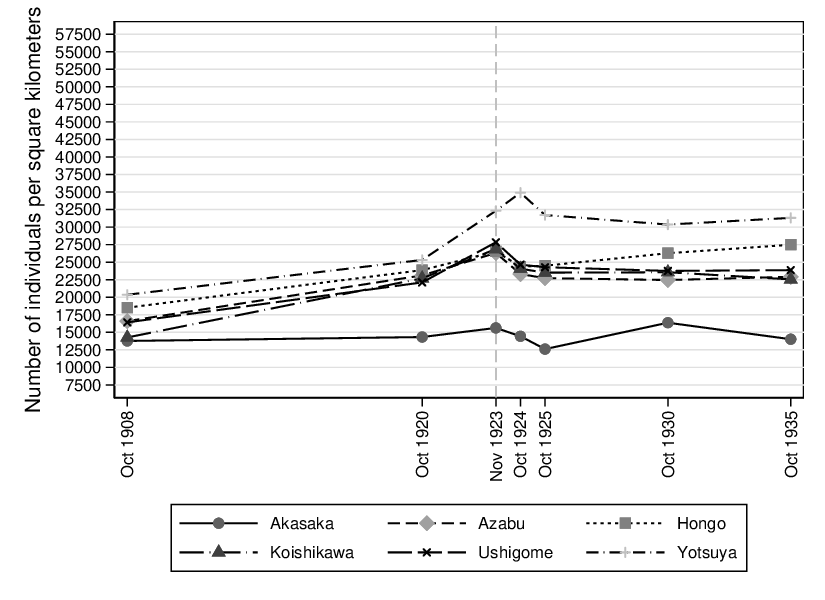}}
\caption{Time-series Plots of the Number of Individuals and Households\\ in Tokyo City}
\label{fig:census_hh_pop}
\scriptsize{\begin{minipage}{450pt}
\setstretch{0.9}
Notes:
Figures~\ref{fig:census_hh_ba} and~\ref{fig:census_hh_uba} show the number of households in burned and unburned wards, respectively.
Figures~\ref{fig:census_pop_ba} and~\ref{fig:census_pop_uba} illustrate the number of people in burned and unburned wards, respectively.
Figures~\ref{fig:ts_ward_pden_ba} and~\ref{fig:ts_ward_pden_uba} illustrate the population density (per square kilometer) in burned and unburned wards, respectively.
The burned wards are those that lost more than a quarter of the ward area due to the fire, whereas the unburned wards are the remaining wards.
The administrative ward area excludes the Imperial Palace grounds, where citizens are unable to live.
The statistics in 1908; 1920; 1924; 1925; 1930; 1935 are surveyed on 1st October, whereas those in 1923 is surveyed on 15th November.\\
Sources:
Created by the author using the ward-level census statistics in Panels A-1 and A-2 of Table~\ref{tab:sum}.
The data on the ward area are obtained from Tokyo City Office (1911; 1922; 1926; 1927; 1932; 1937). 
\end{minipage}}
\end{figure}

Figure~\ref{fig:census_hh_pop} summarizes the time-series plots of the number of individuals and households in Tokyo City measured in these complete surveys by burned and unburned wards.
Figures \ref{fig:census_hh_ba} and \ref{fig:census_pop_ba} show trends in population and households within the burned area, while Figures \ref{fig:census_hh_uba} and \ref{fig:census_pop_uba} show trends in population and households within the unburned area.

First, the number of households and the population in the burned area declined dramatically immediately after the earthquake, while they increased in the unburned area.
This reflects the temporary migration of households and people to the unburned area.
Second, in the burned area, the population remained at a lower level in 1930, but largely recovered by 1935.
In the unburned area, the number of households and the population recovered to pre-earthquake levels by 1930 and increased slightly by 1935.
Third, in most burned areas, the number of households remained below pre-earthquake levels until 1935.

\begin{figure}[htbp]
\centering
\captionsetup{justification=centering,margin=1.5cm}
\subfloat[1920]{\label{fig:map_ward_pden1920}\includegraphics[width=0.33\textwidth]{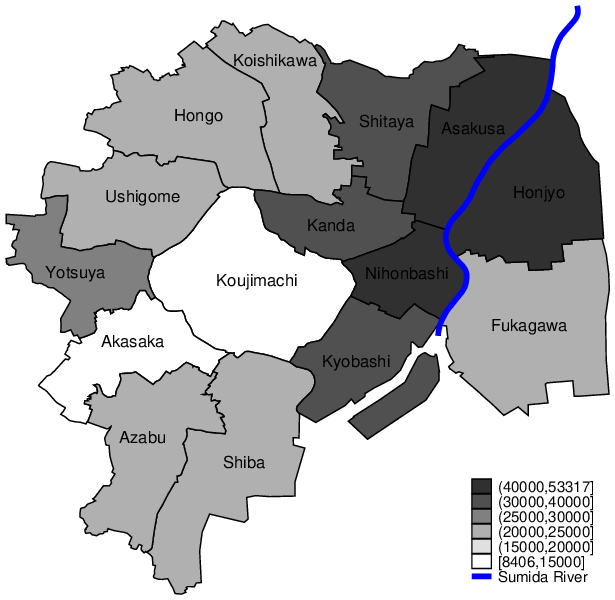}}
\subfloat[1930]{\label{fig:map_ward_pden1930}\includegraphics[width=0.33\textwidth]{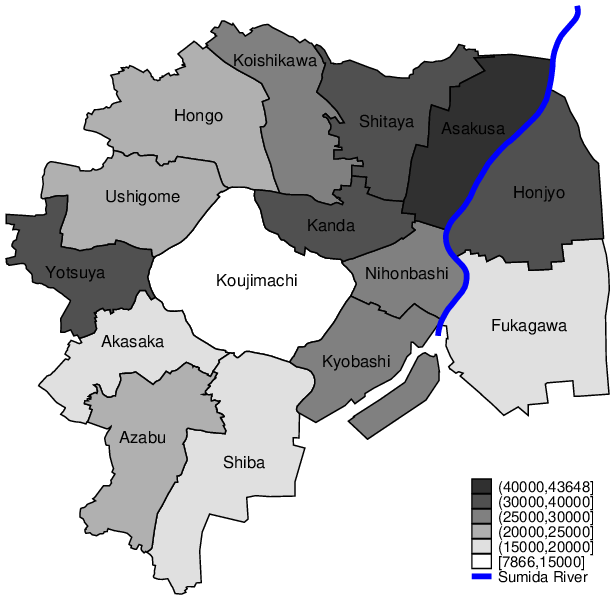}}
\subfloat[1935]{\label{fig:map_ward_pden1935}\includegraphics[width=0.33\textwidth]{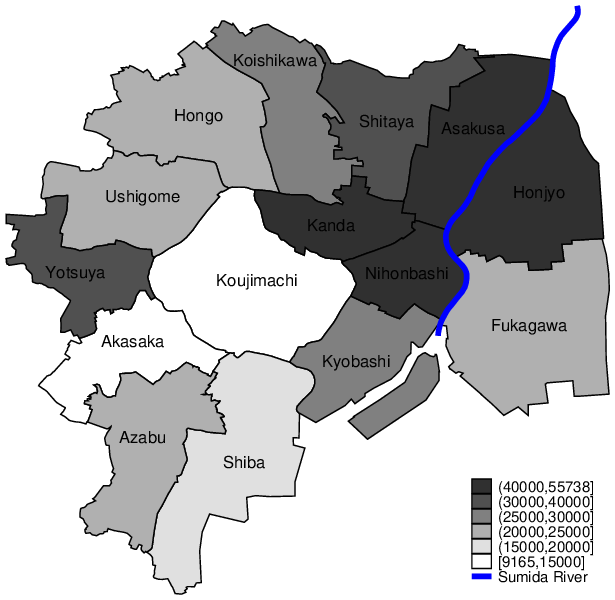}}
\caption{Population Density in the 15 Administrative Wards in Tokyo City}
\label{fig:map_ward_pden}
\scriptsize{\begin{minipage}{450pt}
\setstretch{0.9}
Notes:
Figures~\ref{fig:map_ward_pden1920},~\ref{fig:map_ward_pden1930}, and~\ref{fig:map_ward_pden1935} illustrate the population density in 1920, 1930, and 1935, respectively.
The population statistics are from the population censuses conducted on 1st October each year.
The ward area statistics are from a survey conducted by Tokyo City around the end of each year.\\
Sources:
The census data on population are from Tokyo City Office (1922a; 1932a) and Statistics Bureau of the Cabinet (1937).
The data on the ward area are obtained from Tokyo City Office (1922; 1932; 1937). 
Created by the author using the official shapefile (Ministry of Land, Infrastructure, Transport and Tourism, database).
The location data of the Sumida River are obtained from the Ministry of Land, Infrastructure, Transport, and Tourism (website).
\end{minipage}}
\end{figure}

\subsubsection{Population by Household Size}\label{sec:seca_hh_pop}

\begin{figure}[htbp]
\centering
\captionsetup{justification=centering}
\subfloat[Household in Burned Blocks]{\label{fig:ts_block_hh_abt}\includegraphics[width=0.5\textwidth]{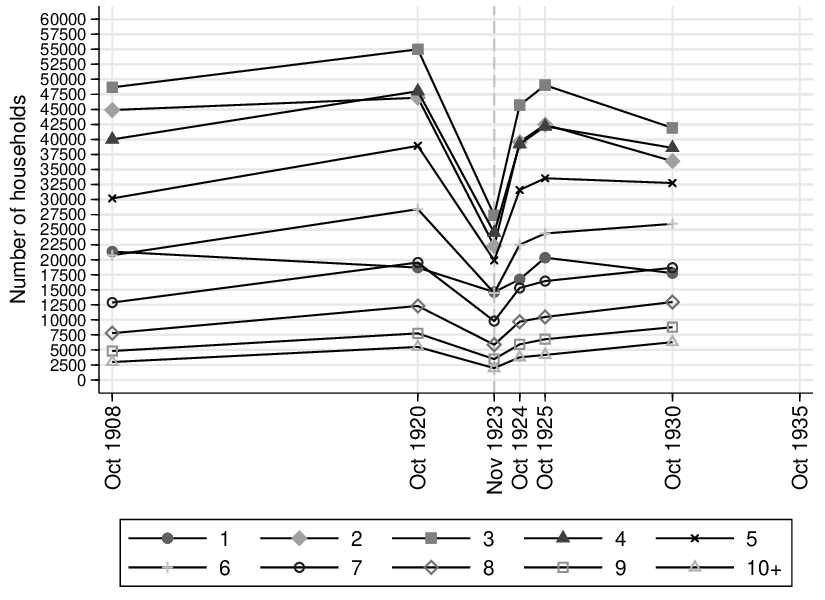}}
\subfloat[Household in Unburned Blocks]{\label{fig:ts_block_hh_abc}\includegraphics[width=0.5\textwidth]{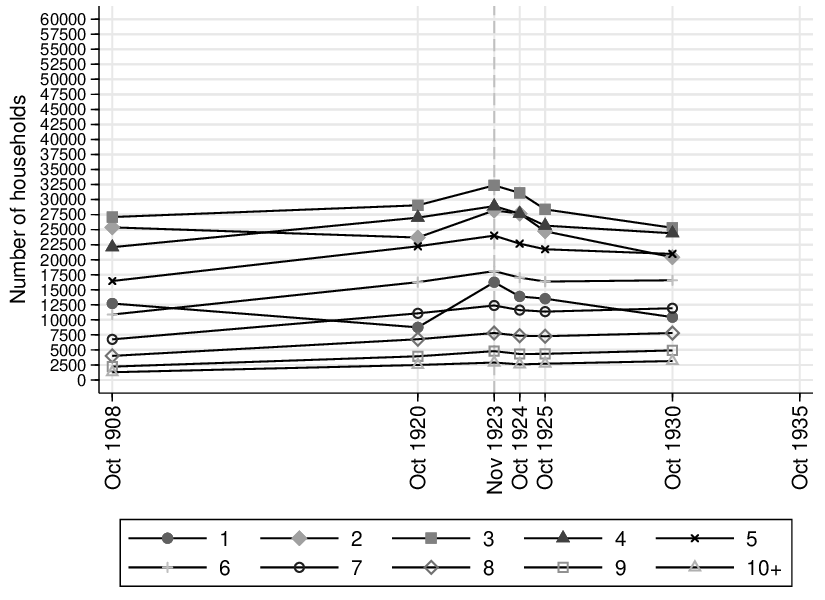}}\\
\subfloat[Population in Burned Blocks]{\label{fig:ts_block_pop_abt}\includegraphics[width=0.5\textwidth]{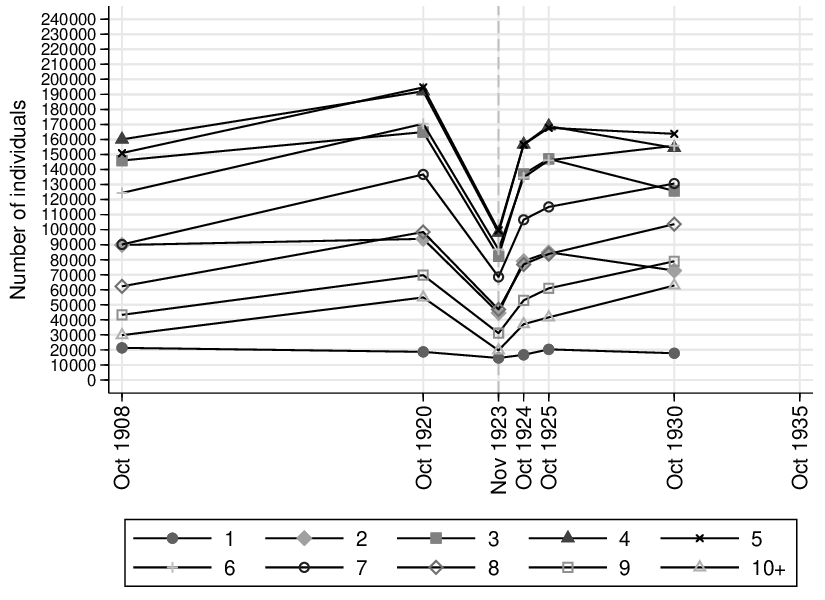}}
\subfloat[Population in Unburned Blocks]{\label{fig:ts_block_pop_abc}\includegraphics[width=0.5\textwidth]{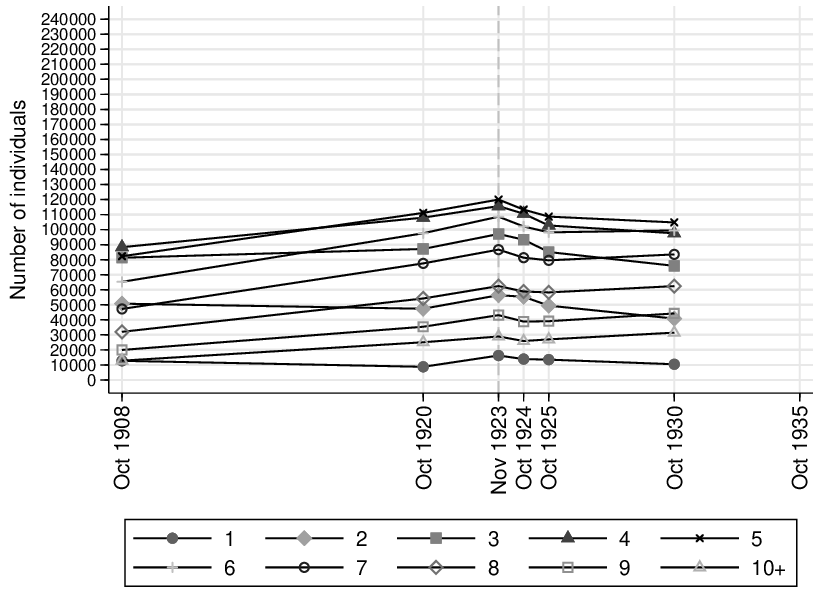}}\\
\caption{Households and Population:\\ Burned v. Unburned Blocks by Household Size Bins}
\label{fig:ts_block_bin}
\scriptsize{\begin{minipage}{450pt}
\setstretch{0.9}
Notes:
This figure illustrates the number of households and individuals in the burned and unburned blocks, grouped by household size, as measured in the censuses of 1908, 1920, 1923, 1924, 1925, and 1930.
The figures in the legend indicate household size.
For example, ``4'' means the bin for a household with four members.
Figures~\ref{fig:ts_block_hh_abt} and~\ref{fig:ts_block_hh_abc} show the number of households.
Figures~\ref{fig:ts_block_pop_abt} and~\ref{fig:ts_block_pop_abc} illustrate the number of individuals.
Sources:
Created by the author using block-level census statistics.
The sample matches that listed in Panel B-2 of Table~\ref{tab:sum}.
\end{minipage}}
\end{figure}

Figure~\ref{fig:ts_block_hh_abt} illustrates the trend in the number of households in burned blocks.
In 1923, the number of households declined sharply across all categories.
From 1923 to 1925, following the earthquake, the number of households recovered to some extent.
Subsequently, the number of households increased in the large-household class ($6+$ people) and decreased in the small-household class (1--5 people) through 1930.
As a result, the number of households in 1930 fell below the 1920 level in the small-household class, whereas that recovered to the 1920 level in the large-household class.
Figure \ref{fig:ts_block_hh_abc} shows the trends in unburned blocks.
In the small household size category (1--5 persons), the number of households decreased slightly from 1920 to 1930, but this decrease was negligible compared to the decline in burned blocks.
Conversely, the number of households increased slightly from 1920 to 1930.

The population trends correspond to changes in the number of households.
In the burned blocks, the population in the small household size category (1--5 people) has decreased, whereas it has increased in the large household size category ($6+$ people) (Figure \ref{fig:ts_block_pop_abt}).
In the unburned blocks, the population in the small household size category (1--5 people) has decreased, whereas it has increased in the large household size category ($6+$ people).
As seen earlier, this indicates that the net population in the unburned area increased due to the rise in the number of large households.
The above is consistent with the fact that the population decreased in the burned area and increased in the unburned area between 1920 and 1930 (Figure~\ref{fig:census_hh_pop}).

\subsection{Additional Quantitative Analyses}\label{sec:secd2}

\subsubsection{Results based on Alternative Treatment Variable Definition: Evidence from Ward-level Census Statistics}\label{sec:secd_r_ward_alt}

One potential concern with the regression model for the ward-level dataset in Equation~\ref{model_es} is the non-linear gradient in the effects of the continuous treatment variable, $R_{i}$.
In this section, I test whether my baseline results from Equation~\ref{model_es} are consistent with those obtained using an alternative binary treatment variable.
Specifically, following the definition of the burned ward used in the descriptive analysis (Figure~\ref{fig:ts_ward_build_rent}), I consider an indicator variable for the ward in which more than a quarter of the administrative area was burned, rather than $R_{i}$ as follows:
	\begin{align}\label{model_es_alt}
	z_{i, t} = \alpha + \sum_{\substack{j \in \{1,...,T\}\\ j \neq F-1}} \beta_{j} \bar{D}_{i, t}^{j} + \nu_{i} + \lambda_{t} + e_{i, t},
	\end{align}
where $\bar{D}_{i, t}^{j}= \bar{B}_{i} \times I(t = j)$ for $j \in \{1,...,T\}$ and $j \neq F - 1$, where $\bar{B}_{i}$ is an indicator variable for the burned wards.

\begin{figure}[htbp]
\centering
\captionsetup{justification=centering}
\subfloat[Impacts on Buildings ($/ \text{km}^{2}$)\\ in Burned Area]{\label{fig:r_ward_build}\includegraphics[width=0.45\textwidth]{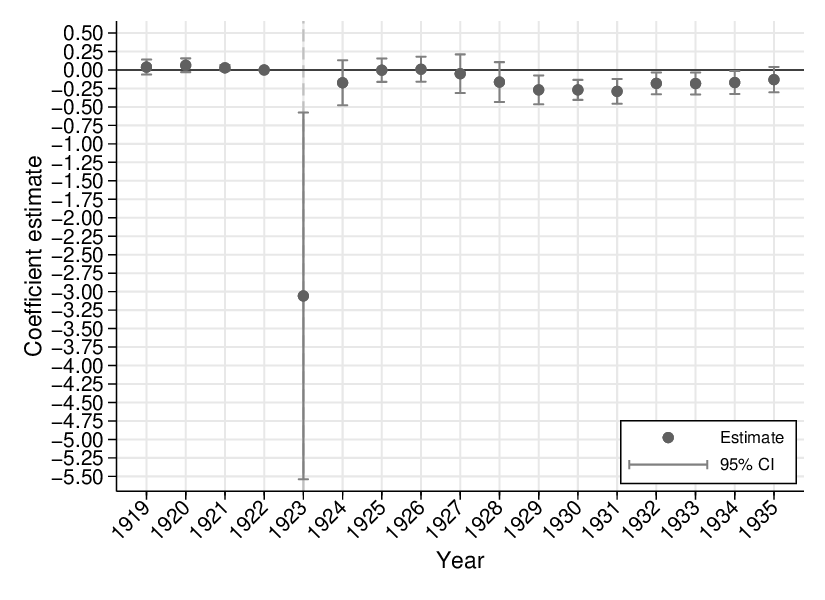}}
\subfloat[Impacts on Multi-story Dwellings ($/ \text{km}^{2}$)\\ in Burned Area]{\label{fig:r_ward_mbuild}\includegraphics[width=0.45\textwidth]{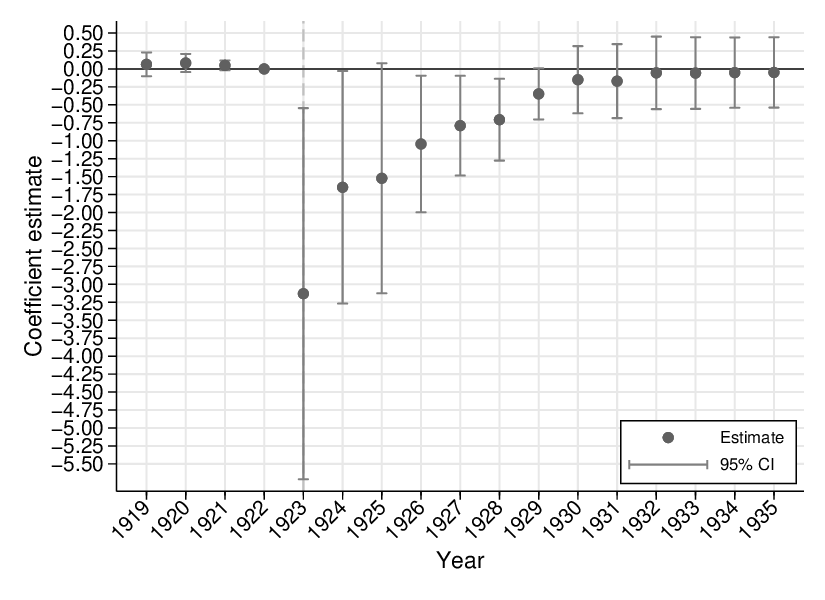}}\\
\subfloat[Impacts on Floor Area ($/ \text{km}^{2}$)\\ in Burned Area]{\label{fig:r_ward_build_area}\includegraphics[width=0.45\textwidth]
{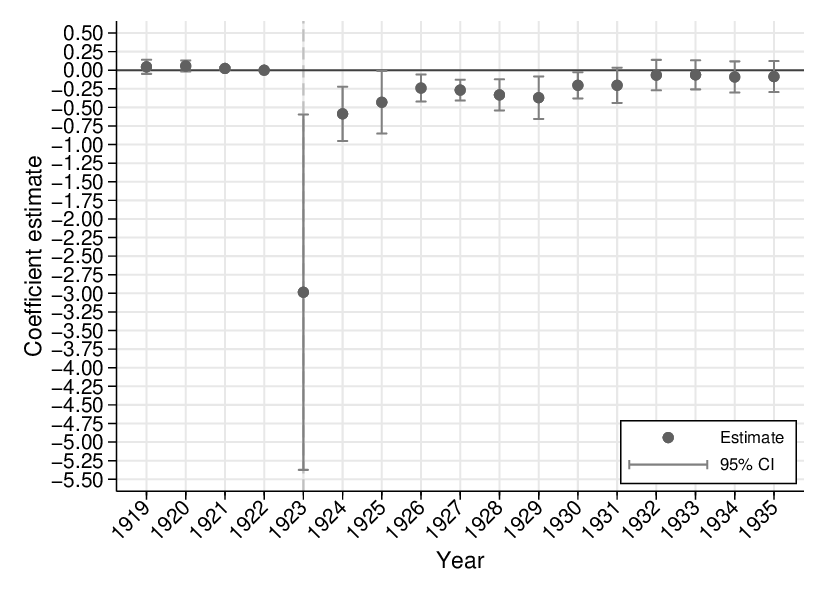}}
\subfloat[Imapcts on Rent ($/ \text{m}^{2}$)\\ in Burned Area]{\label{fig:r_ward_rent}\includegraphics[width=0.45\textwidth]{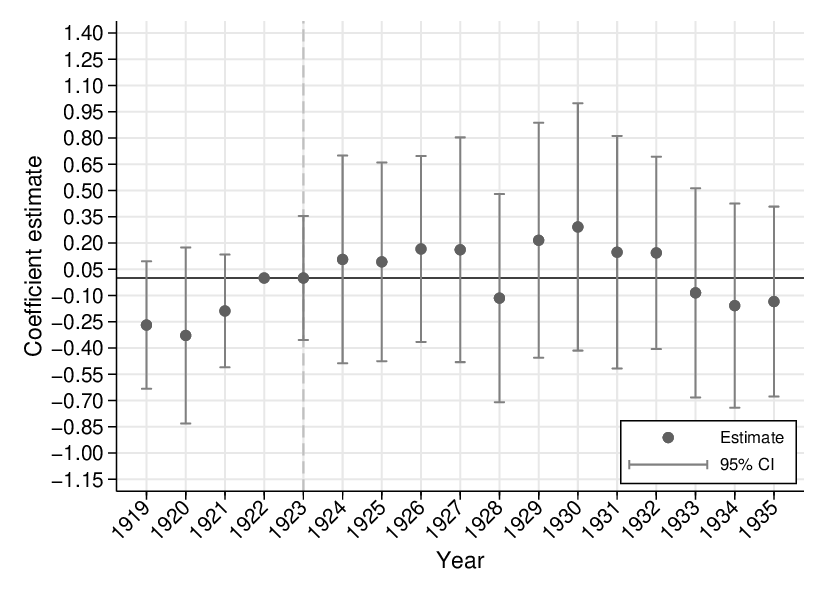}}\\
\subfloat[Impacts on Population\\ in Burned Area]{\label{fig:r_ward_pop}\includegraphics[width=0.45\textwidth]
{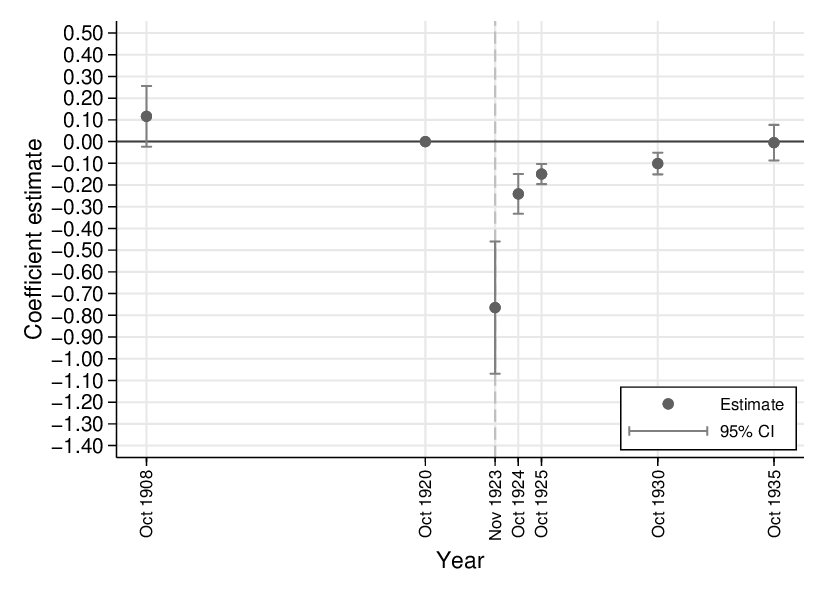}}
\subfloat[Impacts on Households\\ in Burned Area]{\label{fig:r_ward_hh}\includegraphics[width=0.45\textwidth]
{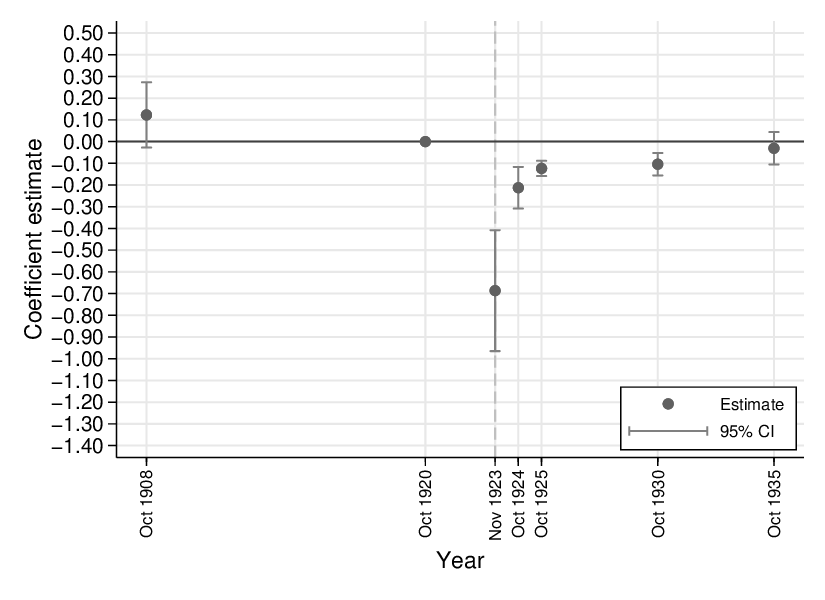}}
\caption{Impacts on the Residences, Rents, and Population in Burned Area:\\ Evidence from the Ward-level Statistics}
\label{fig:r_did}
\scriptsize{\begin{minipage}{450pt}
\setstretch{0.9}
Notes:
This figure shows the estimated coefficients on the interaction terms between the indicator variable for the burned wards and the year dummies in Equation~\ref{model_es_alt}.
Figure~\ref{fig:r_ward_build},~\ref{fig:r_ward_mbuild}, and~\ref{fig:r_ward_build_area} illustrate the results for the log-transformed number of wooden buildings (per $\text{km}^{2}$), multi-story wooden dwellings (per $\text{km}^{2}$), and floor area of the wooden buildings (per $\text{km}^{2}$), respectively.
Figure~\ref{fig:r_ward_rent} shows the result for the log-transformed average rent per square meter.
Rents are deflated by the urban consumer price index excluding rent.
An odd estimate for 1928 is due to an outlier in Koishikawa Ward, which is included in the unburned wards.
All regressions for these building statistics are weighted by ward area, excluding land for the imperial palace, and include ward- and year-fixed effects.
Figures~\ref{fig:r_ward_pop} and~\ref{fig:r_ward_hh} show the results for the log-transformed population and households, respectively.
The 95\% confidence intervals are obtained from the cluster-robust variance-covariance matrix estimator using wards as the clustering unit.\\
Sources: Created by the author using the ward-level census statistics listed in Panel A-1 of Table~\ref{tab:sum}.
\end{minipage}}
\end{figure}

Figure~\ref{fig:r_did} presents the results for this specification.
The results are materially similar to those presented in Figure~\ref{fig:r_did_rate}.
Since the treatment variable is now binary, the estimated coefficient directly suggests the average effect on the burned area.
For example, the estimate for the number of buildings in 1930 is $-0.25$, suggesting that the disaster decreased the number of buildings by about $22$\% in the burned area.
This magnitude is nearly identical to that implied by the baseline estimate from Equation~\ref{model_es}.
Similarly, the estimates suggest an increase of about 29\% in the average rents during the 1929--1930 period, relative to the unburned wards, which is consistent with the magnitude of my baseline estimate (about 28\%).
As explained in the main text, the odd estimate for 1928 in Figure~\ref{fig:r_ward_rent} is due to an outlier in Koishikawa Ward, which is included in the unburned wards.
In other words, this specification would be more sensitive to influential observations in the rent statistics than the baseline specification, which assumes linearity in treatment effects.
The estimated magnitude for population, a decline of about $10$\% in 1930, is also close to my baseline magnitude (about $11$\%).

\subsubsection{Results for Rent under Alternative Specification: Evidence from Ward-level Census Statistics}\label{sec:secd_r_ward_rent}

\def\arraystretch{1.0}
\begin{table}[htbp]
\begin{center}
\captionsetup{justification=centering,margin=1.5cm}
\caption{Results for Rent under Alternative Specification: Evidence from Ward-level Census Statistics}
\label{tab:r_rent_ward}
\scriptsize
\scalebox{1.0}[1]{
\begin{tabular}{lD{.}{.}{-2}D{.}{.}{-2}D{.}{.}{-2}D{.}{.}{-2}}
\toprule[1pt]\midrule[0.3pt]
&\multicolumn{4}{c}{Dependent Variable: ln(Rent)}\\
\cmidrule(rrrr){2-5}
&\multicolumn{1}{c}{(1)}
&\multicolumn{1}{c}{(2)}
&\multicolumn{1}{c}{(3)}
&\multicolumn{1}{c}{(4)}\\\hline
Burned Area (\%) $\times$ $I(\text{Year} \in \left[1923, 1933\right])$		&0.004$**$	&0.005$**$	&			&			\\
														&[0.002]		&[0.002]		&			&			\\
Burned Wards $\times$ $I(\text{Year} \in \left[1923, 1933\right])$		&			&			&0.282$**$	&0.304$**$	\\
														&			&			&[0.131]		&[0.132]		\\\hline
Sample period 		&\multicolumn{1}{c}{1919-1935}	&\multicolumn{1}{c}{1919-1935}	&\multicolumn{1}{c}{1919-1930}		&\multicolumn{1}{c}{1919-1930}	\\
1928 excluded?	&\multicolumn{1}{c}{No}		&\multicolumn{1}{c}{Yes}		&\multicolumn{1}{c}{No}		&\multicolumn{1}{c}{Yes}	\\
Sample mean of the DV		&-1.61		&-1.62		&-1.61		&-1.62		\\
Standard deviation of the DV	&0.62		&0.62		&0.62		&0.62		\\
Ward FE 				&\multicolumn{1}{c}{Yes}		&\multicolumn{1}{c}{Yes}		&\multicolumn{1}{c}{Yes}		&\multicolumn{1}{c}{Yes}	\\
Year FE 				&\multicolumn{1}{c}{Yes}		&\multicolumn{1}{c}{Yes}		&\multicolumn{1}{c}{Yes}		&\multicolumn{1}{c}{Yes}	\\
Observations			&\multicolumn{1}{c}{254}		&\multicolumn{1}{c}{239}		&\multicolumn{1}{c}{254}		&\multicolumn{1}{c}{239}	\\\midrule[0.3pt]\bottomrule[1pt]
\end{tabular}
}
{\scriptsize
\begin{minipage}{400pt}   
\setstretch{0.85}
***, **, and * denote statistical significance at the 1\%, 5\%, and 10\% levels, respectively.
The standard errors in brackets are clustered at the 15-ward level.\\
Notes:
This table summarizes the results of the regressions using ward-level average rent statistics.
The dependent variable is the log-transformed deflated rent ($/\text{m}^{2}$) used in Figure~\ref{fig:r_ward_rentr} and~\ref{fig:r_ward_rent}.
`Burned Area (\%)' indicates the percentage of area burned.
`Burned Wards' is an indicator variable for the wards in which more than a quarter of the administrative area was burned.
$I(\text{Year} \in \left[1923, 1933\right])$ is an indicator variable for the boom years after the disaster.
Columns (1) and (3) show the result for the full sample.
Columns (2) and (4) use the sample excluding 1928, which has an odd observation in the rent statistics in Koishikawa Ward.
All regressions are weighted by ward area, excluding lands for the Imperial Palace.\\
Sources: Created by the author using the ward-level census statistics listed in Panel A-1 of Table~\ref{tab:sum}.
\end{minipage}
}
\end{center}
\end{table}

The regressions presented in Table~\ref{tab:r_rent_ward} aim to capture the average difference in rent between burned and unburned wards, with the boom years (1923-1933) as the treatment period.
This means that Equation~\ref{model_es} is simplified to capture the average effects during the boom as follows:
	\begin{align}\label{model_es_boom}
	z_{i, t} = \alpha + \beta R_{i} \times I(1923 \leq t \leq 1933) + \nu_{i} + \lambda_{t} + e_{i, t},
	\end{align}
where $R_{i}$ is the percentage of burned area used.
I also consider an alternative model of Equation~\ref{model_es_boom} by replacing $R_{i}$ with an indicator variable for burned wards, $\ddot{B}_{i}$, as in Equation~\ref{model_es_alt}.

Columns (1) and (2) use the percentage of burned area as the treatment variable in Equation~\ref{model_es}, whereas Columns (3) and (4) use the binary variable as the treatment variable.
Columns (1) and (3) show the result for the full sample, whereas Columns (2) and (4) exclude 1928, which includes an odd observation in Koishikawa Ward.
The estimate in Column (1) suggests that a one-percentage-point increase in the burned-area share increases the rent by $0.4$\% between 1923 and 1933.
This means that the average rent might have increased by roughly $32$--$49$\% in wards where $70$--$100$\% of the ward area was affected relative to the unburned wards.
Column (2) shows the same estimate, supporting that my baseline specification using a continuous treatment variable is relatively robust against outliers in the dependent variable.
Column (3) shows that the disaster increased the average rent in the burned wards by roughly $32.6$\% relative to the unburned wards.\footnote{Percentage effects are computed as $100 \times (e^{\hat{\beta}} - 1)$, where $\hat{\beta}$ is the estimated coefficient on the binary treatment variable in the log-linear specification.}
This magnitude increased to roughly $35.5$\% if I excluded an influential year of 1928 (Column (4)).
Overall, these results suggest that the average rent in the devastated wards rose by at least $30$\% relative to the unburned wards during the boom years.

\subsubsection{Results for Population and Household Densities: Evidence from Ward-level Census Statistics}\label{sec:secd_r_den}

\begin{figure}[htbp]
\centering
\captionsetup{justification=centering}
\subfloat[Impacts on Population Density\\ in Burned Area]{\label{fig:r_ward_pden}\includegraphics[width=0.45\textwidth]
{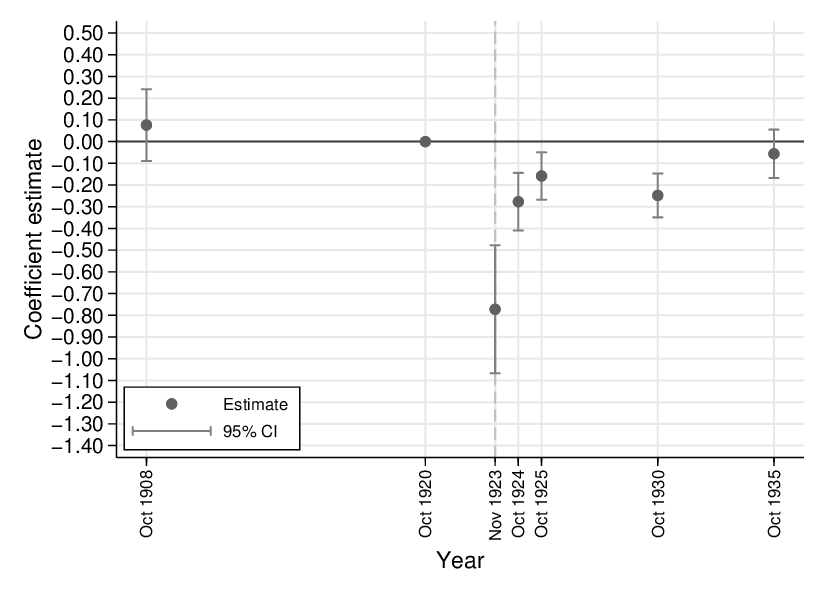}}
\subfloat[Impacts on Household Density\\ in Burned Area]{\label{fig:r_ward_hden}\includegraphics[width=0.45\textwidth]
{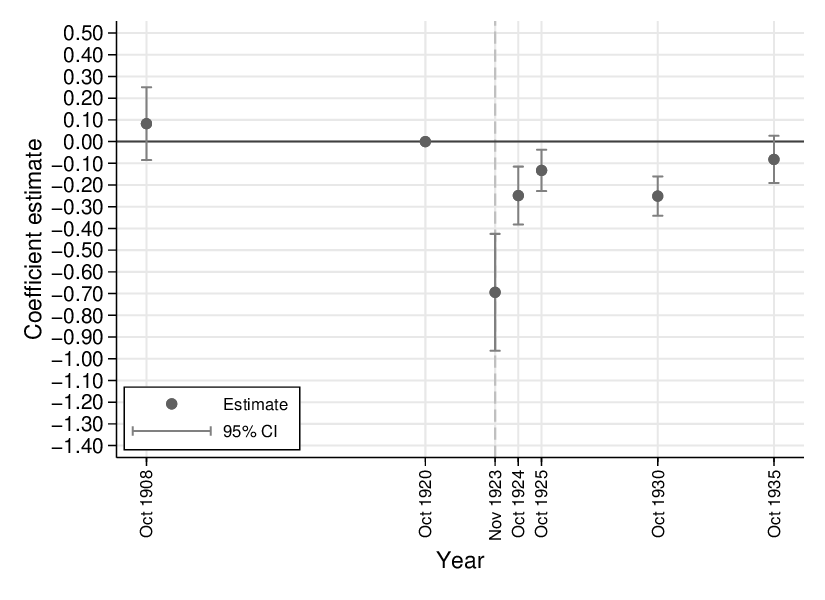}}
\caption{Impacts on the Population and Household Densities:\\ Evidence from the Ward-level Statistics}
\label{fig:r_ward_den}
\scriptsize{\begin{minipage}{450pt}
\setstretch{0.9}
Notes:
This figure shows the estimated coefficients on the interaction terms between the indicator variable for burned wards and the year dummies in Equation~\ref{model_es}.
Figures~\ref{fig:r_ward_pden} and~\ref{fig:r_ward_hden} show the results for the log-transformed population and household densities, respectively.
All the regressions include both ward- and year-fixed effects.
The 95\% confidence intervals are obtained from the cluster-robust variance-covariance matrix estimator using wards as the clustering unit.\\
Sources: Created by the author using the ward-level census statistics listed in Panel A-1 of Table~\ref{tab:sum}.
The ward area used as a denominator is obtained from Tokyo City Office (1911--1932).
\end{minipage}}
\end{figure}

Figures~\ref{fig:r_ward_pden} and~\ref{fig:r_ward_hden} show the results for population and household densities under Equation~\ref{model_es}, respectively.
The results are materially similar to the results for the population and household.
This supports the evidence that the influence of time-series variations in the ward area is negligible during this period, as it is absorbed by the ward-fixed effects.

\subsubsection{Marginal Effects of Industrial Zoning for Population and Households: Evidence from Block-level Census Statistics}\label{sec:secd_r_block_het}

\begin{figure}[htbp]
\centering
\captionsetup{justification=centering}
\subfloat[Impacts of Zoning on Population\\ in Burned and Unburned Blocks]{\label{fig:r_block_het_pop_zone}\includegraphics[width=0.45\textwidth]{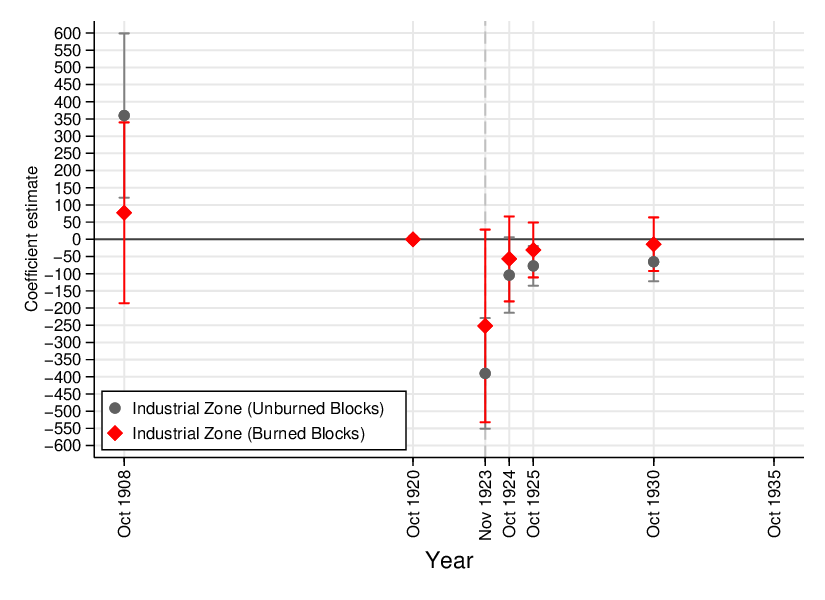}}
\subfloat[Impacts of Zoning on Households\\ in Burned and Unburned Blocks]{\label{fig:r_block_het_hh_zone}\includegraphics[width=0.45\textwidth]{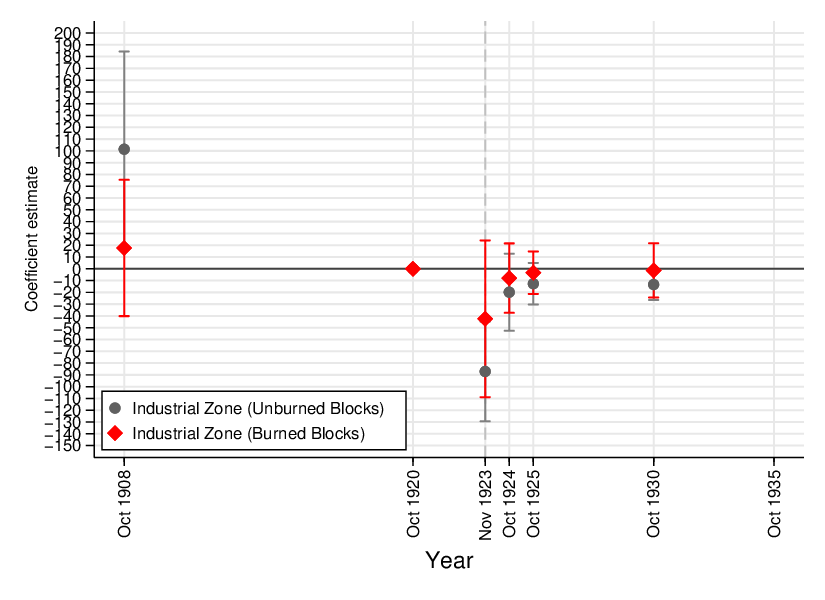}}
\caption{Results for the Population and Households: \\Evidence from the Block-level Census Statistics}
\label{fig:r_did_block_pop_hh}
\scriptsize{\begin{minipage}{450pt}
\setstretch{0.9}
Notes:
The marginal effects of zoning calculated using the estimated coefficients for Equation~\ref{model_es_het} are shown in Figures~\ref{fig:r_block_het_pop_zone}--\ref{fig:r_block_het_hh_zone}.
The estimated effects on the burned (unburned) blocks for each census year are presented in black (red) circles.
`Industrial Zone' includes both commercial and manufacturing zones defined by the zoning system designed prior to the earthquake (Section~\ref{sec:sec_hb_cbd}).
The $95$\% confidence intervals are shown in the black and red caps in each figure.
Standard errors in the regressions based on Equation~\ref{model_es_het} are clustered at the block-level.\\
Sources: Created by the author using the block-level census statistics listed in Panel B-1 in Table~\ref{tab:sum}.
\end{minipage}}
\end{figure}

Figure~\ref{fig:r_block_het_pop_zone} and~\ref{fig:r_block_het_hh_zone} summarize the estimated marginal effects of the industrial zoning on the population and households, respectively.
Each estimate illustrated in both figures shows the partial difference with respect to an indicator variable for industrial zone blocks ($Z_{i}$) in Equation~\ref{model_es_het}.
In both figures, the confidence intervals for the marginal effects include zero in most post-disaster periods.
This suggests that the zoning system did not have a meaningful influence on the population changes after the disaster.

\subsubsection{Testing Heterogeneity with Respect to Industrial Zones: Evidence from Ward-level Labor Statistics}\label{sec:secd_r_ward_het}

\def\arraystretch{0.95}
\begin{table}[htbp]
\begin{center}
\captionsetup{justification=centering,margin=1.5cm}
\caption{Testing Heterogeneity with Respect to Industrial Zones: Evidence from the Ward-level Labor Statistics}
\label{tab:r_het_ward}
\scriptsize
\scalebox{1.0}[1]{
\begin{tabular}{lD{.}{.}{-2}D{.}{.}{-2}D{.}{.}{-2}D{.}{.}{-2}}
\toprule[1pt]\midrule[0.3pt]
&\multicolumn{4}{c}{Dependent Variable}\\
\cmidrule(rrrr){2-5}
&\multicolumn{1}{c}{(1) Workers}
&\multicolumn{1}{c}{(2) Workers}
&\multicolumn{1}{c}{(3) ln(Workers)}
&\multicolumn{1}{c}{(4) ln(Workers)}\\\hline
Burned Wards							&-5.014$**$	&-6.582$*$	&-0.165$***$	&-0.135$**$	\\
$\times$ $I(\text{Year}=1930)$				&[1.816]		&[3.207]		&[0.024]		&[0.049]		\\
Industrial Zone (\%)						&			&-0.093		&			&0.001		\\
$\times$ $I(\text{Year}=1930)$				&			&[0.111]		&			&[0.001]		\\
Burned Wards $\times$ Industrial Zone (\%) 	&			&0.095		&			&-0.001		\\
 $\times$ $I(\text{Year}=1930)$				&			&[0.114]		&			&[0.002]		\\\hline
Sample mean of the DV					&63.19		&63.19		&4.05		&4.05		\\
Standard deviation of the DV				&27.38		&27.38		&0.47		&0.47		\\
Ward FE 					&\multicolumn{1}{c}{Yes}		&\multicolumn{1}{c}{Yes}		&\multicolumn{1}{c}{Yes}		&\multicolumn{1}{c}{Yes}	\\
$I(\text{Year}=1930)$ 		&\multicolumn{1}{c}{Yes}		&\multicolumn{1}{c}{Yes}		&\multicolumn{1}{c}{Yes}		&\multicolumn{1}{c}{Yes}	\\
Observations				&\multicolumn{1}{c}{30}		&\multicolumn{1}{c}{30}		&\multicolumn{1}{c}{30}		&\multicolumn{1}{c}{30}	\\\midrule[0.3pt]\bottomrule[1pt]
\end{tabular}
}
{\scriptsize
\begin{minipage}{420pt}
\setstretch{0.85}
***, **, and * denote statistical significance at the 1\%, 5\%, and 10\% levels, respectively.
The standard errors in brackets are clustered at the 15-ward level.\\
Notes:
This table summarizes the results of the regressions using ward-level labor statistics from the 1920 and 1930 censuses.
The dependent variable in Columns (1) and (2) is the number of gainfully occupied persons in all industries (in 1,000 individuals), whereas the dependent variable in Columns (3) and (4) is the log-transformed version of this variable.
In Columns (1) and (3), the dependent variable is regressed on an interaction term between the indicator variable of the burned wards and the 1930 year dummy, as well as the ward and year fixed effects.
Columns (2) and (4) add an interaction term between the share of industrial blocks and the 1930 year dummy, and a triple interaction term among the indicator variable of the burned wards, the share of industrial blocks, and the 1930 year dummy.
The share of industrial blocks indicates the number of industrial blocks per 100 ward blocks.\\
Sources: Created by the author using the Statistics Bureau of the Cabinet (1929; 1933).
\end{minipage}
}
\end{center}
\end{table}

The ward-level statistics on the number of gainfully occupied persons by industry are available from the 1920 and 1930 censuses.\footnote{For 1920, I use the number of persons whose main occupation was in each industry (\textit{hongy\=osha}), excluding dependents (\textit{j\=uzokusha}). This corresponds to the number of gainfully occupied persons by industry in the 1930 census, which includes all employment statuses (employers, salaried employees, and wage workers). I do not use factory statistics because their definitions were revised in the 1920s and no official statistics provide a consistent definition of the number of factories in Tokyo City.}
The total number of gainfully occupied persons in the city increased from roughly $900$ thousand in 1920 to $996$ thousand in 1930.
By sector, the number of manufacturing workers fell from roughly $380$ thousand to $319$ thousand, whereas the number of commercial workers increased from roughly $303$ thousand to $420$ thousand.
The number of retail stores for daily necessities listed in official statistics remained at around $53,000$ in both years, suggesting that the commercial sector grew in terms of employment rather than the number of stores.
Since these are statistics for the entire city, however, they do not reveal whether the changes in the burned wards were related to the earthquake or to zoning classifications.
I therefore use the total number of gainfully occupied persons in each ward as the dependent variable, which captures the overall labor demand in the ward regardless of the sector.

Table~\ref{tab:r_het_ward} shows the estimation results.
Column (1) reports the result from a two-period version of Equation~\ref{model_es} in which the treatment is an indicator variable for the burned wards.
The estimated coefficient on the interaction term between the burned-ward indicator and the 1930 year dummy is negative and statistically significant ($-5.014$), implying that the number of gainfully occupied persons decreased by about $5,000$ per burned ward relative to the unburned wards.
Column (2) reports the result from a specification based on Equation~\ref{model_es_het}, which adds the interaction term between the share of industrial blocks and the 1930 year dummy, and the triple interaction term among the burned-ward indicator, the share of industrial blocks, and the 1930 year dummy.
The estimated coefficients on both terms are close to zero and statistically insignificant, whereas the coefficient on the burned-ward indicator, which represents the effect for a ward with a zero share of industrial blocks, remains negative and statistically significant ($-6.582$).
The results are materially similar when I use the log-transformed dependent variable (Columns (3) and (4)).
Column (3) suggests that the number of gainfully occupied persons decreased by about $15$\% in the burned wards ($100 \times (e^{-0.165}-1)$).
There is no statistically significant difference in the share of industrial blocks (Column (4)).

\subsubsection{Results for the Sex Ratio: Evidence from Block-level Census Statistics}\label{sec:secd_r_block_sr}

\begin{figure}[htbp]
\centering
\captionsetup{justification=centering}
\subfloat[Impacts on Sex Ratio\\ in Burned Blocks]{\label{fig:r_block_sr}\includegraphics[width=0.45\textwidth]{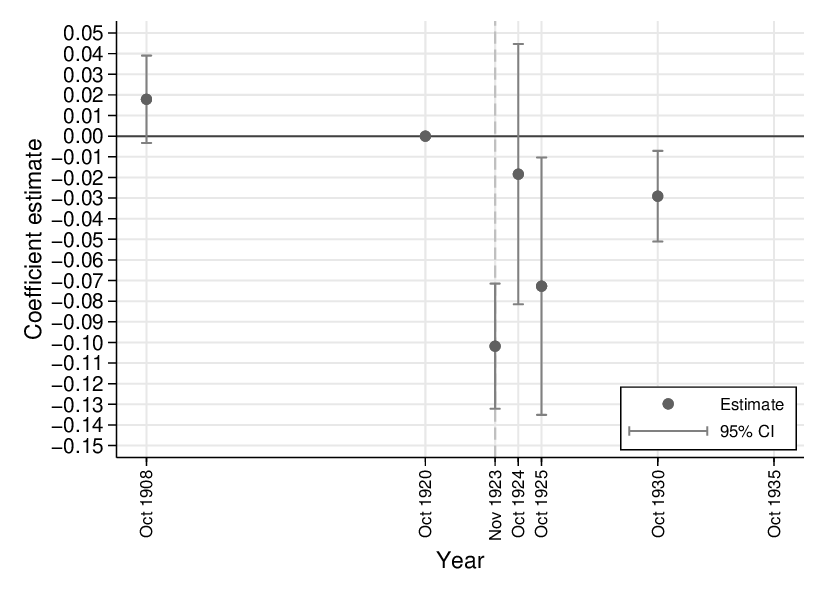}}
\subfloat[Impacts on Sex Ratio\\ in Industrial and Non-Industrial Zones]{\label{fig:r_block_het_sr_burn}\includegraphics[width=0.45\textwidth]{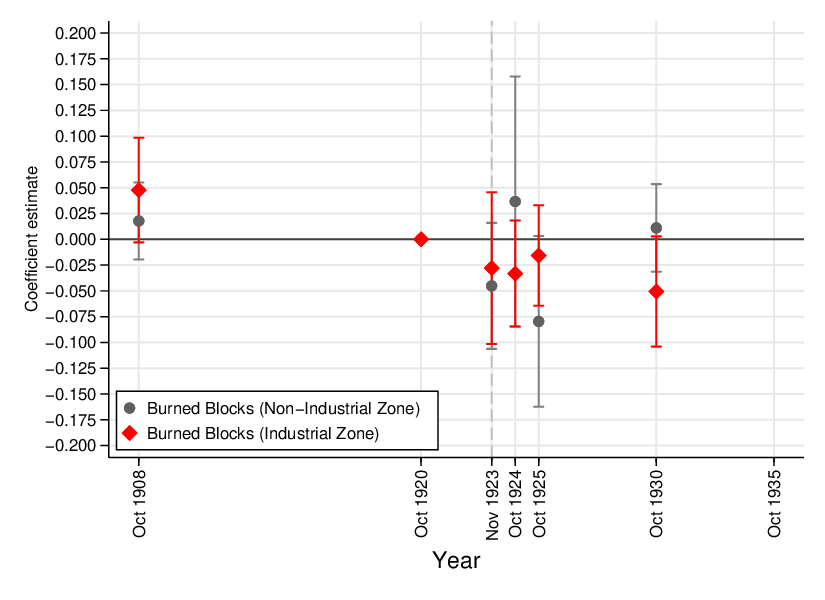}}
\caption{Results for the Sex Ratio: Evidence from the Block-level Census Statistics}
\label{fig:r_did_block}
\scriptsize{\begin{minipage}{450pt}
\setstretch{0.9}
Notes:
Figure~\ref{fig:r_block_sr} shows the results for the sex ratio based on the regression of Equation~\ref{model_es_block}.
The sex ratio is the number of individuals per household and females per male, respectively.
Figure~\ref{fig:r_block_het_sr_burn} shows the estimated marginal effects of disaster for the sex ratio under Equation~\ref{model_es_het}, respectively.
The estimated effects on the non-industrial (industrial) zone are illustrated in black (red) circles.
The `Industrial Zone' includes both commercial and manufacturing zones defined by the zoning system designed prior to the earthquake (Section~\ref{sec:sec_hb_cbd}).
The 95\% confidence intervals are shown in the black and red caps in each figure.
Standard errors in the regressions based on Equation~\ref{model_es_block} and~\ref{model_es_het} are clustered at the block-level.\\
Sources: Created by the author using the block-level census statistics listed in Panel B-1 of Table~\ref{tab:sum}.
\end{minipage}}
\end{figure}

Figure~\ref{fig:r_block_sr} shows the results for the average sex ratio under Equation~\ref{model_es_block}.
Immediately after the earthquake, the sex ratio declined due to high demand for male workers in reconstruction projects (1935 Census Report, p.~24).
It temporarily recovered (particularly in non-industrial areas) in 1924 as the evacuees returned, and declined again in 1925 as land readjustment projects began in earnest.
Unlike the population, the sex ratio should be independent of the changes in residential land area.
The time-series plots of the average sex ratio do not show a clear correlation with household size (not shown).
Thus, the negative estimate for industrial areas in 1930 may indicate that the demand for male labor increased relative to pre-earthquake levels through a mechanism other than land readjustment.

Figure~\ref{fig:r_block_het_sr_burn} summarizes the estimates from Equation~\ref{model_es_het}.
As with the results for average household size, the differences in effects across zones are not statistically significant.\footnote{While the Bureau of Statistics states that ``in industrial areas, due to their regional characteristics, a large number of men are required as part of the labor force, so it is common to see a surplus of men even in small households,'' it also notes that ``in districts such as Kojimachi, Akasaka, Yotsuya, Asakusa, and Fukagawa, a large number of women are required as domestic servants or commercial employees, resulting in a surplus of women even in large households'' (pp.~24--25). In other words, while the sex ratio in the burned area exhibits regional heterogeneity, it declined as a whole (Figure~\ref{fig:r_block_sr}).}
However, the estimate for the industrial zone in 1930 is negative, suggesting that the decline in the sex ratio in burned blocks may have occurred primarily there.
Regarding this point, the Statistics Bureau offers an explanation for the changes in the sex ratio from 1920 to 1930:
\begin{quote}
``Although there were fluctuations caused by the earthquake, the situation in 1930 showed a relative increase in the male population in the \textit{shitamachi} (industrial district) and a corresponding decrease in the \textit{yamanote} (a residential district), when compared to ten years prior.''
\end{quote}
Then it notes that this trend ``reflects the distinct regional cultural tendencies between the production and consumption zones within this city'' (Tokyo City Office 1932a, p.~12).

\subsubsection{Results for the Number of Households by Household Size Bins: Evidence from Block-level Census Statistics}\label{sec:secd_r_hh}

\begin{figure}[htbp]
\centering
\captionsetup{justification=centering,margin=1.5cm}
\subfloat[Households: Raw Estimate for the 15 Household Size Bins]{\label{fig:r_block_hh}\includegraphics[width=0.80\textwidth]{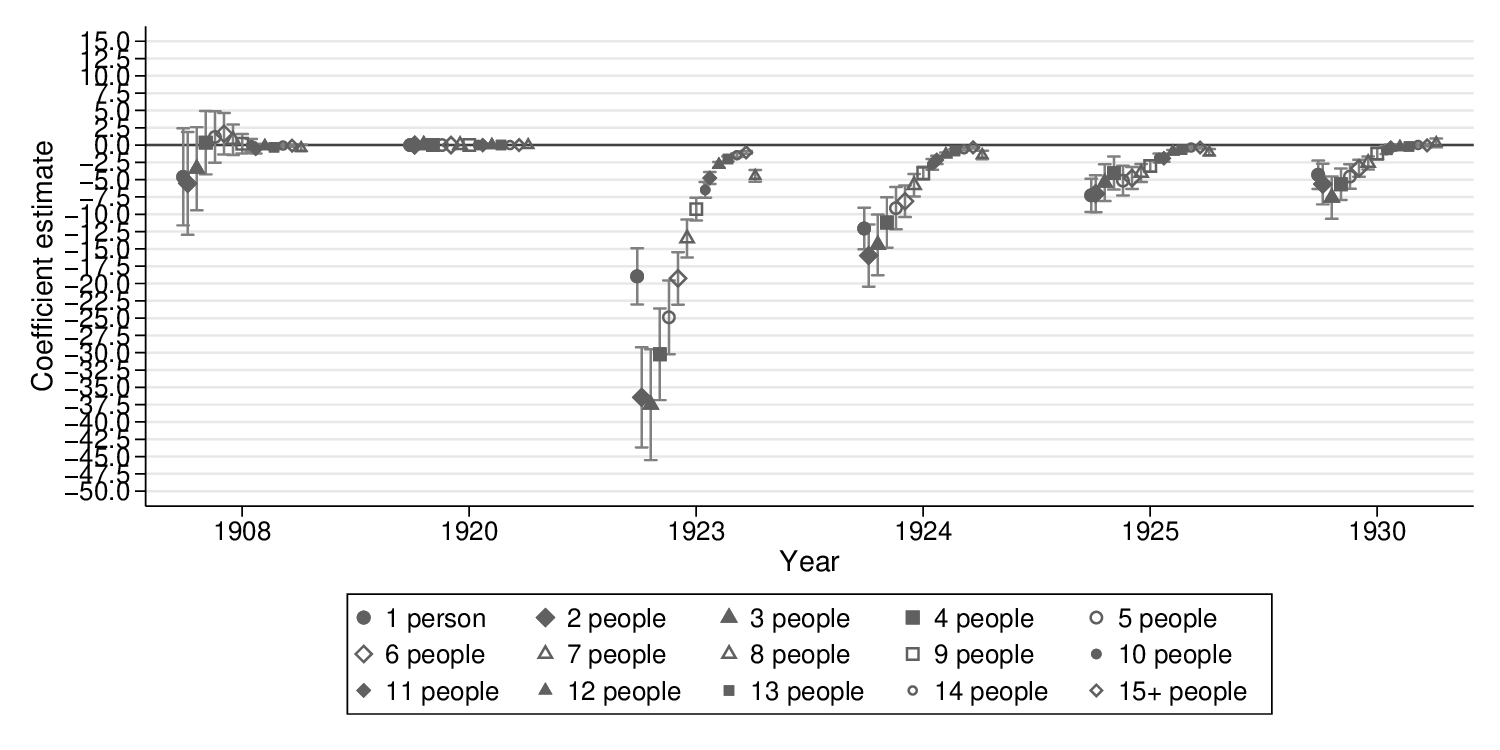}}\\
\subfloat[Households: Estimate / $\hat{\sigma}_{g, 1920}$]{\label{fig:r_block_hh_mag}\includegraphics[width=0.80\textwidth]{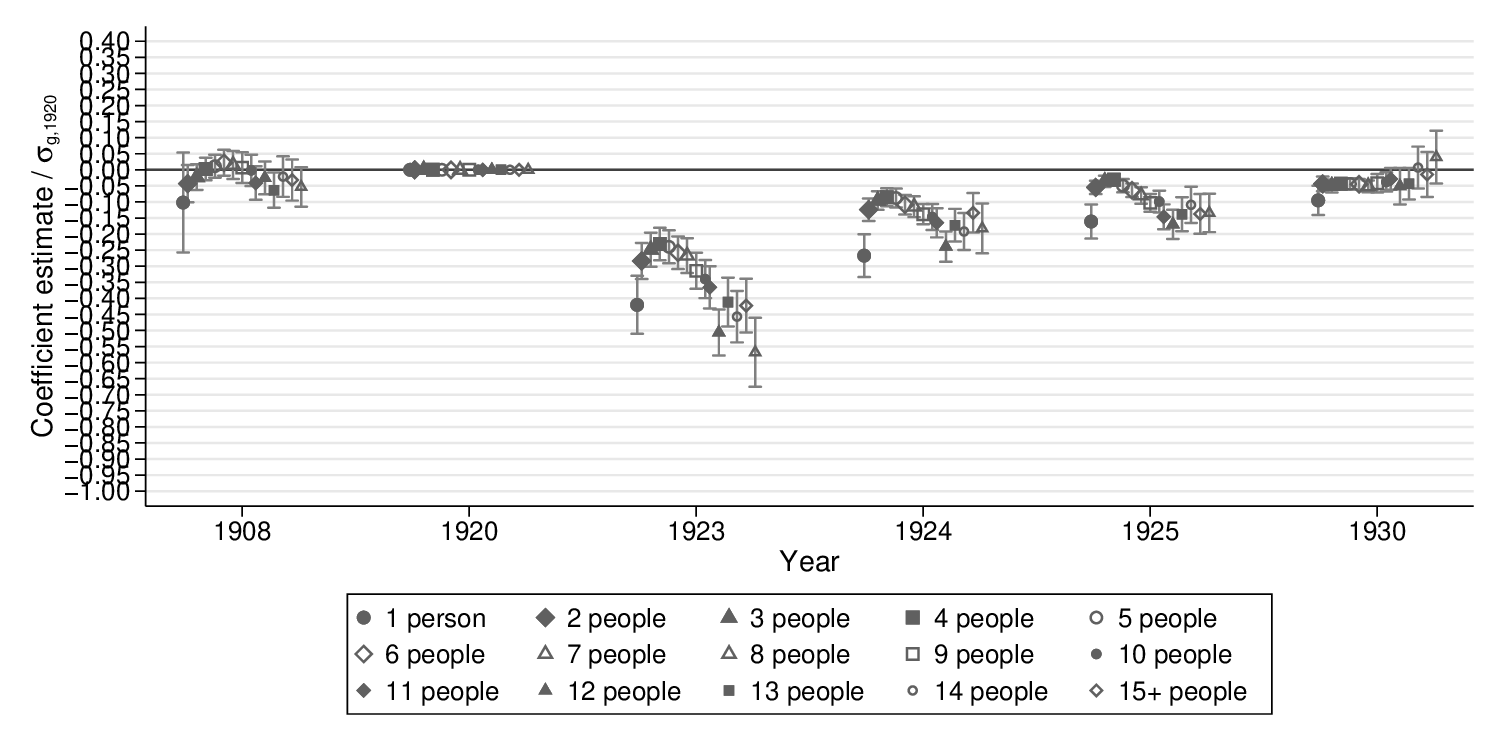}}\\
\caption{Results for the Number of Households by Household Size Bin: Evidence from the Block-level Census Statistics}
\label{fig:r_did_block_hh}
\scriptsize{\begin{minipage}{450pt}
\setstretch{0.9}
Notes:
This figure summarizes the estimated coefficients for the interaction terms between the indicator variable for the burned blocks and the year dummies (Equation~\ref{model_es_block}).
Figure~\ref{fig:r_block_hh} summarizes the results for the number of households in the block-year panel data for each household-size bin (from $1$ to $15+$ individuals).
Figure~\ref{fig:r_block_hh_mag} shows the suggested magnitude of the disaster: the estimates shown in Figure~\ref{fig:r_block_hh} are weighted by the standard deviation of the number of households in each household-size bin measured in the 1920 census.
The $95$\% confidence intervals are obtained from the cluster-robust variance-covariance matrix estimator using the block as the cluster unit.\\
Sources: 
Created by the author using block-size level census statistics on the number of households.
The sample corresponds to that listed in Panel B-2 of Table~\ref{tab:sum}.
\end{minipage}}
\end{figure}

Figure~\ref{fig:r_block_hh} summarizes the results for the number of households for each household-size bin.
Figure~\ref{fig:r_block_hh_mag} shows estimates weighted by the standard deviation of the number of households in each household-size bin, as measured in the 1920 census.
Both figures are similar to the results for the number of individuals shown in the main text (Figure~\ref{fig:r_did_block}).

\subsection{Robustness Appendix}\label{sec:secd3}

\subsubsection{Measurement Error: Are Hidden Households Present in the Census Statistics?}\label{sec:secd_rob_me}

I evaluate the reliability of statistics on the number of households derived from the census.
The Census Enforcement Ordinance (\textit{Kokusei Ch\=osa Shik\=orei}) defines a ``household'' as a group of people who share both a residence and consumption (Article 3, Paragraphs 2 and 3).
In other words, the mere fact of living together is not sufficient to be considered a single ``household.''
The Bureau of Statistics states that in Tokyo City, co-residents were generally understood to be ``those who share only a residence while maintaining separate household budgets'' (Tokyo City Office 1932b, p.~legend).
This means the census rules were consistent with the city's general understanding, and the likelihood of lodger households being miscounted during the survey was reasonably low.

However, the Statistics Bureau (Tokyo City Office 1932b, p.~11) points out the possibility of errors in the identification of households during the survey:
\begin{quote}
``While it is understandable that there were some errors in identifying households during the survey, given the sheer number of households with cohabitants, might not the very nature of these cohabitation arrangements--which gave rise to such identification errors--actually reveal something about the circumstances of those households?''
\end{quote}
This suggests that there were cases in which landlord and lodger households were not included in the statistics on the number of households.
Given this possibility, I assess whether the number of households identified in the census contains any clear measurement errors.

\subsubsection*{Method}

I divide the ordinary households in block $i$ into three groups: landlord households and lodger households, both of which are recorded in the SSDH ([A-1-2-1] and [A-1-2-2] in Figure~\ref{fig:type}), and other households, which comprise all remaining ordinary households.
The number of households in the census ($h_{i}^{Census}$) is expressed as follows using the number of landlord households ($h_{i}^{Landlord}$), the number of lodger households ($h_{i}^{Lodger}$), and the number of other households ($h_{i}^{Other}$):
\begin{align}\label{hh_census}
h_{i}^{Census} = h_{i}^{Landlord} + \rho h_{i}^{Lodger} + h_{i}^{Other},
\end{align}
where $\rho$ is a weight satisfying $0 < \rho \leq 1$ and represents the proportion of lodger households that are correctly counted as separate households in the census; $1-\rho$ is thus the proportion of lodger households that are not identified.
Even if some lodger households are not identified as separate households, their members are still counted in the census population (as members of the landlord's household).
Therefore, no measurement error occurs in the census population:
\begin{align}\label{pop_census}
p_{i}^{Census} = p_{i}^{Landlord} + p_{i}^{Lodger} + p_{i}^{Other}.
\end{align}
Using this property, the average size of other households can be computed from the census and the SSDH as follows:
\begin{align}\label{size_ns}
\tilde{s}_{i}^{Other} \equiv \frac{p_{i}^{Census} - p_{i}^{Landlord} - p_{i}^{Lodger}}{h_{i}^{Census} - h_{i}^{Landlord} - h_{i}^{Lodger}} = \frac{p_{i}^{Other}}{h_{i}^{Other} - (1-\rho) h_{i}^{Lodger}}.
\end{align}
Equation~\ref{size_ns} takes the true value when there is no measurement error ($\rho = 1$) and takes a value greater than the true value when there is measurement error ($0 < \rho < 1$):
\begin{align}\label{size_ns_rel}
\mu^{Other} = \mathbb{E} \left[ \left. \tilde{s}_{i}^{Other} \right|_{\rho = 1} \right] < \mathbb{E} \left[ \left. \tilde{s}_{i}^{Other} \right|_{0 < \rho < 1} \right] = \ddot{\mu}^{Other}.
\end{align}

The test procedure is as follows.
I divide the sample ($\Omega$) into blocks without lodger households (subset $\Xi$) and blocks with lodger households (subset $\Xi^{c}$) ($\Xi^{c}=\Omega \setminus \Xi$).
Note that, in blocks without lodger households (i.e., $i \in \Xi$), the number of other households (population) is equal to the total number of households (population), so that no measurement error can arise.
Using this property, I estimate the average size of other households in the city ($\mu^{Other}$) as follows:
\begin{align}\label{size_ns_est}
\hat{\mu}^{Other} = \frac{\sum_{i \in \Xi} p_{i}^{Other}}{\sum_{i \in \Xi} h_{i}^{Other}} = \frac{\sum_{i \in \Xi} p_{i}^{Census}}{\sum_{i \in \Xi} h_{i}^{Census}}.
\end{align}
In contrast, the average size of other households in blocks with lodger households ($i \in \Xi^{c}$) is estimated as follows:
\begin{align}\label{size_ns_est_me}
\hat{\ddot{\mu}}^{Other} = \frac{\sum_{i \in \Xi^{c}} \left(p_{i}^{Census} - p_{i}^{Landlord} - p_{i}^{Lodger}\right)}{\sum_{i \in \Xi^{c}} \left(h_{i}^{Census} - h_{i}^{Landlord} - h_{i}^{Lodger}\right)}.
\end{align}
From the above, one can consider the following hypothesis test:
\begin{align}\label{size_ns_test}
\mathbb{H}_{0}: \mu^{Other} - \ddot{\mu}^{Other} = 0 \quad \text{v.} \quad \mathbb{H}_{1}: \mu^{Other} - \ddot{\mu}^{Other} < 0.
\end{align}
For this, I perform a two-sample $t$-test.
If the null hypothesis is rejected, $\rho = 1$ is not supported, meaning that one cannot rule out the possibility that the census undercounts lodger households.
On the other hand, if the null hypothesis is not rejected, there is no evidence that the census household count contains measurement error.

\subsubsection*{Results}

\begin{figure}[htbp]
\centering
\captionsetup{justification=centering,margin=1.5cm}
\includegraphics[width=0.55\textwidth]{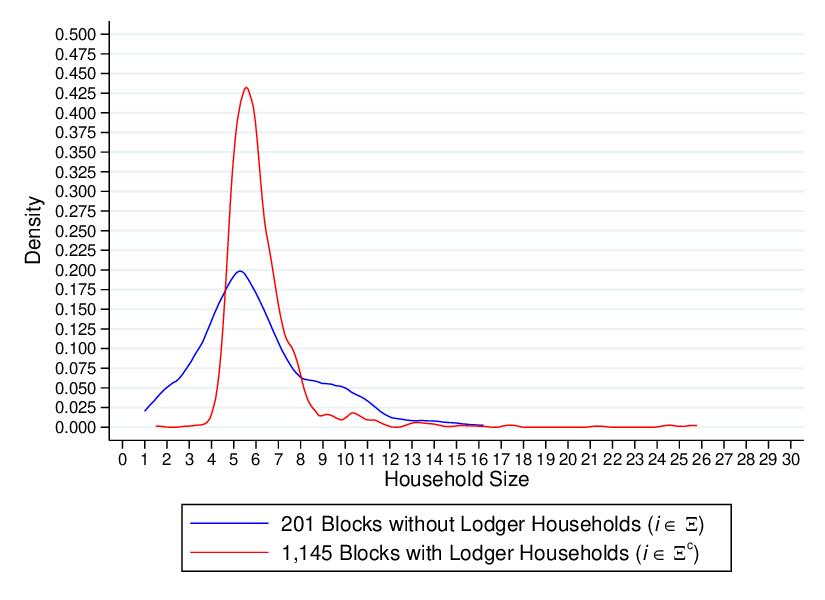}
\caption{Densities of the Average Size of Other Households by Blocks with and without Lodger Households}
\label{fig:kden_size_ns}
\scriptsize{\begin{minipage}{450pt}
\setstretch{0.9}
Notes:
This figure shows the estimated density of the average size of other households (i.e., households other than landlord and lodger households) in 1930.
The blue line shows the density for the $201$ blocks without lodger households ($i \in \Xi$).
The red line shows the density for the $1,145$ blocks with lodger households ($i \in \Xi^{c}$).
Source:
Created by the author using the block-level survey complete statistics listed in Panel B-3 of Table~\ref{tab:sum}.
\end{minipage}}
\end{figure}

Figure~\ref{fig:kden_size_ns} shows the estimated density of the average size of other households.
The average values are very similar for both samples.
However, while the density of households with 0--4 and 8--12 members is high in blocks without lodger households, it is relatively low in blocks with lodger households.
The sample size for blocks with lodger households is more than five times that of blocks without lodger households, which affects the kurtosis.
Thus, this does not directly indicate a systematic shift in the density distribution.

\def\arraystretch{1.0}
\begin{table}[htb]
\begin{center}
\captionsetup{justification=centering,margin=1.5cm}
\caption{Testing Difference in the Average Size of Other Households: Comparison between Blocks with and without Lodger Households}
\label{tab:r_ttest}
\scriptsize
\scalebox{0.95}[1]{
\begin{tabular}{lD{.}{.}{-2}D{.}{.}{-2}D{.}{.}{-2}D{.}{.}{-2}D{.}{.}{-2}D{.}{.}{-2}}
\toprule[1pt]\midrule[0.3pt]
\multicolumn{7}{l}{\textbf{Panel A: Two Sample $t$-Tests for the Average Size of Other Households}}\\
&\multicolumn{3}{c}{(1) Threshold for $\Xi^{c}$: None}&\multicolumn{3}{c}{(2) Threshold for $\Xi^{c}$: Size $< \max \Xi + |\Delta_{\text{Full}}|$}\\
\cmidrule(rrr){2-4}\cmidrule(rrr){5-7}
&\multicolumn{1}{c}{Without ($\Xi$)}
&\multicolumn{1}{c}{With ($\Xi^{c}$)}
&\multicolumn{1}{c}{Difference ($\Delta_{\text{Full}}$)}
&\multicolumn{1}{c}{Without ($\Xi$)}
&\multicolumn{1}{c}{With ($\Xi^{c}$)}
&\multicolumn{1}{c}{Difference ($\Delta_{\text{Trim}}$)}\\\hline
Sample Mean							&6.045	&6.254	&-0.2092	&6.045	&6.155	&-0.1102	\\
$\mathbb{H}_{1}$: $\Delta < 0$ [$p$-value]	&		&		&[0.1448]	&		&		&[0.2847]	\\
Observations							&\multicolumn{1}{c}{201}		&\multicolumn{1}{c}{1,145}	&		&\multicolumn{1}{c}{201}		&\multicolumn{1}{c}{1,138}	&\\
&&&&&&\\
\multicolumn{7}{l}{\textbf{Panel B: Two Sample $t$-Tests for Average Sex Ratio (Testing Random Assignment of $\Xi$)}}\\
&\multicolumn{3}{c}{(1) Threshold for $\Xi^{c}$: None}&\multicolumn{3}{c}{(2) Threshold for $\Xi^{c}$: Size $< \max \Xi + |\Delta_{\text{Full}}|$}\\
\cmidrule(rrr){2-4}\cmidrule(rrr){5-7}
&\multicolumn{1}{c}{Without ($\Xi$)}
&\multicolumn{1}{c}{With ($\Xi^{c}$)}
&\multicolumn{1}{c}{Difference ($\Delta_{\text{Full}}^{SR}$)}
&\multicolumn{1}{c}{Without ($\Xi$)}
&\multicolumn{1}{c}{With ($\Xi^{c}$)}
&\multicolumn{1}{c}{Difference ($\Delta_{\text{Trim}}^{SR}$)}\\\hline
Sample Mean							&0.879	&0.877	&0.0024	&0.879	&0.878	&0.0010	\\
$\mathbb{H}_{1}$: $\Delta \neq 0$ [$p$-value]	&		&		&[0.9604]	&		&		&[0.9833]	\\
Observations							&\multicolumn{1}{c}{200}		&\multicolumn{1}{c}{1,143}	&		&\multicolumn{1}{c}{200}		&\multicolumn{1}{c}{1,136}	&\\\midrule[0.3pt]\bottomrule[1pt]
\end{tabular}
}
{\scriptsize
\begin{minipage}{430pt}
\setstretch{0.85}
Notes:
``Without ($\Xi$)'' and ``With ($\Xi^{c}$)'' denote blocks without and with lodger households, respectively. Other households are ordinary households other than landlord and lodger households.\\
\textbf{Panel A} shows the results of the $t$-tests for the mean difference in the average size of other households between blocks with and without lodger households.
Column (1) shows the result for the entire sample, meaning that several large values in the blocks with lodger households ($\Xi^{c}$) are included in the test.
Column (2) presents the result for the sample excluding $7$ blocks with very large values.
The trimming threshold is set to the maximum value in $\Xi$ plus the absolute mean difference in Column (1), $|\Delta_{\text{Full}}|$.\\
\textbf{Panel B} shows the results of the $t$-tests for the mean difference in the average sex ratio between blocks with and without lodger households.
Columns (1) and (2) show the results for the same sample definitions as Columns (1) and (2) of Panel A, respectively.
A block in $\Xi$ that has no men is omitted from the sample.\\
Source:
Created by the author using the block-level complete survey statistics listed in Panel B-3 of Table~\ref{tab:sum}.
\end{minipage}
}
\end{center}
\end{table}

Panel A in Table~\ref{tab:r_ttest} summarizes the results of the two-sample $t$-tests.
First, Column (1) presents the results for the entire sample, showing that the difference in the average size of other households between the two groups is $-0.2092$ and is not statistically significant ($p$-value $= 0.1448$).
Column (2) shows the results for the sample excluding blocks with very high values.
Specifically, I excluded the $7$ blocks where the value exceeded the sum of the maximum value in blocks without lodger households ($16.2$ people) and the absolute value of the mean difference between the two groups.\footnote{The absolute value of the mean difference between the two groups was added to account for potential measurement errors in the number of households in blocks with lodger households. The results here remain unchanged, even if the absolute value of the difference between the two groups' averages is not taken into account.}
This further excluded cases where the possibility of errors in statistical aggregation or reporting could not be ruled out (Section~\ref{sec:sec_data_census}).
As is evident, the difference in the average size of other households between the two groups is sufficiently close to zero ($-0.1098$).

Panel B in Table~\ref{tab:r_ttest} presents the results of testing the randomness of the assignment of blocks without lodger households ($\Xi$).
As explained in the main text, industrial blocks have a relatively higher proportion of men, while other residential blocks have a relatively higher proportion of women.
Therefore, if blocks without (or with) lodger households were concentrated in specific industrial areas, the average sex ratio would differ significantly from that of blocks with (or without) lodger households.
Columns (1) and (2) both show that there is no significant difference in the average sex ratio between the two groups.
This supports the randomness of the assignment, which is a prerequisite for the hypothesis test in Panel A.

The above results provide no evidence that the number of households reported in the census undercounts lodger households.
As the Bureau of Statistics notes, there may have been ``some errors in the census's household counts,'' but these potential measurement errors are unlikely to materially affect the number of households obtained from the census.

\subsubsection{Additional Analysis using the SSDH}\label{sec:secd_rob_size}

\def\arraystretch{0.95}
\begin{table}[htbp]
\begin{center}
\captionsetup{justification=centering,margin=1.5cm}
\caption{Average Household Size and Sex Ratio: Evidence from the Survey of Shared-Dwelling Households}
\label{tab:r_rob_hs}
\scriptsize
\scalebox{0.90}[1]{
\begin{tabular}{lD{.}{.}{-2}D{.}{.}{-2}D{.}{.}{-2}D{.}{.}{-2}D{.}{.}{-2}}
\toprule[1pt]\midrule[0.3pt]
\multicolumn{6}{l}{\textbf{Panel A: Average Household Size}}\\
&\multicolumn{5}{c}{Dependent Variable: Average Household Size}\\
\cmidrule(rrrrr){2-6}
&\multicolumn{1}{c}{(1) Entire}
&\multicolumn{1}{c}{(2) Entire}
&\multicolumn{1}{c}{(3) Landlord}
&\multicolumn{1}{c}{(4) Lodger}
&\multicolumn{1}{c}{(5) Landlord + Lodger}\\\hline
Burned Blocks				&0.319$***$	&0.142$***$	&0.097$**$	&0.064$*$	&0.073$**$\\
						&[0.093]		&[0.054]		&[0.046]		&[0.038]	&[0.029]\\\hline
Sample mean of the DV		&5.319		&5.191		&3.633		&1.791	&2.607\\
Standard deviation of the DV	&1.416		&0.999		&0.899		&0.586	&0.592\\
Ward FE 					&\multicolumn{1}{c}{Yes}		&\multicolumn{1}{c}{Yes}	&\multicolumn{1}{c}{Yes}	&\multicolumn{1}{c}{Yes}	&\multicolumn{1}{c}{Yes}\\
$F$-statistic $p$-value for	&\multicolumn{1}{c}{0.000}	&\multicolumn{1}{c}{0.000}&\multicolumn{1}{c}{0.000}&\multicolumn{1}{c}{0.000}&\multicolumn{1}{c}{0.000}\\
the Ward FE (zero slope)	&\multicolumn{1}{c}{}	&\multicolumn{1}{c}{}&\multicolumn{1}{c}{}&\multicolumn{1}{c}{}&\multicolumn{1}{c}{}\\
Observations				&\multicolumn{1}{c}{1344}		&\multicolumn{1}{c}{1143}	&\multicolumn{1}{c}{1143}	&\multicolumn{1}{c}{1143}&\multicolumn{1}{c}{1143}\\
&&&&&\\
\multicolumn{6}{l}{\textbf{Panel B: Average Sex Ratio}}\\
&\multicolumn{5}{c}{Dependent Variable: Average Sex Ratio}\\
\cmidrule(rrrrr){2-6}
&\multicolumn{1}{c}{(1) Entire}
&\multicolumn{1}{c}{(2) Entire}
&\multicolumn{1}{c}{(3) Landlord}
&\multicolumn{1}{c}{(4) Lodger}
&\multicolumn{1}{c}{(5) Landlord + Lodger}\\\hline
Burned Blocks				&-0.064$***$			&-0.067$***$			&-0.046$***$		&0.039		&-0.007\\
						&[0.016]				&[0.013]				&[0.016]			&[0.026]		&[0.015]\\\hline
Sample mean of the DV		&0.877				&0.877				&1.063			&0.583		&0.836\\
Standard deviation of the DV	&0.346				&0.253				&0.377			&0.381		&0.292\\
Ward FE 					&\multicolumn{1}{c}{Yes}		&\multicolumn{1}{c}{Yes}	&\multicolumn{1}{c}{Yes}	&\multicolumn{1}{c}{Yes}	&\multicolumn{1}{c}{Yes}\\
$F$-statistic $p$-value for	&\multicolumn{1}{c}{0.000}	&\multicolumn{1}{c}{0.000}&\multicolumn{1}{c}{0.005}&\multicolumn{1}{c}{0.000}&\multicolumn{1}{c}{0.000}\\
the Ward FE (zero slope)	&\multicolumn{1}{c}{}	&\multicolumn{1}{c}{}&\multicolumn{1}{c}{}&\multicolumn{1}{c}{}&\multicolumn{1}{c}{}\\
Observations				&\multicolumn{1}{c}{1343}		&\multicolumn{1}{c}{1143}	&\multicolumn{1}{c}{1138}	&\multicolumn{1}{c}{1138}&\multicolumn{1}{c}{1144}\\
&&&&&\\
\multicolumn{6}{l}{\textbf{Panel C: Average Sex Ratio by Industrial Zone}}\\
&\multicolumn{5}{c}{Dependent Variable: Average Sex Ratio}\\
\cmidrule(rrrrr){2-6}
&\multicolumn{1}{c}{(1) Entire}
&\multicolumn{1}{c}{(2) Entire}
&\multicolumn{1}{c}{(3) Landlord}
&\multicolumn{1}{c}{(4) Lodger}
&\multicolumn{1}{c}{(5) Landlord + Lodger}\\\hline
Burned Blocks				&-0.059$***$	&-0.060$***$	&-0.046$**$		&0.074$**$	&0.008	\\
						&[0.020]		&[0.019]		&[0.018]			&[0.035]		&[0.020]	\\
Burned Blocks $\times$ 		&-0.025		&-0.023		&0.002			&-0.092$**$	&-0.039	\\
Industrial Zone Blocks		&[0.030]		&[0.029]		&[0.027]			&[0.038]		&[0.024]	\\\hline
Sample mean of the DV		&0.877		&0.877		&1.063			&0.583		&0.836	\\
Standard deviation of the DV	&0.346		&0.253		&0.377			&0.391		&0.292	\\
Ward FE 					&\multicolumn{1}{c}{Yes}		&\multicolumn{1}{c}{Yes}	&\multicolumn{1}{c}{Yes}	&\multicolumn{1}{c}{Yes}	&\multicolumn{1}{c}{Yes}\\
Industrial Zone Dummy 	&\multicolumn{1}{c}{Yes}		&\multicolumn{1}{c}{Yes}	&\multicolumn{1}{c}{Yes}	&\multicolumn{1}{c}{Yes}	&\multicolumn{1}{c}{Yes}\\
$F$-statistic $p$-value for	&\multicolumn{1}{c}{0.000}	&\multicolumn{1}{c}{0.000}&\multicolumn{1}{c}{0.013}&\multicolumn{1}{c}{0.000}&\multicolumn{1}{c}{0.000}\\
the Ward FE (zero slope)	&\multicolumn{1}{c}{}	&\multicolumn{1}{c}{}&\multicolumn{1}{c}{}&\multicolumn{1}{c}{}&\multicolumn{1}{c}{}\\
Observations				&\multicolumn{1}{c}{1343}		&\multicolumn{1}{c}{1143}	&\multicolumn{1}{c}{1138}	&\multicolumn{1}{c}{1138}&\multicolumn{1}{c}{1144}\\\midrule[0.3pt]\bottomrule[1pt]
\end{tabular}
}
{\scriptsize
\begin{minipage}{430pt}   
\setstretch{0.85}
***, **, and * denote statistical significance at the 1\%, 5\%, and 10\% levels, respectively.
The standard errors listed in the brackets are based on the HC2 estimator proposed by Horn et al. (1975).
$F$-statistics $p$-values for the null of the zero-slope hypothesis for the ward fixed effects are reported in each column.\\
Notes:
\textbf{Panel A} shows the results for the average household size.
Columns (1) and (2) show the estimates for the average size of the entire households.
Columns (3)--(5) show the estimates for the average size of the landlord, lodger, and landlord and lodger households, respectively.
Column (1) shows the results for the entire sample, whereas the other columns show the results for the blocks with lodger households.
\\
\textbf{Panel B} shows the results for the average sex ratio.
Columns (1) and (2) show the estimates for the average sex ratio of all households.
Columns (3)--(5) show the estimates for the average sex ratio of the landlord, lodger, and landlord and lodger households, respectively.
Column (1) shows the results for the entire sample, whereas the other columns show the results for the blocks with lodger households.\\
\textbf{Panel C} shows the results for the sex ratio under the expanded regression of Equation~\ref{model_cs}, which also includes the industrial zone dummy and the interaction terms between the indicator variable for burned blocks and the industrial zone dummy.
The column layout follows that of Panels A and B.\\
Source: 
Created by the author using the Tokyo City Social Welfare Bureau (1930).
\end{minipage}
}
\end{center}
\end{table}

Panel A of Column (1) presents the results of estimating Equation \ref{model_cs} using all blocks, with average household size as the dependent variable.
The estimated coefficient is positive, suggesting that the average household size in burned blocks was larger than in unburned blocks.
This is consistent with the results for the panel data (Figure~\ref{fig:r_block_size}).
Column (2) shows the estimate is $0.142$, corresponding to $14$\% of the standard deviation ($0.149/0.999$).
Columns (3)--(5) present the results for the same blocks as in Column (2), with the average number of members in landlord households and lodger households (or both) as the dependent variables, respectively.
All coefficients are positive and statistically significant.
Given these, Column (2) reflects that the average number of people in landlord households and lodger households in the burned blocks is larger than that in the unburned blocks.

Column (1) in Panel B presents the results for the sex ratio for the total population.
The sex ratio in burned blocks is significantly negative, and this result remains unchanged when the sample is restricted to blocks containing lodger households (Column (2)).
Columns (3) and (4) present the results regarding the sex ratio of landlord and/or lodger households in blocks where lodger households exist.
Interestingly, the estimated coefficients are negative for landlord households and positive for lodger households.
It follows that the estimated coefficient approaches zero when both are combined (Column (5)).

Panel C summarizes the results of a model that includes an interaction term between burned blocks and industrial zones.
Columns (1) and (2) show that the estimated coefficients on the interaction term are not statistically significant.
This implies that the characteristics of landlord households do not vary across industrial and non-industrial zones.
Columns (3) and (4) present the results for the landlord or lodger households.
While no regional differences are observed in the sex ratio of landlord households, the sex ratio of lodger households tends to be higher in non-industrial zones and lower in industrial zones.
Due to the interplay of differing effects, the estimated coefficient converges toward zero when both types of households are considered (Column (5)).

The differences in results between landlord and lodger households in Panels B and C are intriguing.

First, the average sex ratio for landlord households is relatively balanced (Column (3) in Panel B: $1.06$) and exceeds the average for all households (Column (1) in Panel B: $0.88$).
Given the average household size (Column (3) in Panel A: $3.6$ persons), this implies that landlord households are more likely to include extended family households (i.e., a married couple plus relatives).
Thus, the negative coefficients (Column (3) in Panels B-C) suggest that there were likely many cases where a married couple lived together with the head of household's younger brother or eldest son (not as lodgers) among landlord households in the burned blocks.

Next, the average sex ratio in lodger households is quite low (Column (4) in Panel B: $0.58$).
This is consistent with the Statistics Bureau's report, which states that many single men live in lodger households.
Thus, the coefficient of the interaction term in Column (4) in Panel C indicates that, within lodger households that already have a high proportion of men, the demand for men is particularly high in the industrial zones of the burned blocks.
The main effect in the same column suggests that there are relatively fewer men in the non-industrial zones of the burned blocks.
The Statistics Bureau's description provides insight into this:

\begin{quote}
``The surplus of women in certain large households in areas such as Yotsuya, Asakusa, and Fukagawa is likely due to the fact that many women working as laborers in the entertainment districts are concentrated in these large households'' (p.~23).
\end{quote}

It is likely that lodger households in the non-industrial zones (i.e., residential zones) of the burned blocks included a relatively large number of single women who were lodgers working in the entertainment districts that had formed within the industrial zones.

\subsubsection{Testing Confounding Factor: Accessibility to Railway Station as a Control Variable}\label{sec:secd_rob_trans}

\begin{figure}[htbp]
\centering
\captionsetup{justification=centering,margin=1.5cm}
\subfloat[Railway stations in 1920]{\label{fig:map_station_1920}\includegraphics[width=0.45\textwidth]
{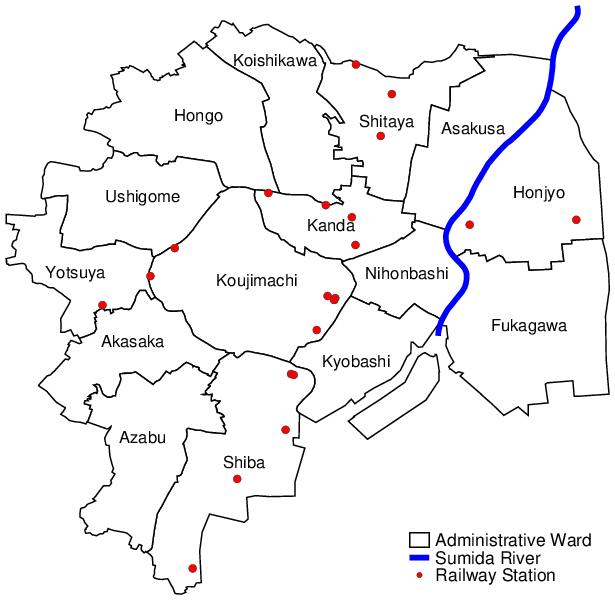}}
\subfloat[Railway stations in 1930]{\label{fig:map_station_1930}\includegraphics[width=0.45\textwidth]
{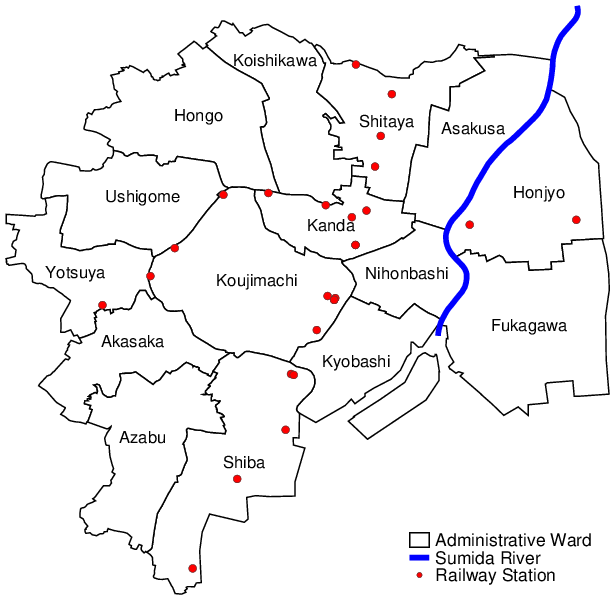}}
\caption{Spatial Distributions of Railway Stations in 1920 and 1930}
\label{fig:map_station}
\scriptsize{\begin{minipage}{450pt}
\setstretch{0.9}
Notes:
Figures~\ref{fig:map_station_1920} and~\ref{fig:map_station_1930} show the spatial distributions of the railway stations in 1920 and 1930, respectively.
Sources:
Created by the author using the official shapefile (Ministry of Land, Infrastructure, Transport and Tourism, database).
The location data of the Sumida River are obtained from the Ministry of Land, Infrastructure, Transport, and Tourism (website).
The location data on the railway stations is from the Ministry of Land, Infrastructure, Transport and Tourism (website).
\end{minipage}}
\end{figure}

This section assesses the robustness of my baseline results for the population and household to the inclusion of accessibility to railway stations.
The location information on the railway station in the sample period is available from the database created by the Ministry of Land, Infrastructure, Transport and Tourism (Section~\ref{sec:secc_ward}).
Figure~\ref{fig:map_station} shows the spatial distribution of the railway stations in 1920 and 1930.
As shown, the number of stations did not increase rapidly in the 1920s.
This means that the accessibility to the railways had not dramatically changed in Tokyo City during this period.
Despite this, there are a few places where accessibility to the stations has improved, such as Hongo and Koishikawa Wards.
Considering this, I calculated the distance from the ward centroid to the nearest railway station using the basemap provided by the Ministry of Land, Infrastructure, Transport and Tourism.

\begin{figure}[htbp]
\centering
\captionsetup{justification=centering,margin=1.5cm}
\subfloat[Impacts on Population\\ in Burned Blocks]{\label{fig:r_block_be_pop_rob}\includegraphics[width=0.45\textwidth]{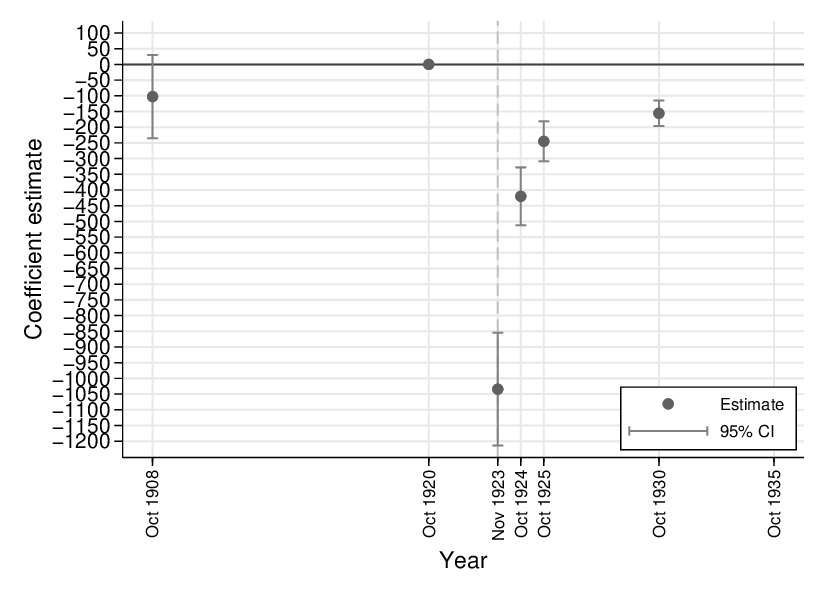}}
\subfloat[Impacts on Households\\ in Burned Blocks]{\label{fig:r_block_be_hh_rob}\includegraphics[width=0.45\textwidth]{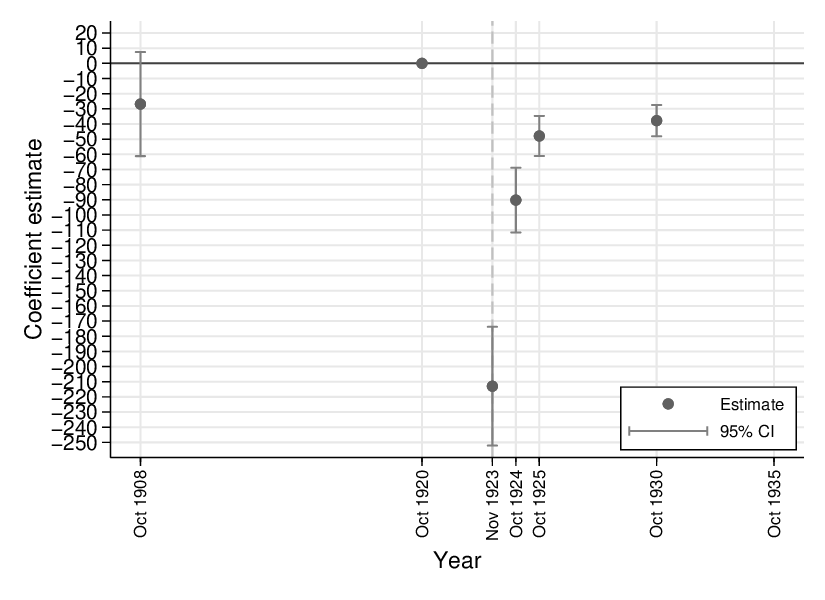}}
\caption{Results for the Population and Households after Controlling for the Distance to the Nearest Railway Station: Evidence from Block-level Census Statistics}
\label{fig:r_did_block_pop_hh_rob}
\scriptsize{\begin{minipage}{450pt}
\setstretch{0.9}
Notes:
Figures~\ref{fig:r_block_be_pop_rob} and~\ref{fig:r_block_be_pop_rob} report the results for block-level population and households, respectively, based on Equation~\ref{model_es}.
Both figures show the estimated coefficients for the interaction terms between the indicator variable for burned blocks and the year dummies.
All regressions include the distance from the ward centroid to the nearest railway station (in $\text{km}$), and block- and year-fixed effects.
The $95$\% confidence intervals are shown in the black and red caps in each figure.\\
Sources: Created by the author using the block-level census statistics listed in Panel B-1 in Table~\ref{tab:sum}.
\end{minipage}}
\end{figure}

Figure~\ref{fig:r_ward_pop_rob} shows the results for the ward-level population and household data under the specification including the distance to the nearest station as a control variable.
Similarly, Figure~\ref{fig:r_did_block_pop_hh_rob} illustrates the estimates for the block-level population and household data after controlling for this distance variable.
Both figures show quite similar results for my baseline results (Figures~\ref{fig:r_ward_pop} and~\ref{fig:r_did_block_pop_hh}).
This result supports the evidence that the changes in accessibility to hub stations are not correlated with the distribution of burned area.

An alternative measure of accessibility to the railway station may be the railway station density, such as the number of railway stations per square kilometer of ward area.
However, the density variable ignores stations located in neighboring wards but close to the ward boundary.
In fact, Figure~\ref{fig:map_station} shows that the stations are located very close to the administrative borders.
For example, although Ushigome Ward has no stations, it has access to a station on the border with Kojimachi Ward.
The same is true of the Hongo and Koishikawa Wards.
The density variable completely ignores this type of case.

\begin{figure}[htbp]
\centering
\captionsetup{justification=centering}
\subfloat[Impacts on Population\\ in burned area]{\label{fig:r_ward_pop_rob}\includegraphics[width=0.45\textwidth]
{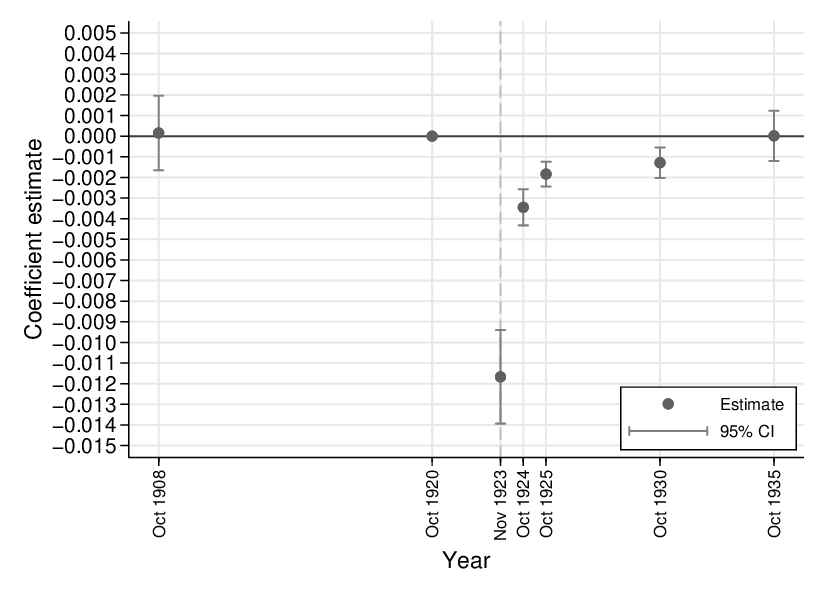}}
\subfloat[Impacts on Household\\ in burned area]{\label{fig:r_ward_hh_rob}\includegraphics[width=0.45\textwidth]
{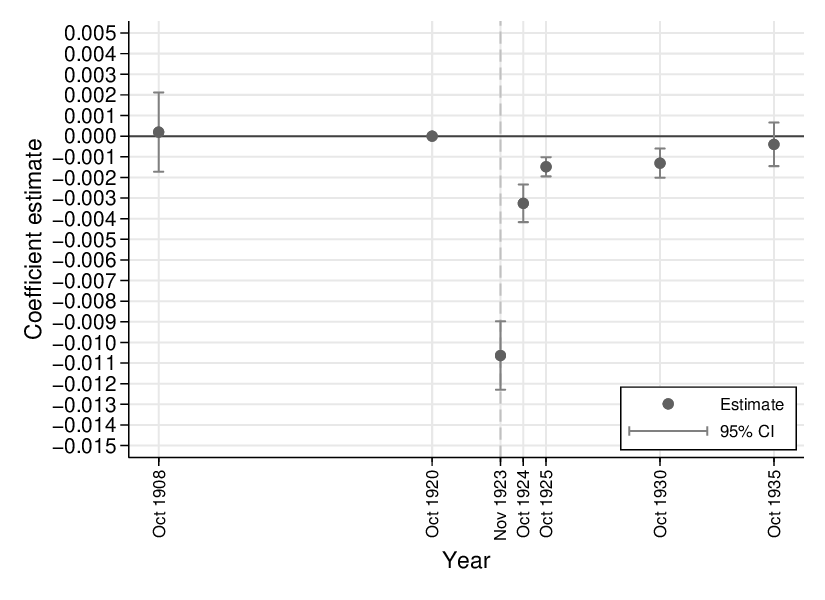}}
\caption{Impacts on the Population and Households after Controlling for the Distance to the Nearest Railway Station: Evidence from Ward-level Statistics}
\label{fig:r_ward_pop_rob}
\scriptsize{\begin{minipage}{450pt}
\setstretch{0.9}
Notes:
This figure shows the estimated coefficients on the interaction terms between the indicator variable for the burned wards and the year dummies (Equation~\ref{model_es}).
Figures~\ref{fig:r_ward_pop_rob} and~\ref{fig:r_ward_hh_rob} show the results for the log-transformed population and household, respectively.
All regressions include the distance from the ward centroid to the nearest railway station (in $\text{km}$) and ward- and year-fixed effects.
The $95$\% confidence intervals are obtained from the cluster-robust variance-covariance matrix estimator using wards as the clustering unit.\\
Sources: Created by the author using the ward-level census statistics listed in Panel A-2 of Table~\ref{tab:sum}.
The location data on the railway stations is from the Ministry of Land, Infrastructure, Transport and Tourism (website).
\end{minipage}}
\end{figure}

\newpage
\renewcommand{\refname}{References used in the Appendices}

\renewcommand{\refname}{Documents used in the Appendices}

\end{document}